\documentclass[rmp,twocolumn,showpacs,amsmath,amssymb,eqsecnum,preprintnumbers,floatfix]{revtex4-1}
\pdfoutput=1
\usepackage{graphicx}
\usepackage{dcolumn}
\usepackage{bm}
\usepackage{color}
\usepackage{enumitem}
\usepackage{afterpage}
\usepackage{hhline}
\usepackage{amsfonts}
\usepackage{epstopdf}
\usepackage{totcount}
\usepackage{refcount}
\def\tr{{\rm tr}\,}

\def\sgn{{\rm sgn\,}}

\def\be{\begin{equation}}
\def\ee{\end{equation}}
\def\bea{\begin{eqnarray}}
\def\eea{\end{eqnarray}}
\def\bml{\begin{mathletters}}
\def\eml{\end{mathletters}}
\def\bse{\begin{subequations}}
\def\ese{\end{subequations}}
\def\me{m_{\text{e}}}
\def\epsilonF{\epsilon_{\text{F}}}

\def\vF{v_{\text{F}}}
\def\Tc{T_{\text{c}}}
\newcommand{\mub}{$\mu_{\textrm{B}}$}
\newcommand{\TC}{$T_{\textrm{C}}$}
\newcommand{\TN}{$T_{\textrm{N}}$}
\newcommand{\pc}{$p_{\textrm{c}}$}
\newcommand{\xc}{$x_{\textrm{c}}$}
\newcommand{\NF}{NbFe$_{2}$}
\newcommand{\NFy}{Nb$_{1-y}$Fe$_{2+y}$}
\newcommand{\CRGS}{CeRu$_{2}$(Ge$_{1-x}$Si$_x$)$_{2}$} 
\newcommand{\CS}{CeSi$_{x}$}
\newcommand{\CPR}{CePd$_{1-x}$Rh$_{x}$}
\newcommand{\ZNZ}{Zr$_{1-x}$Nb$_{x}$Zn$_{2}$}
\newcommand{\CTG}{CeTi$_{1-x}$V$_{x}$Ge$_{3}$}
\newcommand{\CAS}{CeAgSb$_{2}$}
\newcommand{\CRPO}{CeRuPO}
\newcommand{\CFPO}{CeFePO}

\newcommand{\CFPAO}{CeFeAs$_{1-x}$P$_{x}$O}
\newcommand{\CRFPO}{CeRu$_{1-x}$Fe$_{x}$PO}
\newcommand{\UTNS}{U$_{1-x}$Th$_{x}$NiSi$_{2}$}
\newcommand{\UNCS}{UNi$_{1-x}$Co$_{x}$Si$_{2}$}
\newcommand{\NV}{Ni$_{1-x}$V$_{x}$}
\newcommand{\SCRO}{Sr$_{1-x}$Ca$_{x}$RuO$_{3}$}
\newcommand{\SROO}{Sr$_{3}$Ru$_{2}$O$_{7}$}
\newcommand{\NP}{Ni$_{1-x}$Pd$_{x}$}
\newcommand{\YNP}{YbNi$_{4}$P$_{2}$}
\newcommand{\YNPA}{YbNi$_{4}$(P$_{1-x}$As$_{x}$)$_{2}$}
\newcommand{\YRS}{YbRh$_{2}$Si$_{2}$}
\newcommand{\YRCS}{Yb(Rh$_{1-x}$Co$_{x}$)$_{2}$Si$_{2}$}
\newcommand{\YRCSFM}{Yb(Rh$_{0.73}$Co$_{0.27}$)$_{2}$Si$_{2}$}
\newcommand{\SCGP}{SrCo$_{2}$(Ge$_{1-x}$P$_{x}$)$_{2}$}
\newcommand{\YFA}{YFe$_2$Al$_{10}$}
\begin{document}
\title{Metallic Quantum Ferromagnets}
\preprint{Master Copy Version \today}
\author{M. Brando}\thanks{brando@cpfs.mpg.de}
\affiliation{Max Planck Institute for Chemical Physics of Solids, N{\"o}thnitzer Str. 40, D-01187 Dresden, Germany}
\author{D.Belitz}\thanks{dbelitz@uoregon.edu}
\affiliation{Department of Physics, and Institute of Theoretical Science, and Materials Science Institute, University of Oregon, Eugene, Oregon 97403, USA}
\author{F. M. Grosche}\thanks{fmg12@cam.ac.uk}
%\author{G. G. Lonzarich}\thanks{Electronic address: gl238@cam.ac.uk}
\affiliation{University of Cambridge, Cavendish Laboratory, CB3 0HE Cambridge, UK}
\author{T.R.Kirkpatrick}\thanks{tedkirkp@umd.edu}
\affiliation{Institute for Physical Science and Technology, and Department of Physics, University of Maryland, College Park, Maryland 20742, USA}
\date{\today}
\begin{abstract}
This review gives an overview of the quantum phase transition (QPT) problem
in metallic ferromagnets, discussing both experimental and theoretical
aspects. These QPTs can be classified with respect to the presence and strength of quenched disorder:
Clean systems generically show a discontinuous, or first-order, QPT from the ferromagnetic state
to a paramagnetic one as a function of some control parameter, as predicted by
theory. Disordered systems are much more complicated, depending on
the disorder strength and the distance from the QPT. In many disordered materials the 
QPT is continuous, or second order, and Griffiths-phase effects
coexist with QPT singularities near the transition.
In other systems the transition from the ferromagnetic state at low temperatures is to a different type of long-range
order, such as an antiferromagnetic or a spin-density-wave state. In still other materials
a transition to a state with glass-like spin dynamics is suspected. The review provides
a comprehensive discussion of the current understanding of these various transitions, and 
of the relation between experimental and theoretical developments.
\end{abstract}
\maketitle
\makeatletter
\def\l@paragraph{\@dottedtocline{4}{5.3em}{2.1em}}
\makeatother
\tableofcontents
%
%********************************************************************************************
\section{Introduction}
\label{sec:I}
%This review summarizes the experimental results for, and the theoretical understanding of, the quantum phase transition in metallic ferromagnets. In this section we start with some general remarks about quantum phase transitions and then turn to the metallic ferromagnetic quantum phase transition in particular.
%
Metallic ferromagnets have been studied since ancient times, as this class of materials includes elemental iron, which gave ferromagnetism its name, 
as well as nickel and cobalt. Detailed studies in the early 1900s led to one of the first examples of mean-field theory~\cite{Weiss_1907}.
A more elaborate version of mean-field theory by \textcite{Stoner_1938} 
explained how a nonzero magnetization can arise from a spontaneous splitting of the conduction band. When it became clear, 30 years later, that 
mean-field theory does not correctly describe the behavior close to the phase transition, ferromagnetism became one of the testing grounds for the 
theory of critical phenomena \cite{Stanley_1971, Wilson_Kogut_1974}. More recently metallic ferromagnets with low Curie temperatures, ranging 
from tens of 
degrees to a few degrees, or even lower, have attracted much attention. One motivation for the study of these materials is that many of them allow 
for decreasing the Curie temperature even further, by applying pressure or by changing the chemical composition. This allows the study of the 
quantum phase transition that occurs at zero temperature and for fundamental reasons must be quite different in nature from the thermal phase 
transition observed at a nonzero Curie temperature. Over the years it again became clear that the quantum version of mean-field theory does not 
correctly describe the behavior close to the transition, contrary to early suggestions.

This review summarizes the experimental and theoretical understanding of this quantum phase transition. In line with this goal, our exposition of 
the experimental results is restricted to materials that have a ferromagnetic ground state in some part of the phase diagram, and for which the phase 
transition that marks the instability of the ferromagnetic phase at low temperatures is clearly observed and reasonably well characterized. In parallel 
to this discussion we 
describe the relevant theoretical ideas and the extent to which they explain, and in some cases predicted, the experimental observations. In this section 
we start with some general remarks about quantum phase transitions and then turn to the one in metallic ferromagnets. % Introduction
\subsection{General remarks on quantum phase transitions}
\label{subsec:I.A}

Quantum phase transitions (QPTs) have been discussed for many years and remain 
a subject of great interest \cite{Hertz_1976, Sachdev_1999}.
Whereas classical or thermal phase transitions occur
at a nonzero transition temperature and are driven by thermal fluctuations,
QPTs occur at zero temperature, $T=0$, as
a function of some non-thermal control parameter (typical examples are pressure,
composition, or an external magnetic field) and are driven by
quantum fluctuations. The ways in which the description of QPTs
differs from that of their classical counterparts are subtle and took a
long time to understand.
Early on it was realized that at a mean-field level the description is the same
for both quantum and classical phase transitions. Indeed, the earliest
theory of a QPT was the Stoner theory of ferromagnetism \cite{Stoner_1938}.
Stoner considered the case of itinerant ferromagnets, where the conduction electrons are responsible
for the ferromagnetism,%
\footnote{We refer to systems where the conduction electrons are the sole source
of the magnetization as ``itinerant ferromagnets'', and to ones where part or all of the magnetization is due to
localized spins as ``localized-moment ferromagnets''.} 
and developed a mean-field theory that describes
both the classical and the quantum ferromagnetic transition. 

Important mathematical developments were the proof of the Trotter 
formula \cite{Trotter_1959}, and the coherent-state formalism \cite{Casher_Lurie_Revzen_1968},
which proved useful for representing the partition
function of quantum spin systems in terms of a functional 
integral \cite{Suzuki_1976a, Suzuki_1976b}. It implied, at least for certain
spin models, that a quantum phase transition in a system with $d$ spatial dimensions 
could be described in terms of the corresponding classical phase transition in 
an effective dimension $d_{\text{eff}}=d+1$. An example is the Ising model in a transverse 
field \cite{DeGennes_1963, Stinchcombe_1973}. The crucial observation behind this conclusion
was that the functional-integral representation of the partition function contains an
integration over an auxiliary variable (usually referred to as imaginary time) that extends from zero to the inverse 
temperature $1/T$. At $T=0$, this integration range becomes infinite and mimics
an additional spatial integration in the thermodynamic limit.
If space and time scale in the same way, then $d_{\text{eff}}=d+1$
follows. In particular, since the upper critical dimension $d_{\text{c}}^+$, above which 
mean-field theory provides an exact description of the transition, for the classical Ising 
model is $d_{\text{c}}^+ = 4$, it follows that the critical behavior of the quantum Ising model
in a transverse field in $d>3$ is mean-field like \cite{Suzuki_1976a}. More generally, it
also implied that the statics and the dynamics are 
intrinsically coupled at QPTs. This is unlike the case of classical 
phase transitions, where the dynamic critical phenomena are decoupled from the 
statics \cite{Ferrell_et_al_1967, Ferrell_et_al_1968, Halperin_Hohenberg_1967,
Halperin_Hohenberg_1969, Hohenberg_Halperin_1977}. 

This leads to the following general conceptual point: In the context of classical
critical phenomena, the dynamic universality classes are much smaller
(and therefore more numerous) than the static ones.
Physically, this is due to the fact that the order-parameter
fluctuations that determine the universality class can be conserved (such as in, e.g., a ferromagnet)
or non-conserved (such as in, e.g., an antiferromagnet), 
and they can couple to any number of other slow or soft modes or excitations,
with each of these cases realizing a different universality class \cite{Hohenberg_Halperin_1977}. 
By the same argument one expects quantum phase transitions in, for instance, metals
to be very different from those in insulators 
because the respective dynamical processes are very different.%
\footnote{To date, no
comprehensive classification of QPTs, at a level of the 
classification of classical critical dynamics given by \textcite{Hohenberg_Halperin_1977},
exists.}

In an important paper, \textcite{Hertz_1976}, among other things, generalized
the Trotter-Suzuki formulation to the case where space and time do not scale the
same way.
\footnote{Initially, mathematical results for specific spin models that yielded
$z=1$ had been applied more broadly than their validity warranted, which led to considerable
confusion.}  
He showed that if the slow order-parameter time 
scale $t_{\xi}$ at a continuous QPT diverges as $t_{\xi} \propto \xi^z$, with
$\xi$ the correlation length and $z$ the dynamical scaling exponent (which in general is
not equal to unity), then the imaginary-time
integral is analogous to a spatial integration over an additional $z$ spatial dimensions.
For such a class of problems the critical behavior at the continuous QPT is
equivalent to that at the corresponding classical transition in
$d_{\text{eff}}=d+z$ dimensions. At
this point it seemed that QPTs were, in fact,
not fundamentally different from their classical counterparts.
The statics and the dynamics couple, leading to an effective
dimension different from the physical spatial dimension, and the number of universality classes is different,
but the technical machinery that had been developed to solve the classical phase
transition problem \cite{Wilson_Kogut_1974, Ma_1976, Fisher_1983} could be generalized to treat QPTs 
as well and map them onto classical transitions in a different dimension.%
\footnote{\label{MIT_footnote} There are important QPTs that
have no classical analogs; examples include various metal-insulator transitions in disordered
electron systems with or without the electron-electron interaction taken into
account \cite{Anderson_1958, Evers_Mirlin_2008, Finkelstein_1984a, Belitz_Kirkpatrick_1994,
Lee_Ramakrishnan_1985, Kramer_MacKinnon_1993}.  While they do not allow for
a mapping onto a classical counterpart, their theoretical descriptions still use the
same concepts that were developed for classical transitions.}

The above considerations assume that the phase transition separates an ordered phase
from a disordered one, with the ordered phase characterized by a local order parameter. For
the ferromagnetic transition that is the subject of this review, this is indeed the case. It should be
mentioned, however, that there are very interesting phase transitions, both classical and quantum,
that do not allow for a description in terms of a local order parameter. One example is provided by spin
liquids \cite{Balents_2010}, others, by the quantum Hall effects \cite{von_Klitzing_Dorda_Pepper_1980,
Tsui_Stormer_Gossard_1982} and topological insulators \cite{Hasan_Kane_2010, Qi_Zhang_2011}. Other interesting 
cases are the Anderson and Anderson-Mott metal insulator transitions.$^{\ref{MIT_footnote}}$ It has been 
proposed that for these transitions, and indeed for all QPTs, the von Neumann 
entanglement entropy $S_{\text{e}}$ is a useful concept %, and that $S_{\text{e}}$ shows nonanalyticities 
%even at QPTs where the ground state energy is known to be analytic, such as the Anderson 
%transition 
since it displays nonanalyticities characteristic of the QPT \cite{Kopp_Jia_Chakravarty_2007}. $S_{\text{e}}$ is defined as
the entropy of a subsystem of a larger system, and it is a measure of correlations in the ground state.
It tends to scale with the area of the subsystem rather than its volume, and provides interesting
connections between correlated electrons, quantum information theory, and the thermodynamics of
black holes \cite{Eisert_Cramer_Plenio_2010}.

\subsection{Quantum ferromagnetic transitions in metals}
\label{subsec:I.B}

The prime example studied by \textcite{Hertz_1976} was the same as that considered by Stoner,
namely, an itinerant ferromagnet. Here the magnetization serves as an order parameter,
and Hertz derived a dynamical Landau-Ginzburg-Wilson (LGW) functional for this transition
by considering a model of itinerant electrons that interact only through a contact potential in 
the particle-hole spin-triplet channel. He analyzed this LGW functional by means of
renormalization-group (RG) methods. He concluded that in this case the dynamical
critical exponent has the value $z=3$, and that the QPT for an itinerant Heisenberg
ferromagnet hence maps onto the corresponding classical transition in $d_{\text{eff}} = d + 3$ dimensions.
Since the upper critical dimension for classical ferromagnets is $d_{\text{c}}^+ = 4$, 
this seemed to imply that Stoner theory was exact,
as far as the critical behavior at the transition was concerned, in the physical spatial dimensions $d=2$ and $d=3$.
This in turn implied that the transition was generically continuous or second order, with
mean-field static critical exponents. Preceding Hertz's work, Moriya and collaborators in the early 1970s 
had developed a comprehensive
description of itinerant quantum ferromagnets that one would now classify as a self-consistent one-loop
theory (historically, it was often referred to as self-consistently renormalized or SCR spin-fluctuation theory);
this work was summarized by \textcite{Moriya_1985}. \textcite{Millis_1993} used Hertz's RG framework to study 
the behavior at small but nonzero temperature and the crossover between 
the quantum and classical scaling behaviors. Most of the explicit results obtained via the RG confirmed the
earlier results of the spin-fluctuation theory. This combined body of work is often referred to as
Hertz-Millis-Moriya or Hertz-Millis theory. We will discuss its basic features and results in Sec.\ \ref{subsubsec:III.C.2}.

Apart from these developments, which were aimed at understanding the critical behavior at
the (presumed second-order) ferromagnetic quantum phase transition, a related but separate
line of investigations dealt with quantitative issues regarding the strength of the magnetism,
and the properties of the ordered phase, in itinerant ferromagnets. It was
realized early on that Stoner theory and its extension to finite temperature
\cite{Edwards_1968} leaves key questions unanswered, especially for
metals with low Curie temperatures \TC: Firstly, why
is the exchange energy, which can be extracted from band structure
probes or from careful analysis of the magnetic equation of state,
typically at least an order of magnitude larger than $k_{\text{B}}$\TC? If the
order was destroyed solely by a thermal smearing of the Fermi function, the two
would be expected to be of similar magnitude. Secondly, why is the
ordered moment in the low-temperature limit only a small fraction of
the fluctuating moment as extracted from the Curie constant in the
temperature dependent susceptibility? Thirdly, why is the temperature
dependence of the magnetization at low temperature proportional to
$T^2$ rather than $T^{3/2}$, as would be expected from including
spin-wave excitations?

The key to answering these questions, and to achieving a quantitative
description of band ferromagnets with low ordering temperatures, was to
include the effect of fluctuations of the local order parameter, the
magnetization, as was demonstrated by \textcite{Murata_Doniach_1972}. More
comprehensive models were developed in the 1970s by Moriya and
collaborators \cite{Moriya_1985} in the spin-fluctuation-theory work already mentioned above. 
As inelastic neutron scattering became feasible, which
demonstrated the existence of magnetic fluctuations and allowed for their
quantitative parameterization \cite{Ishikawa_1982, Bernhoeft_1983},
it became possible to accurately model key material properties such as
\TC, the low-temperature ordered moment and its temperature
dependence, as well as the
temperature dependence of the magnetic susceptibility and the
associated fluctuating moment, in a further development of the SCR spin-fluctuation approach \cite{Lonzarich_Taillefer_1985}. 

Returning to the statistical-mechanics description of the phase transition itself,
a key result of both the SCR theories and Hertz's RG description was the value
of the dynamical exponent, $z=3$. This
result in a clean metallic ferromagnet can be made plausible by
general arguments that are independent of the technical details 
of Hertz's theory, and, more importantly, independent of whether or not the conduction electrons
themselves are responsible for the magnetism. In the absence of soft modes other than the
order-parameter fluctuations, the bare order-parameter susceptibility $\chi_{\text{OP}}$ at
criticality as a function of the frequency $\omega$ and the wave number $k$ has the 
form \cite{Hohenberg_Halperin_1977}
\bse
\label{eqs:1.1}
\be
\chi_{\text{OP}}^{-1}(k,\omega) = -i\omega/\gamma + k^2
\label{eq:1.1a}
\ee
if the order parameter is not a conserved quantity, or
\be
\chi_{\text{OP}}^{-1}(k,\omega) = -i\omega/\lambda\,k^2 + k^2
\label{eq:1.1b}
\ee
\ese
if it is, with $\gamma$ and $\lambda$ kinetic coefficients. At $T>0$, or at $T=0$ in the 
presence of quenched disorder, $\gamma$ and
$\lambda$ are weakly $k$-dependent and approach constants as $k\to 0$. However, in clean
systems at $T=0$ these coefficients do not exist in the limit of zero frequency and wave number,
and in metallic systems their effective behavior is 
$\gamma \propto \lambda \propto 1/k$.%
\footnote{Throughout this review, $a\propto b$ means
``$a$ is proportional to $b$'', $a\approx b$ means ``$a$ is approximately equal to $b$",
$a\sim b$ means ``$a$ scales as $b$'', and $a\cong b$ means ``$a$ is isomorphic to $b$".}%
$^,$\footnote{More generally,
$\gamma$ and $\lambda$ scale as $\gamma \sim 1/(\omega + k^{z_{\gamma}})$ and
$\lambda \sim 1/(\omega + k^{z_{\lambda}})$, with $z_{\gamma}$ and $z_{\lambda}$ dynamical
exponents characteristic of the respective transport coefficient. In the non-conserved case, $z_{\gamma}\geq 1$ 
leads to $\omega \sim k$, i.e., $z=1$. In the conserved case, $z_{\lambda} \geq 2$ leads to
$\omega \sim k^2$, i.e., $z=2$, and $z_{\lambda}<2$ leads to $\omega \sim k^{4-z_{\lambda}}$, i.e.,
$z=4-z_{\lambda}$. For the conduction electrons in a metal, $\omega \sim k$, and as long as the
order parameter couples to the conduction electrons one therefore expects $z_{\lambda} = 1$,
or $z=3$.}
For a non-conserved order parameter this leads to
$z=1$, as in the case of a quantum antiferromagnet \cite{Chakravarty_Halperin_Nelson_1989},
or an Ising model in a transverse field \cite{Suzuki_1976a}. For a ferromagnet, where the
order parameter is conserved, we find from Eq.\ (\ref{eq:1.1b}) $z=3$ in the clean case,
and $z=4$ in the disordered case. This is consistent with Hertz's explicit calculation for
a specific model. It is important to recognize that the Eqs.\ (\ref{eqs:1.1}) do not 
get qualitatively changed by renormalizations, provided $d_{\text{eff}} = d+z$ is greater than
the upper critical dimension: The coupling between the statics and the dynamics ensures that
the critical exponent $\eta$
\footnote{\label{exponents_footnote} For a definition of critical exponents, see Appendix\ 
 \ref{app:B}.}
is zero and the exponents in Eqs.\ (\ref{eqs:1.1}) remain unchanged.
Simple mean-field arguments, including Eqs.\ (\ref{eqs:1.1}), are therefore self-consistently
valid for all $d > d_{\text{c}}^+ - z$, the static critical behavior is mean-field like, and the
dynamical critical exponent is the one that follows from Eqs.\ (\ref{eqs:1.1}). However, it is
important to remember that all of these considerations are valid only under the assumption
that there are no other soft modes that couple to the order parameter. In metallic ferromagnets
this assumption is {\em not} valid, as we will explain in detail in Sec.\ \ref{sec:III}.

The experimental situation through the 1990s was confusing: In some materials a second-order
or continuous transition was observed, but many others showed a first-order or discontinuous
transition. Within mean-field theory, the standard explanation (if one can call it that) for a first-order transition is that the
coefficient of the quartic term in the Landau expansion happens to be negative \cite{Landau_Lifshitz_V_1980}.
While this can always be the case in some specific materials, for reasons related to the band structure, 
there is no reason to believe that it will be the case in whole classes of materials. 
A much more general mechanism for a first-order transition was proposed in 1999,
when two of the present authors, together with Thomas Vojta,
showed theoretically that the quantum phase transition in two-dimensional and three-dimensional metallic systems from
a paramagnetic (PM) phase to a homogeneous ferromagnetic one
is generically first order, provided the material is sufficiently clean
(\onlinecite{Belitz_Kirkpatrick_Vojta_1999}, to be referred to as BKV). The physical reason
underlying this universal conclusion is a coupling of the magnetization to electronic
soft modes that exist in any metal, which leads to a fluctuation-induced first-order
transition. The same conclusion was reached by other groups \cite{Chubukov_Pepin_Rech_2004, Rech_Pepin_Chubukov_2006,
Maslov_Chubukov_Saha_2006}. This theoretical work was later generalized to include the effects of an external magnetic field, 
which leads to tricritical wings in the phase diagram \cite{Belitz_Kirkpatrick_Rollbuehler_2005}, 
and to cases where both itinerant and localized electrons are important%
\footnote{\label{itinerant_localized} Systems in which both localized and itinerant carriers
are important raise interesting questions regarding spin conservation. This has been
discussed in the context of observed anomalous damping of paramagnets in uranium-based
systems by \textcite{Chubukov_Betouras_Efremov_2014}.}\footnotetext{\label{CEP_footnote} A critical end point (CEP) is defined as a point where a line of second-order transitions terminates at a line of first-order transitions, with the first-order line continuing into an ordered region, see, e.g., \textcite{Chaikin_Lubensky_1995} and references therein. In the recent literature the term CEP is often misused.}
% A critical end point (CEP) is usually defined as a point where a line of second-order transitions intersects, and terminates on, a line of first-order transitions, see, e.g., \textcite{Fisher_Barbosa_1991}. In the literature the term is sometimes used in a broader sense; we employ it as used in the original papers.
%
or where the magnetic order may be ferrimagnetic or magnetic-nematic rather than 
ferromagnetic \cite{Kirkpatrick_Belitz_2011, Kirkpatrick_Belitz_2012b}.
Since the role of the electronic soft modes diminishes with increasing temperature, 
this theory predicts that in clean systems there necessarily exists a tricritical point
in the phase diagram, i.e., a temperature that separates a line of first-order transitions
at low temperatures from a line of second-order transitions at higher temperatures as
the control parameter is varied. In addition, BKV showed that non-magnetic quenched
disorder suppresses the tricritical temperature, and that the transition remains 
second order down to zero temperature if the disorder strength exceeds a critical
value. 
                      
\begin{figure}[t]
\begin{center}
\includegraphics[width=0.95\columnwidth,angle=0]{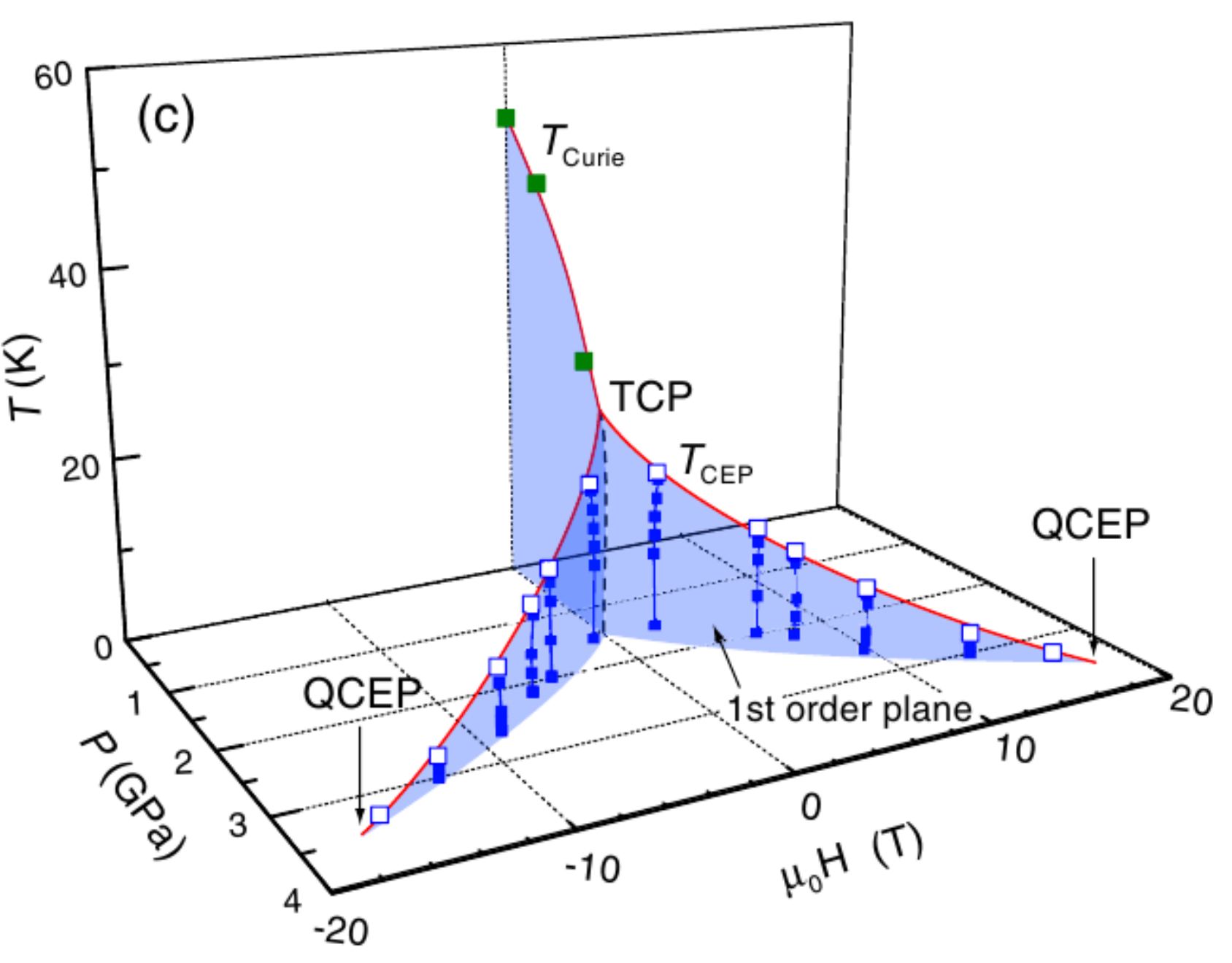}
\end{center}
\caption{Observed phase diagram of UGe$_2$ in the space spanned by temperature ($T$), pressure ($P$) and magnetic field ($H$). Solid red curves represent lines of second-order transitions, blue planes represent first-order transitions. Also shown are the tricritical point (TCP), and the extrapolated ``quantum critical end points'' (QCEP)$^{\ref{CEP_footnote}}$ at the wing tips. From \textcite{Kotegawa_et_al_2011b}.}
\label{figure:UGe_2_phase_diagram}
\end{figure}

Many experiments are consistent with these predictions, and over time experiments
on cleaner samples, or at lower temperatures, or both, showed a first-order transition
at sufficiently low temperatures even in cases where previously a continuous transition
had been found. The predicted tricritical point and associated tricritical wings
have also been observed in many systems. A representative example of this
type of phase diagram is shown in Fig.~\ref{figure:UGe_2_phase_diagram}.
Strongly disordered materials, on the other hand, almost always show a continuous transition,
also in agreement with the theoretical prediction. There are, however, exceptions from these
general patterns, which we will discuss in Sec.\ \ref{subsubsec:II.C.1}.

These predictions and observations are for systems where the transition is to a homogeneous
ferromagnetic state; the schematic phase diagrams for the discontinuous and continuous cases,
respectively, are shown in Fig.~\ref{fig:schematic_phase_diagrams} a) and b). In other materials, magnetic order of a 
different kind is found to compete with homogeneous
ferromagnetism at low temperatures, as schematically illustrated in Fig.~\ref{fig:schematic_phase_diagrams} c). In
strongly disordered systems, spin-glass freezing and quantum Griffiths effects may occur
at low temperatures and augment or compete with critical behavior, see Fig.~\ref{fig:schematic_phase_diagrams} d).
These effects will be discussed in detail in Secs.\ \ref{subsec:II.D}, \ref{subsec:II.E} and \ref{subsec:III.D}, 
\ref{subsec:III.E}. 

The striking difference between the predictions of BKV and Hertz theory is due to a coupling
of the order-parameter fluctuations to electronic degrees of freedom. Hertz
theory treats this coupling in too simple an approximation to capture all of its qualitative effects. In metals at $T=0$ there
are soft or gapless two-fermion excitations that couple to the
magnetic order-parameter fluctuations in important ways. In effect, the combined fermionic
and bosonic (order-parameter) fluctuations determine
the quantum universality class in all spatial dimensions $d<3$. As
a result of this coupling, the upper critical dimension is $d_{\text{c}}^+=3$,
rather than $d_{\text{c}}^+=1$ as predicted by Hertz theory, and the transition
is first order, rather than continuous with mean-field exponents. The mechanism behind 
this phenomenon is similar to what is known as a fluctuation-induced first-order transition in
classical phase transitions \cite{Halperin_Lubensky_Ma_1974, Chen_Lubensky_Nelson_1978}, but it is different in at least 
one crucial way, cf. Secs.\ \ref{subsubsec:III.B.2} and \ref{subsec:IV.A}. Two well-known classical examples of a 
fluctuation-induced first-order transition are the superconducting (BCS) transition,
and the nematic-to-smectic-A transition in liquid crystals. In
these cases, soft fluctuations of the electromagnetic vector potential
(in superconductors) or the nematic order parameter (in liquid crystals) couple to the order parameter 
and lead to a cubic term in the equation of state, which in turn leads
to a discontinuous phase transition. As we will discuss in Sec.\ \ref{sec:IV}, this
type of mechanism is even more important and efficient in the quantum 
case.%
\footnote{\label{superconductors_liquid_crystals_foonote} In superconductors, the first-order transition occurs so close to the
critical point that it is unobservable \cite{Chen_Lubensky_Nelson_1978}, and in liquid crystals it took a long time until
the weakly first-order transition was observed \cite{Anisimov_et_al_1990}.
We will discuss in Secs.\ \ref{subsubsec:III.B.2} and \ref{subsec:IV.A} why the fluctuation-induced first-order
transition in quantum ferromagnets is so much more robust.}

We add some remarks about the relative strength of fluctuations at second-order and certain first-order
transitions. At a second-order transition above the upper critical dimension, treating the fluctuations in a Gaussian 
approximation suffices to obtain the exact critical behavior; this is what Hertz theory concluded for the
ferromagnetic QPT. At a critical point below the upper critical dimension this is not true; fluctuations are
strong enough to modify the critical exponents, although they do not
change the continuous nature of the transition. At a fluctuation-induced first-order transition,
the combined effects of order-parameter fluctuations and other soft modes are so strong that they
change the order of the transition predicted by mean-field theory.%
\footnote{\label{first_order_footnote}  It is often thought that at first-order transitions, as opposed to
second-order ones, fluctuations are not important. In the case of
a fluctuation-induced first-order transition this notion is obviously misleading, as the name implies. 
Less obviously, and more generally, all first-order transitions can be understood as a limiting case of second-order
transitions where the critical exponents (including $\beta = 0$) can be determined exactly. In
particular, the scaling and renormalization-group concepts familiar from 
second-order transitions, properly interpreted, still apply at any first-order transition 
\cite{Nienhuis_Nauenberg_1975, Fisher_Berker_1982}.}
The prediction of BKV was that this will happen at the ferromagnetic QPT in clean systems.

The continuous quantum ferromagnetic transition in disordered metals, in systems where 
the disorder is strong enough to suppress the tricritical temperature to zero, has also been 
studied in detail theoretically \cite{Kirkpatrick_Belitz_1996, Belitz_et_al_2001a,
Belitz_et_al_2001b}. In this case the itinerant electrons are moving diffusively, rather than ballistically.
Because this is a slower process, there is an effective diffusive
enhancement of the exchange interaction that causes ferromagnetism, and some crucial
signs in the theory are changed compared to the clean case. The net
result is that the second-order transition predicted by Hertz theory becomes, so to speak,
even more continuous by the coupling to the electronic soft modes: For example, the theory 
predicts that in $d=3$ the critical exponent\footnotemark[\getrefnumber{exponents_footnote}] $\beta$ is equal to $2$, compared to $\beta=1/2$ in Hertz theory.%
\footnote{\label{logs_footnote} The asymptotic critical behavior in this case actually consists
of power laws multiplied by log-normal terms, see Sec.\ \ref{subsubsec:III.C.3}.}
This large value of $\beta$ may give the impression of a ``smeared transition",
even though there still is a sharp critical point. This, as well as the predicted values of other
exponents, is consistent with numerous experiments in disordered systems, as we will discuss. 
In related developments, much work has been done on Griffiths singularities and Griffiths phases in 
disordered metallic magnets. Depending on the nature and symmetry of the order parameter, 
these theories predict that in some systems the Griffiths-phase effects are very weak, while in others they
lead to strong power-law singularities with continuously varying exponents, and in yet others they
completely destroy the sharp quantum phase transition (for a review, see \onlinecite{Vojta_2010}). 
If these effects are important, they will be superimposed on the critical behavior. 

%a transition from a paramagnet to a homogeneous ferromagnetic state is not possible in metallic systems at $T=0$, and some %inhomogeneous magnetic phase appears instead.
%
Finally, there are theories that suggest that in some metallic systems an inhomogeneous magnetic phase may form in between the paramagnetic and the homogeneous ferromagnetic state at low $T$. This was first suggested by \textcite{Belitz_Kirkpatrick_Vojta_1997},
and has been explored in detail by others. Spiral phases, spin nematics, and spin-density waves
have been proposed to appear between the uniform ferromagnet and the paramagnetic 
phase \cite{Chubukov_Pepin_Rech_2004, Rech_Pepin_Chubukov_2006, 
Maslov_Chubukov_Saha_2006, Efremov_Betouras_Chubukov_2008, Chubukov_Maslov_2009, Conduit_Green_Simons_2009, 
Karahasanovic_Kruger_Green_2012}. We will discuss these and related theories in Sec.\ \ref{subsec:III.E}.

\section{Experimental results}
\label{sec:II}
In this section we discuss experimental results on quantum ferromagnets, organized with respect to the
observed structures of the phase diagram as shown in Fig.\ \ref{fig:schematic_phase_diagrams}. % Experimental results
\subsection{General remarks}
\label{subsec:II.A}

\begin{figure}[t]
\begin{center}
\includegraphics[width=0.95\columnwidth,angle=0]{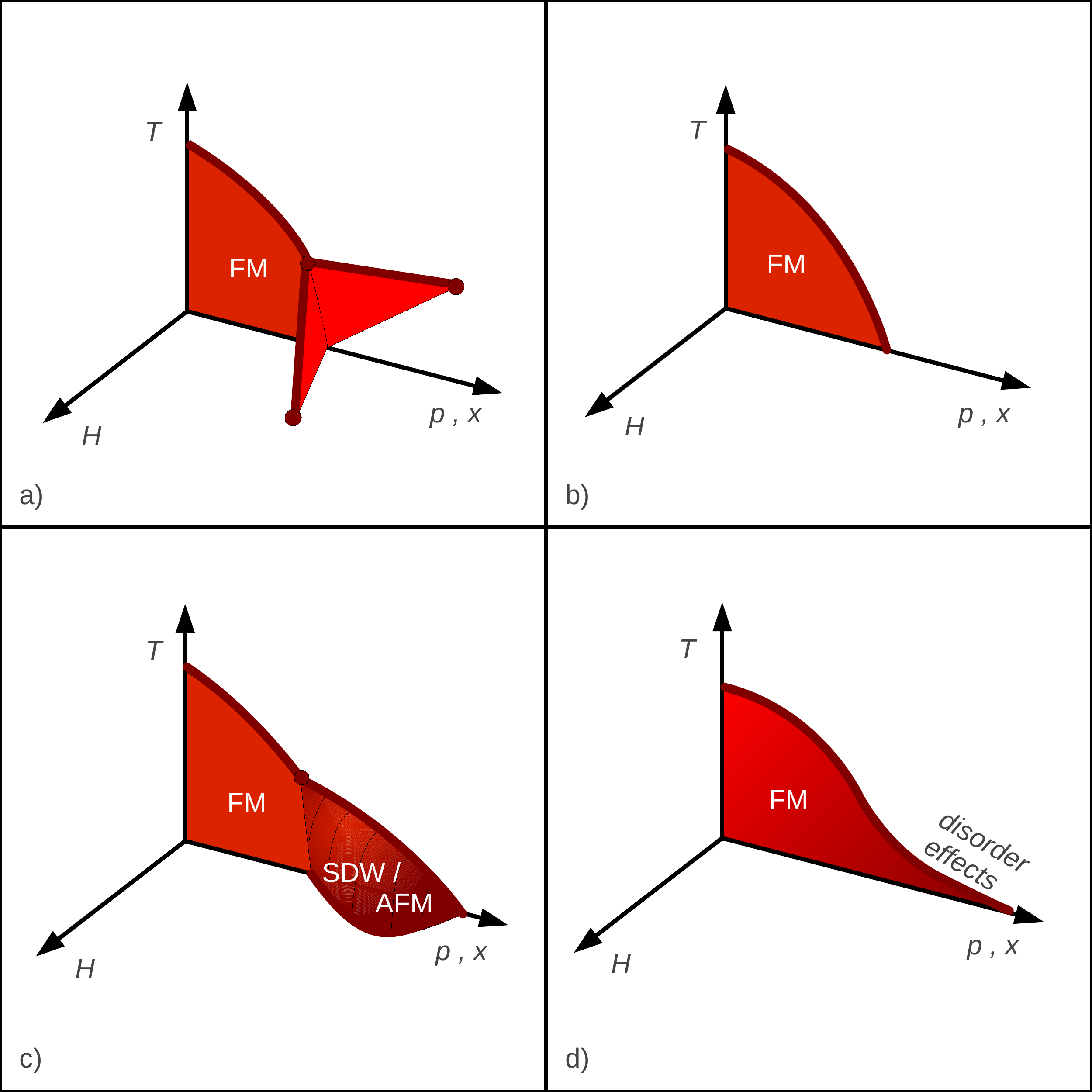}
\end{center}
\caption{(Color online) Schematic phase diagrams observed in ferromagnetic (FM) systems
               that show, at the lowest temperatures realized, a) a discontinuous
               transition and tricritical wings in a magnetic field, b) a continuous
               transition, c) a change to spin-density-wave (SDW) or
               antiferromagnetic (AFM) order, d) a continuous transition in strongly
              disordered systems that may be accompanied by quantum Griffiths
              effects or spin-glass freezing in the tail of the phase diagram.}
\label{fig:schematic_phase_diagrams}
\end{figure}

During the last two decades a large number of ferromagnetic (FM) metals have been found that (1) have a low Curie temperature,
and (2) can be driven across a ferromagnet-to-paramagnet quantum phase transition. The control parameter is often either
hydrostatic pressure or uniaxial stress, but the transition can also be triggered by composition, or an external magnetic field.
The initial motivation for many of these experiments was to look for a ferromagnetic quantum critical point (QCP), and
possibly novel states of matter in its vicinity, as had been found in many antiferromagnetic (AFM)
metals \cite{Grosche_et_al_1996, Mathur_et_al_1998, Gegenwart_et_al_2008, von_Loehneysen_et_al_2007, Park_et_al_2006}. 
It soon became clear, however, that the FM
case is quite different from the AFM one. Instead of displaying a quantum critical point, many systems were found to undergo a
first-order quantum phase transition, with a tricritical point in the phase diagram separating a line of second-order transitions at
relatively high temperatures from a line of first-order transitions at low temperatures. In several of these materials the existence of a 
tricritical point has been confirmed by the observation of tricritical wings upon the application of an external magnetic field $H$,
as shown schematically in Fig.~\ref{fig:schematic_phase_diagrams}\,a. Some systems,
such as ZrZn$_2$, were initially reported to have a QCP, but with increasing sample quality the transition at low temperatures was
found to be first order. The first-order transition occurs across a large variety of materials, including transition metals in which the
magnetism is due to $3d$ electrons, as well as $4f$- and $5f$-electron systems, see Tables~\ref{table:1a},~\ref{table:1b}.
Some systems do show
a continuous quantum phase transition to the lowest temperatures observed, see Tables~\ref{table:2_pt_1}, \ref{table:2_pt_2}, \ref{table:2_pt_3} and 
Fig.~\ref{fig:schematic_phase_diagrams}\,b. Several of these are either strongly 
disordered, as judged by their residual resistivities,%
\footnote{\label{resistivity_footnote} Throughout this review we will use the residual resistivity, denoted by $\rho_0$, as a measure of 
                quenched disorder. One needs to keep in mind that $\rho_0$ is a very rough and incomplete measure of disorder, that
                many transport theories make very simple assumptions regarding the scattering process, and that relating the measured 
                value of $\rho_0$ to theoretical considerations can therefore be difficult.
                Unfortunately, more extensive experimental characterizations
                of disorder as well as more sophisticated theoretical treatments are rarely available.}
or their crystal structure makes them quasi-one-dimensional. Finally, the 
expectation of additional phases was borne out. In some systems the long-range order changes from ferromagnetic to modulated 
spin-density-wave (SDW) or AFM order, see Fig.~\ref{fig:schematic_phase_diagrams}\,c), and strongly disordered systems often show a
spin-glass-like phase in the tail of the phase diagram, Fig.~\ref{fig:schematic_phase_diagrams}\,d). Accordingly, we distinguish four categories
of metallic quantum ferromagnets, namely: (1) Systems that display a first-order quantum phase transition; (2) systems that display, 
or are suspected to display, a quantum critical point; (3) systems that undergo a phase transition to a different type of magnetic 
order before the FM quantum phase transition is reached; and (4) systems with spin-glass-like characteristics or other manifestations
of strong disorder at very low 
temperatures. This phenomenological classification, which is independent of the microscopical origin of the magnetism, is 
reflected in Fig.~\ref{fig:schematic_phase_diagrams} and Tabs. \ref{table:1a}--\ref{table:4}. 
For each of these categories we discuss a number of representative materials in which the QPT has been reasonably well characterized. 
This list of materials is not exhaustive.

We also mention that superconductivity has been found to coexist with itinerant ferromagnetism in four U-based FM metals: 
UGe$_{2}$ (\textcite{Saxena_et_al_2000}), URhGe (\textcite{Aoki_et_al_2001,Yelland_et_al_2011}), 
UCoGe (\textcite{Huy_et_al_2007a}), and UIr (\textcite{Kobayashi_et_al_2006}).
%
%\footnote{\label{ZrZn_2_SC_footnote} Coexistence of superconductivity and ferromagnetism has also been reported to occur in ZrZn$_2$ \cite{Pfleiderer_et_al_2001}, but was later attributed to non-stoichiometric samples with a higher Zr content than ZrZn$_2$ \cite{Yelland_et_al_2005}.}
%
While very interesting, this topic is outside the scope of this review and will be mentioned only in passing. Another very interesting
class of materials that we do not cover are ferromagnetic semiconductors which have recently been reviewed by
\textcite{Jungwirth_et_al_2006}.

\subsection{Systems showing a discontinuous transition}
\label{subsec:II.B}
We first discuss systems in which there is strong evidence for a first-order transition at low temperatures. 
These include the transition-metal compounds MnSi and ZrZn$_2$, several uranium-based compounds, and 
some other materials; their properties are summarized in Tables~\ref{table:1a},~\ref{table:1b}. The wide spread pattern of 1st order transitions near the QPT is consistent with fundamental arguments such as the BKV theory~\cite{Belitz_Kirkpatrick_Vojta_1999, Belitz_Kirkpatrick_Rollbuehler_2005},  which for clean ferromagnets predicted a first-order quantum phase transition at $T=0$, a tricritical point in the phase diagram, and associated tricritical wings in an external magnetic field. This theory will be reviewed in Sec.\ \ref{sec:III}, where we will give a detailed discussion of the relation between theory and experiment.
%
%\onecolumngrid
%\begin{absolutelynopagebreak}
\begin{table*}[t]
\caption{Systems showing a first-order transition I: Transition-metal and uranium-based compounds. FM = ferromagnetism, SC = superconductivity.
               \TC\ = Curie temperature, $T_{\text{tc}} =$ tricritical temperature.
                      $\rho_0 =$ residual resistivity. n.a. = not available}
%\smallskip
\begin{ruledtabular}
\begin{tabular}{lllllllll}
System \ \ \ 
   & Order of 
      & \TC/K$\,^b$
         & magnetic 
            & tuning  
               & $T_{\text{tc}}$/K 
                  & wings 
                     & Disorder$\,^d$  
                        & Comments \\
   & Transition$\,^a$ 
      &                            
         & moment/$\mu_{\text{B}}^c$
            &  parameter 
               &                              
                  &  observed  
                     & ($\rho_0/\mu\Omega$cm)\ \ 
                        & \\
%\\
\hline\\[-7pt]
%%%%%%%%%%%%%%%%%%%%%%%%%%%%%%%
MnSi
   & 1st$\,^{1,2}$ 
      & $29.5\,^3$
         & $0.4\,^3$
            & pressure$\,^1$ 
               & $\approx 10\,^{1,e}$  
                  & yes$\,^4$
                     & $0.33\,^4$
                        & weak helimagnet$\,^{5,6}$\\
\\[-5pt]
%%%%%%%%%%%%%%%%%%%%%%%%%%%%%%%%%
ZrZn$_2$
   & 1st$\,^{7}$
      & $28.5\,^7$ 
         & $0.17\,^7$
            & pressure$\,^7$
               & $\approx 5\,^7$
                  & yes$\,^7$
                     & $\geq 0.31\,^8$ 
                        & long history$\,^{9}$\\
 \\[-5pt]
%%%%%%%%%%%%%%%%%%%%%%%%%%%%%%%%%%%%
CoS$_2$
   & 1st$\,^{10,11}$ 
      & $122\,^{10}$  
         & $0.84\,^{12}$  
            & pressure$\,^{10}$  
               & $\approx 118\,^{10}$ 
                  & (yes)$\,^{f}$ 
                     & $0.2$ -- $0.6\,^{13}$
                        & high \TC\ and $T_{\text{tc}}$\\
\\[-5pt]
%%%%%%%%%%%%%%%%%%%%%%%%%%%%%%%%%%%%%%%%%%
Ni$_3$Al
   & (1st?)$\,^{g}$
      & 41 -- 15$\,^{h}$
         & 0.075$\,^{i}$
            & pressure$\,^{14}$ 
               & n.a. 
                  & no 
                     & 0.84$\,^{15}$
                        & 1st order trans-\\
& & & & & & & & ition suspected \\[2pt]
\hline\\[-7pt]
%%%%%%%%%%%%%%%%%%%%%%%%%%%%%%%%%%%%%%%%%%
UGe$_2$
   & 1st$^{\,16,17}$ 
      & $52^{\,18}$
         & $1.5^{\,18}$
            & pressure$\,^{18,19}$
               & $24\,^{20}$
                  & yes$\,^{18,20}$
                     & $0.2\,^{19}$
                        & easy-axis FM \\
& & & & & & & & coex. FM+SC$\,^{19}$ \\                    
\\[-5pt]
%%%%%%%%%%%%%%%%%%%%%%%%%%%%%%%%%%%%%%%%%%
U$_3$P$_4$
   & 1st$^{\,21}$ 
      & 138$^{\,22}$
         & 1.34$^{\,23,j}$
            & pressure$^{\,21}$
               & 32$^{\,21}$
                  & yes$\,^{21,k}$
                     & 4$^{\,21,l}$
                        & canted easy-axis FM \\
\\[-5pt]
%%%%%%%%%%%%%%%%%%%%%%%%%%%%%%%%%%%%
URhGe
   & 1st$\,^{17,24}$
      & $9.5^{\,25}$
         & $0.42\,^{25}$
            &$\perp B$-field$\,^{24,26}$
               & $\approx 1\,^{24}$
                  & yes$\,^{24}$
                     & $8\,^{27}$
                        & easy-plane FM\\
& & & & & & & & coex. FM+SC$\,^{25}$ \\                          
\\[-5pt]
%%%%%%%%%%%%%%%%%%%%%%%%%%%%%%%%%%%%%%%
UCoGe
   & 1st$\,^{17,28}$
      & $2.5\,^{28}$
         & $0.03\,^{29}$
            & none 
               & $>2.5\,?$ $^m$
                  & no 
                     & $12\,^{29}$ 
                        & very weak FM \\
& & & & & & & & coex. FM+SC$\,^{29}$ \\                         
\\[-5pt]
%%%%%%%%%%%%%%%%%%%%%%%%%%%%%%%%
UCoAl 
   & 1st$\,^{30,n}$
      & $\approx 0\,^{30,31,o}$
         & $0\,^{30,31,o}$
            & pressure$\,^{30,31}$
               & $> 11$\,K$\,^{30}$
                 & yes$\,^{30}$
                   & 24$\,^{30}$
                     & easy-axis FM\\
\\[-5pt]
%%%%%%%%%%%%%%%%%%%%%%%%%%%%%%%%%%%%%%%%%%
URhAl 
   & 1st$\,^{33}$
      & $34$ -- $25\,^{32,33}$
         & $\approx 0.9\,^{32,33}$
            & pressure$\,^{33}$
               & $\approx 11\,^{33}$
                  & yes$\,^{33}$
                     & $\approx$ 65$\,^{33}$
                        & weakly 1st order \\ [2pt]
%& & & & & & & & suspected\\[2pt]
\hline\hline\\[-5pt]
%\vskip -3pt
%\hline\\
\multicolumn{9}{l} {$^a$ At the lowest temperature achieved.}\\
\multicolumn{9}{l} { $^b$ A single value of \TC\ at the default value of the tuning parameter (ambient pressure,
                               zero field) is given if $T_{\text{tc}}$ has also been}\\ 
\multicolumn{9}{l} {\hskip 7pt measured; a range of \TC\ for a range of control parameters in all other cases. \quad $^c$ Per formula unit unless otherwise noted.}\\
\multicolumn{9}{l} {$^d$ For the highest-quality samples.\quad $^e$ Disputed by \textcite{Stishov_et_al_2007}; see text.}\\
\multicolumn{9}{l} {$^{f}$ Metamagnetic behavior in 1st-order region indicative of wings.}\\
\multicolumn{9}{l} {$^{g}$ Suspected 1st order transition near $p=80$\,kbar  \cite{Niklowitz_et_al_2005, Pfleiderer_2007}.}\\
\multicolumn{9}{l} {$^{h}$ For pressures $p = 0$ -- $60$ kbar.\quad 
                              $^{i}$ Per Ni at $p=0$ \cite{Niklowitz_et_al_2005}.\quad $^j$ Per U.}\\ 
\multicolumn{9}{l} {$^k$ Via a metamagnetic transition; wings have not been mapped out.\quad $^l$ At the critical pressure $p_c \approx 4$\,GPa.}\\
\multicolumn{9}{l} {$^m$ Pressure decreases \TC\ \cite{Slooten_et_al_2009}; TCP not accessible. \TC\ increases nonmonotonically
                                       upon doping with Rh }\\
\multicolumn{9}{l} {\hskip 7pt \cite{Sakarya_et_al_2008}; order of transition for URh$_x$Co$_{1-x}$Ge not known except for 
                              $x=1$ (2nd order with \TC\ = 9.5\,K).}\\
\multicolumn{9}{l} {$^n$ Inferred from existence of tricritical wings.}\\
\multicolumn{9}{l} { $^o$ PM at zero pressure. Uniaxial pressure induces FM, so does doping, see \textcite{Ishii_et_al_2003} and references therein.}\\
%\multicolumn{9}{l} { $^p$ 1st order transition suspected near $p_c \approx 5.3\,$GPa \cite{Combier_2013}.}\\   
                              [-5pt]
                                 \\\hline\\[-5pt]
\multicolumn{2}{l}{$^1$ \textcite{Pfleiderer_et_al_1997}}  
      & \multicolumn{3}{l}{$^2$ \textcite{Uemura_et_al_2007}} 
            & \multicolumn{3}{l}{$^3$ \textcite{Ishikawa_et_al_1985}} 
                  & \multicolumn{1}{l}{$^4$ \textcite{Pfleiderer_Julian_Lonzarich_2001}} 
                      \\
\multicolumn{2}{l}{$^5$ \textcite{Ishikawa_et_al_1976}}  
      & \multicolumn{3}{l}{$^6$ \textcite{Muehlbauer_et_al_2009}} 
            & \multicolumn{3}{l}{$^7$ \textcite{Uhlarz_Pfleiderer_Hayden_2004}} 
                  & \multicolumn{1}{l}{$^8$ \textcite{Sutherland_et_al_2012}} 
                      \\
\multicolumn{2}{l}{$^{9}$ \textcite{Pfleiderer_2007}}  
      & \multicolumn{3}{l}{$^{10}$ \textcite{Goto_et_al_1997}} 
            & \multicolumn{3}{l}{$^{11}$ \textcite{Goto_Fukamichi_Yamada_2001}} 
                  & \multicolumn{1}{l}{$^{12}$ \textcite{Adachi_Sato_Takeda_1969}} 
                       \\
\multicolumn{2}{l}{$^{13}$ \textcite{Sidorov_et_al_2011a}}  
      & \multicolumn{3}{l}{$^{14}$ \textcite{Niklowitz_et_al_2005}} 
            & \multicolumn{3}{l}{$^{15}$ \textcite{Steiner_et_al_2003}} 
                  & \multicolumn{1}{l}{$^{16}$ \textcite{Huxley_et_al_2001}}
                         \\ 
\multicolumn{2}{l}{$^{17}$ \textcite{Aoki_et_al_2011b}}  
      & \multicolumn{3}{l}{$^{18}$ \textcite{Kotegawa_et_al_2011b}} 
            & \multicolumn{3}{l}{$^{19}$ \textcite{Saxena_et_al_2000}} 
                  & \multicolumn{1}{l}{$^{20}$ \textcite{Taufour_et_al_2010}} 
                        \\ 
\multicolumn{2}{l}{$^{21}$ \textcite{Araki_et_al_2015}}
      & \multicolumn{3}{l}{$^{22}$ \textcite{Trzebiatowski_Troc_1963}}  
            & \multicolumn{3}{l}{$^{23}$ \textcite{Wisniewski_Gukasov_Henkie_1999}} 
                    & \multicolumn{1}{l}{$^{24}$ \textcite{Huxley_et_al_2007}}
                        \\ 
\multicolumn{2}{l}{$^{25}$ \textcite{Aoki_et_al_2001}} 
      & \multicolumn{3}{l}{$^{26}$ \textcite{Levy_et_al_2005}}  
            & \multicolumn{3}{l}{$^{27}$ \textcite{Miyake_Aoki_Flouquet_2009}} 
                  & \multicolumn{1}{l}{$^{28}$ \textcite{Hattori_et_al_2010}} 
                        \\     
\multicolumn{2}{l}{$^{29}$ \textcite{Huy_et_al_2007b}}           
      &  \multicolumn{3}{l}{$^{30}$ \textcite{Aoki_et_al_2011a}}           
            & \multicolumn{3}{l} {$^{31}$ \textcite{Ishii_et_al_2003}}                 
                 & \multicolumn{1}{l} {$^{32}$ \textcite{Veenhuizen_el_al_1988} }
                      \\    
\multicolumn{2}{l} {$^{33}$ \textcite{Shimizu_et_al_2015a} }
      & 
            & 
                  & 
                        \\     
%%%%%%%%%%%%%%%%%%%%%%%%%%%%%%%%%%%%
\end{tabular}
\end{ruledtabular}
\vskip -5mm
\label{table:1a}
\end{table*}

%
%\onecolumngrid
\begin{table*}[t]
\caption{Systems showing a first-order transition II: Lanthanide-based compounds, and strontium ruthenates.
               \TC\ = Curie temperature, $T_{\text{tc}} =$ tricritical temperature. $\rho_0 =$ residual resistivity. n.a. = not available.}
\smallskip
\begin{ruledtabular}
\begin{tabular}{lllllllll}
System 
   & Order of 
      & \TC/K
         & magnetic 
            & tuning  
               & $T_{\text{tc}}$/K 
                  & wings 
                     & Disorder$\,^c$  
                        & Comments \\
   & Transition$\,^a$ 
      &                            
         & moment/$\mu_{\text{B}}\,^b$
            &  parameter 
               &                              
                  & observed
                     & ($\rho_0/\mu\Omega$cm)
                        & \\
%\\
\hline\\[-7pt]
%%%%%%%%%%%%%%%%%%%%%%%%%%%%%%%
La$_{1-x}$Ce$_x$In$_2$
   & 1st$\,^{1}$
      & 22 -- 19.5 $\,^{1,d}$ 
         & n.a.
            & composition$\,^{1}$
               & $>22$ ?$\,^e$
                  & no 
                     &  \hskip -15pt n.a. 
                        & third phase?$\,^{1}$\\
\\[-5pt]
%%%%%%%%%%%%%%%%%%%%%%%%%%%%%%%%%%%%%
SmNiC$_2$
   & 1st$\,^{2}$
      & 17 -- 15$\,^{2,f}$
         & 0.32$\,^{2}$
            & pressure$\,^{2}$
               & $> 17$ ?
                  & no
                     &  \hskip -15pt 2
                        & other phases$\,^{2}$\\
\\[-5pt]
%%%%%%%%%%%%%%%%%%%%%%%%%%%%%%%%%%%%
YbCu$_2$Si$_2$
   & 1st$\,^{3}$ 
      & 4.7 -- 3.5$\,^{3,2,g}$ 
         & 0.16 -- 0.42$\,^{3,h}$ 
             & pressure$\,^{3-5}$
                & n.a.
                   & no
                      &  \hskip -15pt $<$ 0.5$\,^{6}$
                         & strong Ising\\
& & &  & & & & & anisotropy$\,^{3}$\\
%%%%%%%%%%%%%%%%%%%%%%%%%%%%%%%%%%%%%%%%%%
YbIr$_2$Si$_2$
   & 1st$\,^{7}$ 
      & 2.3 -- 1.3$\,^i$ 
         & n.a. 
            & pressure$\,^{7}$
               & n.a. 
                  & no 
                     &  \hskip -15pt $\approx 22$ $^j$  
                        & FM order \\
& & & & & & & & suspected$\,^{7}$ \\
%%%%%%%%%%%%%%%%%%%%%%%%%%%%%%%%%%%%%%%%%%
CePt
   & (1st?)$\,^{8}$
      & 5.8 -- 0$\,^{9,8}$
         & n.a.
            & pressure$\,^{8}$
               & n.a.
                  & no
                     &  \hskip -15pt $\approx$ 11$\,^{9}$                  
                        & 1st order trans-\\
& & & & & & & & ition suspected \\[2pt]
\hline\\[-7pt]
%%%%%%%%%%%%%%%%%%%%%%%%%%%%%%%%%
\SCRO
   & 1st$\,^{10}$
      & 160 -- 0$\,^k$
         & 0.8 -- 0$\,^k$
            & composition$\,^{10}$
               & n.a.
                  & no
                     &  \hskip -15pt n.a.
                         & ceramic samples\\
%%%%%%%%%%%%%%%%%%%%%%%%%%%%%%%%%
Sr$_3$Ru$_2$O$_7$
   & 1st$\,^l$
      & 0$\,^m$
         & 0$\,^m$
            & pressure$\,^m$
                & n.a.
                  & yes $^{11}$ 
                     &  \hskip -15pt $< 0.5$$\,^{11}$
                        & foliated wings,\\
& & & & & & & & exotic phase$\,^{11}$ \\[2pt]
\hline\hline\\[-5pt]
%\vskip -3pt
%\hline\\
\multicolumn{9}{l} {$^a$ At the lowest temperature achieved.\quad $^b$ Per formula unit unless otherwise noted.\quad $^c$ For the highest-quality samples.}\\
\multicolumn{9}{l} { $^d$ For $x = 1.0$ -- $0.9$.\quad 
                               $^e$ 1st order for $x=1$, TCP not accessible.\quad
                               $^f$ For $p = 0$ -- 2\,GPa.\quad
                               $^g$ For pressures $p \approx 11.5$ -- $9.4$\,GPa.}\\
\multicolumn{9}{l} {$^h$ For pressures $p = 9.4$ -- $11.5$\,GPa.\quad
                              $^i$ For pressures $p \approx 10$ -- $8$\,GPa.}\\  
\multicolumn{9}{l} {$^j$ For a magnetic sample at pressures $p \approx 8$ -- $10$\,GPa. Samples with
                                       $\rho_0$ as low as 0.3$\mu\Omega$cm at ambient pressure have been}\\
\multicolumn{9}{l} {\hskip 7pt prepared \cite{Yuan_et_al_2006}.\quad $^k$ For $x = 0$ to $x\agt 0.7$.}\\
\multicolumn{9}{l} {$^l$ Phase diagram not mapped out completely; the most detailed measurements show tips of wings. 
                                     See \textcite{Wu_et_al_2011}.}\\
\multicolumn{9}{l} {$^m$ Paramagnetic at ambient pressure. Hydrostatic pressure drives the system away from FM, 
                                       uniaxial stress drives it towards}\\
\multicolumn{9}{l} {\hskip 7pt FM. See \textcite{Wu_et_al_2011} and references therein, especially \textcite{Ikeda_et_al_2000}.}\\                                                         
                              [-5pt]
                                 \\\hline\\[-5pt]
\multicolumn{3}{l}{$^{1}$ \textcite{Rojas_et_al_2011}}  
      &  \multicolumn{2}{l}{ \hskip -15pt $^{2}$ \textcite{Woo_et_al_2013}} 
            & \multicolumn{2}{l}{ \hskip -15pt $^{3}$ \textcite{Tateiwa_et_al_2014}} 
                  & \multicolumn{2}{l}{ \hskip -0pt $^{4}$ \textcite{Winkelmann_et_al_1999}} 
                      \\
\multicolumn{3}{l}{$^{5}$ \textcite{Fernandez-Panella_et_al_2011}}  
      & \multicolumn{2}{l}{ \hskip -15pt $^{6}$ \textcite{Colombier_et_al_2009}} 
            & \multicolumn{2}{l}{ \hskip -15pt $^{7}$ \textcite{Yuan_et_al_2006}} 
                  & \multicolumn{2}{l}{ \hskip -0pt $^{8}$ \textcite{Larrea_et_al_2005}} 
                      \\
\multicolumn{3}{l}{$^{9}$ \textcite{Holt_et_al_1981}}  
      & \multicolumn{2}{l}{ \hskip -15pt $^{10}$ \textcite{Uemura_et_al_2007}} 
            & \multicolumn{2}{l}{ \hskip -15pt $^{11}$ \textcite{Wu_et_al_2011}} 
                  & \multicolumn{2}{l}{ } 
                       \\
%%%%%%%%%%%%%%%%%%%%%%%%%%%%%%%%%%%%
\end{tabular}
\end{ruledtabular}
\vskip -5mm
\label{table:1b}
\end{table*}

\subsubsection{Transition-metal compounds}
\label{subsubsec:II.B.1}
\begin{figure}[b]
\vskip 0pt
\begin{center}
\includegraphics[width=0.8\columnwidth,angle=0]{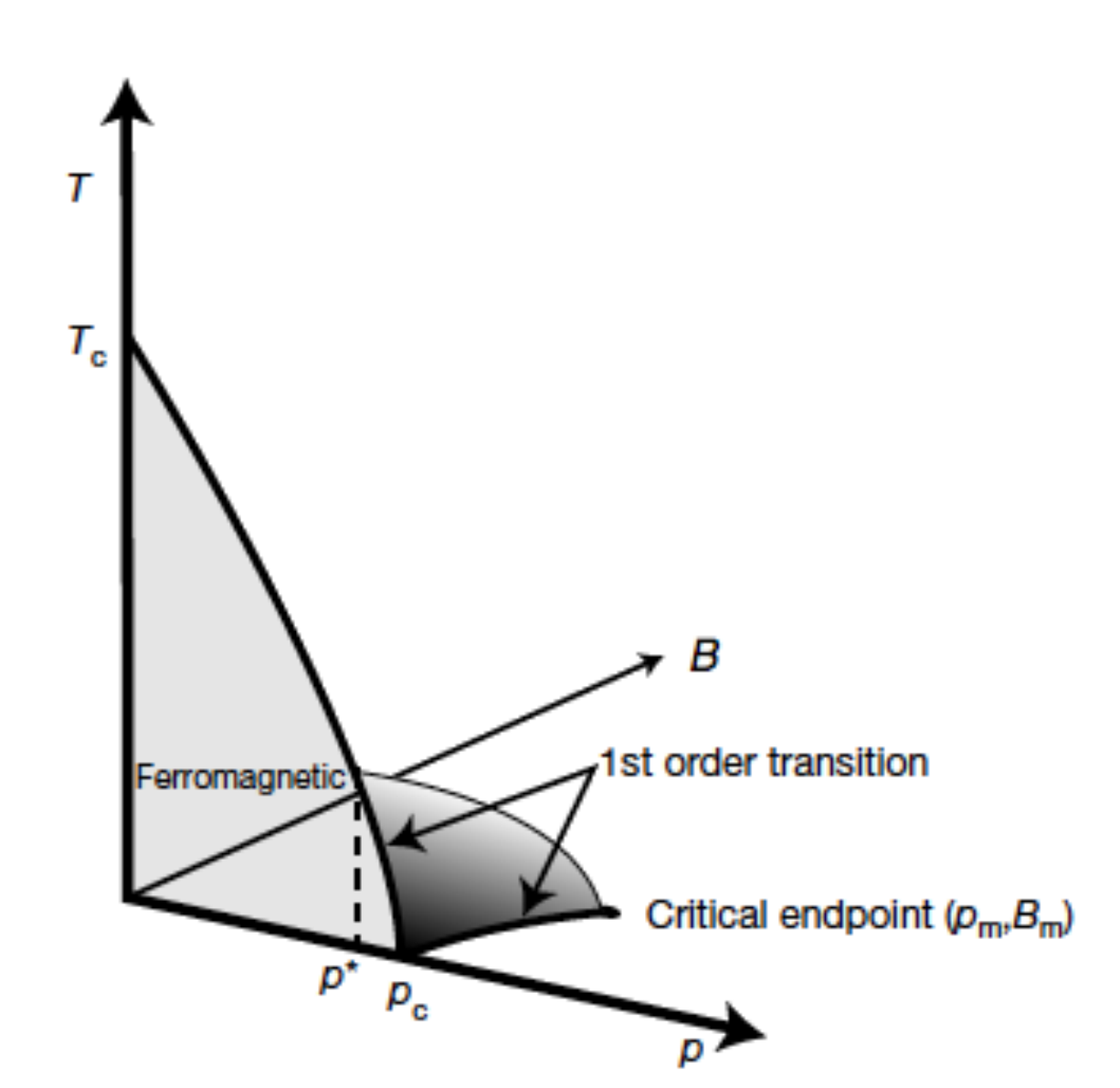}
\end{center}
\vskip -10pt
\caption{Phase diagram of MnSi. In the temperature - pressure ($T$-$p$) plane the transition temperature drops from
 \TC\ $ = 29.5\,$K at ambient pressure and changes from second to first order at $p^* = 12\,$kbar where 
 \TC\ $ \approx 12\,$K. \TC\ vanishes at $p_{\text{c}} = 14.6$ kbar. In the magnetic field - pressure ($B$-$p$) 
 plane at $T=0$, and everywhere across the shaded wing, the transition is first order up to a ``critical endpoint''$^{\ref{CEP_footnote}}$ estimated to be located at $B_{\text{m}} = 0.6\,$T and $p_{\text{m}} = 17\,$kbar. From \textcite{Pfleiderer_Julian_Lonzarich_2001}.}
\label{figure:MnSi_phase_diagram}
\end{figure}

\begin{figure}[b]
\vskip -0pt
\begin{center}
\includegraphics[width=0.65\columnwidth,angle=0]{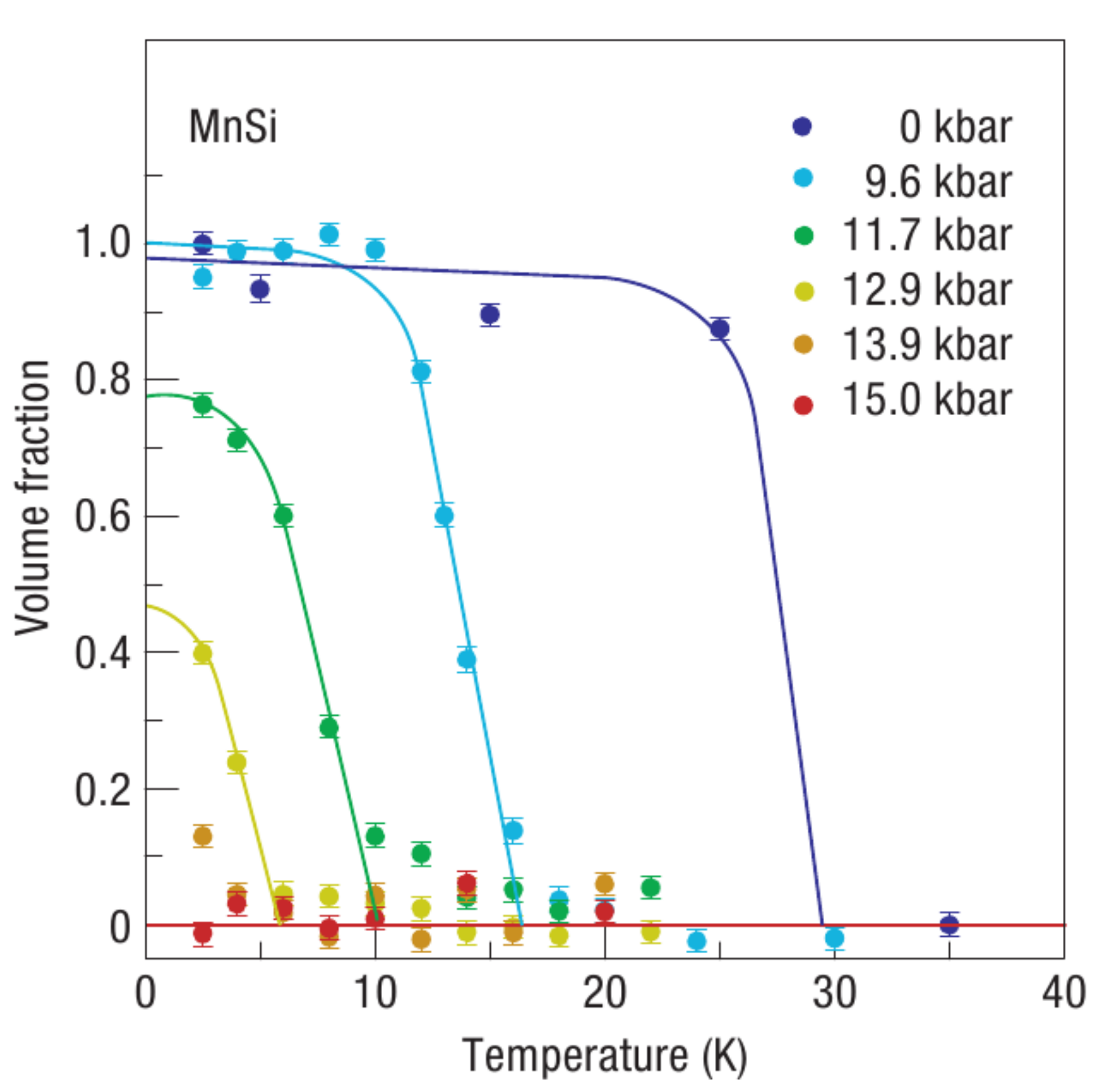}
\end{center}
\vskip -10pt
\caption{$\mu$SR results for the volume fraction with static magnetic order. The nonzero volume fraction less then unity at
              $T=0$ for intermediate pressures indicates phase separation, which in turn is indicative of a first-order transition.
              From \textcite{Uemura_et_al_2007}.}
\label{figure:MnSi_volume_fraction}
\end{figure}

\paragraph{MnSi}
\label{par:II.B.1.a}
MnSi is a very well-studied material in which the search for a FM QCP resulted in the observation of a first-order quantum
phase transition. The transition temperature at ambient pressure is \TC\, $\approx 29.5\,$K, 
\footnote{\label{Tc_footnote} We denote the ferromagnetic transition temperature by \TC\ irrespective of the order
of the transition. In parts of Sec.\ \ref{sec:III}, where we want to emphasize that a transition is second order, we denote the critical
temperature by $\Tc$.}
and the application of hydrostatic pressure suppresses \TC\ to zero at a critical pressure 
$p_c \approx 14.6\,$kbar \cite{Pfleiderer_MeMullan_Lonzarich_1994,Pfleiderer_et_al_1997}. This compound is actually a weak helimagnet \cite{Ishikawa_et_al_1976},
due to its B21 crystal structure that lacks inversion symmetry,
with a complicated phase diagram (see \onlinecite{Muehlbauer_et_al_2009} and references therein). 
However, the long wavelength of 
the helix, about 180\ \AA, allows one to approximate the system as a ferromagnet. The helical order implies that the transition 
should be very weakly first order even at ambient pressure \cite{Bak_Jensen_1980}. This has indeed been
observed \cite{Stishov_et_al_2007, Stishov_et_al_2008, Janoschek_et_al_2013}.%
\footnote{\label{Brazovskii_footnote} The first-order transition at ambient pressure was found by \textcite{Janoschek_et_al_2013} 
 to be of a type that was first predicted by \textcite{Brazovskii_1975} for different systems. It differs slightly from the type
 predicted by \textcite{Bak_Jensen_1980} for helical magnets.}
\textcite{Pfleiderer_et_al_1997} found evidence of a strongly first-order transition for pressures 
$p^* < p < p_c$ with $p^* \approx 12\ \text{kbar}$. The tricritical temperature (i.e., the transition temperature at $p = p^*$) is 
$T_{\text{tc}} \approx 12\,$K.%
\footnote{\label{pseudo_TCP_footnote} Since the transition is likely to be weakly first order for all $p<p^*$, the
                                                               observed apparent tricritical point separates a very weakly first-order                                                              transition from one that is more strongly first order.}
These results were later corroborated by the observation of tricritical wings (\onlinecite{Pfleiderer_Julian_Lonzarich_2001}, the observed phase diagram is shown in Fig.~\ref{figure:MnSi_phase_diagram}), and by $\mu$SR data that show, for $p^*<p<p_c$, phase separation indicative of a first-order transition \cite{Uemura_et_al_2007}, see Fig.~\ref{figure:MnSi_volume_fraction}. Moreover, this has been confirmed by neutron Larmor diffraction experiments under pressure~\cite{Pfleiderer_et_al_2007}. Conversely, data presented by \textcite{Stishov_et_al_2007}, \textcite{Petrova_et_al_2009}, and \textcite{Petrova_Stishov_2012}, suggests that the quantum phase transition at $p = p_c$ is either continuous or very weakly first order. Although the evidence for a pressure induced first-order transition appears convincing in the purest crystals, no agreement has been reached~\cite{Otero-Leal_et_al_2009a, Stishov_2009, Otero-Leal_et_al_2009b}.

For $p < p_c$ the properties of MnSi are in good semiquantitative agreement with the SCR theory~\cite{Pfleiderer_et_al_1997}. Specifically, $T^{4/3}_{\textrm{C}}$ is a linear function of the pressure. This agreement fails at $p \rightarrow p_c$ due to the presence of the first-order transition. Also, a striking $T^{3/2}$ power law for the resistivity was observed in a broad range of $p \gtrsim p_c$~\cite{Pfleiderer_et_al_2001} where we would expect a Fermi-liquid $T^{2}$ behavior. The nature of this power law in resistivity, which seems to be a common feature in itinerant magnets near their QPT (cf. Sec.~\ref{subsec:IV.B}), is still unclear. 

\paragraph{ZrZn$_2$}
\label{par:II.B.1.b}
Another transition-metal compound with very similar behavior across the quantum ferromagnetic transition, but without
the complications resulting from helical order, is ZrZn$_2$. It crystallizes in the cubic C15 structure and
is a true ferromagnet \cite{Matthias_Bozorth_1958,Pickart_et_al_1964} with a small magnetic anisotropy and an ordered 
moment of $0.17\,$\mub\  per formula unit \cite{Uhlarz_Pfleiderer_Hayden_2004}. The material can be tuned across the transition by means of hydrostatic pressure. While early experiments \cite{Smith_Mydosh_Wohlfarth_1971,Huber_Maple_Wohlleben_1975, Grosche_et_al_1995} suggested the existence of a quantum critical point, an increase in sample quality led to the realization that the transition becomes first order at high $p$, causing a first-order quantum phase transition with a critical pressure $p_c \approx 16.5\,$kbar \cite{Uhlarz_Pfleiderer_Hayden_2004}. The transition temperature at ambient pressure is \TC\ $\approx 28.5\,$K, and the tricritical temperature is $T_{\text{tc}} \approx 5\,$K. The phase diagram is qualitatively the same as that shown in Fig.~\ref{figure:MnSi_phase_diagram}; the observation of tricritical wings by \textcite{Uhlarz_Pfleiderer_Hayden_2004} confirmed an earlier suggestion by \textcite{Kimura_et_al_2004}.
The first-order nature of the QPT was confirmed by \textcite{Kabeya_et_al_2012,Kabeya_et_al_2013}, who also
studied crossover phenomena above the tricritical wings. However, the transition is weakly first order and, for $p < p_c$, ZrZn$_{2}$ can be reasonably well undestood within the SCR theory~\cite{Grosche_et_al_1995,Smith_et_al_2008}. Surprisingly, the resistivity exponent shows an abrupt change from 5/3 for $p < p_c$ to 3/2 at $p > p_c$ and remains 3/2 up to higher pressures (about 25\,kbar), similarly to MnSi. In ZrZn$_{2}$ as well as in other itinerant magnets such a NFL behavior is still not understood (cf. Sec.~\ref{subsec:IV.B}).

\paragraph{CoS$_2$}
\label{par:II.B.1.c}
Cobalt disulphide crystallizes in a cubic pyrite structure. It is an itinerant ferromagnet with \TC\ $\approx 124$\,K, an ordered moment 
of 0.84\,\mub/Co, and an effective moment of 1.76\,\mub/Co~\cite{Jarrett_et_al_1968, Adachi_Sato_Takeda_1969}. Density-functional 
calculations concluded that CoS$_{2}$ is a half-metallic ferromagnet \cite{Zhao_et_al_1993,Mazin_2000}. The spin polarization is high at about
56\% \cite{Wang_et_al_2004}, and the transport coefficients and the thermal expansion coefficient show unusual behavior in the vicinity of 
the transition \cite{Adachi_Ohkohchi_1980, Yomo_1979}. Magnetization measurements indicate that the transition is almost first order at ambient 
pressure \cite{Wang_et_al_2004}. Hydrostatic pressure decreases \TC, and at a pressure of about 0.4\,GPa the nature of the transition
changes from second order to first order, with a tricritical temperature $T_{\text{tc}} \approx 118\,$K \cite{Goto_et_al_1997}. A much lower
value for the tricritical pressure was found by \textcite{Otero-Leal_et_al_2008}; however, this analysis depended on a specific model
equation of state. \textcite{Sidorov_et_al_2011a} confirmed a strongly first order QPT at a critical pressure of about 4.8\,GPa. \TC\, is
also suppressed if selenium is substituted for sulphur, and the transition again becomes first order at a small selenium concentration,
with 1\% of selenium roughly equivalent to a pressure of 1\,GPa \cite{Hiraka_Endoh_1996}. 

Two groups have investigated the $p\,$-$T$ phase diagram at higher pressures up to the QPT:~\textcite{Barakat_et_al_2005} observed a monotonically 
decreasing \TC\, with increasing pressure. They inferred a first-order quantum phase transition at $p_{\text{c}} \approx 6$\,GPa from a
change of the temperature dependence of the resistivity ($\rho(T) = \rho_{0} + AT^{n}$) from $n = 2$ in the FM phase to $n \approx 1.6$ 
for $p > p_{\text{c}}$. Their samples had a residual resistivity $\rho_0 \approx 2\,\mu\Omega$cm and a residual resistance ratio (RRR) of about 60. 
\textcite{Sidorov_et_al_2011a} performed experiments on a better sample ($\rho_0 \approx 0.7\,\mu\Omega$cm) and concluded that 
$p_{\text{c}} = 4.8$\,GPa. They found that the temperature dependence of the resistivity does not change across the transition, with
$n=2$ both below and above $p_{\text{c}}$, while the residual resistivity drops by about a factor of 3 as the transition is crossed.

These discrepancies notwithstanding, all experiments agree on the first-order nature of the quantum phase transition. This makes
the phase diagrams of CoS$_2$, ZrZn$_2$, and MnSi qualitatively the same. It is worthwhile noting that the tricritical temperatures
correlate with the size of the ordered moment, with the largest moment corresponding to the highest $T_{\text{tc}}$.

\paragraph{Ni$_3$Al}
\label{par:II.B.1.d}

Ni$_3$Al crystallizes in the simple cubic Cu$_3$Au structure. Its magnetic properties depend on the exact composition; the stoichiometric
compound at ambient pressure is a ferromagnet with a Curie temperature \TC\,= 41\,K and a small ordered moment of 0.075\,$\mu_{\text{B}}$/Ni
\cite{de_Boer_et_al_1969, Niklowitz_et_al_2005}. \TC\, decreases upon the application of hydrostatic pressure and vanishes at a critical 
pressure of 8.1\,GPa \cite{Niklowitz_et_al_2005}. The resistivity of stoichiometric Ni$_3$Al shows a pronounced non-Fermi-liquid (NFL) temperature 
dependence on either side of the transition, $\Delta\rho \propto T^n$, with $n$ somewhere between 3/2 and 5/3
\cite{Fluitman_et_al_1973, Steiner_et_al_2003, Pfleiderer_2007}. At ambient pressure and in zero magnetic field \textcite{Steiner_et_al_2003}
found $n = 1.65$ for temperatures between about 0.5 and 3.5\,K. The prefactor is comparable with that of the $T^{3/2}$ behavior of the
resistivity in ZrZn$_2$ \cite{Pfleiderer_et_al_2001, Yelland_et_al_2005}. 

The transition at ambient pressure is second order, and the overall form of the phase diagram is consistent with the results of the
spin-fluctuation theory described in Sec.\ \ref{subsubsec:III.C.2}, as is the logarithmic temperature dependence of the specific
heat \cite{Sato_1975, Yang_et_al_2011, Niklowitz_et_al_2005}. However, studies of the temperature dependence of the resistivity under
pressure suggest that the quantum phase transition at the critical pressure is first order \cite{Niklowitz_et_al_2005, Pfleiderer_2007}.
This would be analogous to the behavior of MnSi, Sec.\ \ref{par:II.B.1.a}, where overall behavior consistent with spin-fluctuation theory
also gives way to a first-order transition at low temperatures. Since the magnetic moment in Ni$_3$Al is smaller than in MnSi, the 
theory discussed in Sec.\ \ref{subsubsec:III.B.2} predicts the tricritical temperature in the former to be lower than in the latter, see the
discussion in Sec.\ \ref{subsubsec:II.B.5}, which is consistent with the experimental evidence.

\TC\ also decreases upon doing of Ni$_3$Al with Pd \cite{Sato_1975} or Ga \cite{Yang_et_al_2011}; these systems are discussed in
Sec.~\ref{NiAlGa}.

\subsubsection{Uranium-based compounds}
\label{subsubsec:II.B.2}

Ferromagnetism with a first-order transition at low transition temperatures has been observed in the uranium-based
heavy-fermion compounds UGe$_2$ \cite{Huxley_Sheikin_Braithwaite_2000,Taufour_et_al_2010, Kotegawa_et_al_2011b}, URhGe \cite{Huxley_et_al_2007}, and UCoGe \cite{Hattori_et_al_2010}. UCoAl is paramagnetic at ambient pressure, but very close
to a first-order quantum phase transition \cite{Aoki_et_al_2011a}. The ferromagnetism is due to
$5f$ electrons. The extent to which these electrons are localized or itinerant, and the consequences for neutron-scattering
observations, have been investigated in some detail \cite{Yaouanc_et_al_2002, Fujimori_et_al_2012, Chubukov_Betouras_Efremov_2014}. 
Coexistence of ferromagnetism and superconductivity (SC) has been found in UGe$_2$ \cite{Saxena_et_al_2000, Huxley_et_al_2001},
URhGe \cite{Aoki_et_al_2001}, and UCoGe \cite{Huy_et_al_2007b}; for a recent overview, see \textcite{Aoki_Flouquet_2014}. 
%This topic is beyond the scope of this review and will be mentioned only cursorily.

\paragraph{UGe$_{2}$}
\label{par:II.B.2.a}

\begin{figure}[t]
\vskip 0pt
\begin{center}
\includegraphics[width=0.85\columnwidth,angle=0]{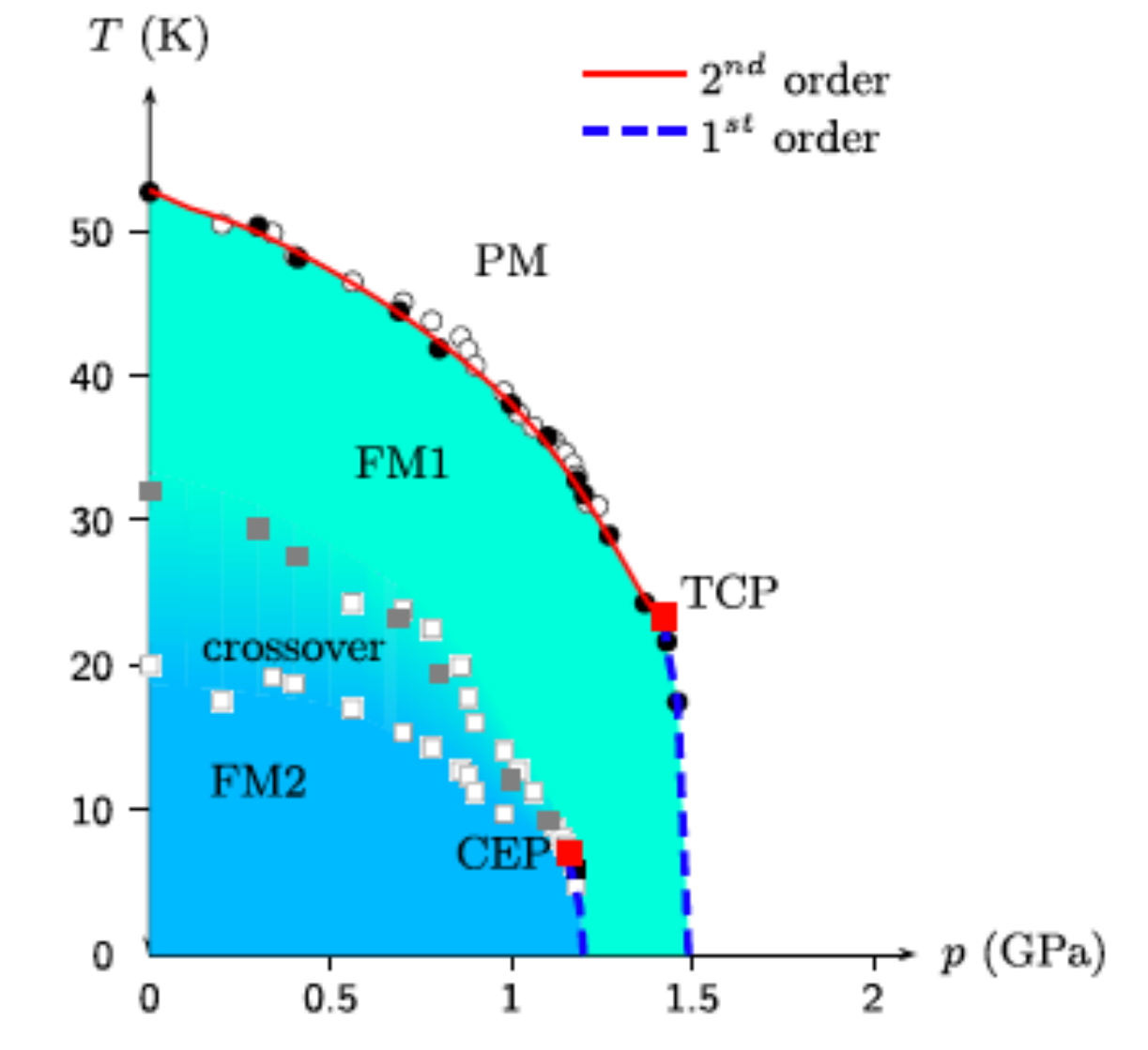}
\end{center}
\vskip -10pt
\caption{Phase diagram of UGe$_2$ in the temperature-pressure plane. Shown are the paramagnetic (PM) phase,
              two ferromagnetic phases (FM1 and FM2), and the tricritical point (TCP). The critical point marked CEP $^{\ref{CEP_footnote}}$ is related to the transition between the phases FM1 and FM2. From \textcite{Taufour_et_al_2010}.}
\label{figure:UGe_2_p-T_phase_diagram}
\end{figure}

UGe$_2$ has received much attention since superconductivity coexists with ferromagnetism in part of the ordered 
phase \cite{Saxena_et_al_2000}. It crystallizes in an inversion-symmetric orthorhombic structure, and the best samples
have been reported to have residual resistivities as low as $0.2\,\mu\Omega$cm \cite{Saxena_et_al_2000}. \textcite{Taufour_et_al_2010} found the residual resistivity to be strongly pressure dependent. The Curie temperature at 
ambient pressure is \TC\ $\approx 52\,$K \cite{Saxena_et_al_2000, Huxley_et_al_2001, Aoki_et_al_2001, 
Aoki_Flouquet_2012}. \TC\ decreases with increasing hydrostatic pressure and vanishes at $p \approx 16$ kbar, 
which coincides with the pressure where the superconductivity disappears. Within the ferromagnetic phase a further transition is observed, across which the magnitude of the magnetic moment changes discontinuously. The associated transition line starts near the peak in the superconducting transition temperature, ends in a critical point at a temperature of about 4\,K, and is replaced by a crossover at higher temperatures
\cite{Huxley_et_al_2007, Taufour_et_al_2010}, see Fig.~\ref{figure:UGe_2_p-T_phase_diagram}.
The tricritical temperature has been measured to be $T_{\text{tc}} \approx 24\,$K \cite{Taufour_et_al_2010, Kotegawa_et_al_2011b},
but values as high as $T_{\text{tc}} \approx 31\,$K have been reported \cite{Huxley_et_al_2007} with a tricritical pressure
$p_{\text{tc}} \approx 13$ kbar. \textcite{Kabeya_et_al_2010} found a somewhat smaller value of $\approx 12.5$~kbar from
measurements of the linear thermal expansion coefficient.
The tricritical wings have been mapped out in detail, see Fig.~\ref{figure:UGe_2_phase_diagram}.

\paragraph{U$_3$P$_4$}
\label{U3P4}

U$_3$P$_4$ at ambient pressure is a ferromagnet with \TC\ = 138\,K~\cite{Trzebiatowski_Troc_1963}. It crystallizes in a bcc
structure with no inversion symmetry, and the magnetic structure is canted with a FM component along 
$\langle 111\rangle$~\cite{Burlet_et_al_1981, Wisniewski_Gukasov_Henkie_1999, Heimbrecht_Zumbusch_Biltz_1941, Zumbusch_1941}.
Pressure reduces \TC\ until a QPT is reached at $p_{\text{c}} \approx 4\,$GPa. From measurements of the resistivity and the magnetic susceptibility at $p \approx 1.5\,$GPa,~\textcite{Araki_et_al_2015} concluded that the transition changes from second order to first order with a tricritical temperature $T_{\text{tc}} = 32\,$K. Consistent with this, the pressure-dependence of \TC\ changes from a Hertz-type \TC\ $\propto (p - p_{\text{c}})^{3/4}$ behavior to \TC\ $\propto (p - p_{\text{c}})^{1/2}$. In a magnetic field, metamagnetic behavior has been observed that is indicative of tricritical wings, although the wings have not been mapped out.

\paragraph{URhGe and UCoGe}
\label{par:II.B.2.b}

Both of these materials belong to the ternary UTX intermetallic U-compounds where T is one of the late transition metals and 
X a $p$-electron element. They crystallize in the orthorhombic TiNiSi structure (space group $P_{nma}$). For lattice parameters, see \textcite{Troc_Tran_1988} and \textcite{Canepa_et_al_1996}. Because the $5f$-electrons, which carry the magnetic moments, are partially delocalized in these materials, the ordered moment is often reduced compared to the free ion value and an enhanced electronic specific heat is observed. In addition, they are characterized by a strong Ising anisotropy \cite{Sechovsky_Havela_1998}. Two main mechanisms control the delocalization of the $5f$ electrons and thus 
the magnetism: the direct overlap of neighbouring U $5f$ orbitals and the hybridization of those with the $d$-electrons. For 
inter-U distances smaller than the so-called Hill limit ($d_{U-U} \approx 3.4 - 3.6$\,\AA)~\cite{Hill_1970} the strong direct overlap of the $5f$ orbitals results in a non-magnetic ground state. Larger values yield a FM or AFM ordered ground state. For values close to this limit the $f-d$ hybridization strength controls the magnetic properties. There is a clear tendency of these systems to show magnetic order with increasing $d$-electron filling of the T element~\cite{Sechovsky_Havela_1998}. The strongest electronic correlations are therefore found in UTX compounds with intermediate values of $d_{U-U}$ and 
$d$-electron filling.
\begin{figure}[b]
\vskip 0pt
\begin{center}
\includegraphics[width=0.85\columnwidth,angle=0]{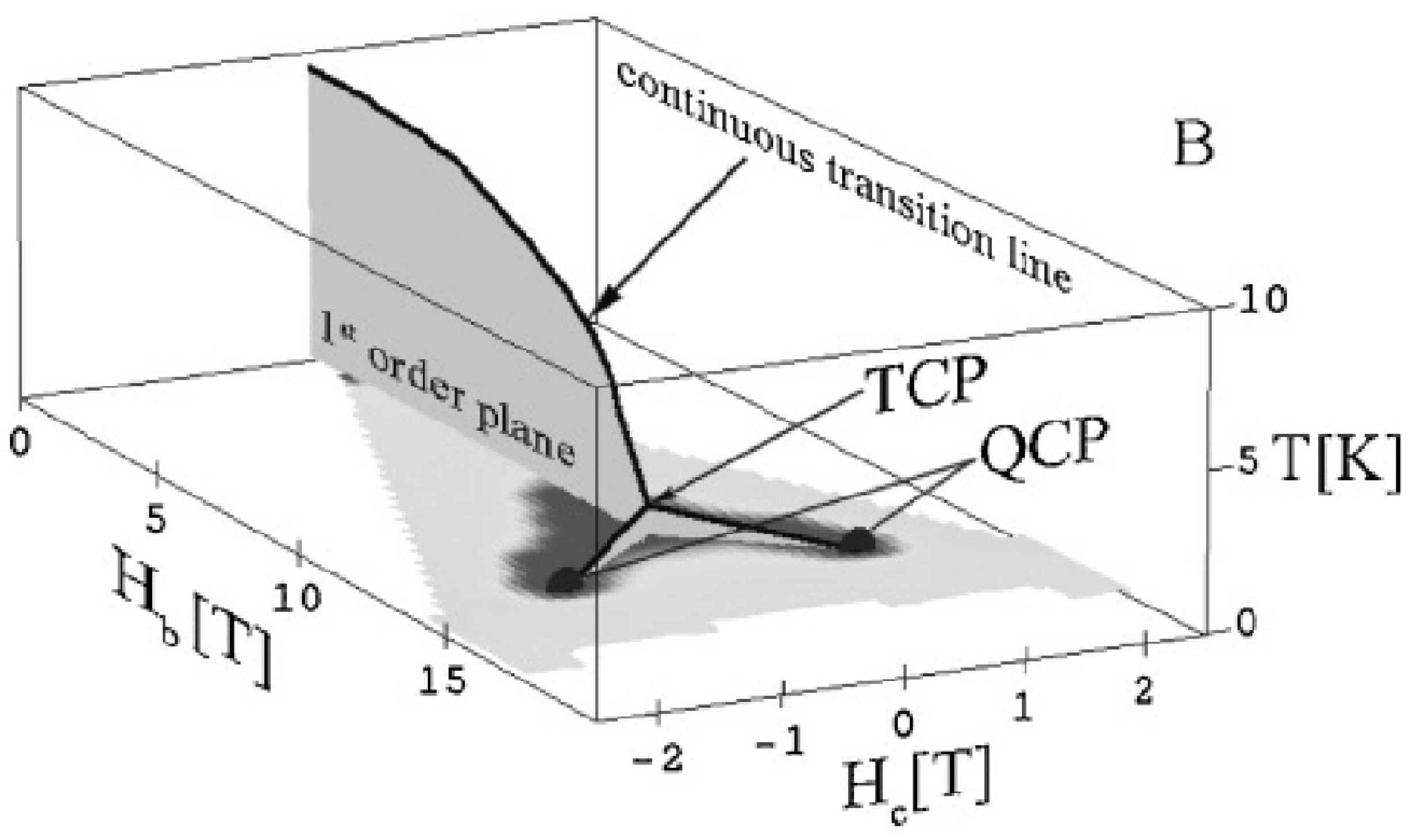}
\end{center}
\vskip -10pt
\caption{Phase diagram of URhGe in the space spanned by temperature and magnetic fields in the b- and c-directions. 
              The dark shaded regions indicate the presence of superconductivity. From \textcite{Huxley_et_al_2007}.}
\label{fig:URhGe_phase_diagram}
\end{figure}

URhGe has a $d_{U-U} = 3.5$\,\AA\ close to the Hill limit. It is ferromagnetic with a Curie temperature \TC\ $= 9.5\,$K
and an ordered moment of 0.42\,\mub, oriented along the c-axis. It was the second U-based compound (after UGe$_2$) for
which coexistence of superconductivity and ferromagnetism was found in high-quality samples \cite{Aoki_et_al_2001}. It
is unique in that a magnetic field parallel to the b-axis suppresses \TC\ and leads to a tricritical point at $T\approx 1\,$K and
$H_b \approx 12\,$T \cite{Huxley_et_al_2007}. With an additional field in the c-direction, tricritical wings appear, see Fig.~\ref{fig:URhGe_phase_diagram}. 
The superconductivity is absent at intermediate fields, but reappears at low
temperatures in the vicinity of the tricritical wings \cite{Levy_et_al_2005, Huxley_et_al_2007}. 
\begin{figure}[b]
\vskip 0pt
\begin{center}
\includegraphics[width=0.85\columnwidth,angle=0]{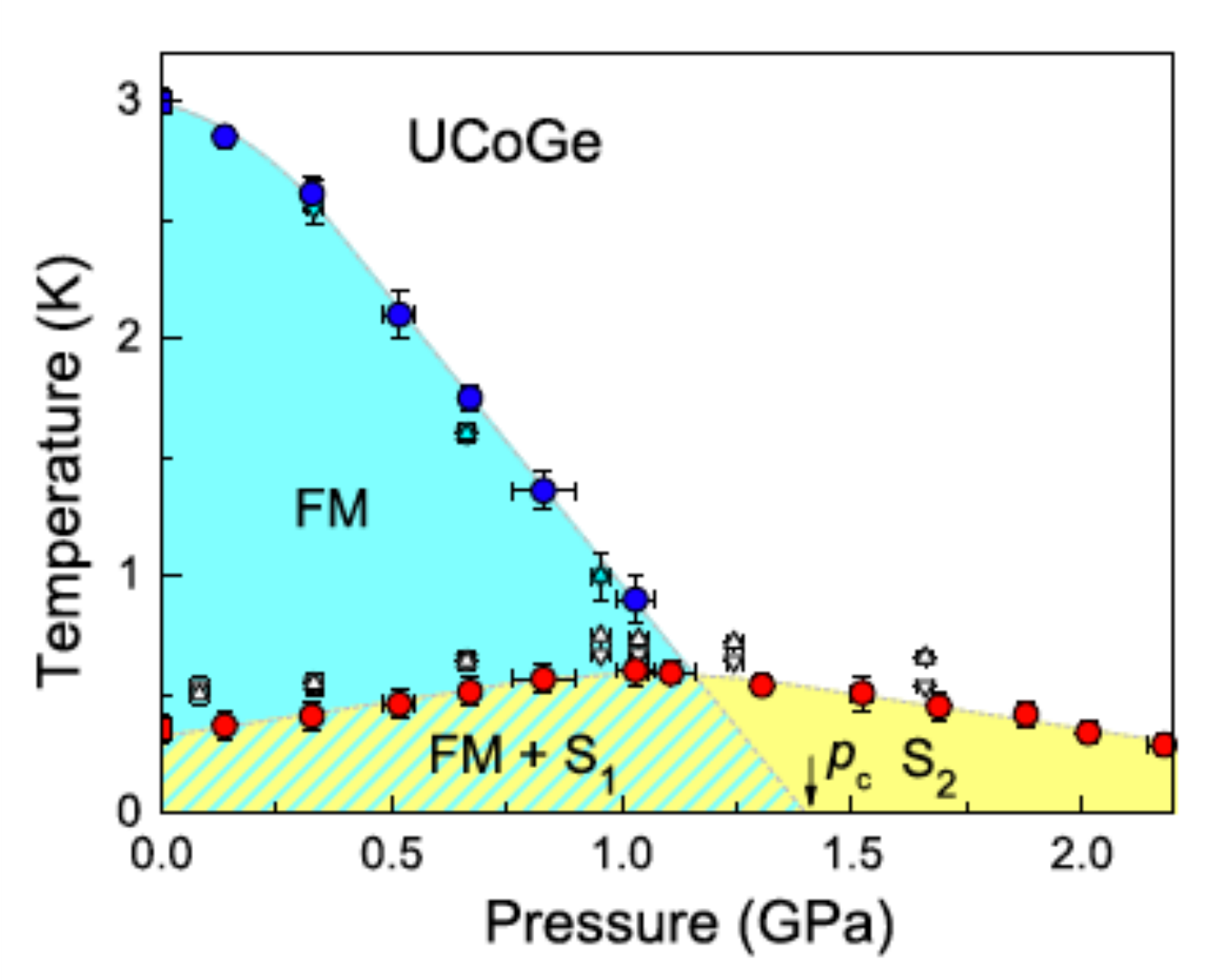}
\end{center}
\vskip -10pt
\caption{Phase diagram of UCoGe in the temperature-pressure plane, showing the ferromagnetic and superconducting
              phases. The magnetic transition is first order for all pressure values \cite{Hattori_et_al_2010}.
After~\textcite{Slooten_et_al_2009}.}
\label{fig:UCoGe_T_p_phase_diagram}
\end{figure}
\begin{figure}[t,b]
\vskip 0pt
\begin{center}
\includegraphics[width=0.85\columnwidth,angle=0]{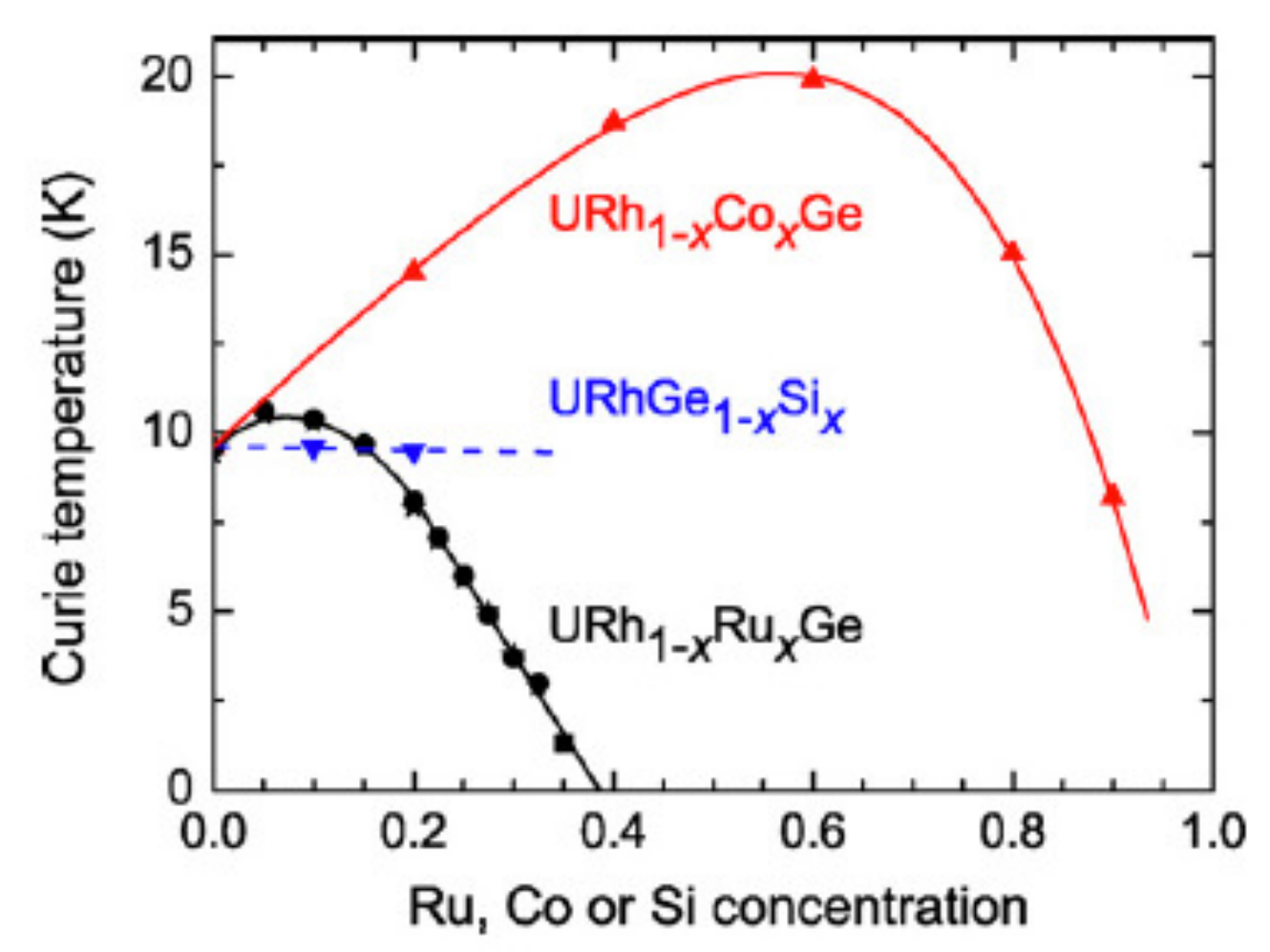}
\end{center}
\vskip -10pt
\caption{Curie temperature of URhGe doped with Co, Si, or Rh as a function of the dopant concentration. The transition in pure
              URhGe is 2nd order, in pure UCoGe, 1st order. From \textcite{Sakarya_et_al_2008}.}
\label{fig:URhGe_doped_phase_diagram}
\end{figure}

The nature of the magnetic order in UCoGe, ferromagnetic or otherwise, was initially unclear. This, together with the observation that URhGe is
ferromagnetic, prompted the study of URh$_{1-x}$Co$_x$Ge alloys \cite{Sakarya_et_al_2008}, and the final conclusion 
was that UCoGe is indeed a weak ferromagnet with a Curie temperature near 3\,K and a small ordered moment of 
0.03\,\mub\ \cite{Huy_et_al_2007b}. The transition was found to be weakly first order by means of nuclear quadrupole resonance measurements \cite{Hattori_et_al_2010}. Hydrostatic pressure decreases \TC\ \cite{Hassinger_et_al_2008,Slooten_et_al_2009} which vanishes near the maximum of the superconducting dome, see Fig.~\ref{fig:UCoGe_T_p_phase_diagram}. A tricritical point must appear as \TC\ increases upon doping with Rh, see Fig.~\ref{fig:URhGe_doped_phase_diagram}, but the order of the transition has not been studied as a function of the Rh concentration. Similarly, in pure UCoGe tricritical wings should appear in a magnetic field, analogously to what is observed in UCoAl, see Fig.~\ref{fig:UCoAl_phase_diagram}. A recent study has reported that \TC\ is suppressed by doping with Ru, with an extrapolated critical Ru concentration of about 31\% \cite{Valiska_et_al_2014}. The order of the transition has not been determined.

UCoGe displays coexistence of superconductivity and ferromagnetism below 0.8\,K \cite{Huy_et_al_2007b, Slooten_et_al_2009}. 
In contrast to both UGe$_2$ and URhGe the superconductivity is observed in both the ferromagnetic and paramagnetic phases,
see Fig.~\ref{fig:UCoGe_T_p_phase_diagram}.

\paragraph{UCoAl}
\label{par:II.B.2.d}

\begin{figure}[b]
\begin{center}
\includegraphics[width=0.95\columnwidth,angle=0]{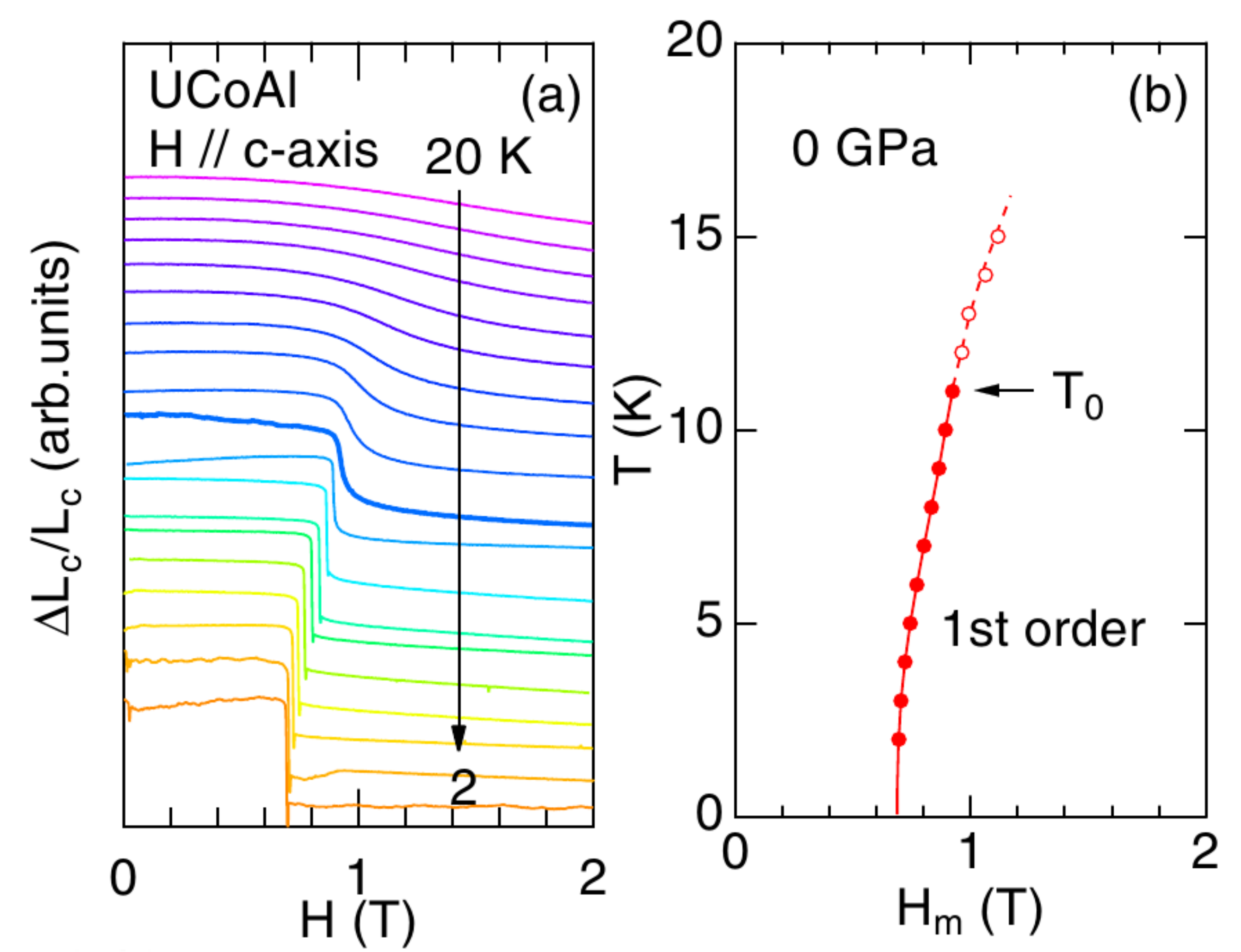}
\end{center}
\caption{a) Magnetostriction measured along the $c$-axis with $H \parallel c$ for temperatures between 2 and 21\,K 
              (every 1\,K). b) Temperature vs field evolution of the metamagnetic transition which changes from first order to a crossover for 
              $T > T_{0} = 11$\,K. From~\textcite{Aoki_et_al_2011a}.}
\label{Aoki_et_al_2011b}
\end{figure}
At ambient pressure and zero field, UCoAl is a paramagnet with a strong uniaxial magnetic anisotropy \cite{Sechovsky_et_al_1986}. 
It crystallizes in the hexagonal ZrNiAl structure consisting of U-Co and Co-Al layers that alternate along the c-axis. 
The inter-U distance is $d_{U-U} \approx 3.5$\,\AA\ (same value as in URhGe, see \ref{par:II.B.2.b}), but a large $d$-filling leads to 
UCoAl being paramagnetic~\cite{Sechovsky_Havela_1998}. Its isoelectronic analog URhAl is ferromagnetic with 
$d_{U-U} \approx 3.63$\,\AA,  (cf. Sec.~\ref{par:II.B.2.e}).
These observations suggest that UCoAl is close to a FM instability, which is indeed the case: Application of a magnetic field 
along the easy magnetization axis (the crystallographic $c$-axis) induces a first-order metamagnetic phase transition at 
$H_{m} \approx 0.7$\,T at low temperature with an induced moment of about 
0.3\,$\mu_{\textrm{B}}$~\cite{Andreev_et_al_1985, Mushnikov_et_al_1999}. Moreover, uniaxial stress induces 
ferromagnetism~\cite{Ishii_et_al_2003, Shimizu_et_al_2015b}. The susceptibility shows Curie-Weiss behavior for $T > 40$\,K with a fluctuating moment 
of about 1.6\,$\mu_{\textrm{B}}$, much larger than the induced moment of 0.3\,$\mu_{\textrm{B}}$~\cite{Havela_et_al_1997}. 
The magnetism is believed to be itinerant with the U $5f$-electrons providing the main contribution \cite{Eriksson_Johansson_Brooks_1989,
Wulff_et_al_1990, Mushnikov_et_al_1999}; polarized-neutron diffraction experiments have found  the magnetic moment 
exclusively at the U sites with the orbital moment being twice as large as (and antiparallel to) the spin
 moment~\cite{Wulff_et_al_1990,Javorsky_et_al_2001}.

\begin{figure}[t]
\begin{center}
\includegraphics[width=0.95\columnwidth,angle=0]{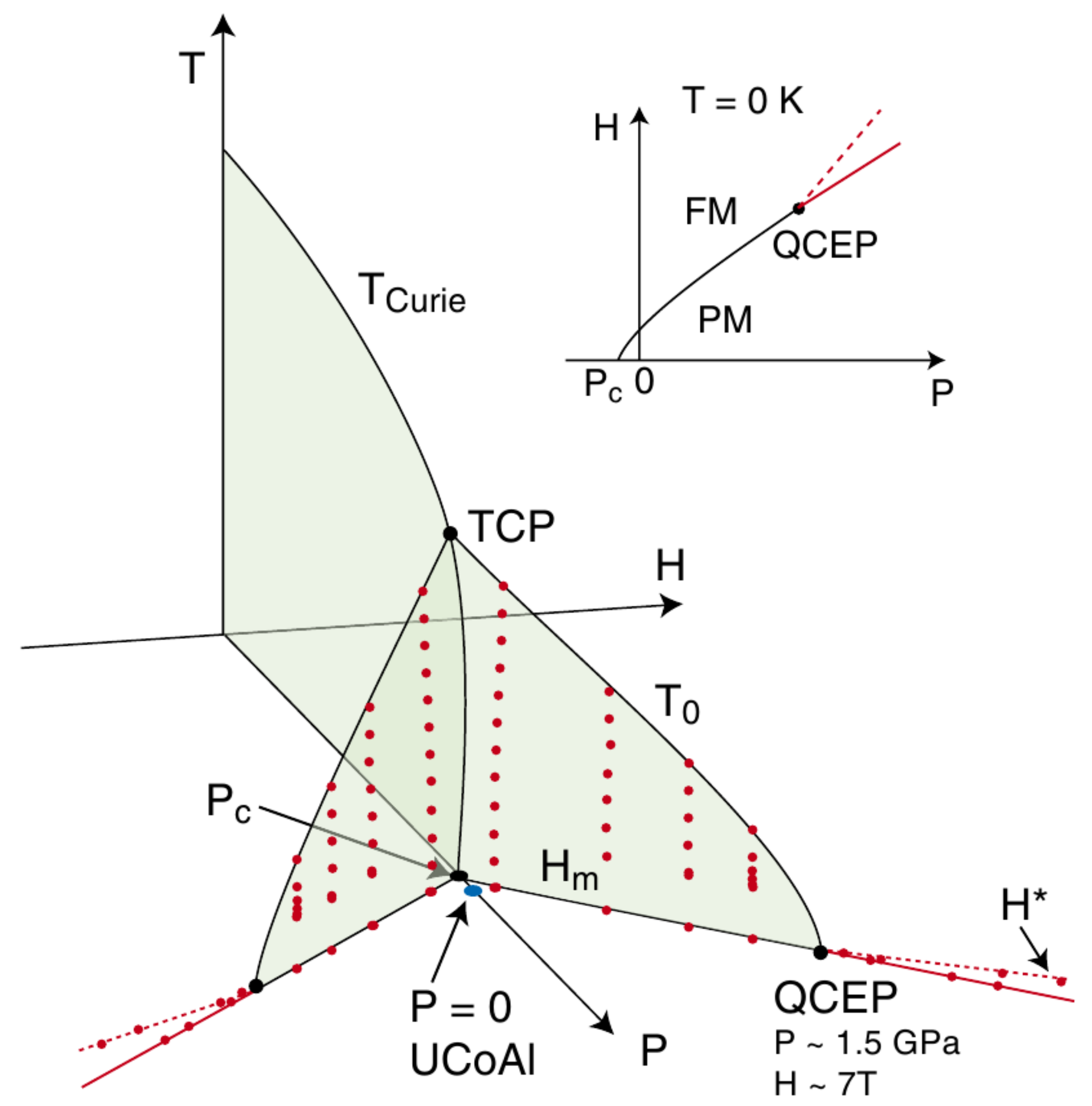}
\end{center}
\caption{Semi-schematic $T$-$P$-$H$ phase diagram with $H \parallel c$ showing the existence of tricritical wings in UCoAl. 
              The wings are determined by the observation of a first-order metamagnetic transition at $H_m$ (red dots); they are bounded by lines of second-order transitions at $T_0$ and end in quantum ``critical end points'' (QCEPs)$^{\ref{CEP_footnote}}$. The critical pressure $P_{\text{c}}$ is negative and the tricritical point (TCP) is not accessible. See the text for more information.  From~\textcite{Aoki_et_al_2011a}. 
}
\label{fig:UCoAl_phase_diagram}
\end{figure}
Studies of the magnetostriction, magnetoresistivity~\cite{Aoki_et_al_2011a}, nuclear magnetic resonance~\cite{Karube_et_al_2012} and thermopower~\cite{Palacio-Morales_et_al_2013} indicate that the field-induced first-order transition terminates in a critical point at a temperature $T_{0} = 11$\,K at ambient pressure, as illustrated in Fig.~\ref{Aoki_et_al_2011b}: $\Delta L(H)/L$ shows a step-like jump at $H_{m}$ for $T < T_{0}$ which becomes smooth for $T \ge T_{0}$. A determination of critical exponents suggests that the transition at $T_0$ is in the three-dimensional Ising universality class~\cite{Karube_et_al_2012}. $H_m$ increases with pressure and each wing terminates in a quantum critical point (denoted by QCEP in the figure)$^{\ref{CEP_footnote}}$ at $P \approx 1.5$\,GPa and $\mu_{0}H \approx 7$\,T. At the wing-tip point a pronounced enhancement of the effective mass (derived from the coefficient of the $T^{2}$ term in the electrical resistivity) is observed~\cite{Aoki_et_al_2011a}. 

The resulting $T$-$P$-$H$ semi-schematic phase diagram is shown in Fig.~\ref{fig:UCoAl_phase_diagram}, which demonstrates the 
presence of tricritical wings in UCoAl. The red dots represent the experimental values for $H_{m}$ determined by magnetoresistivity 
(with $J \perp H$) and magnetostriction measurements. Since $T_{0} = 11$\,K at the ambient pressure, the tricritical point (TCP) must 
be located at $T > 11$\,K. At pressures higher than 1.5\,GPa, the first-order character of the metamagnetic transition disappears and 
new features in the form of kinks in the magnetoresistivity and Hall effect are observed at $H_{m}$ and $H^{*}$~\cite{Combier_et_al_2013}. 
Very recent investigations of the transverse and longitudinal resistivities and of the magnetization under pressure~\cite{Combier_2013} point to a much richer 
phase diagram, where the exact location of the QCEP remains uncertain, with possible changes of the Fermi surface as well as the appearance of new phases around the QCEP.

The substitution of Fe for Co was found to lead to FM ground state in zero field and ambient pressure by \textcite{Karube_et_al_2015}.
By nuclear quadrupole resonance measurements these authors found a first-order transition in U(Co$_{1-x}$Fe$_x$)Al with a transition
temperature of about 10\,K and about 17\,K for $x=0.1$ and $x=0.02$, respectively.

\paragraph{URhAl}
\label{par:II.B.2.e}

URhAl belongs to the same UTX compound family as URhGe, UCoGe, and UCoAl. 
It has the same layered hexagonal ZrNiAl-type crystal structure as UCoAl, but with $d_{U-U} = 3.63$\,\AA, larger than the 
Hill limit (cf. Sec. \ref{par:II.B.2.b}).
\begin{figure}[t]
\begin{center}
\includegraphics[width=0.95\columnwidth,angle=0]{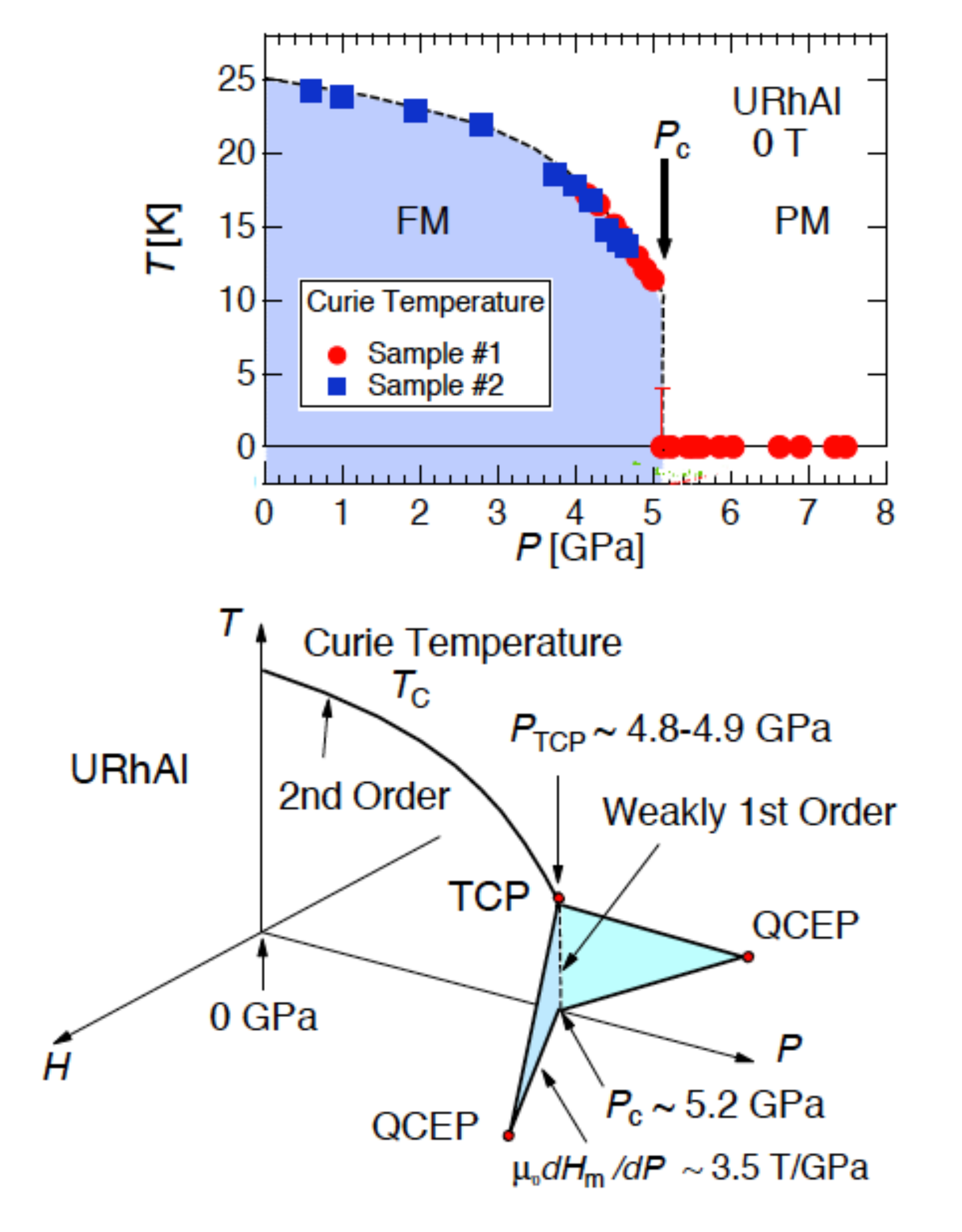}
\end{center}
\caption{Upper panel: Temperature-pressure phase diagram of URhAl in zero field determined from resistivity measurements. Lower panel: 
              Temperature-pressure-field diagram inferred from metamagnetic behavior observed in an external field. From \textcite{Shimizu_et_al_2015a}.}
\label{fig:URhAl_phase_diagram}
\end{figure}
Consistent with this, and contrary to UCoAl which has a nonmagnetic ground state, URhAl orders ferromagnetically via a second-order transition. Values of the Curie temperature between \TC\, = 27\,K and \TC\,= 34\,K have been reported, with strong Ising-like ordered moments of 0.9\,$\mu_{\textrm{B}}$/U along the $c$-axis~\cite{Veenhuizen_el_al_1988, Combier_2013, Shimizu_et_al_2015a}. 

The itinerant vs localized nature of magnetism in URhAl is controversial, as it is in many other UTX compounds. A peak at 380\,meV in inelastic neutron scattering 
experiments~\cite{Hiess_et_al_1997} was interpreted as indication of an intermultiplet transition, suggesting $5f$-electron localization. X-ray magnetic circular 
dichroism (XMCD) experiments also indicate a high degree of localization of the $5f$-orbitals~\cite{Grange_et_al_1998}. On the other hand, polarized neutron 
studies point to a rather strong delocalization of the $5f$ electrons~\cite{Paixao_et_al_1992}. Moreover, band structure calculations based on an itinerant approach 
can reproduce quite well most of the experimental findings, i.e. magneto-optical Kerr effect~\cite{Kucera_et_al_1998}, equilibrium volume, bulk modulus, 
magnetocrystalline anisotropy energy, magnetocrystalline anisotropy of the U moments and the shape of the XMCD lines~\cite{Kunes_et_al_2001}.

Pressure experiments were performed on a rather clean single crystal with a RRR $\approx 14$ and \TC\, = 28\,K~\cite{Combier_2013}.  At ambient 
pressure the phase transition is mean-field-like characterized by a single peak in $C/T$ and a kink in the thermal expansion ratio $\Delta L/L$. The magnetization 
with $H \parallel c$ shows a clear hysteresis at 2\,K with a remanent magnetization of 0.9\,$\mu_{\textrm{B}}$/U. 
%This sample was investigated under hydrostatic pressure by AC-calorimetry. Measurements down to 2\,K and up to 5.8\,GPa tracked the FM phase transition under pressure; the resulting phase diagram is shown in Fig.~\ref{fig:Combier_2013_Fig_4.13}. The nature of the transition at very low temperature is not clear yet. In the calorimetry data the signature at \TC\, was visible for pressures up to 5.1\,GPa and did not show any change in shape when approaching the quantum phase transition at $P_{\text{c}} \approx 5.2$\,GPa, suggesting that the transition remains second order. Also, around $P_{\text{c}}$ the Sommerfeld coefficient at 3\,K increases, possibly indicating strong spin fluctuations. However, the fact that the ground state of URhAl just above $P_{\text{c}}$ is similar to that of UCoAl at zero pressure suggests a first-order transition at the lowest temperatures and the existence of a tricritical point \cite{Combier_2013}. Future experiments in a magnetic field, and at temperatures below 1\,K, are 
desirable.
Recent transport experiments on moderately disordered samples ($\rho_0 \approx  65\,\mu\Omega$cm near the transition) have mapped out the phase 
diagram in more detail \cite{Shimizu_et_al_2015a, Shimizu_et_al_2015c}. The QPT at a critical pressure $p_c \approx 5.2$\,GPa
is weakly first order, and metamagnetic signatures in an applied magnetic field imply the existence of tricritical wings. Due to the weakly first-order nature of the
transition strong spin fluctuations are still observable in the behavior of the electrical resistivity and the specific heat. 

\subsubsection{Lanthanide-based compounds}
\label{subsubsec:II.B.3}

\paragraph{La$_{1-x}$Ce$_x$In$_2$}
\label{par:II.B.3.LaCeIn}

CeIn$_2$ crystallizes in the orthorhombic CeCu$_2$ structure and undergoes a first-order transition to a ferromagnetic state
at a Curie temperature \TC\,= 22\,K \cite{Rojas_et_al_2009}. This conclusion on the basis of discontinuities at \TC\, in the
resistivity, the thermal expansion, and the magnetic entropy was later corroborated by $\mu$SR measurements \cite{Rojas_et_al_2011}.
Application of hydrostatic pressure increases \TC\, \cite{Mukherjee_Iyer_Sampathkumaran_2012}, but
upon doping with lanthanum \TC\, decreases, to about 19.4\,K in La$_{1-x}$Ce$_x$In$_2$ with $x = 0.9$, and the
transition remains first order \cite{Rojas_et_al_2011}. The same $\mu$SR measurements indicated the existence of a second magnetic 
phase with long-range order in between the FM and PM phases. The nature of this phase is not known. Doping with Ni decreases
\TC\, sharply, and the transition in Ce(In$_{1-x}$Ni$_x$)$_2$ has been reported to be second order to a FM for $x = 0.025$, $0.05$, and 
$0.15$ \cite{Rojas_Espeso_Fernandez_2013}. However, an earlier experiment by \textcite{Sung_et_al_2009} concluded that the 
ground state for $x = 0.15$ is AFM.

\paragraph{SmNiC$_2$}
\label{par:SmNiC}

The ferromagnetic charge-density-wave (CDW) compound SmNiC$_2$ has a \TC\ of about 17\,K which is weakly susceptible to 
pressure \cite{Woo_et_al_2013}. The polycrystalline samples measured are rather clean, with a residual resistivity of less than 
2\,$\mu\Omega$cm for pressures below about 3\,GPa.
The PM-FM transition is first order and remains first order as the pressure is increased from zero to 2\,GPa, with \TC\ dropping to 15\,K. At higher
pressure, there is a second or weakly first-order transition from the FM to a phase of unclear nature, and at least two other phases appear at low
temperature. Since the nonmagnetic phase in this material is a CDW state below $T\approx 150\,$K, the phase diagram may fall outside 
the classification provided by Fig.~\ref{fig:schematic_phase_diagrams} and the first-order transition may be of different origin than in other materials, 
see Sec.~\ref{subsec:III.F}. If the theory discussed in Sec.~\ref{subsubsec:III.B.2} applies, one expects a tricritical point at negative pressure. In this 
case a metamagnetic transition corresponding to the tricritical wings should be visible in an applied magnetic field, provided the zero-pressure line is not already past the wing tips.

 \paragraph{Yb-based systems}
\label{par:II.B.3.YbXSi}

YbCu$_2$Si$_2$ crystallizes in the body-centered ThCr$_2$Si$_2$ structure and does not order magnetically at ambient pressure. 
A transition to a magnetically ordered phase under pressure was suggested early on the basis of transport measurements 
\cite{Alami-Yadri_Jaccard_1996,Alami-Yadri_Wilhelm_Jaccard_1998}, and later 
confirmed by means of M{\"o}ssbauer data \cite{Winkelmann_et_al_1999}. \textcite{Fernandez-Panella_et_al_2011} concluded from susceptibility measurements that 
the nature of the order is FM, and the transition is likely first order \cite{Winkelmann_et_al_1999, Colombier_et_al_2009, Fernandez-Panella_et_al_2011}. The FM 
nature of the ordered phase was confirmed by \textcite{Tateiwa_et_al_2014}, who also found evidence for phase separation indicative of a first-order transition. 

YbIr$_2$Si$_2$ crystallizes in either the ThCr$_2$Si$_2$ structure, or the $P$-type CaBe$_2$Ge$_2$ structure, depending on the synthesis conditions 
\cite{Hossain_et_al_2005}. The former is magnetically (presumably AFM) ordered below 0.7\,K, whereas the latter is a paramagnet at ambient pressure. 
\textcite{Yuan_et_al_2006} found that by applying pressure the system in its $P$-type structure undergoes a first-order QPT to an ordered phase at a critical 
pressure $p_{\text{c}} \approx 8\,$GPa. The nature of the ordered phase is suspected to be FM, but additional investigations are needed. Recent measurements 
of the resistivity under hydrostatic pressure as high as 15\,GPa found NFL behavior in a rather broad range in pressure, $3 \leq p \leq 8$\,GPa, and confirmed the 
sudden appearence of magnetic ordering at 8.3\,GPa suggestig a first-order QPT~\cite{Macovei_2010}. The transition temperature shifts to higher values and shows 
a weak maximum around 11\,GPa, a behavior very similar to that of \YRS\, under pressure~\cite{Mederle_et_al_2001,Knebel_et_al_2006}. \YRS\, evolves from an AFM to a FM ordered state under chemical pressure (Co substitution)~\cite{Lausberg_et_al_2013} and possibly even under hydrostatic 
pressure~\cite{Knebel_et_al_2006}. This suggests that the nature of the magnetic ordered phase in YbIr$_2$Si$_2$ could also be AFM, but more investigations are 
needed.

\paragraph{\rm{CePt}}
\label{par:II.B.3.CePt}
CePt under pressure has also been reported to display a ferromagnetic quantum phase transition at $p_{\textrm{c}} \approx 12.1\,$GPa~\cite{Larrea_et_al_2005}. The phase transition at $p = 0$ is second order~\cite{Holt_et_al_1981}. No magnetization measurements have been performed under pressure. The FM signature is strongly weakened under pressure well before $p_{\textrm{c}}$ and transport experiments indicate a sudden drop of the phase boundary line close to $p_{\textrm{c}}$ suggesting the presence of a first-order transition. 
\subsubsection{Strontium Ruthenates}
\label{subsubsec:SrRuO}

The perovskite ruthenates, which include Sr$_2$RuO$_4$ and Sr$_4$Ru$_3$O$_{10}$ in addition to SrRuO$_3$ and Sr$_3$Ru$_2$O$_7$, belong to a class of 
materials known as the 
Ruddlesden-Popper series; for a historical overview, see~\textcite{Mackenzie_Grigera_2004}. In SrRuO$_3$ a QPT can be triggered by means of doping
with Calcium, whereas the phase diagram of Sr$_3$Ru$_2$O$_7$ has been explored by applying pressure and an external magnetic field. In \SCRO\ a
variety of very different behaviors has been observed, which is likely due to different sample preparation methods (bulk ceramic, bulk powder, and thin films).
We therefore discuss this material both in the present section and in Secs.\ \ref{subsubsec:II.C.1} and \ref{subsubsec:II.E.3}. 

\paragraph{\SCRO\ (bulk ceramic samples)}
\label{SCRO_ceramic}

\begin{figure}[b]
\begin{center}
\includegraphics[width=0.65\columnwidth,angle=0]{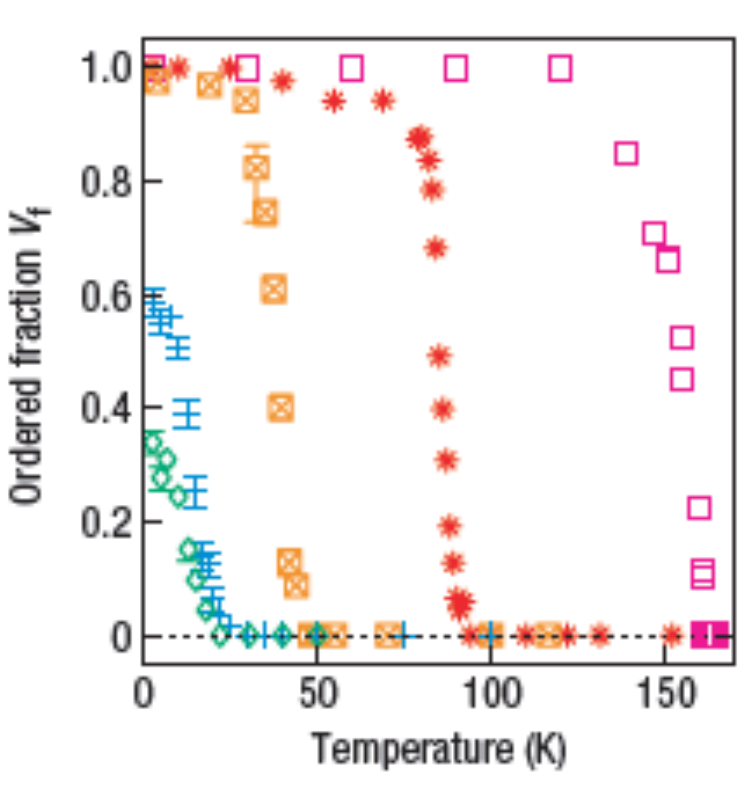}
\end{center}
\caption{$\mu$SR results for the volume fraction with static magnetic order for ceramic \SCRO\ samples. The symbols from right (open squares) to left (open
              diamonds) correspond to Ca concentrations $x = 0$, $0.3$, $0.5$, $0.65$, and $0.7$, respectively. Note the similarity with the corresponding results
              for MnSi in Fig.~\ref{figure:MnSi_volume_fraction}. From \textcite{Uemura_et_al_2007}.}
\label{fig:SCRO_volume_fraction}
\end{figure}

\SCRO\, is a metallic system that crystallizes in an orthorhombically distorted perovskite structure. SrRuO$_{3}$ is an itinerant ferromagnet with a second-order transition 
at \TC\,$\approx 160$\,K~\cite{Kim_et_al_2003} and an ordered moment of about 1\,\mub/Rh, while CaRuO$_{3}$ is a strongly exchange-enhanced Pauli paramagnet 
with no sign of metamagnetism and a Fermi-liquid ground state with an anomalously low coherence scale~\cite{Schneider_et_al_2014}. Long-range FM order
disappears for a Ca concentration around $x_c \approx 0.7$, and NMR experiments established the presence of FM spin fluctuations for all concentrations,
the Curie-Weiss behavior of the susceptibility with a Weiss temperature that changes sign at $x_c$ notwithstanding \cite{Yoshimura_et_al_1999}. This, and the
large effective moment (compared to the ordered one) of about 3\,\mub/Ru seemed to make \SCRO\  a good candidate for the SCR theory of itinerant ferromagnetism 
(cf. Sec.~\ref{subsec:I.B}). However, a $\mu$SR study by \textcite{Uemura_et_al_2007} of ceramic samples with $x = 0.65$ and $x=0.7$ found a finite 
volume fraction of ferromagnetic order, see Fig.~\ref{fig:SCRO_volume_fraction}, and a suppression of the critical dynamics visible at smaller values of $x$.
These results are very similar to the corresponding ones in MnSi (Fig.~\ref{figure:MnSi_volume_fraction}) and
are indicative of a first-order transition. No information about the disorder strength in these samples is available. For bulk powder samples and epitaxial 
thin films of the same material rather different results have been obtained, see Secs.\ \ref{SCRO_powder} and \ref{subsubsec:II.E.3}, respectively. 
\paragraph{Sr$_3$Ru$_2$O$_7$}
\SROO\ is not a simple ferromagnet, and its properties are incompletely understood. Here we discuss some aspects that are analogous to the properties of simpler quantum ferromagnets.

\begin{figure}[t]
\begin{center}
\includegraphics[width=0.95\columnwidth,angle=0]{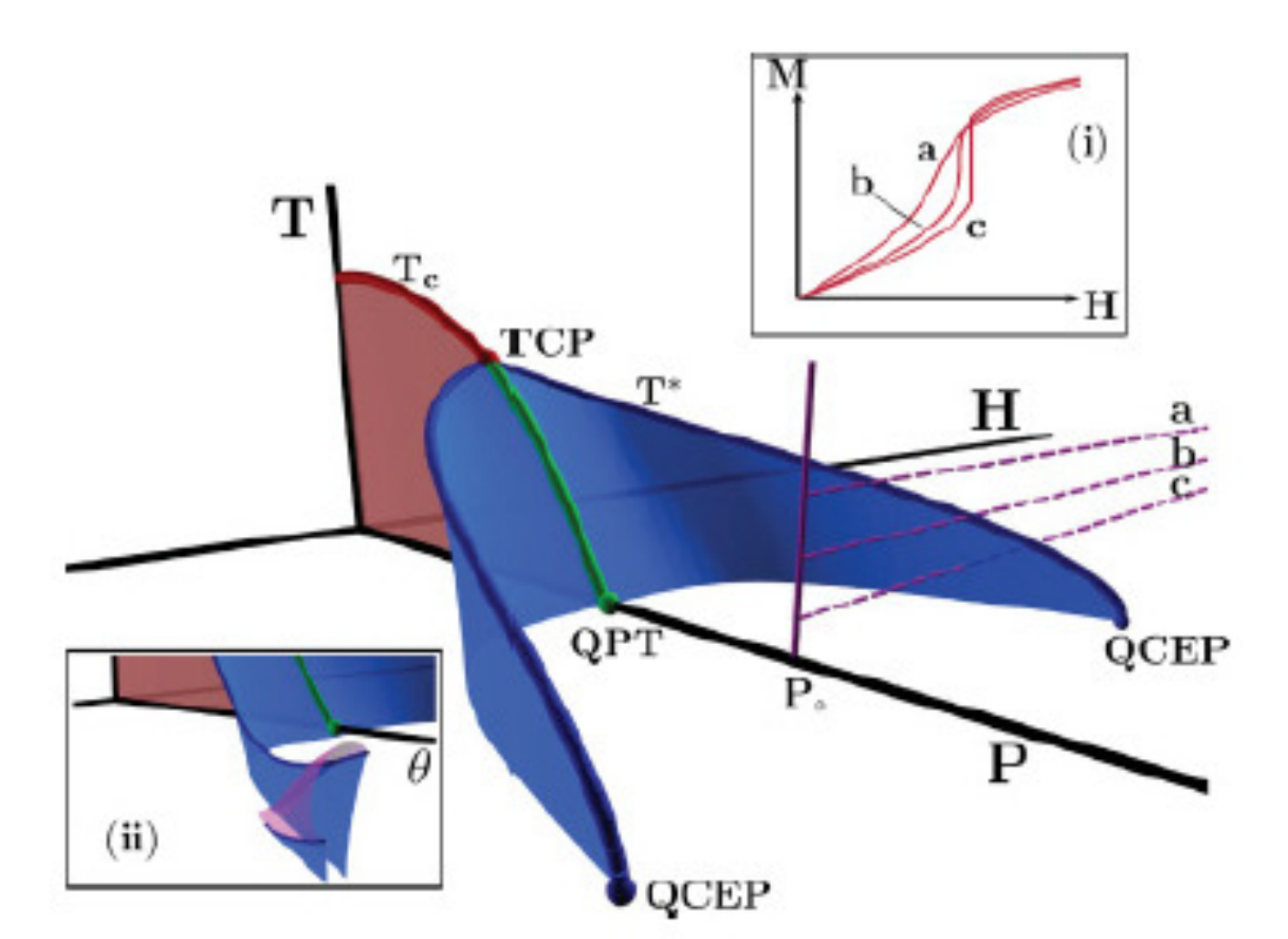}
\end{center}
\caption{Schematic phase diagram of \SROO\ in the space spanned by temperature ($T$), hydrostatic pressure ($P$), and magnetic field ($H$), with 
              $H \perp c$. Ambient pressure ($P_0$) is indicated by the solid purple line. The tricritical point (TCP) is not accessible, but the tricritical wings are observed by 
              sweeping the magnetic field at fixed temperature (dashed purple lines and inset (i)). The critical temperature $T^*$ can be tuned by rotating the magnetic field 
              by an angle $\theta$ out of the magnetically easy $ab$-plane. As $T^*$ decreases, the wing tips split and a phase displaying a strong transport anisotropy is 
              found in between two sheets, with a second-order phase boundary on top (inset (ii) and Fig.~\ref{fig:SrRuO_phase_diagram_2}).
              From~\textcite{Wu_et_al_2011}.}
\label{fig:SrRuO_phase_diagram}
\end{figure}

Very pure samples of Sr$_3$Ru$_2$O$_7$ have been prepared, with residual resistivities of less than 0.25\,$\mu\Omega$cm~\cite{Perry_Maeno_2004}. The ground state in zero field and at ambient pressure is paramagnetic close to a ferromagnetic instability~\cite{Ikeda_et_al_2000}. This places the system in the generic phase diagram 
of Fig.~\ref{fig:schematic_phase_diagrams} a) between the tricritical wings (see Fig.~\ref{fig:SrRuO_phase_diagram}), as is the case for UCoAl, 
Fig.~\ref{fig:UCoAl_phase_diagram}.
\begin{figure}[h,t]
\begin{center}
\includegraphics[width=0.75\columnwidth,angle=0]{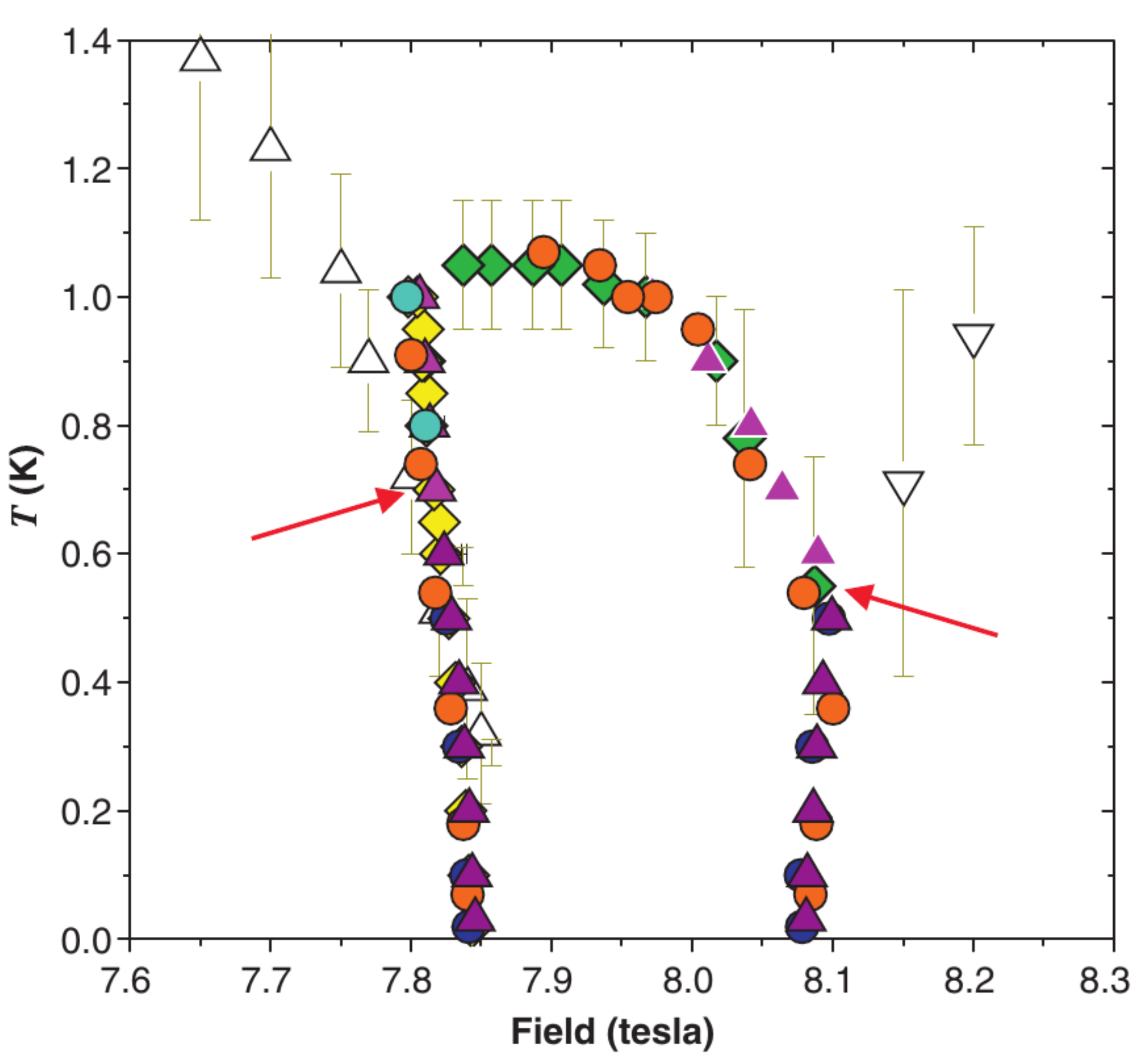}
\end{center}
\caption{$T$-$H$ phase diagram of \SROO\ at ambient pressure with $H \parallel c$ inferred from measurements of the resistivity, magnetic susceptibility, magnetostriction, and thermal expansion coefficient. The near-vertical lines are the two sheets of the tricritical wings, with the nematic phase in between. The transition below (above) the red arrows is first (second) order. From~\textcite{Grigera_et_al_2004}.}
\label{fig:SrRuO_phase_diagram_2}
\end{figure}
A magnetic field applied in the magnetically easy $ab$-plane takes the system through the metamagnetic wings at about 5\,T if the temperature is low enough, see Fig.~\ref{fig:SrRuO_phase_diagram} inset (i). Hydrostatic pressure and uniaxial stress drive the system away from and towards ferromagnetism, respectively~\cite{Ikeda_et_al_2001,Chiao_et_al_2002,Ikeda_et_al_2004,Wu_et_al_2011}.~\textcite{Wu_et_al_2011} investigated the ac susceptibility under pressure across the metamagnetic transition. They found a quantum critical point (denoted by QCEP in Fig.~\ref{fig:SrRuO_phase_diagram})$^{\ref{CEP_footnote}}$ at $p_{\textrm c} \approx 13.6$\,kbar but no divergence of the 
susceptibility at this point as would be expected for the generic model of quantum critical metamagnetism~\cite{Millis_et_al_2002}, implying that the metamagnetism cannot solely be explained by field-induced ferromagnetism. 

Another way to navigate the phase diagram is to change the field direction out of the magnetically easy $ab$-plane: Changing the field tilt angle $\theta$ allows to follow the wings and suppress the critical temperature $T^*$ that marks the top of the wing~\cite{Grigera_et_al_2001,Grigera_et_al_2003}. As $T^*$ goes to zero, a second sheet of the wing appears, and instead of the quantum critical point that is observed in simpler systems (see Sec. \ref{subsubsec:II.B.2}) 
a more complicated phase structure emerges~\cite{Perry_et_al_2004,Grigera_et_al_2004,Rost_et_al_2011}. Such a bifurcation has been modeled phenomenologically by means of a Landau theory~\cite{Green_et_al_2005}. The phase between the two sheets has been interpreted as a magnetic nematic (i.e., a non-s-wave ferromagnet)~\cite{Grigera_et_al_2004, Borzi_et_al_2007, Raghu_et_al_2009, Stingl_et_al_2011,Rost_et_al_2011}, or an inhomogeneous phase analogous to the 
Fulde-Ferrell-Larkin-Ovchinnikov phase in superconductors~\cite{Berridge_et_al_2009,Berridge_et_al_2010}, but the details are not well understood. The novel phase is observed with field tuning, but not with pressure tuning \cite{Wu_et_al_2011}. 

Recent magnetic neutron scattering experiments have identified an incommensurate SDW order with an ordered moment of about 0.1\,\mub/Ru and wavevector ${\bm q} = (0.233,0,0)$ in the phase shown in Fig.~\ref{fig:SrRuO_phase_diagram_2} and an additional phase at slightly higher magnetic fields with a different ordering wavevector ${\bm q} = (0.218,0,0)$~\cite{Lester_et_al_2014}. More work on this interesting system is clearly warranted.
\subsubsection{Discussion, and comparison with theory}
\label{subsubsec:II.B.5}
A striking aspect of the phase diagrams discussed in this section is their universality. As illustrated in Tables~\ref{table:1a},~\ref{table:1b} and discussed above, phase diagrams that are qualitatively the same as the one shown in Fig.~\ref{figure:UGe_2_phase_diagram} are observed in a wide variety of systems with very different electronic structures and different symmetries of the order parameter. Their only commonality is that they are metallic ferromagnets with reasonably small amounts of disorder.
\footnote{\label{measure_of_disorder_footnote} Throughout this review we use the residual resistivity as a convenient
               measure of the amount of quenched disorder in a material. We note, however, that it is possible that different 
               manifestations of disorder affect the resistivity differently than they affect magnetism. This may be relevant for
               interpreting certain observations in nominally rather clean systems, such as Ni$_x$Pd$_{1-x}$, see 
               Sec.\ \ref{NP}.}
This universal behavior calls for an equally universal explanation of the first-order nature of the quantum phase transition. Although detailed quantitative modelling of the phase diagram is still lacking, the theory described in Sec.\ \ref{subsubsec:III.B.2} can explain the qualitative structure of the phase diagram in terms of a fluctuation-induced first-order transition, with generic soft modes in the conduction-electron system playing the role of the extraneous (to the order parameter) soft modes that drive the transition first order. 

There are large quantitative differences between the systems listed in Tables~\ref{table:1a},~\ref{table:1b}. \textcite{Sang_Belitz_Kirkpatrick_2014} have shown that the sizes of the tricritical wings, which vary dramatically from material to material, correlate with the saturation magnetization as expected from the theory discussed in Sec.\ \ref{subsubsec:III.B.2}. In this context we add a remark on the shape of the wings. Theory and all experiments agree that the wings point in the ``forward'' direction, i.e., the wing tips are located at a larger value of the control parameter (for most systems, pressure) than the first-order transition in zero field. However, the curvature of the wings is not expected to be universal; it depends on the relation between the experimental control parameter and the theoretical one (i.e., the mass term in a LGW theory), which in turn depends on microscopic details. For instance, the wings in UGe$_2$, Fig.~\ref{figure:UGe_2_phase_diagram}, have a pronounced ``backward'' 
curvature, whereas the ones in UCoAl, Fig.~\ref{fig:UCoAl_phase_diagram}, are almost flat. 
Similarly, the shape of the lines of second-order transitions that connect the tricritical point with the wing tips is not universal. \textcite{Wysokinski_Abram_Spalek_2014, Wysokinski_Abram_Spalek_2015} have considered a model containing $f$-electrons in addition to conduction electrons that couple to the magnetization and have achieved good agreement with the shape of the wings in UGe$_2$. The physical mechanism that leads to a first-order QPT in their theory is the same as the one discussed in Sec.~\ref{subsubsec:III.B.2}.

There also is a clear correlation between the size of the ordered moment and the value of $T_{\text{tc}}$, see Tables~\ref{table:1a},~\ref{table:1b}. This is consistent with the theory, which predicts that the tricritical temperature is proportional to the ordered magnetic moment (for given microscopic temperature and magnetic-moment scales, which one would expect to be similar for systems that are chemically similar) \cite{Belitz_Kirkpatrick_Rollbuehler_2005}. For instance, within the uranium-based systems there is a rough correlation between $T_{\text{tc}}$ and the ordered moment. A U-based system in which no first-order transition has been found is UIr, see Sec.\ \ref{subsubsec:II.C.1} and Table \ref{table:2_pt_1}. Since the ordered moment in the phase FM3 of UIr, from which the QPT to the PM phase occurs, is smaller than the one in UGe$_2$ by more than a factor of 30, and smaller than the one in URhGe by a factor of more than 10 \cite{Kobayashi_et_al_2006}, one expects the tricritical temperature to be 
smaller by a similar factor. This would put $T_{\text{tc}}$ well below 1\,K, and possibly lower than 100\,mK, which is less than the lowest Curie temperature observed in UIr. Similarly, in the first group of materials in Tables~\ref{table:1a},~\ref{table:1b} the tricritical temperature correlates with the size of the ordered moment: CoS$_2$ has the largest moment and the highest $T_{\text{tc}}$, while in Ni$_3$Al, which has the smallest moment, a first-order transition at very low temperatures is suspected but has not yet been convincingly observed. More generally, it is conceivable that the tricritical temperature in several weakly disordered systems is rather low and for that reason has not been observed so far. A related issue is the robustness of the first-order transition described by the renormalized Landau theory; this is discussed in Secs.\ \ref{subsubsec:III.B.2} and \ref{subsec:IV.A}.
%
%There also has been theoretical work aimed at quantitatively understanding the low transition temperature in these systems, as well as other features. \textcite{Lonzarich_Taillefer_1985} have used spin-fluctuation theory \cite{Moriya_1985} to express the magnetic equation of state in terms of four parameters that can be obtained from neutron-scattering experiments,or from band-structure calculations. They obtained good quantitative agreement with data on Ni$_3$Al and MnSi. This theory does not, however, reflect the first-order nature of the transition at low temperatures. 

\subsection{Systems showing a continuous transition.}
\label{subsec:II.C}

We now discuss systems that show a continuous transition at low temperatures; their properties are summarized in Tables \ref{table:2_pt_1}, 
\ref{table:2_pt_2}, \ref{table:2_pt_3}. Most of these materials are composition-tuned, which introduces 
various amounts of disorder, and they can be classified with respect to the disorder strength. The first group is known or suspected to be 
relatively weakly disordered as judged by the residual resistivity, see Table~\ref{table:2_pt_1} (see, however, footnote \ref{resistivity_footnote}).
Consistent with this, their phase diagrams have the shape shown in panel b) of Fig.~\ref{fig:schematic_phase_diagrams}. In the
second group the disorder is strong, see Table~\ref{table:2_pt_2}, and the phase diagrams are of the form shown schematically in 
panel d) of Fig.~\ref{fig:schematic_phase_diagrams}. YbNi$_4$P$_2$ falls into a separate category due to its quasi-one-dimensional electronic 
structure which sets it apart from all other materials discussed in this review, see Table~\ref{table:2_pt_3}.

\begin{table*}[t]
\caption{Systems showing a second-order transition: Weakly disordered bulk systems. \TC\ =
              Curie temperature, $\rho_0 =$ residual resistivity. FM = ferromagnetism, SC = superconductivity. n.a. = not available. }
\smallskip
\begin{ruledtabular}
\begin{tabular}{lllllll}
System 
   & Order of 
      & \TC/K 
         & magnetic 
            & tuning  
               & Disorder  
                  & Comments \\

   & Transition $^a$ 
      &                            
         & moment/$\mu_{\text{B}}$ $^b$
            &  parameter  
               & ($\rho_0/\mu\Omega$cm) $^c$
                  & \\
%\\
\hline\\[-7pt]
%%%%%%%%%%%%%%%%%%%%%%%%%%%%%%%
Ni$_x$Pd$_{1-x}$ 
   & 2nd$\,^1$
      & 600 -- 7$\,^d$
         & 0.2 -- 2.45$\,^e$ 
            & composition$\,^1$
                & 1.5 (?)$\,^f$
                   & low-$T$ behavior unclear\\
%& & & & & & lowest \TC\ rather high \\
\\[-5pt]
%%%%%%%%%%%%%%%%%%%%%%%%%%%%%%%%%%%%%%%%%%
(Ni$_{1-x}$Pd$_x$)$_3$Al 
   & 2nd$\,^{2}$
      & 42 -- 4$\,^{g}$
         & 0.075 -- 0$\,^{h}$
            & composition$\,^{2}$
                & 10$\,^{2,i}$
                   & moderate disorder\\
\\[-5pt]
%%%%%%%%%%%%%%%%%%%%%%%%%%%%%%%%%%%%%%%%%%
Ni$_3$Al$_{1-x}$Ga$_x$ 
   & 2nd$\,^{3}$
      & 41 -- 5$\,^{3,j}$
         & 0.075 -- 0.02$\,^{3,j}$
            & composition$\,^{3}$
                & n.a.
                   & disorder unknown\\
\\[-5pt]
%%%%%%%%%%%%%%%%%%%%%%%%%%%%%%%
UIr
   & 2nd$\,^{4}$
      & 46 -- 1$\,^{4}$
         & 0.5$\,^{4}$
            & pressure$\,^{4}$
               & n.a.$\,^{k}$
                  & three FM phases, \\
   & & & & & & coex. FM+SC \\
%%%%%%%%%%%%%%%%%%%%%%%%%%%%%%%%%%%%%%%%%%
UNiSi$_2$
  & 2nd$\,^{5}$
     & 95$\,^{5,6,7}$
         & 1.2$\,^{6}$
            & pressure$\,^{5}$
               & $\approx 25\,^{5}$
                  & two FM phases\\
\\[-5pt]
%%%%%%%%%%%%%%%%%%%%%%%%%%%%%%%%%%%%%%%%%%
(Cr$_{1-x}$Fe$_x$)$_2$B
   & 2nd$\,^{8}$
      & 45 -- 8$\,^{8,l}$
         & 1.4 -- 0.25$\,^l$
            & composition
               & 35\,$^{8}$
                  & moderate disorder\\
\\[-5pt]
%%%%%%%%%%%%%%%%%%%%%%%%%%%%%%%%%%%
Zr$_{1-x}$Nb$_{x}$Zn$_{2}$ 
   & 2nd$\,^{9}$
      & 18 -- 0$\,^{9,m}$
         & 0.08 -- 0$\,^{9,m}$
            & composition$\,^{9}$
               & n.a.
                  & disorder unknown \\
\\[-5pt]                  
%%%%%%%%%%%%%%%%%%%%%%%%%%%%%%%%
\SCRO
   & 2nd$\,^{10}$
      & 160 -- 0$\,^n$
         & 0.8 -- 0$\,^n$
            & composition
               & n.a.
                  & bulk powder samples\\
\\[-5pt]
%%%%%%%%%%%%%%%%%%%%%%%%%%%%%%%%
SrCo$_{2}$(Ge$_{1-x}$P$_{x}$)$_{2}$ 
   & 2nd$\,^{11}$  
     & 35 -- 2$\,^{11,o}$
        & 0.1 -- 0.02$\,^{11,o}$ 
           & composition$\,^{11}$
              & n.a.
                 & FM induced by\\
& & & & & & dimer-breaking \\
%%%%%%%%%%%%%%%%%%%%%%%%%%%%%%%%
CeSi$_{1.81}$
   & 2nd$\,^{12}$
      & 9.5 -- 3$\,^{12,p}$
         & $0.2$ -- $0\,^{12,q}$
            & pressure
               & 12 (30)$\,^r$
                  & moderate disorder\\
\\[-5pt]
%%%%%%%%%%%%%%%%%%%%%%%%%%%%%%%%
CePd$_{1-x}$Ni$_x$
   & 2nd$\,^{13}$
      & 10.5 -- 6.1$\,^{13,s}$
         & n.a.
            & composition$\,^{13}$
               & $\approx 15\,^{13}$
                  &\TC\ nonmonotonic\\
\\[-5pt]
%%%%%%%%%%%%%%%%%%%%%%%%%%%%%%%%
U$_4$Ru$_7$Ge$_6$
   & 2nd$\,^{14}$
      & 11.2 -- 3$\,^t$
         & 0.2$\,^{u}$
            & pressure$\,^{14}$
                & 58$\,^{t}$
                   & intermediate disorder\\
\\[-5pt]
%%%%%%%%%%%%%%%%%%%%%%%%%%%%%%%%%
U$_4$(Ru$_{1-x}$Os$_x$)$_7$Ge$_6$
   & n.a.
      & 12 -- 1$\,^{v}$
         & 0.2$\,^u$
            & composition$\,^{15}$
                & n.a.
                   & disorder unknown\\
\\[-5pt]
%%%%%%%%%%%%%%%%%%%%%%%%%%%%%%%%%
(Sc$_{1-x}$Lu$_x$)$_{3.1}$In
   & 2nd$\,^{16}$
      & 4 -- 1$\,^{w}$
         & 0.13 -- 0$\,^{w}$
            & composition$\,^{16}$
               & n.a.
                  & quasi-1-$d$ chains of Sc-In\\
\\[-5pt]
%%%%%%%%%%%%%%%%%%%%%%%%%%%%%%%%%
\hline\hline\\[-5pt]
\multicolumn{7}{l} {$^a$  At the lowest temperature achieved.\quad 
                              $^b$ Per formula unit unless otherwise noted.\quad 
                              $^c$ For the highest-quality samples.}\\
\multicolumn{7}{l} {$^d$ For $x = 1$ -- $0.027$~\cite{Nicklas_et_al_1999}. $^e$ For $0.018 \lesssim x \lesssim 0.1$~\cite{Nicklas_et_al_1999}.}\\        
\multicolumn{7}{l} {$^f$ \textcite{Tari_Coles_1971, Nicklas_2000}; \textcite{Ikeda_1987} reported $\rho_0$ as small as 
                                      $0.01\,\mu\Omega$cm for the relevant Ni concentrations.}\\    
\multicolumn{7}{l} {$^{g}$ For $x=0$ -- 0.9.\quad
                              $^{h}$ For $x=0$~\cite{Niklowitz_et_al_2005} to $x = 0.1$~\cite{Sato_1975}.\quad
                              $^{i}$ For $x=0.1$.\quad
                              $^{j}$ For $x=0$ -- 0.33.}\\       
\multicolumn{7}{l} {$^k$ RRR up to 250 \cite{Kobayashi_et_al_2006}.\quad
                              $^l$ For $x = 0.05$ -- $0.02$.\quad
                              $^{m}$ For $x = 0$ -- $0.08$.\quad
                              $^n$ For $x = 0$ -- $x\agt 0.7$.}\\
\multicolumn{7}{l} {$^{o}$ Per Co for $x = 0.55$ -- $0.35$.\quad
                               $^p$ For $p = 0$ -- $13\,$kbar.\quad
                               $^q$ At $T=1.7\,$K for $p = 0$ -- $14\,$kbar.}\\
\multicolumn{7}{l} { $^r$ For CeSi$_{1.86}$ at $p=0$ with a current in a- (c-) direction \cite{Sato_et_al_1988}.\quad
                               $^s$ For  $x \approx 0.5$ -- $0.94$.}\\
\multicolumn{7}{l} {$^t$ For $p = 0$ -- $2\,$GPa.\quad
                              $^u$ per U.\quad
                              $^{v}$ For $x=0$ -- 0.3 \cite{Colineau_et_al_2001}.\quad
			      $^{w}$ For $x = 0$ -- 0.03.} \\
                              \\[-5pt]
\hline\\[-5pt]
\multicolumn{2}{l}{$^1$ \textcite{Nicklas_et_al_1999} }
      & \multicolumn{2}{l}{$^{2}$ \textcite{Sato_1975}}
            & \multicolumn{2}{l}{$^{3}$ \textcite{Yang_et_al_2011} }
                  & $^{4}$ \textcite{Kobayashi_et_al_2006} \\ 
\multicolumn{2}{l}{$^{5}$ \textcite{Sidorov_et_al_2011b} }
      & \multicolumn{2}{l}{$^{6}$ \,\textcite{Das_et_al_2000}}
            & \multicolumn{2}{l}{$^{7}$ \textcite{Pikul_Kaczorowski_2012} }
                  & $^{8}$ \textcite{Schoop_et_al_2014} \\                  
\multicolumn{2}{l}{$^{9}$ \textcite{Sokolov_et_al_2006}}
      & \multicolumn{2}{l}{$^{10}$ \textcite{Itoh_Mizoguchi_Yoshimura_2008}}
            & \multicolumn{2}{l}{ $^{11}$ \textcite{Jia_et_al_2011} }
                  & $^{12}$ \textcite{Drotziger_et_al_2006} \\   
\multicolumn{2}{l}{$^{13}$ \textcite{Kappler_et_al_1997}}
      & \multicolumn{2}{l}{$^{14}$ \textcite{Hidaka_et_al_2011}}
            & \multicolumn{2}{l}{$^{15}$ \textcite{Colineau_et_al_2001}}
                  & $^{16}$ \textcite{Svanidze_et_al_2014} \\      
%%%%%%%%%%%%%%%%%%%%%%%%%%%%%%%%%%%%%%%%%%%
\end{tabular}
\end{ruledtabular}
\vskip -3mm
\label{table:2_pt_1}
\end{table*}
\subsubsection{Weakly disordered systems}
\label{subsubsec:II.C.1}
\paragraph{{\rm Ni}$_{x}${\rm Pd}$_{1-x}$}
\label{NP}
NiPd alloys, which crystallize in an fcc structure, form a series of solid solutions whose composition can be varied continuously from pure Pd to pure Ni. The 
alloying procedure can produce very little disorder as measured by the residual resistivity $\rho_0$, which has been reported not to exceed 0.1\,$\mu\Omega$cm for any concentration~\cite{Ikeda_1987}. A small concentration (about 2.5\%) of Ni induces ferromagnetic order~\cite{Murani_Tari_Coles_1974}. This
composition-induced quantum phase transition was studied by \textcite{Nicklas_et_al_1999} by means of heat capacity, electrical resistivity and magnetization 
measurements. For Ni concentrations up to 10\% above the critical concentration $x_{\text{c}} \approx 0.026$ they found a critical temperature
\TC\ $\propto (x - x_{\text{c}})^{3/4}$ and a $T\ln T$ contribution to the specific heat down to 0.3\,K. The $T$-dependence of the resistivity shows a 
power-law behavior
\be
\rho (T\to 0) = \rho_0 + A\,T^n\ .
\label{eq:2.C.1}
\ee
The exponent $n$ displays a sharp minimum of $n = 5/3$ near the critical concentration, while the prefactor $A$ shows an equally sharp maximum. 
% In addition, the dc susceptibility $\chi \propto T^{-4/3}$ for samples with $x \approx x_{\text{c}}$. 
These results are all consistent with the predictions of 
Hertz-Millis-Moriya theory (cf. Sec.~\ref{subsubsec:III.C.2}).
\begin{figure}[t]
\begin{center}
\includegraphics[width=0.95\columnwidth,angle=0]{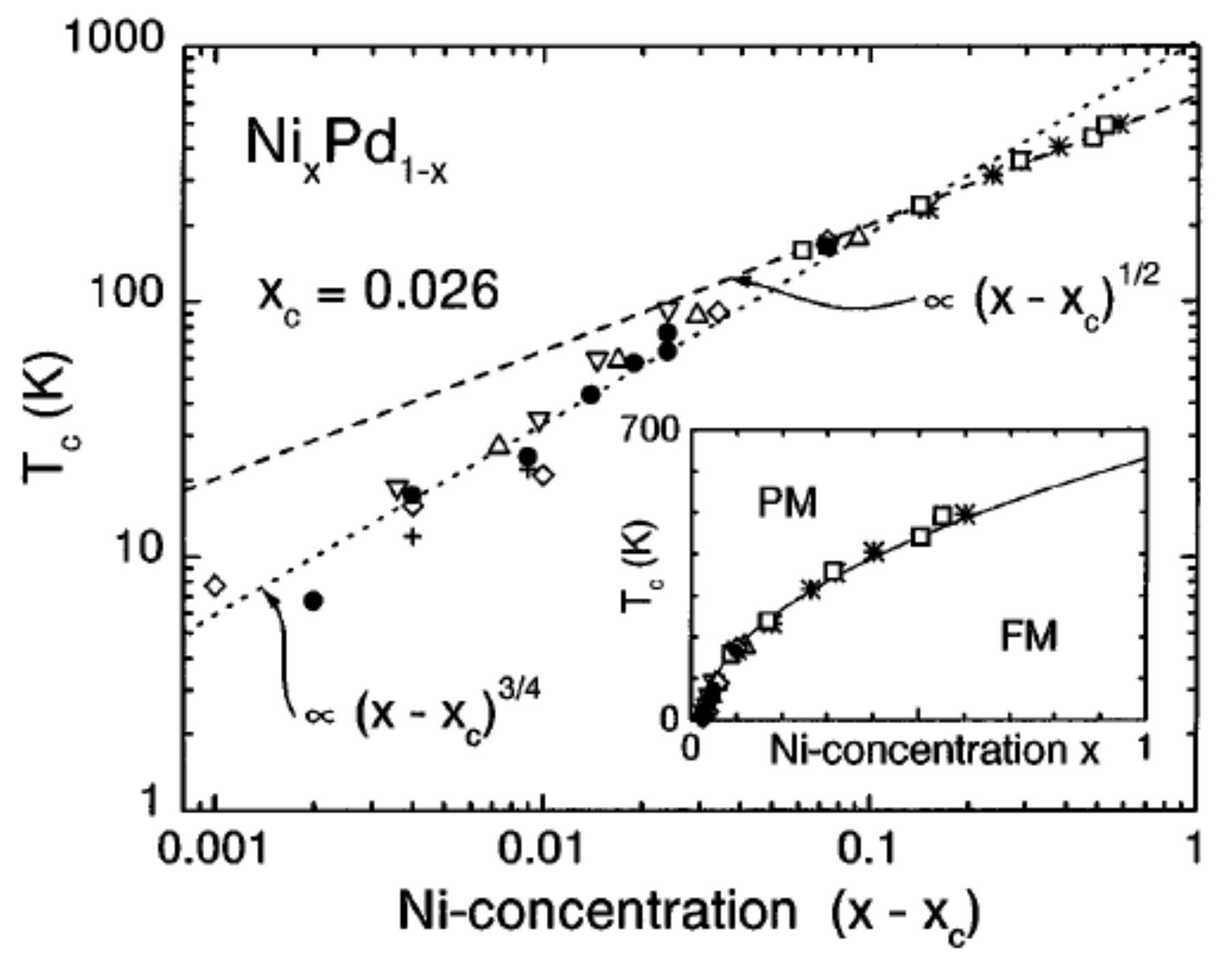}
\end{center}
\caption{Observed phase diagram of Ni$_{x}$Pd$_{1-x}$. Filled dots are data taken by~\textcite{Nicklas_et_al_1999}, the other symbols represent earlier data. 
              From~\textcite{Nicklas_et_al_1999}.}
\label{fig:NiPd_phase_diagram}
\end{figure}
\begin{figure*}[ht]
\begin{center}
\hspace*{-25pt}
\includegraphics[width=2.30\columnwidth,angle=0]{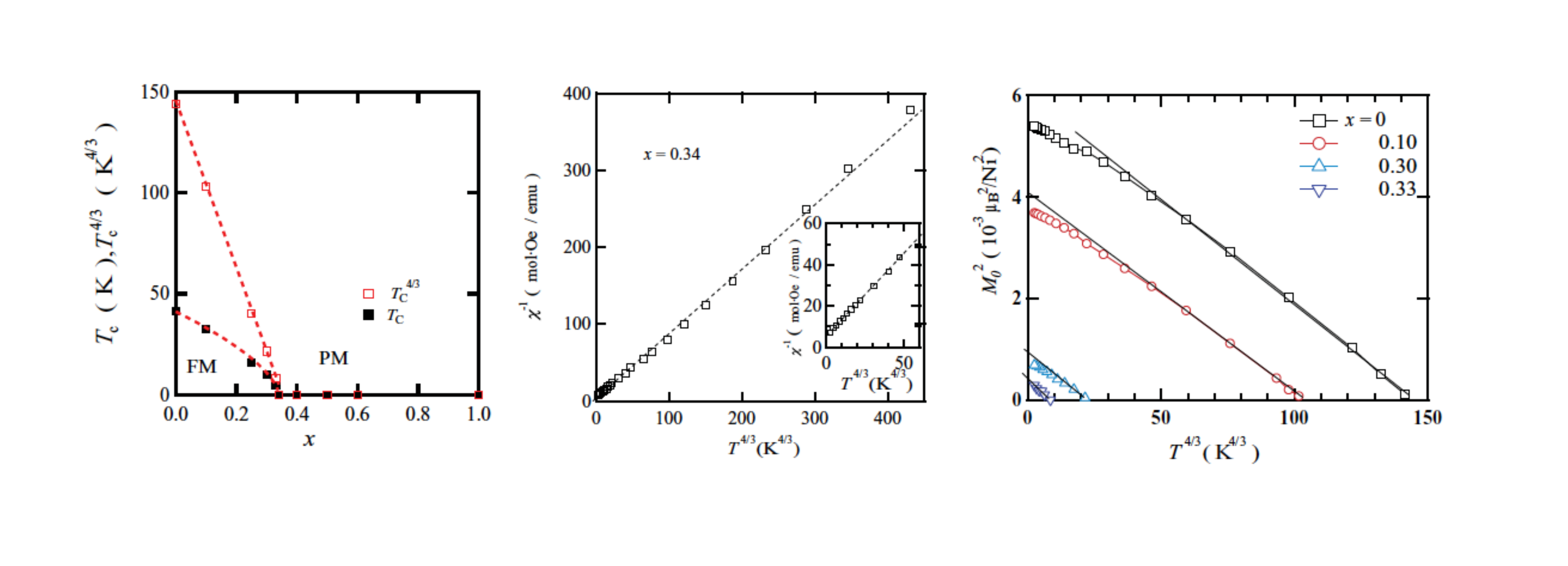}
\end{center}
\vskip -40pt
\caption{Hertz-type scaling behavior as observed in Ni$_3$Al$_{1-x}$Ga$_2$. From left to right: \TC\ vs $x$ phase diagram, inverse magnetic susceptibility 
              vs. $T^{4/3}$ for the critical sample, magnetization squared vs. $T^{4/3}$ for various concentrations. From~\textcite{Yang_et_al_2011}.}
\label{fig:Ni3AlGa}
\end{figure*}

The lowest transition temperature achieved in these experiments was \TC\ $\approx 7\,$K at $x - x_{\text{c}} \approx 0.002$, see Fig.~\ref{fig:NiPd_phase_diagram}. 
This is on the same order as the temperature above which, e.g., MnSi displays behavior consistent with Hertz theory even though the behavior at low $T$ is very
different. Subsequent ac susceptibility and zero-field-cooled/field-cooled magnetization measurements on the same NiPd samples at temperatures as low as 2\,K 
found evidence for spin-glass freezing in a small region of the phase diagram ($0.025 \leq x \leq 0.028$)~\cite{Nicklas_2000}. To corroborate this observation a 
measurement of the thermal expansion was performed on the same $x = 0.024$ polycrystal studied by~\textcite{Nicklas_et_al_1999}. \textcite{Kuechler_et_al_2006}
found that the Gr{\"u}neisen ratio (i.e., the thermal expansion coefficient divided by the specific heat) does not increase with decreasing $T$, but remains constant below 3\,K, in contrast to what is expected at a QCP~\cite{Kuechler_et_al_2006}. 
Single-crystalline samples investigated by~\textcite{Franz_et_al_2010} show similar transport and thermodynamic properties as those seen in polycrystals studied by~\textcite{Nicklas_et_al_1999}, but a detailed analysis of the magnetization indicates that at low fields and low temperatures strong deviations emerge from the conventional mean field predictions of a FM QCP in the clean limit. Considering that neutron-depolarization imaging experiments have shown that polycrystalline samples are much more homogeneous than the single-crystalline samples~\cite{Pfleiderer_et_al_2010}, these results all indicate that the behavior at asymptotically low temperatures is intrinsic and different from what is expected for a clean ferromagnet, which raises the question of disorder present in the samples. The strength of the disorder, or how to characterize it, is not quite clear. One measure is the residual resistivity, 
$\rho_0$. The data obtained by \textcite{Ikeda_1987} 
suggest $\rho_0 \approx 5\,\mu\Omega$cm for $x$ around the critical concentration.~\textcite{Tari_Coles_1971} reported a low-temperature ($< 4.2$\,K) resistivity of 
about 1\,$\mu\Omega$cm for a sample with $x = 0.025$. The residual resistivity $\rho_0$ of the samples studied by \textcite{Nicklas_2000} is about 
$0.5\,\mu\Omega$cm for pure Pd (RRR = 40) and for the sample with $x = 0.1$,  $1.5\,\mu\Omega$cm for $x \approx x_{\text{c}}$, and it reaches a maximum of 
$3\,\mu\Omega$cm at $x \approx 0.04$. These results suggest that there is an intrinsic and substantial amount of disorder even in the best samples as reflected in the 
value of $\rho_0$. It would be desirable to revisit the QPT in NiPd, while carefully characterizing the amount of disorder in the samples. A $\rho_0$ of 
1 $\mu\Omega$cm would put the sample marginally in the intermediate Regime II of the theory discussed in Sec.~\ref{subsubsec:III.B.3}, where the theory predicts 
a continuous transition with effectively mean-field exponents. Substantially cleaner samples would have to show a first-order transition at low temperatures in order 
to be consistent with the theory. However, if the spin-glass effects found by \textcite{Nicklas_2000} were to be corroborated this theory would be inapplicable
and the system would have to be classified with the materials discussed in Sec.~\ref{subsec:II.E}.
\paragraph{Ni$_3$Al$_{1-x}$Ga$_x$ and (Ni$_{1-x}$Pd$_x$)$_3$Al}
\label{NiAlGa}
The FM order in Ni$_3$Al with \TC\ = 41\,K (see Sec.\ \ref{par:II.B.1.d}) can be suppressed by substitution of Pd for Ni~\cite{Sato_1975}, or by doping with 
Ga~\cite{Yang_et_al_2011}. In the former system, a quantum critical point is reached at $x \approx 0.095$, at which concentration the samples measured
by \textcite{Sato_1975} had a residual resistivity $\rho_0 \approx 10\,\mu\Omega$cm, indicating moderate disorder. The observed critical behavior is consistent 
with the Hertz-Millis-Moriya theory, as one would theoretically expect for systems in this disorder regime, see Sec.\ \ref{subsubsec:III.B.3}.

\begin{figure}[b]
\begin{center}
\includegraphics[width=0.9\columnwidth,angle=0]{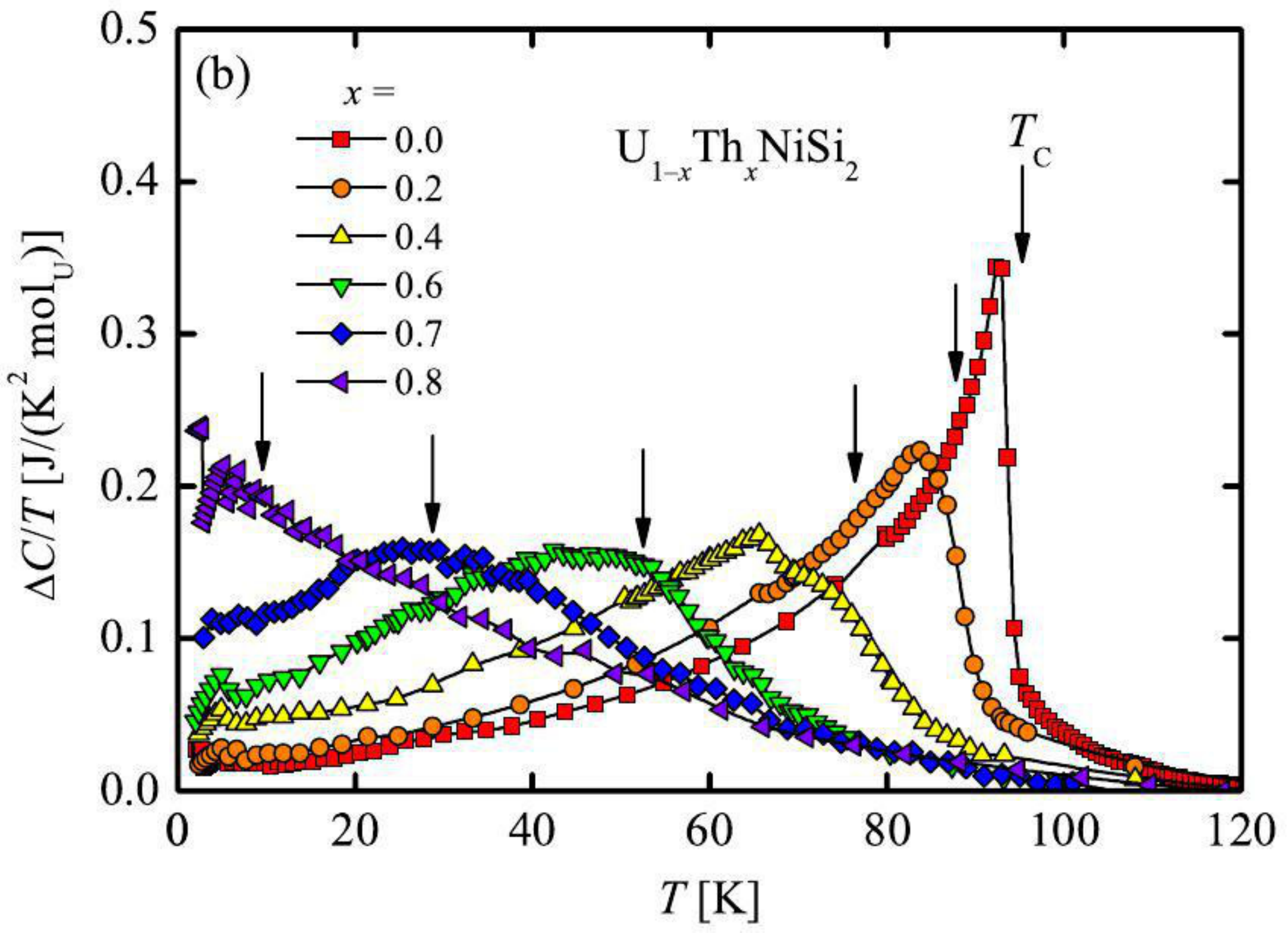}
\end{center}
\caption{$5f$-electron Sommerfeld coefficient $\Delta C(T)/T$ of \UTNS. For $x=0$ the FM ordering is visible at \TC\ = 95\,K in the form of a $\lambda$-like peak. For the behavior with increasing Th doping see Sec.\ \ref{par:UThNiSi}. From~\textcite{Pikul_et_al_2012a}.}
\label{pikul2012a1}
\end{figure}

\begin{figure}[b]
\begin{center}
\includegraphics[width=0.9\columnwidth,angle=0]{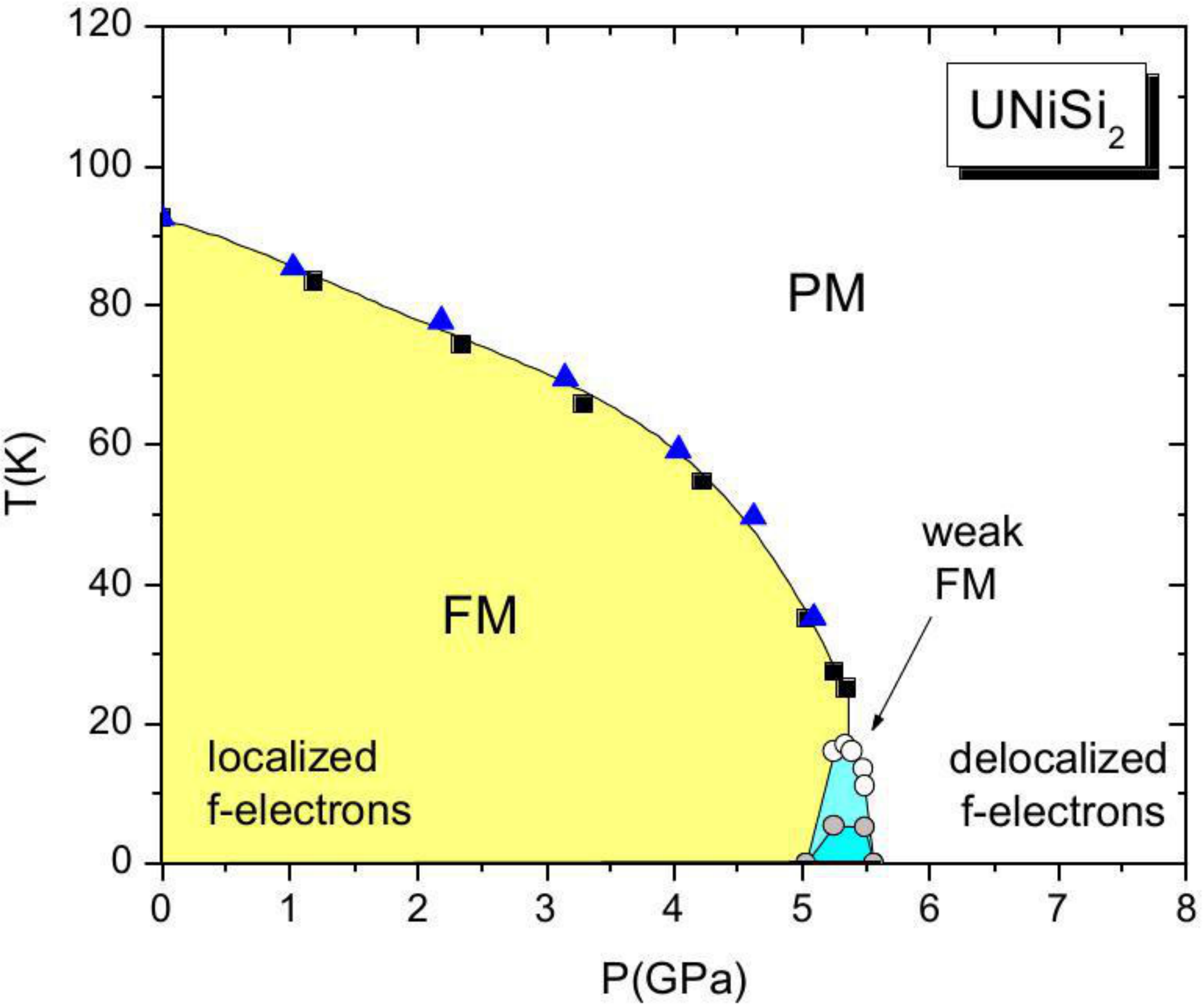}
\end{center}
\caption{Temperature-pressure ($T$-$P$) phase diagram of UNiSi$_2$ from resistivity, ac susceptibility, and specific heat measurements. The turquoise region 
              represents a separate phase characterized by weak ferromagnetism and a large Sommerfeld coefficient. In the paramagnetic (PM) phase the Sommerfeld 
              coefficient remains large as does the resistivity. This suggests a delocalization of the $f$-electrons across the FM-to-PM quantum phase transition. 
              From~\textcite{Sidorov_et_al_2011b}.}
\label{sidorov2011a1}
\end{figure}

\TC\, also decreases monotonically  upon doping with Ga, leading to a QPT in Ni$_3$Al$_{1-x}$Ga$_x$ at $x=x_c \approx 0.34$
\cite{Yang_et_al_2011}; Ni$_3$Ga is paramagnetic \cite{Hayden_Lonzarich_Skriver_1986}. The disorder strength in the polycrystalline samples investigated
by \textcite{Yang_et_al_2011} is not known, but assuming that this system is in the same moderate-disorder regime as Ni$_3$Al$_{1-x}$Ga$_2$ one would
expect to observe Hertz-type critical behavior according to the theoretical analysis reviewed in Sec.\ \ref{subsubsec:III.B.3}. This is indeed borne out by the
experiment, see Fig.~\ref{fig:Ni3AlGa}: The temperature-concentration phase diagram obeys Eq.~(\ref{eq:3.52}), the susceptibility at the critical concentration
diverges as $T^{-4/3}$, Eq.~(\ref{eq:3.56}), and the magnetization as a function of temperature near \TC\ obeys Eq.~(\ref{eq:3.59'}). The first result reflects
the combination of critical exponents $\nu\,z/(1+2\nu) = 3/4$, see Eq.~(\ref{eq:3.48}). The second one reflects the exponent $\gamma_T = 4/3$, and if combined 
with the first one it also implies $\gamma = 1$, in agreement with Eq.~(\ref{eq:3.48}), since $\chi \sim T^{-4/3} \sim \vert x - x_c\vert^{-1}$.
The third one reflects $\beta = 1/2$ in addition to the combination $\nu\,z/(1+2\nu)$. See Appendix~\ref{app:B} for the definitions of the critical exponents, 
and Sec.~\ref{par:III.C.2.b} for the
scaling considerations underlying the above statements. As emphasized in Sec.~\ref{sec:III}, this behavior is expected to hold, strictly speaking, only in
a pre-asymptotic regime. However, for moderate disorder strengths the true asymptotic regime is expected to be unobservably small.
\paragraph{UIr}
\label{par:UIr}
UIr at ambient pressure is an Ising-like ferromagnet with a $T_{\text{C}} \approx 46$\,K. High-quality samples with a RRR $\approx 250$ have been investigated 
under hydrostatic pressure~\cite{Akazawa_et_al_2004, Kobayashi_et_al_2006, Kobayashi_et_al_2007}. The overall phase diagram is qualitatively similar to that 
of UGe$_2$, but the details are different. With increasing pressure there are three distinct FM phases labeled FM1, FM2, and FM3~\cite{Kobayashi_et_al_2006}, 
and strain and resistivity measurements suggest that these three phases have slightly different crystal structures \cite{Kotegawa_et_al_2011a}. FM1 has an ordered 
moment of 0.5\,$\mu_{\text{B}}$/U. There is a first-order metamagnetic transition between FM1 and FM2 at $p \approx 1.7$\,GPa (at $T=0$). The ordered moment 
in FM2 and FM3 is less than 0.5 $\mu_{\text{B}}$/U. \TC\, goes to zero, and FM3 gives way to paramagnetism, at a critical pressure $p_{\text{c}} \approx 2.8$\,GPa. 
Near $p_{\text{c}}$ in the FM3 phase, superconductivity is observed at temperatures below 140\,mK \cite{Akazawa_et_al_2004}. The absence of metamagnetic 
behavior in the PM phase is indicative of the FM3-PM transition remaining second order to the lowest observed \TC\,$\approx 0.8$\,K.
\paragraph{UNiSi$_2$}
\label{par:UNiSi}
UNiSi$_{2}$ is a collinear ferromagnet with \TC\,$\approx 95$\,K and U moments of 1.2\,$\mu_{\textrm{B}}$ directed along the crystallographic $c$-axis of the 
orthorhombic structure~\cite{Geibel_et_al_1990, Kaczorowski_1996, Taniguchi_et_al_1998, Das_et_al_2000,Pikul_et_al_2012a}.%
\footnote{\textcite{Pikul_et_al_2012a} gives a different value, $\mu_{sp} \approx 0.8$\,\mub/U.}
Although the amount of magnetic entropy below \TC, $\Delta S \approx 11.3$\,J/mole-K, suggests that the uranium moments are mostly localized, this value is lower 
than the value $\Delta S = R\ln(10) = 19.1$\,J/mole-K expected for fully localized U$^{3+}$ moments \cite{Sidorov_et_al_2011b}.%
\footnote{\textcite{Pikul_et_al_2012a} finds an even lower value, $\Delta S \approx 8$\,J/mole-K}
This is possibly due to the crystalline electric field and the Kondo effect which is seen in the $\rho \propto -\ln T$ behavior of the resistivity above 
\TC~\cite{Kaczorowski_1996,Sidorov_et_al_2011b}. In polycrystalline samples as well as in single crystals of good quality (RRR $\approx$ 7) the FM phase transition 
in zero field is clearly second order, characterized by a $\lambda$-like peak in $C(T)/T$ (see Fig.~\ref{pikul2012a1}). Partial substitution of Th for U suppresses
\TC\ and leads to pronounced disorder effects; this system is discussed in Sec.\ \ref{par:UThNiSi}.
\textcite{Sidorov_et_al_2011b} investigated single crystals of UNiSi$_{2}$ under hydrostatic pressure, up to about 6\,GPa. With increasing pressure the Curie 
temperature decreases, moderately for pressures up to about 4\,GPa, and then more sharply, vanishing above 5.5\,GPa, see the phase diagram in 
Fig.~\ref{sidorov2011a1}.
The FM phase transition remains second order (from ac calorimetry measurements) in the pressure range $0 \leq p \leq 5.1$\,GPa where the transition could be detected. 
In the pressure range near the QPT, between 5.1 and 5.5\,GPa, another phase appears (turquoise region in Fig.~\ref{sidorov2011a1}), which is characterized by weak 
ferromagnetism. This feature, which is reminiscent of the distinct FM phases in UIr (see Sec. \ref{par:UIr}) is indicated by an upturn in the ac susceptibility and signatures in 
the resistivity and the specific heat~\cite{Sidorov_et_al_2011b}. The magnetic entropy is strongly reduced under pressure. This effect and the enhanced Sommerfeld 
coefficient (as well as the resistivity) at 5.5\,GPa led \textcite{Sidorov_et_al_2011b} to suggest that at the FM QPT a delocalization of the $f$-electrons takes place with 
the formation of a hybridized $f$-$d$ band of heavy electrons.  
\paragraph{{\rm (Cr}$_{1-x}${\rm Fe}$_x$)$_2${\rm B}}
\label{par:II.C.1.c}
Ferromagnetism can be induced in the paramagnetic compound Cr$_2$B by doping
with Fe. The resulting system (Cr$_{1-x}$Fe$_x$)$_2$B undergoes a QPT near
$x_{c} = 0.02$~\cite{Schoop_et_al_2014}. Doping introduces a substantial
amount of disorder resulting in a residual resistivity near the critical
concentration of $\rho_0 \approx 35$\,$\mu\Omega$cm. The exponent $n$ in Eq.\
(\ref{eq:2.C.1}) falls from its Fermi-liquid value $n = 2$ at $x = 0$ to $n =
1$ at $x_{c}$, and remains there for larger values of $x$. The prefactor $A$
peaks around $x_{c}$. However, the absolute change of $\rho$ with temperature in the linear-in-$T$ range is extremely small. 
For instance, for $x = 0.025$ between 8 and 20\,K $\Delta\rho = \rho - \rho_0 \approx 0.2$\,$\mu\Omega$cm, which is very small compared
to the rather large value of $\rho_0 = 40$\,$\mu\Omega$cm. This results in a tiny
value of $A \approx $\,18\,n$\Omega$cm/K; for other values of $x$ the value is even smaller. The
application of a magnetic field of 14\,T restores the Fermi-liquid value $n = 2$. The specific heat shows a $T\ln T$ term similar to that observed in \NP,
with a prefactor that is largest around $x_{c}$, but again the maximal change
is very small, $\Delta C/T \approx 2$\,mJ/K$^{2}$mol. These observations are
in principle consistent with the existence of a QCP at $x_{c}$, but the NFL properties characteristic of
critical behavior are extremely weak. Nevertheless, the systems is promising and the nature and properties of the
transition warrant further investigation. Given the disorder strength, the
theory reviewed in Sec.\ \ref{subsec:III.B} predicts a continuous transition.
\paragraph{\ZNZ}
\label{ZNZ}
Itinerant ferromagnetism in ZrZn$_{2}$ (\TC\, = 28.5\,K,~\textcite{Matthias_Bozorth_1958,Pickart_et_al_1964}) can be tuned to zero by a small amount of Nb 
substituting for Zr.~\textcite{Sokolov_et_al_2006} investigated the magnetization of several polycrystalline samples of the series \ZNZ\, with $0 \leq x \leq 0.14$ 
down to 1.8\,K. From an Arrott-plot analysis they found that \TC\, is suppressed to zero at $x_{\text{c}} \approx 0.08$. The dependence of \TC\, on the Nb 
concentration can be well described by $T_{\text{C}} \propto (x - x_{\text{c}})^{3/4}$, and the spontaneous moment vanishes linearly with \TC. Furthermore, 
the inverse magnetic susceptibility could be fitted to $\chi^{-1} = a\,T^{4/3} + b(x - x_{\text{c}})$, with $a$ and $b$ constants. All of this is consistent with the
results of Hertz-Millis-Moriya theory, see Sec.\ \ref{par:III.C.2.b}. There were no indications of a first-order transition or metamagnetism. 

These results are reminiscent of those for \NP\ (cf. Sec.~\ref{NP}). However, in the present case little is known about the disorder present in the samples.
X-ray diffraction experiments revealed single-phase specimens with Laves phase C15 structure, but no resistivity data are available since the solid-state reaction 
used produced powder unsuitable for resistivity measurements~\cite{Sokolov_2014}.
\paragraph{\SCRO\ (bulk powder samples)}
\label{SCRO_powder}
Early studies of polycrystalline \SCRO\ samples showed that \TC\ goes to zero linearly as $x$ approaches $x_{c} \approx 0.7$
\cite{Kanbayasi_1978, Kiyama_et_al_1999}, which led to the proposal of a QCP in this material. \textcite{Itoh_Mizoguchi_Yoshimura_2008} studied powder 
samples and concluded from magnetization measurements that there is indeed a QCP. From Arrott plots 
for $x$ near $x_c$ they inferred a field-dependence of the magnetization $M \propto H^{2/3}$, i.e., a critical exponent $\delta = 3/2$. This 
% is inconsistent with the mean-field value $\delta = 3$ of SCR theory, but 
agrees with the prediction of the generalized Landau theory for disordered systems described in 
Sec.\ \ref{subsubsec:III.B.3}. Unfortunately, no information is available about the disorder strength in these samples. Recently, \textcite{Huang_et_al_2015} have studied the dynamical scaling of the magnetization and specific heat and found $\delta = 1.6$ in agreement with~\textcite{Itoh_Mizoguchi_Yoshimura_2008}.
Their scaling analysis yields a very unusual temperature dependence of the specific-heat coefficient $\gamma \propto \text{const.} - T^{0.7}$.
% but an unusual small dynamic critical exponent $z = 1.76$ which is likely governed by strong disorder~\cite{Huang_et_al_2015}. 
The behavior is markedly different from that of ceramic samples (see Sec.\ \ref{SCRO_ceramic}) and epitaxial thin films (see Sec.\ \ref{subsubsec:II.E.3}). 
\paragraph{{\rm SrCo}$_{2}$({\rm Ge}$_{1-x}${\rm P}$_{x}$)$_{2}$}
\label{SCGP}
Pressure, doping, or an external magnetic field are standard external parameters for inducing a quantum phase transition. For the ferromagnetism that develops
in \SCGP\, at $x \approx 0.325$, \textcite{Jia_et_al_2011} have proposed a new tuning mechanism: the breaking of bonds in Ge-Ge dimers, which the authors
argue is more important than the simple change in carrier concentration with $x$.
SrCo$_{2}$Ge$_{2}$ forms in the ThCr$_{2}$Si$_{2}$ structure, with Co$_{2}$Ge$_{2}$ layers separated by a Ge-Ge 
interlayer bond, i.e., a dimer. This causes the layers to be pulled together and to form a collapsed tetragonal cell with a three-dimensional electronic structure. 
SrCo$_{2}$Ge$_{2}$ 
is a simple Pauli paramagnet. The lack of such a dimer in SrCo$_{2}$P$_{2}$ causes the same structure to be uncollapsed and, therefore, 
to have a more two-dimensional electronic structure and a shorter Co-Co separation within the layers, which increases the Co-Co interaction. From measurements of the magnetization, the susceptibility, and the specific heat of polycrystalline samples~\textcite{Jia_et_al_2011} concluded that at 
$x \approx 0.325$ the system develops bulk ferromagnetism via a QCP. The Curie temperature increases with increasing $x$, reaches a maximum of
\TC\ $\approx$ 35\,K at $x \approx 0.55$, and then decreases. For $x \agt 0.75$ \textcite{Jia_et_al_2011} found that the 
%ferromagnetism disappears via a crossover rather than a phase transition. 
ground state is a Stoner-enhanced paramagnet rather than a ferromagnet. The ferromagnetism is of the itinerant type, characterized by a 
Curie-Weiss behavior with an effective moment much 
larger than the saturation moment. This is in agreement with band-structure calculations which show a maximum in the density of states at 
$x \approx 0.5$ where 
\TC\, reaches its maximal value~\cite{Cuervo-Reyes_Nesper_2014}. \TC\, increases linearly from $x$ = 0.3 to 0.5 having a value of 
about 5\,K for $x = 0.35$. No sign 
of a first-order transition or spin-glass behavior was detected. For a sample with $x = 0.325$ and no \TC\, the susceptibility was found to 
behave as $\chi \propto T^{-4/3}$ 
(down to 2\,K) and the specific heat $C/T \propto -\ln T$ (down to 0.4\,K), in agreement with Hertz-Millis-Moriya theory (cf. Sec.~\ref{subsubsec:III.C.2}). No
resistivity measurements have been reported, and the role of disorder in this material is unknown. %
\paragraph{\CS}
\label{par:II.C.1.e}
\CS\, can be considered the first example of a FM dense Kondo system~\cite{Yashima_et_al_1982, Sato_et_al_1988}. It crystallizes in the $\alpha$-ThSi$_{2}$ 
structure with a broad homogeneity range, $1.7 \leq x \leq 2$~\cite{Ruggiero_Olcese_1964, Yashima_Satoh_1982, Yashima_et_al_1982}. Within this range, the 
system shows a paramagnetic ground state for $x \geq 1.85$, while a magnetically ordered state was found for $x \leq 1.8$ with a transition temperature around 10\,K. The 
nature of the magnetic order is not entirely clear~\cite{Drotziger_et_al_2006}. Susceptibility measurements suggest that the magnetic structure depends on the Si vacancy distribution~\cite{Shaheen_Mendoza_1999}, and magnetization 
measurements indicate that the ground state may not be pure ferromagnet, but rather a ferrimagnet or canted ferromagnet resulting from the RKKY interaction 
between the Ce $4f$ local moments on two different lattice sites~\cite{Drotziger_et_al_2006}.

\textcite{Drotziger_et_al_2006} studied the magnetization of a single crystal of CeSi$_{1.81}$ as a function of temperature and hydrostatic pressure. At ambient
pressure the transition temperature was \TC\ = 9.5K; pressure was found to suppress \TC\ monotonically. \TC\ vanishes at a critical pressure 
$p_{\text{c}} \approx 13.1$\,kbar, with a concomitant continuous suppression of the magnetic moment from 0.2\,\mub/Ce to zero. The transition at $T_{\textrm{C}}$ 
was found to always be of second order down to the lowest observed value of \TC\ $\approx$ 3K. 
The electrical resistivity of the $x = 1.81$ sample is not known, but for $x = 1.86$ a residual resistivity of 12\,$\mu\Omega$cm (with current along the $a$-axis) and 
30\,$\mu\Omega$cm (with current along the $c$-axis) has been reported~\cite{Sato_et_al_1988}. If these values are representative for the $x = 1.81$ critical sample 
as well, they put the system \CS\, in a moderately disordered regime where a continuous transition 
is expected theoretically, see Sec. \ \ref{subsubsec:III.B.3}. However, questions about both the nature of the ordered phase and the nature of the transition in the
zero-temperature limit remain~\cite{Drotziger_et_al_2006}.
%

%\paragraph{CeAgSb$_2$}
%\label{par:II.C.1.f}
%
%\CAS\, a Kondo-lattice systems with a planar magnetic anisotropy. It is one example of a ferromagnet where the moments are unexpectedly aligned along the 
%magnetic hard direction; others include YbNiSb~\cite{Bonville_et_al_1992}, \CRPO~\cite{Krellner_et_al_2007}, \YNP~\cite{Steppke_et_al_2013} and 
%\YRCSFM~\cite{Lausberg_et_al_2013}. 
%Neutron scattering experiments have shown that the ordered moment of 0.41\mub/Ce aligns uniaxially along the tetragonal $c$-axis, whereas magnetization 
%measurements indicate a strong Ising-like magnetocrystalline anisotropy with the basal plane as the magnetic easy plane~\cite{Araki_et_al_2003}. 
%For this reason, quantum criticality in \CAS\, has been investigated by transversal-field tuning with $H \perp c$. The critical field was found to be
%$H_{c} \approx 2.8$\,T~\cite{Strydom_et_al_2008, Zou_et_al_2013, Logg_et_al_2013}. The transition is suspected to remain second order to the
%lowest \TC\, measured, about 2\,K. Under pressure, an AFM phase has been detected, see Sec.\ \ref{subsubsec:II.D.3}.

\paragraph{CePd$_{1-x}$Ni$_x$}
\label{par:CePdNi}
The FM \TC\ of CePd (\TC\ = 6.5\,K) initially increases upon alloying with nickel, then decreases for $x\agt 0.8$, and vanishes at a Ni concentration 
$x_c \approx 0.95$ \cite{Kappler_et_al_1997}.
Measurements of the specific heat, magnetization, and resistivity have shown NFL behavior of the resistivity for $3 < T < 30$\,K, and logarithmic behavior of the
specific-heat coefficient in a temperature window between about 1 and 10\,K. 
\paragraph{(Sc$_{1-x}$Lu$_x$)$_{3.1}$In}
\label{ScLuIn}
Upon doping of the non-stoichiometric FM compound Sc$_{3.1}$In with lutetium, evidence for a QCP with unusual values of the critical exponents in (Sc$_{1-x}$Lu$_x$)$_{3.1}$In has been found near a critical concentration $x_c \approx 0.035$ \cite{Svanidze_et_al_2014}. NFL behavior has been observed in both the FM and PM phases, in addition to the vicinity of the QCP. This material may be characterized by a reduced dimensionality due to the one-dimensional nature of the Sc-In chains \cite{Jeong_Kwon_2007, Svanidze_et_al_2014}.
\paragraph{U$_4$Ru$_7$Ge$_6$ and U$_4$(Ru$_{1-x}$Os$_x$)$_7$Ge$_6$}
\label{par:URuGe}
U$_4$Ru$_7$Ge$_6$ is ferromagnetic below \TC\ $\approx 12\,$K; it is a metal with Kondo-like and heavy-fermion features,
while U$_4$Os$_7$Ge$_6$ is a paramagnet. The system U$_4$(Ru$_{1-x}$Os$_x$)$_7$Ge$_6$ has been investigated by
\textcite{Colineau_et_al_2001}, who found that \TC\ is suppressed to zero for $x \approx 0.3$. 

Hydrostatic pressure applied to U$_4$Ru$_7$Ge$_6$ also suppresses \TC, with a FM-PM QPT at $p \approx 2.6$\,GPa.
Resistivity measurements by \textcite{Hidaka_et_al_2011} are consistent with the existence of a QCP with Hertz-type behavior at that pressure. The
residual resistivity of their polycrystalline samples at $p=2.36$\,GPa was about 58\,$\mu\Omega$cm. The resistivity of the
U$_4$(Ru$_{1-x}$Os$_x$)$_7$Ge$_6$ samples is not known, but is presumably higher. This places this system in the
disorder Regime II (intermediate disorder) discussed in Sec.\ \ref{subsubsec:III.B.3}, which is consistent with the experimental
observations.
\subsubsection{Strongly disordered systems}
\label{subsubsec:II.C.2}
\begin{table*}[t]
\caption{Systems showing a second-order transition: Strongly disordered bulk systems. \TC\ =
              Curie temperature, $\rho_0 =$ residual resistivity. FM = ferromagnetism, SC = superconductivity. n.a. = not available. }
\smallskip
\begin{ruledtabular}
\begin{tabular}{lllllll}
System 
   & Order of 
      & \TC/K 
         & magnetic 
            & tuning  
               & Disorder  
                  & Comments \\

   & Transition $^a$ 
      &                            
         & moment/$\mu_{\text{B}}$ $^b$
            &  parameter  
               & ($\rho_0/\mu\Omega$cm) $^c$
                  & \\
%\\
\hline\\[-7pt]
%%%%%%%%%%%%%%%%%%%%%%%%%%%%%%%
LaV$_x$Cr$_{1-x}$Ge$_3$
   & n.a.
      & 55 -- 20$\,^{d}$
         & 1.4$\,^{e}$
            & comp. + press.$\,^{1}$
               & 100$\,^{f}$
                  & lowest \TC\ rather high\\
\\[-5pt]
%%%%%%%%%%%%%%%%%%%%%%%%%%%%%%%%%
URu$_{2-x}$Re$_x$Si$_2$ 
   & 2nd$\,^{2,3}$
      & 25 -- 2$\,^{g}$
         & 0.4 -- 0.03$\,^{3}$
            & composition$\,^{2}$
               & $\approx 100$$\,^{h}$
                  & strong disorder\\
\\[-5pt]
%%%%%%%%%%%%%%%%%%%%%%%%%%%%%%%%%%%
URh$_{1-x}$Ru$_{x}$Ge 
   & 2nd$\,^{4}$
      & $\approx 10$ -- 0$\,^{4,i}$
         & $\approx 0.1$ -- 0$\,^{4,i}$
            & composition$\,^{4}$
               & n.a.$\,^j$
                  & disorder unclear \\
\\[-5pt]
%%%%%%%%%%%%%%%%%%%%%%%%%%%%%%%%%
Th$_{1-x}$U$_{x}$Cu$_2$Si$_2$
   & 2nd$\,^{5}$
      & 101 -- 12$\,^{k}$
         & 0.92 -- 0.09$\,^{k}$
            & composition$\,^{5}$
               & 235$\,^{l}$
                  & disorder unclear \\
\\[-5pt]
%%%%%%%%%%%%%%%%%%%%%%%%%%%%%%%%%
UCo$_{1-x}$Fe$_x$Ge
   & 2nd$\,^{6}$
      & 8.5 -- 3$\,^{m}$
         & 0.1 -- 0.02$\,^{m}$
            & composition$\,^{6}$
               & 430$\,^{n}$
                  & extremely high $\rho_0$ \\
\\[-5pt]
%%%%%%%%%%%%%%%%%%%%%%%%%%%%%%%%%
\hline\hline\\[-5pt]
\multicolumn{7}{l} {$^a$  At the lowest temperature achieved.\quad 
                              $^b$ Per formula unit unless otherwise noted.\quad 
                              $^c$ For the highest-quality samples.}\\
\multicolumn{7}{l} {$^{d}$ For $x = 0.16$ and $p = 0$ -- 3\,GPa. \quad
                              $^{e}$ For $x=0$.\quad
                              $^{f}$ For $x = 0.16$.\quad
                              $^{g}$ For $x=0.6$ -- $0.2$ \cite{Butch_Maple_2009}.}\\
\multicolumn{7}{l} {$^{h}$ For $x=0.1$ \cite{Butch_Maple_2010}.\quad                                                                              
                              $^i$ For $x = 0$ -- $0.4$ \cite{Huy_et_al_2007a}.}\\
\multicolumn{7}{l} {$^j$ Large nominal $\rho_0 \approx 200 - 300\,\mu\Omega$cm due to cracks; not indicative of the intrinsic disorder  \cite{Huy_et_al_2007a}.}\\
\multicolumn{7}{l} {$^{k}$ For $x = 1$ -- $0.15$.\quad 
                              $^{l}$ High $\rho_0$ due to microcracks.\quad 
                              $^{m}$ For $x \approx 0.75$ -- 0.22.\quad 
                              $^{n}$ For $x = 0.22$.}\\
\\[-5pt]
\hline\\[-5pt]
\multicolumn{2}{l}{ $^{1}$ \textcite{Lin_et_al_2013} }
      & \multicolumn{2}{l}{$^{2}$ \textcite{Bauer_et_al_2005}}
            & \multicolumn{2}{l}{$^{3}$ \textcite{Butch_Maple_2009}}
                  &  $^{4}$ \textcite{Huy_et_al_2007a} \\      
\multicolumn{2}{l}{ $^{5}$ \textcite{Lenkewitz_et_al_1997}}
      & \multicolumn{2}{l}{$^{6}$ \textcite{Huang_et_al_2013}}    
            & \multicolumn{2}{l}{}
                   & \\
%%%%%%%%%%%%%%%%%%%%%%%%%%%%%%%%%%%%%%%%%%%
\end{tabular}
\end{ruledtabular}
\vskip -3mm
\label{table:2_pt_2}
\end{table*}
\paragraph{LaV$_x$Cr$_{1-x}$Ge$_3$}
\label{LaVCrGe}
Upon substitution of vanadium for chromium in LaCrGe$_3$, the FM \TC\ drops from 88\,K to 36\,K for $x = 0.21$, which is the
highest Cr concentration for which single crystals could be grown \cite{Lin_et_al_2013}. Pressure applied to a sample with $x = 0.16$
leads to a further decrease of \TC, with a QPT at $p \approx 3\,$GPa. The lowest \TC\ achieved was about 20K, and the order of the
QPT is not known. 
\paragraph{URu$_{2-x}$Re$_x$Si$_2$}
\label{URuReSi}
The parent compound of URu$_{2-x}$Re$_x$Si$_2$, URu$_2$Si$_2$, is a heavy-fermion superconductor (superconducting $T_{\text{c}} \approx 1.5\,$K)
that has an ordered phase of unknown nature, usually referred to as the ``hidden-order'' phase, below about 17K; see \textcite{Mydosh_Oppeneer_2013}
for a recent overview. Substitution of Re, Tc, or Mn leads to the destruction of the hidden-order phase and the emergence of ferromagnetism past a certain dopant 
concentration \cite{Dalichaouch_et_al_1990}, but only URu$_{2-x}$Re$_x$Si$_2$ has been studied in detail. In this system, the hidden-order phase disappears for 
$x \approx 0.1$ and the system develops a FM ground state for $x \agt 0.15$, but the critical Re concentration has proven hard to determine \cite{Butch_Maple_2010}. 
\TC\ increases monotonically with increasing $x$ and reaches a maximum of almost 40\,K at $x \approx 0.8$, above which the material does not remain in a single phase.
The existence of FM long-range order has been ascertained by neutron scattering for $x=0.8$ \cite{Torikachvili_et_al_1992} and by $^{29}$Si NMR for
$x \geq 0.4$ \cite{Kohori_et_al_1993}. Pronounced NFL behavior has been observed in the specific heat and the electrical resistivity for a large concentration
range $0.15 \alt x \alt 0.8$ \cite{Bauer_et_al_2005}, and the dynamical magnetic susceptibility shows unusual behavior for $0.2 < x < 0.6$
\cite{Krishnamurthy_et_al_2008}. The system is highly disordered as judged from the residual resistivity, which is on the order of 100 $\mu\Omega$cm
\cite{Butch_Maple_2010}, and the magnetic moment appears to go to zero continuously. Attempts to determine critical exponents have been hampered
by difficulties in determining the critical concentration precisely \cite{Bauer_et_al_2005, Butch_Maple_2009, Butch_Maple_2010}. The latest results from
scaling plots yield exponents $\delta$ and $\gamma$ that vary continuously with $x$ and approach $1$ and $0$, respectively, for $x$ approaching the
critical value $x_{\text{c}} \approx 0.15$, while $\beta \approx 0.8$ is independent of $x$ \cite{Butch_Maple_2009, Butch_Maple_2010}. These results are 
hard to understand within any phase-transition scenario, even if one interprets the exponents as effective ones in a pre-asymptotic region. $\gamma = 0$
in particular contradicts the very notion of a FM order parameter. There currently is no resolution of this problem. The uncertainty about $x_{\text{c}}$ may
be to blame, and the strong disorder may lead to unusual effects. For instance, it is conceivable that there is a Re concentration region that represents a
quantum Griffiths phase (see Secs. \ref{subsec:II.E} and \ref{subsubsec:III.D.1}) rather than true long-range FM order. It has also been speculated that an 
interplay between remnants of the hidden order and ferromagnetism leads to unusual behavior near the onset of ferromagnetism \cite{Butch_Maple_2010}.
\paragraph{URh$_{1-x}$Ru$_x$Ge}
\label{par:II.C.2.b}
Doping URhGe with Ru decreases the Curie temperature after a small initial increase and suppresses it
to zero at a Ru concentration close to $x = 0.38$ \cite{Huy_et_al_2007a, Sakarya_et_al_2008}, see
Fig.~\ref{fig:URhGe_doped_phase_diagram}. The quantum phase transition was studied by \textcite{Huy_et_al_2007a}.
It was found to be second order with a pronounced $T\ln T$ contribution to the specific heat at the critical
concentration. The $T$-dependence of the electrical resistivity shows a non-Fermi-liquid $T^n$ behavior, with
$n<2$ over a wide range of concentrations, with a minimum of $n = 1.2$ at the critical concentration. Such NFL
behavior has been interpreted as indicative of critical fluctuations and corroborating the existence of a quantum
critical point. The continuous
nature of the transition is consistent with theoretical expectations, assuming that the large critical Ru concentration
leads to a substantial amount of disorder. Unfortunately, the strength of the microscopic disorder is hard to determine
experimentally, since cracks in the brittle system lead to an artificially high residual resistivity of 200 - 300 $\mu\Omega$cm. The Gr{\"u}neisen parameter $\Gamma$ is observed to stay finite at the transition, in disagreement with the theoretical result by \textcite{Zhu_et_al_2003} that predicts a diverging $\Gamma$.

URhGe$_{1-x}$Si$_x$ has also been studied. Si doping up to $x \approx 0.2$ has little effect on \TC,
see Fig.~\ref{fig:URhGe_doped_phase_diagram},
and no quantum phase transition has been observed in this material \cite{Sakarya_et_al_2008}.
\paragraph{UCo$_{1-x}$Fe$_x$Ge}
\label{UCoFeGe}
Doping of the weak FM UCoGe (see Sec.\ \ref{par:II.B.2.b}) with Fe initially increases \TC\ to a maximum of \TC\ $\approx 8.5\,$K around
$x = 0.075$. With further increasing $x$, \TC\ decreases and vanishes at an extrapolated $x_c \approx 0.22$ \cite{Huang_et_al_2013}. The
QPT is believed to be second order, and there is some evidence for quantum critical behavior in the transport and specific-heat data. Since
the transition in UCoGe is first order this implies the existence of a tricritical point in the phase diagram, but this has not been investigated.
The origin of the very large residual resistivity is not clear.
\paragraph{Th$_{1-x}$U$_{x}$Cu$_2$Si$_2$}
\label{ThUCuSi}

UCu$_2$Si$_2$ orders ferromagnetically below \TC\ $\approx 101\,$K, ThCu$_2$Si$_2$ is paramagnetic. In Th$_{1-x}$U$_{x}$Cu$_2$Si$_2$,
\cite{Lenkewitz_et_al_1997} have found a QPT for $x$ close to $0.15$ with the specific-heat coefficient displaying a logarithmic temperature
dependence. The large residual resistivity ($\rho_0 > 200 \mu\Omega$cm) is due to microcracks in the samples and not a measure of
intrinsic disorder.
\subsubsection{Quasi-one-dimensional systems}
\label{subsubsec:II.C.3}
\begin{table*}[t]
\caption{Systems showing a second-order transition: Quasi-one-dimensional (1-$d$) materials. \TC\ = Curie temperature, \\
              $\rho_0 =$ residual resistivity. n.a. = not available. }
\smallskip
\begin{ruledtabular}
\begin{tabular}{lllllll}
System 
   & Order of 
      & \TC/K 
         & magnetic 
            & tuning  
               & Disorder  
                  & Comments \\

   & Transition$\,^a$ 
      &                            
         & moment/$\mu_{\text{B}}\,^b$
            &  parameter  
               & ($\rho_0/\mu\Omega$cm)$\,^c$
                  & \\
%\\
\hline\\[-7pt]
%%%%%%%%%%%%%%%%%%%%%%%%%%%%%%%%%%
YbNi$_4$P$_2$ 
   & 2nd$\,^{1}$  
      & 0.14$\,^{2}$
         & $\approx 0.035\,^{2}$ 
            & none 
               & 2.6$\,^{1}$
                  & quasi-1-$d$\\
\\[-5pt]
%%%%%%%%%%%%%%%%%%%%%%%%%%%%%%%%%%
YbNi$_4$(P$_{1-x}$As$_{x}$)$_2$ 
   & 2nd$\,^{3}$  
      & 0.15 -- 0.025$\,^{3,d}$
         & $\approx 0.05\,^{3}$ 
            & composition 
               & $\approx 5.5$, 15$\,^{3,e}$
                  & quasi-1-$d$, disordered\\
\\[-5pt]
%%%%%%%%%%%%%%%%%%%%%%%%%%%%%%%%%
\hline\hline\\ [-5pt]   
\multicolumn{7}{l} {$^a$  At the lowest temperature achieved.\quad 
                              $^b$ Per formula unit unless otherwise noted.\quad 
                              $^c$ For the highest-quality samples.}\\                                                        
\multicolumn{7}{l} {$^{d}$ For $x = 0$ -- $0.08$ \cite{Steppke_et_al_2013}.\quad
                              $^{e}$ 5.5 for $J \parallel c$, 15 for $J \perp c$ \cite{Steppke_et_al_2013}.}\\
\\[-5pt]                              
\hline\\[-5pt]
\multicolumn{2}{l}{$^{1}$ \textcite{Krellner_et_al_2011}}        
   & \multicolumn{2}{l}{$^{2}$ \textcite{Spehling_et_al_2012}} 
        & \multicolumn{2}{l}{$^{3}$ \textcite{Steppke_et_al_2013} } \\
%%%%%%%%%%%%%%%%%%%%%%%%%%%%%%%%%%%%%%%%%%%
\end{tabular}
\end{ruledtabular}
\vskip -0mm
\label{table:2_pt_3}
\end{table*}
\paragraph{\YNP}
\label{par:II.C.3.a}
\YNP\, is a unique system with respect to FM quantum criticality~\cite{Steppke_et_al_2013}. It is the stoichiometric metallic ferromagnet with the lowest Curie 
temperature ever observed, namely, \TC\, = 0.15\,K. In this compound Ni is not magnetic~\cite{Deputier_et_al_1997, Krellner_et_al_2011}, and the Yb atoms are 
arranged in chains along the $c$-direction and located between edge-connected Ni tetrahedra, forming a ZrFe$_{4}$Si$_{2}$ structure type with a lattice constant 
ratio $c/a \approx 0.5$. The quasi-one-dimensional structure corresponds to a strong anisotropy of the transport properties: Non-correlated band structure calculations 
have identified Fermi surfaces with strong one-dimensional character~\cite{Krellner_et_al_2011}, which is indeed reflected in the anisotropy of the resistivity, 
$\rho_{a}/\rho_{c} \approx 5$ at 1.8\,K~\cite{Krellner_Geibel_2012}.
\begin{figure}[ht]
\begin{center}
\includegraphics[width=0.9\columnwidth,angle=0]{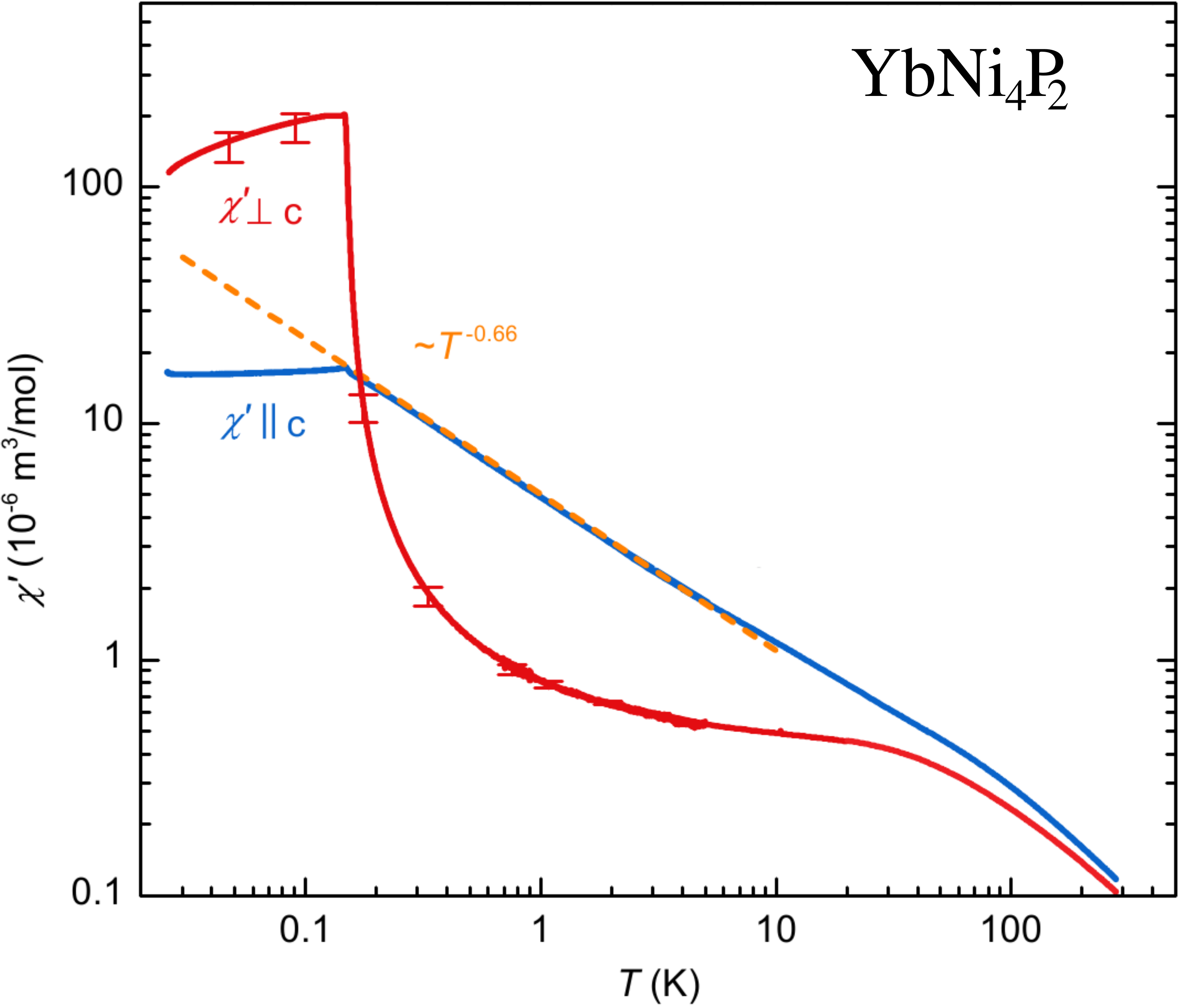}
\end{center}
\caption{Temperature dependence of the ac susceptibility $\chi'(T)$ measured with a modulation field amplitude $\mu_{0}H = 15\,\mu$T parallel and perpendicular, respectively, to the crystalline $c$-axis. The large absolute value $\chi'(T_{\textrm{C}}) \approx 200 \times 10^{-6}$\,m$^{3}$/mol indicates a FM phase transition. The log-log plot 
              emphasizes the strong divergence of $\chi'_{\perp}(T)$ just above \TC. The dashed line indicates that $\chi'_{\perp}(T) \propto T^{-2/3}$ 
              above \TC. From~\textcite{Steppke_et_al_2013}.} 
\label{fig:YbNi4P2_susceptibility}
\end{figure}

The Yb$^{3+}$ ion is located in an orthorhombic crystalline electric field (CEF) which splits the $J = 7/2$ energy levels leaving a Kramers doublet as the 
ground state~\cite{Huesges_et_al_2013} 
and causes the crystalline $c$-axis to be the magnetic easy axis~\cite{Krellner_Geibel_2012}. This can be clearly seen in Fig.~\ref{fig:YbNi4P2_susceptibility}, which
shows the ac susceptibility $\chi'(T)$ in a small field $H$ parallel and perpendicular, respectively, to the $c$-axis. Although \YNP\, is a heavy-fermion system with a 
Kondo temperature of 8\,K, a small unscreened moment of about 0.05\,\mub/Yb orders ferromagnetically at 
0.15\,K~\cite{Krellner_et_al_2011,Spehling_et_al_2012,Steppke_et_al_2013,Gegenwart_et_al_2015}. In addition, despite the strong CEF anisotropy the moments align within the $ab$-plane, 
the magnetic hard direction (see in Fig.~\ref{fig:YbNi4P2_susceptibility} how $\chi'_{\parallel}(T)$ crosses $\chi'_{\perp}(T)$ just above \TC). \YNP\, shares this 
uncommon behavior with just a few other ferromagnetic Kondo-lattice systems, such as \CRPO\, (Sec.~\ref{CRPO}), \CAS\, (Sec.~\ref{CAS}), 
YbNiSn~\cite{Bonville_et_al_1992}, and
\YRCSFM\, (Sec.~\ref{par:YRCS}). This switch in magnetic response can be explained within a local-moment Heisenberg model with competing exchange 
interactions~\cite{Andrade_et_al_2014}. However, this explanation does not work for \YNP\  where quantum effects are strong. For instance, $\chi'(T)$ in 
classical Ising or Heisenberg systems is characterized by a power-law behavior at the transition with well-known universal exponents, while the divergence of
${\chi'~\perp~c}$ in Fig.~\ref{fig:YbNi4P2_susceptibility} just above \TC\, is much stronger than a power law. It is even stronger than what is expected 
for a pure one-dimensional Ising ferromagnet, where $\chi'(T) \propto T^{-1}\exp(2J/k_{\textrm{B}}T)$, with $J$ the coupling constant~\cite{Ising_1925}. 
Another interesting way to explain the switch in 
response is to have strong transverse spin fluctuations which at sufficiently low temperature dominate the magnetic 
anisotropy~\cite{Krueger_Pedder_Green_2014}. This model, however, implies a first-order phase 
transition at \TC, which is not observed in \YNP~\cite{Steppke_et_al_2013}.

In addition to the unique behavior displayed by the susceptibility, \YNP\, presents a number of other unconventional properties. Strong non-Fermi liquid behavior has
been found in the resistivity, $\rho(T) \propto T$, the specific heat $C/T \propto T^{-0.42}$, and the NMR relaxation rate, $1/T_{1}T \propto T^{-3/4}$, in polycrystals in a 
broad $T$-range above \TC~\cite{Krellner_et_al_2011,Sarkar_et_al_2012}.
\paragraph{\YNPA}
\label{par:II.C.3.b}

\begin{figure}[ht]
\begin{center}
\includegraphics[width=0.95\columnwidth,angle=0]{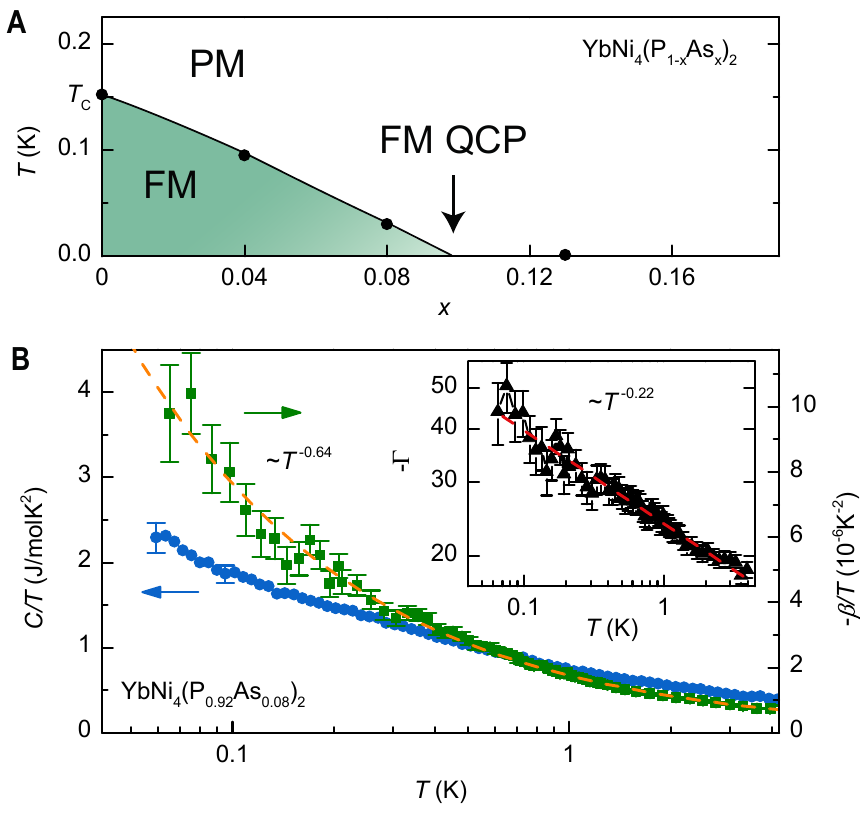}
\end{center}
\caption{Phase diagram of \YNPA. The lowest \TC\, of about 25\,mK was observed at $x = 0.08$. B: Specific heat (blue, left arrow) and volume thermal expansion 
              coefficient (green, right arrow) for this sample. Inset: $T$-dependence of the Gr\"uneisen ratio $\Gamma(T) = \beta(T)/C(T) \propto T^{-0.22}$. 
              From~\textcite{Steppke_et_al_2013}.}
\label{fig:YbNi4P2_grueneisen}
\end{figure}

The results reviewed in Sec.\ \ref{par:II.C.3.a} motivated the growth of single crystals with phosphorus substituted by arsenic (which amounts to negative chemical 
pressure) in order to reduce \TC\, and look for FM quantum criticality. Four single crystals of the series \YNPA\, were grown with a minimum value of 
$\rho_{0} = 5.5\,\mu\Omega$cm for the stochiometric \YNP. \textcite{Steppke_et_al_2013} 
investigated the magnetic and thermodynamic properties of those down to 20\,mK and in particular measured the Gr\"uneisen ratio $\Gamma(T) = \beta(T)/C(T)$, where 
$\beta(T)$ is the volume thermal expansion coefficient. According to~\textcite{Zhu_et_al_2003}, this quantity should diverge as $\Gamma(T) \propto T^{-\lambda}$ at any QCP, where $\lambda = 1/\nu z$ is given in terms of the correlation-length exponent $\nu$ and the dynamical exponent $z$ (cf. Sec.~\ref{sec:III}). 
\textcite{Steppke_et_al_2013} found that the FM phase transition is suppressed at $x_{\textrm{c}} \approx 0.1$ (Fig.~\ref{fig:YbNi4P2_grueneisen}A) and that it remains 
second order even in the sample with $x = 0.08$ with a \TC\ of about 25\,mK. Both $C(T)/T$ and $\beta(T)/T$ diverge (see Fig.~\ref{fig:YbNi4P2_grueneisen}B) with
exponents that are approximately independent of the As concentration, which rules out a possible quantum Griffiths phase (cf.~\ref{subsubsec:III.D.1}).
Most importantly, $\Gamma(T) \propto T^{-0.22}$ in the sample with $x = 0.08$ which is located almost at $x_{\textrm{c}}$ (see Fig.~\ref{fig:YbNi4P2_grueneisen}B). 
This  provides evidence that in \YNPA\, a FM QCP exists.
The nature of this QCP is still unclear. The exponent $\lambda = 0.22$ yields a value of $\nu z \approx 5$ which is rather large. For instance, within Hertz-Millis-Moriya 
theory one has $\nu = 1/2, z = 3$ (cf.~\ref{par:III.C.2.b}). This is not surprising, as no existing theory is expected to apply to this material. Any theoretical framework
will have to take into account the local nature of the Yb $4f$-states with spin-orbit coupling and a strong Kondo effect and, perhaps most importantly, the quasi-one-dimensional
electronic structure. The absence of a first-order phase transition is likely due to the latter, since the soft fermionic modes that are crucial for the first-order mechanism
described in Sec.\ \ref{sec:III} are absent in a one-dimensional system, and in a quasi-one-dimensional one they will be present only in an asymptotically low temperature range the size of which
depends on the anisotropy.
\subsubsection{Discussion, and comparison with theory}
\label{subsubsec:II.C.4}
The materials in which a continuous transition is observed to the lowest temperatures achieved have been grouped into three distinct classes, see Tables \ref{table:2_pt_1}, \ref{table:2_pt_2}, \ref{table:2_pt_3}: Weakly
disordered, strongly disordered, and quasi-one-dimensional. For the last group, the conduction-electron system is 
%not a Fermi liquid, or crosses over to a Fermi liquid only at a very low temperature that depends on the electronic anisotropy. The theories discussed in Sec.~\ref{sec:III}, which all depend on an underlying Fermi liquid therefore do not apply, and no theoretical treatment is currently available.
expected to be a Fermi liquid at asymptotically low temperatures, but it will cross over to a Luttinger liquid at a temperature that depends on the
electronic anisotropy. A determination of the temperature range where the theories discussed in Sec.~\ref{sec:III}, which all depend on an underlying 
Fermi liquid, still apply requires detailed theoretical considerations that are currently not available. For the strongly disordered bulk 
systems, the theory discussed in Sec.\ \ref{subsubsec:III.C.3} predicts a continuous transition, and Griffiths effects may also be present, see
Sec.\ \ref{subsubsec:III.D.1}. Of the two systems in this category, URu$_{2-x}$Re$_x$Si$_2$ is the more thoroughly studied one. As discussed
in Sec.\ \ref{URuReSi}, the current experimental results cannot be easily interpreted with any existing theory. A major obstacle is the uncertainty
about the critical concentration $x_{\text{c}}$, and additional studies about the onset of long-range FM order would be desirable.

In the weakly disordered group, CeSi$_x$ and Ni$_x$Pd$_{1-x}$ come with open questions regarding the nature of the transition, or the
presence of phases other than the FM one, at low temperatures. Given the residual resistivities of these materials, theoretically one would 
expect a continuous transition in CeSi$_x$, and a first-order one in Ni$_x$Pd$_{1-x}$, provided the transition is not pre-empted by a
different phase. In (Cr$_{1-x}$Fe$_x$)$_2$B one expects a second-order transition, and the disorder is in a range where the observed
mean-field critical behavior is consistent with theoretical expectations, see Sec.\ \ref{subsubsec:III.B.3}. For the remaining systems no
information about the disorder strength is available, which makes a comparison with theoretical predictions difficult.

These somewhat inconclusive results may well have to be revisited if cleaner samples and/or measurements at lower temperatures should
become available in the future. The history of ZrZn$_2$, Sec.\ \ref{par:II.B.1.b}, shows that improving sample quality can change the conclusion
about the order of the transition. One also needs to keep in mind that the tricritical temperatures listed in Tables~\ref{table:1a},~\ref{table:1b} span a substantial
range, and the $T_{\text{tc}}$ in, for instance, URhGe is barely higher than the lowest temperature at which UIr has been measured. In summary,
additional low-temperature studies of the materials in this section are highly desirable. 

\subsection{Systems changing to spin-density-wave or antiferromagnetic order}
\label{subsec:II.D}
In some systems the FM phase undergoes a transition to a spatially modulated magnetic phase as the Curie temperature decreases, see the schematic phase diagram in Fig.~\ref{fig:schematic_phase_diagrams}c.%
\footnote{\label{AFM_FM_transition_footnote} There also are cases of transitions from a
                metallic AFM state to a metallic FM state, with the FM being the ground state, see Sec.\ \ref{subsec:IV.B} point 3. We
                do not discuss these materials.}
This produces a Lifshitz point, where the FM, modulated, and PM phases meet, as well as two QPTs, one from the FM phase to the modulated one, and one from the modulated phase to the PM. They are discussed below, and their properties are summarized in Table~\ref{table:3}. In some of these materials the evidence for a modulated phase is stronger than in others, and in some cases there are conflicting experimental results. The classification of some of these systems within our scheme should thus be considered tentative.

\subsubsection{Simple ferromagnets}
\label{subsubsection:II.D.1}

\paragraph{An itinerant magnet: \NFy}
\label{NF}

The Laves phase compound \NF\ shows itinerant antiferromagnetism on the border of a FM phase. First indications of a low-$T$ AFM ordered state with \TN\ $\simeq 10$\,K were found in magnetization and NMR experiments on polycrystals~\cite{Shiga_Nakamura_1987,Yamada_Sakata_1988,Crook_Cywinski_1995}. This has recently been confirmed by a microscopic study with electron spin resonance, muon spin relaxation and M\"ossbauer spectroscopy on single crystals~\cite{Rauch_et_al_2014}. However, this state is characterized by an unusually high magnetic susceptibility, $\chi \approx 0.02$ in SI units, which corresponds to a large Stoner enhancement factor of the order of 180~\cite{Brando_et_al_2008}. These authors speculated that the magnetic order in \NF\ is a long-wavelength modulated state with a small ordering wave number $Q \approx 0.05\,\mathrm{\AA}^{-1}$. This has been corroborated by recent neutron scattering experiments~\cite{Niklowitz_2014}.

\begin{figure}[b]
\begin{center}
\includegraphics[angle=0,width=0.9\columnwidth]{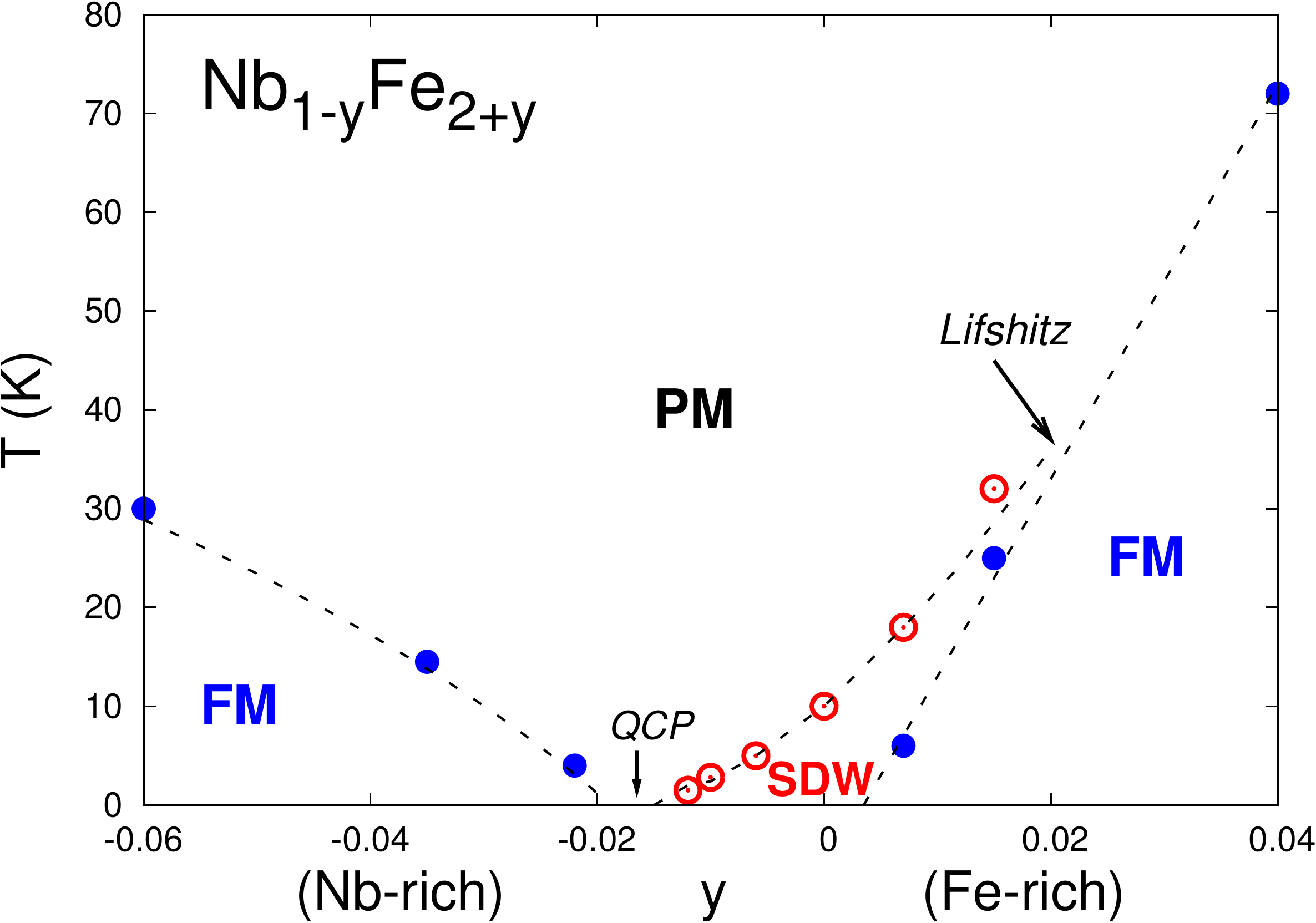}
\end{center}
\caption{Composition vs temperature phase diagram  of \NFy. Within a narrow homogeneity range, \NF\ can be tuned from ferromagnetism (FM) ($y \gtrsim 0.01$) 
via an intermediate spin-density-wave (SDW) modulated state to a quantum critical point (QCP) ($y \simeq -0.015$) and a second FM phase ($y < -0.015$). 
From~\textcite{Moroni_et_al_2009}.}
\label{NFy}
\end{figure}

The ratio of the low-$T$ susceptibility over the Sommerfeld coefficient of the specific heat capacity yields a Wilson ratio of 60, which indicates that the susceptibility is far more strongly enhanced than the carrier effective mass and thereby suggests that stoichiometric \NF\ is very close to the border of ferromagnetism. Indeed, a ferromagnetic
ground state is found in iron-rich samples~\cite{Yamada_Sakata_1988,Crook_Cywinski_1995,Moroni_et_al_2009}. By varying the iron content in \NFy\ within a narrow homogeneity range, \NF\ can be tuned
from ferromagnetism for $y \gtrsim 0.01$ via an intermediate SDW
modulated state around $y \simeq 0$ to a quantum
critical point at $y \approx -0.015$. For $y < -0.015$ the ground
state becomes FM again (Fig.~\ref{NFy}). The fact that both iron and
niobium-rich samples are ferromagnetic at low temperature is
noteworthy. It has been linked to the peculiar electronic structure of
this material~\cite{Subedi_Singh_2010,Tompsett_et_al_2010,Neal_et_al_2011,
  Alam_Johnson_2011}. Part of the phase diagram can also be reproduced
by measurements under hydrostatic pressure~\cite{Duncan_et_al_2010}:
Starting with a FM sample with $y = 0.015$, increasing pressure is
equivalent to moving the system to the left in the phase diagram of
Fig.~\ref{NFy}. A pressure of 2.5\,GPa roughly corresponds to a shift
in composition from $y = 0.015$ to $0.007$.  For $y \agt -0.015$,
\NFy\ is thus a rare example of an itinerant system in which a SDW
state with a small wave vector connects continuously to the FM state
at a Lifshitz point, cf. Fig.~\ref{NFy}.

Signatures of quantum critical behavior have been observed near the QPT where the SDW order disappears:
the electrical resistivity displays a $T^{3/2}$ power-law behavior, and the specific-heat coefficient $\gamma$ shows
a logarithmic temperature dependence down to 0.1\,K \cite{Brando_et_al_2008}. The latter is consistent with
Hertz-Millis-Moriya theory for a clean FM,$^{\ref{HMM_validity_footnote}}$ but the former disagrees with the $T^{5/3}$ behavior expected for
this case, see Eqs.~(\ref{eq:3.62}, \ref{eqs:3.64}) in Sec.~\ref{par:III.C.2.a}. A $T^{3/2}$  behavior of the
resistivity has also been observed in other systems, e.g., in MnSi and ZrZn$_2$, even far from the QPT,
which suggests a more general phenomenon that remains incompletely understood, see Sec.~\ref{subsec:IV.A}.
In contrast to MnSi and ZrZn$_2$, the observed behavior in NbFe$_2$ might be attributable to scattering by
critical fluctuations with a nonzero wave number. The apparent inconsistency between the specific heat and
the resistivity might be due to fluctuations with different wave vectors dominating the transport and
thermodynamic properties, respectively. 
%This issue has also been addressed by \textcite{Oliver_Schofield_2015}, but this treatment ignores the important soft-mode effects discussed in Sec.~\ref{sec:III}.
\footnote{\label{HMM_validity_footnote} As is explained in Sec.~\ref{subsubsec:III.C.2}, in the FM case this does not represent true critical behavior,
               but it may be observable in a sizable pre-asymptotic region.}

The existence of the QCP indicates that the PM-to-SDW transition is second order, at least at low temperature. 
There are indications that the FM-to-SDW transition at $y=0.015$ is first order \cite{Friedemann_et_al_2014}. On the Nb-rich side of the phase diagram,
the transition was found to be second order for all samples investigated by \textcite{Moroni_et_al_2009}.
\begin{table*}[!]
\caption{Systems showing change into a spin-density-wave (SDW) or antiferromagnetic (AFM) order. \TC\ = Curie temperature, $T_{\text{N}} =$
 N\'eel temperature, $\rho_0 =$ residual resistivity. QC = quantum critical, n.a. = not available.}
\smallskip
\begin{ruledtabular}
\begin{tabular}{llllllll}
System 
   & Order of 
      & $T_{\text{C}}$/K 
         & $T_{\text{N}}$/K 
            & magnetic 
               & tuning  
                  & Disorder  
                     & Comments \\

   & Transition$\,^a$ 
      &        
         &                    
            & moment/$\mu_{\text{B}}$$^b$
               &  parameter  
                  & ($\rho_0/\mu\Omega$cm)$^c$
                     & \\
%\\
\hline\\[-7pt]
%%%%%%%%%%%%%%%%%%%%%%%%%%%%%%%%%%%%%%%%%%
Nb$_{1-y}$Fe$_{2+y}$ 
   & 1st (?)$\,^{1,d}$
      & 72 -- 6$\,^e$
         & 32 -- 2.8$\,^f$
            & $\approx 0.02\,^{2}$ 
               & composition$\,^{3}$
                  & 5.5 -- 17$\,^{3,4}$
                     & SDW phase,  \\
   & & & & & pressure$\,^{5}$ & & Lifshitz point \\[2pt]
\\[-7pt]
%%%%%%%%%%%%%%%%%%%%%%%%%%%%%%%%%%%%%%%%%%
PrPtAl 
   & 1st$\,^{6,g}$
      & 4.7$\,^{6}$
         & 5.85, 5.5$\,^{6,h}$
            & $\approx 1\,^{6}$
               & none
                  & n.a.
                     & spiral phase  \\
& & & & & & & two SDW phases \\[2pt]
\hline\\[-7pt]
%%%%%%%%%%%%%%%%%%%%%%%%%%%%%%%%%%%%%%%%%%%
CeRuPO 
   & 1st$\,^{7,i}$
      & 15 -- 11$\,^{7,8,j}$
         & 10 -- 1$\,^{7,8,j}$
            & 1.2$\,^{8}$
               & pressure$\,^{7,8}$
                  & $\approx 30\,^{7,k}$
                     & AFM phase \\
& & & & & & & \\[-5pt]
%%%%%%%%%%%%%%%%%%%%%%%%%%%%%%%%%%%%%%%%%%%
CeFeAs$_{1-x}$P$_{x}$O 
   & n.a.
      & $\approx 10$ -- 6$\,^{9,10,l}$
         & $\approx 3\,^{11,m}$
            & 0.95 -- 0$\,^{9}$
               & composition$\,^{9,10}$
                  & n.a.
                     & conflicting results \\
 \\[-5pt]
 %%%%%%%%%%%%%%%%%%%%%%%%%%%%%%%%%%%%%%%%%%%
CeRu$_{1-x}$Fe$_{x}$PO 
   & n.a.   
     & 15 -- 0.3$\,^{12}$
        & $\approx 0.5$ -- $0 \,^{12}$
          & 1.2$\,^{13}$
            & composition$^{\,12}$
              & n.a.
                 & QC flucts., pos-\\
& & & & & & & sible AFM phase \\[2pt]
\hline\\[-7pt]
%%%%%%%%%%%%%%%%%%%%%%%%%%%%%%%%%%%%%%%%%%
CeAgSb$_{2}$ 
   & 1st$\,^{14}$
      & $9.6$ -- 2$\,^{14}$
         & $\approx $ 6 -- 4$\,^{14}$
            & 0.41$\,^{15}$
               & pressure$\,^{14,16}$
                  & 0.2$\,^{14}$
                     & AFM phase not\\
& & & & & $\perp$ field$\,^{16}$ & & always observed\\[2pt]
\\[-7pt]
%%%%%%%%%%%%%%%%%%%%%%%%%%%%%%%%%%%%%%%%%%%
CeRu$_{2}$(Ge$_{1-x}$Si$_x$)$_{2}$ 
   &n.a.
      & 8 -- 2.5$\,^{17}$
         & 10 -- 1$\,^{17}$
            & n.a.
               & comp./press.$\,^{1,18}$
                  & 0.3$^{\,18}$
                     & hybridization \\
  & & & & & & & suppresses FM \\[2pt]
\hline\\[-7pt]
%%%%%%%%%%%%%%%%%%%%%%%%%%%%%%%%%%%%%%%%%%%
Yb(Rh$_{1-x}$Co$_{x})_{2}$Si$_2$ 
   &n.a.
      & 1.3$\,^{19,n}$
         & 1.2 -- 0.07$\,^{20,o}$
            & 0.1 --
               & composition$\,^{20}$
                  & 0.5 -- 10$\,^{20,21}$
                     & field-induced \\
  & & & & 0.002$\,^{22,19,p}$ & pressure$\,^{23,24}$ & & AFM QPT\\
\\[-5pt]
%%%%%%%%%%%%%%%%%%%%%%%%%%%%%%%%%%%%%%%%%%
\hline\hline\\[-5pt]
\multicolumn{8}{l} {$^a$ For the FM-to-AFM or SDW transition at the lowest temperature achieved.\quad 
                              $^b$ Per formula unit unless otherwise noted.}\\
\multicolumn{8}{l} {$^c$ For the highest-quality samples.}\\
\multicolumn{8}{l} {$^d$ For the FM-SDW transition. The FM-PM transition on the Nb-rich side is second order to the lowest \TC\ measured ($\approx 2\,$K).}\\                              \multicolumn{8}{l} {$^e$ For $0.04 > y > 0.007$ \cite{Brando_et_al_2008, Moroni_et_al_2009}.}\\
\multicolumn{8}{l} {$^f$ For $0.015 > y > -0.01$ \cite{Brando_et_al_2008, Moroni_et_al_2009}.}\\
\multicolumn{8}{l} {$^g$ For the FM-SDW2 and SDW2-SDW1 transitions. The SDW1-PM is 2nd order~\cite{Abdul-Jabbar_et_al_2015}.}\\
\multicolumn{8}{l} {$^h$ For the SDW1-PM and SDW2-SDW1 transitions, respectively~\cite{Abdul-Jabbar_et_al_2015}.}\\
\multicolumn{8}{l} {$^i$ For the FM-AFM transition at $T \agt 9\,$K. The order of the transition at low $T$ is not known.}\\
\multicolumn{8}{l} {$^j$ FM for $0 \leq p \alt 0.7$\,GPa and AFM for $0.7 \alt p \leq 2.8$\,GPa \cite{Lengyel_et_al_2015, Kotegawa_et_al_2013}. }\\
\multicolumn{8}{l} {$^k$ Near the FM-AFM transition.\quad 
                              $^l$ For $x = 0.4$ -- $0.8$ \cite{Luo_et_al_2010,Jesche_et_al_2012}.
                              $^m$ For $x = 0.9$ \cite{Jesche_2011}.}\\
\multicolumn{8}{l} {$^n$ For Yb(Rh$_{0.73}$Co$_{0.27}$)$_2$Si$_2$~\cite{Lausberg_et_al_2013}.\quad $^o$ For $x = 0.27$ -- $0$ \cite{Klingner_et_al_2011}.}\\  
\multicolumn{8}{l} {$^p$ 0.002 $\mu_{\text{B}}$ for YbRh$_2$Si$_2$ \cite{Ishida_et_al_2003};
                                       0.1 $\mu_{\text{B}}$ for Yb(Rh$_{0.73}$Co$_{0.27}$)$_2$Si$_2$ \cite{Lausberg_et_al_2013}.}\\
\\[-5pt]                                       
\hline\\[-5pt]
\multicolumn{2}{l} {$^{1}$ \textcite{Friedemann_et_al_2014}}
      & \multicolumn{2}{l} {\hskip -25pt $^2$ \textcite{Brando_et_al_2008}}
            & \multicolumn{2}{l} {\hskip -15pt $^3$ \textcite{Moroni_et_al_2009}}
                  & \multicolumn{2}{l} {$^4$ \textcite{Friedemann_et_al_2013}}\\
\multicolumn{2}{l} {$^{5}$ \textcite{Duncan_et_al_2010}}
      & \multicolumn{2}{l} {\hskip -25pt $^{6}$ \textcite{Abdul-Jabbar_et_al_2015}}
            & \multicolumn{2}{l} {\hskip -15pt $^{7}$ \textcite{Kotegawa_et_al_2013}}
                  & \multicolumn{2}{l} {\hskip -0pt $^{8}$ \textcite{Lengyel_et_al_2015}}\\
\multicolumn{2}{l} {$^{9}$ \textcite{Luo_et_al_2010}}
      & \multicolumn{2}{l} {\hskip -25pt $^{10}$ \textcite{Jesche_et_al_2012}}
            & \multicolumn{2}{l} {\hskip -15pt $^{11}$ \textcite{Jesche_2011}}
                  & \multicolumn{2}{l} {\hskip -0pt $^{12}$ \textcite{Kitagawa_et_al_2012}}\\
\multicolumn{2}{l} {\hskip -0pt $^{13}$ \textcite{Krellner_et_al_2007}}
      & \multicolumn{2}{l} {\hskip -25pt $^{14}$ \textcite{Sidorov_et_al_2003}}
            & \multicolumn{2}{l} {\hskip -15pt $^{15}$ \textcite{Araki_et_al_2003}}
                  & \multicolumn{2}{l} {$^{16}$ \textcite{Logg_et_al_2013}}\\
\multicolumn{2}{l} {\hskip -0pt $^{17}$ \textcite{Suellow_et_al_1999}}
      & \multicolumn{2}{l} {\hskip -25pt $^{18}$ \textcite{Wilhelm_Jaccard_1998}}
            & \multicolumn{2}{l} {\hskip -15pt $^{19}$ \textcite{Lausberg_et_al_2013}}
                  & \multicolumn{2}{l} {\hskip -0pt $^{20}$ \textcite{Klingner_et_al_2011}}\\
\multicolumn{2}{l} {\hskip -0pt $^{21}$ \textcite{Krellner_et_al_2009}}
      & \multicolumn{2}{l} {\hskip -25pt $^{22}$ \textcite{Ishida_et_al_2003}}
             & \multicolumn{2}{l} {\hskip -15pt $^{23}$ \textcite{Mederle_et_al_2001}}
                       & \multicolumn{2}{l} {\hskip -0pt $^{20}$ \textcite{Knebel_et_al_2006}}\\
\end{tabular}
\end{ruledtabular}
\vskip -3mm
\label{table:3}
\end{table*}

\paragraph{An induced-moment magnet: PrPtAl}
\label{PrPtAl}
PrPtAl is another interesting example of a system in which a modulated
magnetic state has been observed over a narrow temperature range above
the ferromagnetic ordering temperature. Neutron
scattering experiments in conjunction with a theoretical analysis have
been interpreted as indicating that this change is due to quantum
critical fluctuations, in accord with the mechanism reviewed in
Sec.~\ref{subsec:III.E}~\cite{Abdul-Jabbar_et_al_2015}.

%PrPtAl crystallizes in the orthorhombic TiNiSi-type structure.% , in
% which the Pr atoms form a network of chains parallel to the $a$-axis
% and parallel to the $b$-axis.
Because the orthorhombic TiNiSi-type structure of PrPtAl is inversion
symmetric, it does not allow for a Dzyaloshinski-Moriya interaction
(as opposed to the case of MnSi). Initial experiments on polycrystals
identified a second-order phase transition into a FM state at \TC\ =
5.8\,K with an ordered saturation moment of
1\,\mub/Pr~\cite{Hulliger_1993,Kitazawa_et_al_1998}. As in other rare
earth systems, the origin of magnetism is subtle in PrPtAl. Although the
ground state of the Pr$^{3+}$ ion is a non-magnetic singlet, magnetic
moments are induced, when higher lying crystalline electric field
levels are admixed via the inter-site exchange
interaction. This causes strong short-range
correlations, which are responsible for the small entropy release (only
about 15\% of $R\ln2$) below \TC~\cite{Kitazawa_et_al_1998}. The
interplay between induced moments and conduction electrons could be
seen as similar to band magnetism, but with the Pr states boosting the
magnetic response of the conduction electrons and causing significant
magnetic anisotropy. 

Detailed neutron scattering experiments on single crystals have
revealed two SDW phases just above \TC: Below $T_{1} =
5.85$\,K~\textcite{Abdul-Jabbar_et_al_2015} found a doubly modulated
incommensurate SDW (SDW1) followed by another single incommensurate
modulation (SDW2) at a different ordering vector below $T_{2} =
5.5$\,K and eventually the phase transition into the FM state at \TC\
= 4.7\,K. Both SDW phases are suppressed by weak magnetic field;
this is possibly the reason why these phases were not seen before. The
magnetic structure in the SDW2 phase is found to be an elliptical
spiral. Spiral order preceding a FM transition is not uncommon in rare
earth magnets such as, for instance Tb and Dy, and is usually
attributed to a complex interplay between anisotropy energy and
exchange interaction \cite{Miwa_1961}. However, in PrPtAl
\textcite{Abdul-Jabbar_et_al_2015} point at the lack of apparent
nesting vectors which would favour spiral order, the temperature
dependence of the ordering wavevector in the spiral phase SDW2, the
low critical magnetic fields required to tune between SDW2 phase and FM, and the
second-order nature of the transition at $T_{1}$, which contrasts with
the first-order nature at $T_{2}$ and \TC. They argue that the
mechanism behind the spiral formation in SDW2 must involve the strong
magnetic fluctuations in the competing ordered states.
Starting from the model proposed
by~\textcite{Karahasanovic_Kruger_Green_2012} and adding local
moments, the strong anysotropy and weak disorder, the authors derived
a theory that can describe key experimental results
observed in neutron scattering, magnetoresistivity and specific
heat. Although the full phase diagram has not been accessed experimentally, the proposed
phase diagram is similar to that in
Fig.~\ref{figure:nematic_spiral_uniform_phase_diagram} without the
nematic phase. This material may thus represent a rare
case in which
the mechanism discussed in Sec.~\ref{subsec:III.E} is realized.
\subsubsection{Ferromagnetic Kondo-lattice systems: CeTPO}
\label{subsubsec:II.D.2}

Ferromagnetic Kondo-lattice systems are rare, and therefore quantum criticality in such systems has received relatively little attention either experimentally
or theoretically. This is in contrast to a large number of AFM Kondo lattices which have been studied extensively; see, e.g., \textcite{Gegenwart_et_al_2008} and
references therein. Only recently have Ce and Yb-based systems been studied that show a FM transition at low temperature. Quite often these materials possess peculiar
crystal structures, such as the quasi-one-dimensional heavy-fermion material \YNP, see Sec.\ \ref{subsubsec:II.C.3}, or the quasi-two-dimensional cerium transition-metal 
(T) phosphide oxides CePTO, which are the topic of this section. For other Kondo-lattice systems, see Sec.\ \ref{subsubsec:II.D.3}.

The quasi-two-dimensional tetragonal crystal structure of ZrCuSiAs-type of the CePTO systems is quite simple and related to the iron-based superconductor 
LaFePO~\cite{Kamihara_et_al_2006}. It consists of alternating layers of TP$_{4}$ and OCe$_{4}$ along the crystallographic $c$-axis~\cite{Zimmer_et_al_1995}. 
Importantly, the Ce-Ce interatomic distance is in the range where the RKKY interaction is ferromagnetic~\cite{Sereni_1991, Chevalier_Malaman_2004}. However,
not all CeTPO systems are FM, for instance, CeOsPO is an AFM~\cite{Krellner_et_al_2007}. Two compounds that have been studied with respect to FM 
quantum criticality are \CRPO\, and \CFPO. \CRPO\, is a ferromagnet with \TC\, = 15\,K~\cite{Krellner_et_al_2007}, and \CFPO\, 
is a paramagnet with very strong in-plane FM fluctuations~\cite{Bruening_et_al_2008}. In what follows we review studies of \CRPO\, under hydrostatic 
pressure, and of \CRFPO\, and \CFPAO. The special case of stoichiometric \CFPO\, is discussed in Sec.~\ref{CeFePO}.

\paragraph{\CRPO}
\label{CRPO}

The low-temperature properties of \CRPO\, polycrystalline samples (RRR = 50) were first investigated by~\textcite{Krellner_et_al_2007}. These 
authors found a Curie-Weiss behavior of the susceptibility at high $T$ (Ce is trivalent in this compound and Fe is non-magnetic) with a positive 
Weiss temperature $\Theta_{\textrm{W}} = 8$\,K and FM order below \TC\, = 15\,K. The phase transition at \TC\, is second order, indicated by 
the $\lambda$-like shape of the specific heat. The resistivity shows a distinctive drop below about 50\,K which is a signature of coherent Kondo 
scattering. The Kondo temperature $T_{\textrm{K}} \approx 10$\,K, estimated from an analysis of the entropy, is comparable with \TC.

\begin{figure}[ht]
\begin{center}
\includegraphics[width=0.95\columnwidth,angle=0]{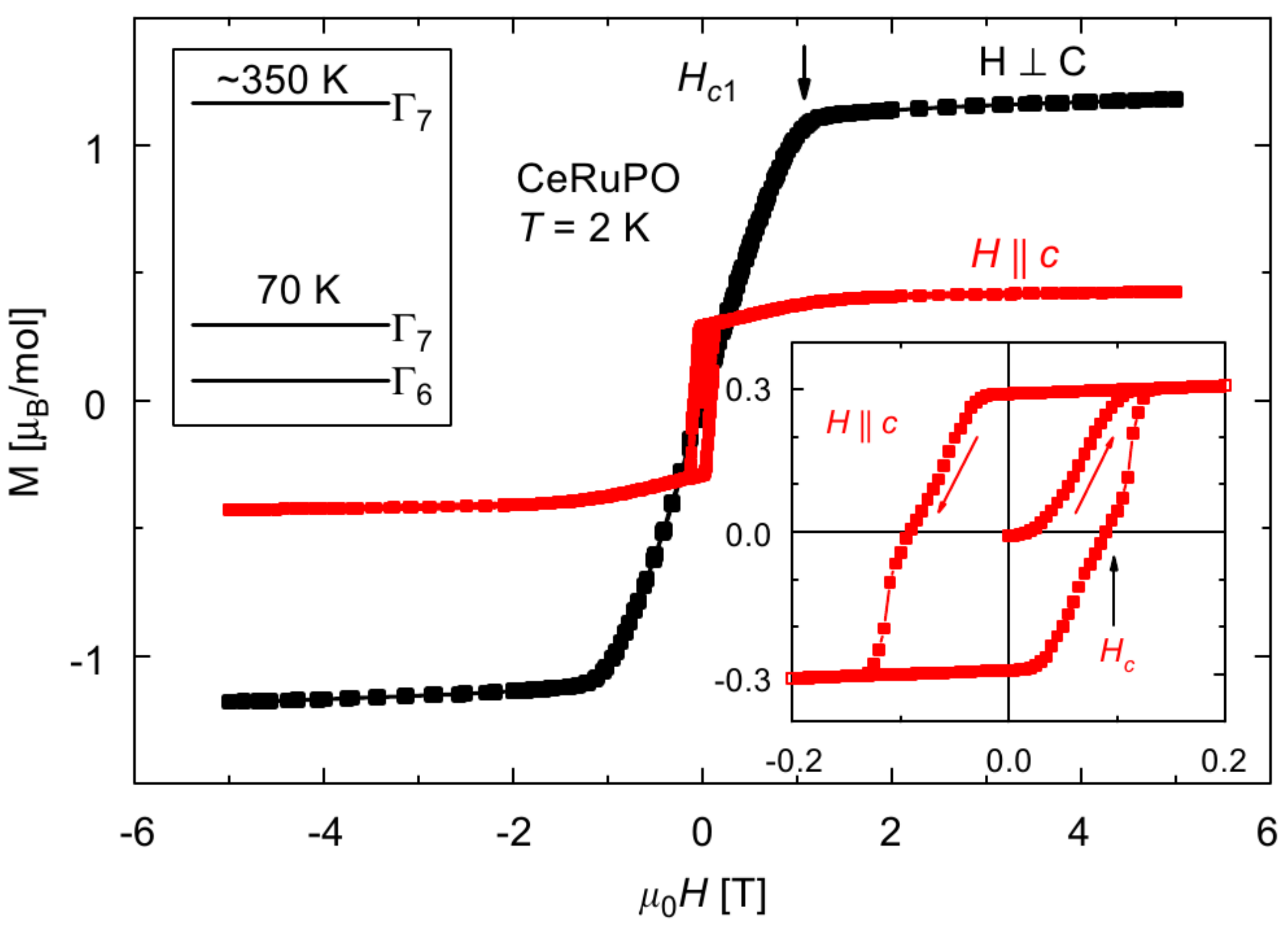}
\end{center}
\caption{Magnetization isotherms at 2\,K with $H \parallel c$ and $H \perp c$. From~\textcite{Krellner_Geibel_2008}.}
\label{CeRuPO_magnetization}
\end{figure}

Using a Sn-flux method, the same authors grew high-quality single crystals  with $\rho_{0} = 5$\,$\mu\Omega$cm (RRR = 30)~\cite{Krellner_Geibel_2008}, and studied the magnetic anisotropy. At high $T$ the susceptibility measured  with $H \parallel c$ and $H \perp c$ shows a Curie-Weiss behavior, but with very different Weiss temperatures for the two cases: $\Theta_{\textrm{W}}^{ab} \approx 4$\,K and $\Theta_{\textrm{W}}^{c} = -250$\,K. These temperatures (specifically their difference) can be expressed on the basis of molecular field theory~\cite{Bowden_et_al_1971} as a function of the first CEF parameter which is a measure of the strength of the magnetocrystalline anisotropy. Such a difference in the Weiss temperatures indicates that the CEF anisotropy favors the moments to be aligned within the $ab\,$-plane. However, the magnetic moments below \TC\, align along the $c$-axis. This is shown in Fig.~\ref{CeRuPO_magnetization}, which displays the magnetization isotherms at 2\,K.
The magnetization measured with $H \parallel c$ shows a clear hysteresis with a spontaneous moment of 0.3\,$\mu_{\textrm{B}}$ and a saturation magnetization 
of 0.43\,$\mu_{\textrm{B}}$. On the other hand, with $H \perp c$ the magnetization is zero at zero field, and increases up to the critical field of 1\,T where it reaches 
the saturation value of about 1.2\,$\mu_{\textrm{B}}$. This suggests that the magnetic order is rather collinear with moments along the $c$-axis. Ordering of the 
moments along the magnetic hard direction is quite rare, but also found in: YbNiSn~\cite{Bonville_et_al_1992}, \YRCSFM~\cite{Lausberg_et_al_2013}, \YNP~\cite{Steppke_et_al_2013} and CeAgSb$_{2}$~\cite{Araki_et_al_2003}.

Quantum criticality in \CRPO\ has been looked for by means of resistivity measurements under pressure, with the current in the $ab\,$-plane. Pressure was found to decrease \TC, to 5.9\,K at 2.1\,GPa, which by extrapolation suggested a QCP at about 3.2\,GPa \cite{Macovei_et_al_2009}. \textcite{Lengyel_et_al_2015} also investigated the ac susceptibility under pressure and performed resistivity experiments
at pressures up to 7.5\,GPa. These experiments found that the FM ground state changed into an AFM one (with unknown structure) at a pressure
of about 0.87\,GPa. At $p \geq 3$\,GPa the resistivity data no longer show a phase transition, and the observations suggested a first-order QPT
at a critical pressure \pc\, $\approx 3$\,GPa, with no QCP. Above \pc\, the ground state was proposed to be a Fermi liquid, due to a $T^{2}$ behavior of the resistivity. 
Remarkably, the coefficient of the $T^{2}$ term in the resistivity shows a maximum at about 4\,GPa, well inside the FL paramagnetic state. 
%Quantum criticality in \CRPO\, was investigated almost simultaneously by two groups with hydrostatic pressure. Both groups measured resistivity with current in the $ab\,$-plane and a magnetic field $H \parallel c$. The first report on pressure studies on \CRPO\, indicated that \TC\, decreases with increasing pressure down to 5.9\,K at 2.1\,GPa, whose extrapolation suggests a QCP at about 3.2\,GPa~\cite{Macovei_et_al_2009}. In her Ph.D. thesis, M. Macovei investigated also the ac-susceptibility under pressure and pushed the resistivity experiments up to 7.5\,GPa. She describes how the FM transition changes into an AFM one at about 0.87\,GPa whose structure is unknown~\cite{Macovei_2010}.
%
\begin{figure}[t]
\begin{center}
\includegraphics[width=0.95\columnwidth,angle=0]{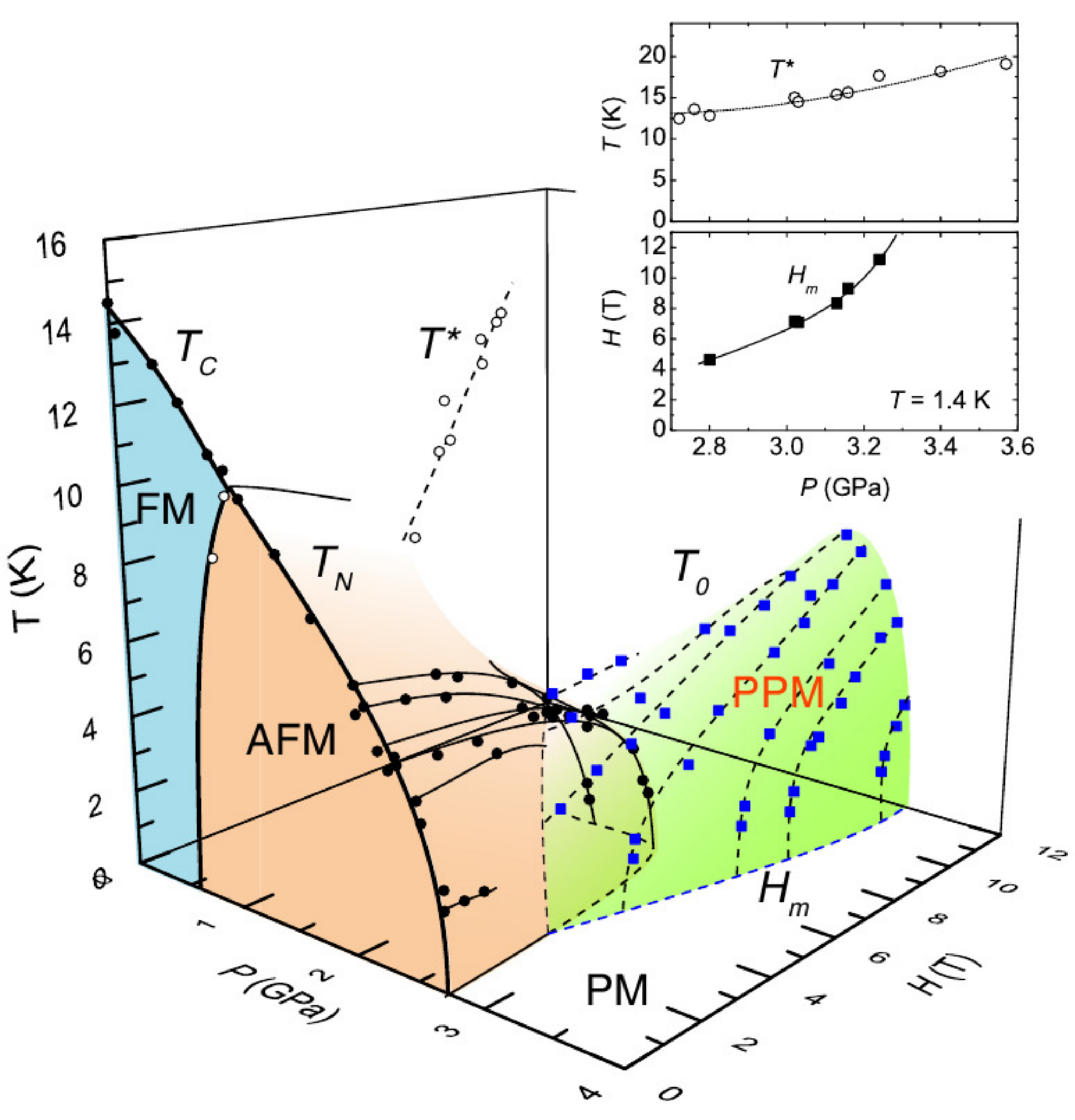}
\end{center}
\caption{Pressure-temperature-magnetic field phase diagram of \CRPO\, derived from resistivity measurements. At ambient pressure and $H = 0$ the ground state is 
              FM, but changes into an AFM one at about 0.7\,GPa. Magnetism is suppressed at \pc\, $\approx 2.8$\,GPa. The AFM state is identified by the transition 
              temperature decreasing with increasing field (black solid lines). At even larger fields ($H_m$) a metamagnetic crossover occurs from a PM state to a 
              polarized PM (PPM) state. $T^{*}$ indicated the Kondo coherence temperature which increases with increasing pressure. From~\textcite{Kotegawa_et_al_2013}.}
\label{CeRuPO_phase_diagram}
\end{figure}
%
%At $p \geq 3$\,GPa the resistivity data do not show any phase transition anymore, suggesting that the quantum phase transition is first order at \pc\, $\approx 3$\,GPa with no QCP. Above \pc\, the ground state is proposed to be a Fermi liquid (FL) from the $T^{2}$ behavior of the resistivity. Remarkably, the $A$ coefficient of the $T^{2}$ resistivity shows a maximum at about 4\,GPa, well inside the FL paramagnetic state. 

The complete $T$-$p\,$-$H$ phase diagram of \CRPO, up to $p=\ $3.5\,GPa, was investigated in great detail by \textcite{Kotegawa_et_al_2013}; it is 
shown in Fig.~\ref{CeRuPO_phase_diagram}. 
%It contains neither a FM QCP nor a FM tricritical point; rather, its structure is that shown schematically in Fig.~\ref{fig:schematic_phase_diagrams}c. 
The observed sensitivity of the magnetic order to a magnetic field ascertained that the FM state changes into an AFM one above $p \approx 0.7$\,GPa, and the 
magnetic order was found to be completely suppressed at \pc\, = 2.8\,GPa. \CRPO\, is thus another case where an anticipated FM QPT is not realized because
a modulated phase intervenes. Although the authors could not determine whether the transition at \pc\, is first or second order, the coefficient 
of the $T^2$ term in the resistivity shows a maximum around \pc, which suggests the presence of an AFM QCP. This is in disagreement with 
\textcite{Lengyel_et_al_2015}. $^{31}$P-NMR experiments have revealed that the magnetic correlations are three-dimensional over the entire pressure range
investigated~\cite{Kitagawa_et_al_2014}, in contrast to what was found in Ce(Rh$_{1-x}$Fe$_x$)PO, see Sec.~\ref{CRFPO}.
% This is different from 
% what has been seen by M. Macovei \textit{et al.} and from 
% the results on \CRFPO\, where the FM ordered state seems to be continuously suppressed to $T = 0$~\cite{Kitagawa_et_al_2012}, see Sec.\ \ref{CRFPO}. 
%This demonstrates that \CRPO\, is one of the cases where the FM QCP is 
% preempted by another phase, as in \NF\, (see Sec.~\ref{NF}), or \YRCS\, (see Sec.~\ref{par:YRCS}). 
% Interestingly, at 0.7 and 0.98\,GPa the system undergoes, with
% decreasing temperature, first a phase transition into the FM state at \TC\, and then a second one into the AFM state across the FM-AFM coexistence line, see 
The FM-AFM transition was found to be first order at the two points denoted by open circles in 
Fig.~\ref{CeRuPO_phase_diagram}; the order of the transition at lower temperatures is not known. We note that the 
AFM-to-FM transition in  \YRCS\ with $x=0.215$ (see Fig.~\ref{YRS_PD} in Sec. \ref{par:YRCS}) has also 
been reported to be first order \cite{Klingner_et_al_2011,Hamann_et_al_2014}, as has the SDW-to-FM transition in \NFy\ (see Fig.~\ref{NFy} in Sec.\ \NF)
\cite{Friedemann_et_al_2014}. 

At pressures close to \pc, \textcite{Kotegawa_et_al_2013} observed another resistivity feature at a magnetic field $H_m$ and a temperature $T_0$
(see Fig.~\ref{CeRuPO_phase_diagram}), which they ascribed to a metamagnetic crossover from a PM state to a polarized PM (PPM) state. This is similar
to what has been observed in doped CeRu$_{2}$Si$_{2}$~\cite{Flouquet_et_al_2002, Flouquet_et_al_2010, Shimizu_et_al_2012}. Interestingly, $T_0$ increases with 
increasing field.
% In fact, the temperature below which this metamagnetic crossover is observed increases with increasing field. Hence, \CRPO\, does not have neither a FM-QCP nor a tricritical point.
%
\paragraph{\CFPAO}
\label{CFPAO}

Historically, \CFPO\, was the first CeTPO system that attracted attention in the context of quantum criticality since it was shown that it is a 
paramagnet very close to a FM instability~\cite{Bruening_et_al_2008}. This motivated searches for an FM ordered phase nearby. The substitution of As at the 
P site seemed promising, since it acts as negative chemical pressure and favors magnetic ordering in Ce-based systems. It also permits to study the evolution of Fe 
and Ce magnetism from the AFM CeFeAsO~\cite{Zhao_et_al_2008} to \CFPO. This was done independently by two groups~\cite{Luo_et_al_2010, Jesche_et_al_2012}.

The first complete phase diagram of \CFPAO\, polycrystalline samples was published by \textcite{Luo_et_al_2010}, who confirmed that the commensurate AFM 
order of the Fe sublattice below $T_N^{\text{Fe}} \approx 140$\,K in CeFeAsO is suppressed by P substitution. With increasing $x$, they found that the unit
cell shrinks along the $c$-axis substantially faster than along the $a$-axis, which is important for the evolution of the $f$-$d$ hybridization strength. 
The AFM order disappears at $x \approx 0.4$, and an AFM QCP was suspected~\cite{Luo_et_al_2010, De_la_Cruz_et_al_2010}. 
No evidence of superconductivity was found in these studies. Moreover, \textcite{Luo_et_al_2010} found that the Ce sublattice also orders 
antiferromagnetically at $T_N^{\text{Ce}} = 4.16$\,K, which is very weakly $x$-dependent for small $x$. At $x \approx 0.37$ the ground state of the Ce sublattice 
changes from AFM to FM, and FM order was found to persist up to a P content of about 0.9. Eventually, in a small concentration region at $x \alt 1$ the system 
was found to be a heavy-fermion (HF) paramagnet with strong FM fluctuations, in agreement with \textcite{Bruening_et_al_2008}, and a second QCP, for the FM-HF
transition, was suggested.
\begin{figure}[t]
\begin{center}
\includegraphics[width=0.95\columnwidth,angle=0]{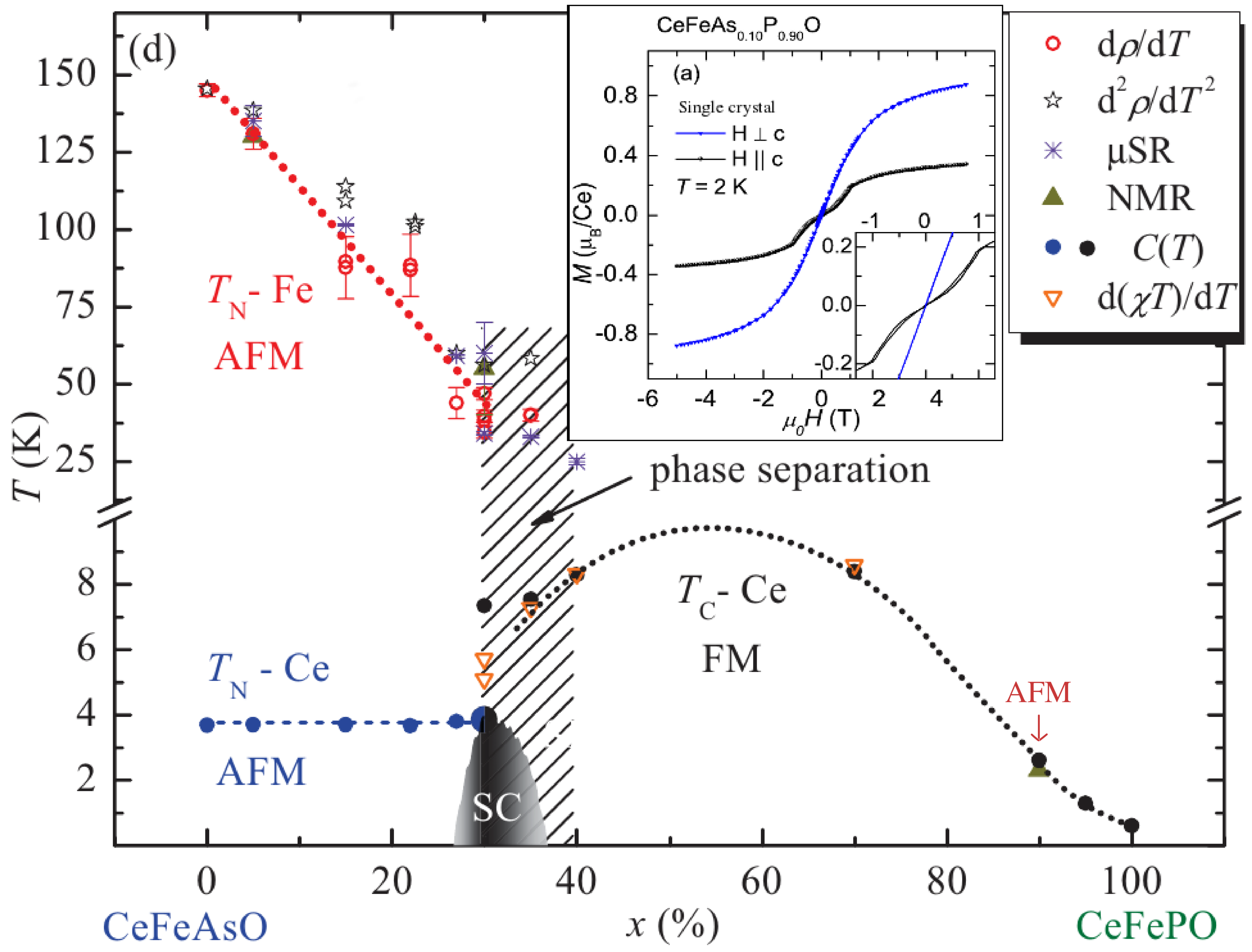}
\end{center}
\caption{Temperature ($T$) - concentration ($x$) phase diagram of \CFPAO\ obtained by a variety of techniques (list in the right inset) for single- and polycrystalline
              samples. Red and blue dotted lines indicate the AFM ordering temperature of the Fe and Ce sublattices, respectively. Superconductivity (SC) is found
              near $x = 0.3$. The black dotted line is the Curie temperature for the FM order of the Ce atoms. The magnetization isotherms in the left inset indicate that
              at $x = 0.9$ the order is no longer FM, but rather AFM. See the text for more information. After~\textcite{Jesche_et_al_2012, Jesche_2011}).}
\label{CeFeAsPO_phase_diagram_new}
\end{figure}

Later studies by \textcite{Jesche_2011} and \textcite{Jesche_et_al_2012} found several additional features in the phase diagram, and disagree in some respects
with \textcite{Luo_et_al_2010}. Their phase diagram (obtained by means of a variety of techniques) is shown in Fig.~\ref{CeFeAsPO_phase_diagram_new}.
Instead of an AFM QCP they found that the AFM order terminates at a nonzero $T_N^{\text{Fe}} \approx 30\,$K and is followed by a region of phase separation,
indicating a possible tricritical point. They also found superconductivity, possibly coexisting with Ce ferromagnetism, in a small dome around $x = 0.3$ at temperatures
up to 4\,K. On the large-$x$ side of the phase diagram, which is of most interest for this review, \textcite{Jesche_2011} and \textcite{Krellner_Jesche_2014} found
that a single-crystal sample with $x = 0.9$ had an AFM ground state rather than a FM one. This evidence is shown in the inset of 
Fig.~\ref{CeFeAsPO_phase_diagram_new}, which shows the isothermal magnetization at $T = 2$\,K, i.e. below the transition temperature of about 2.7\,K for this
concentration. There is no remanent magnetization at $B = 0$ in both field directions, and a metamagnetic transition at $B = 1$\,T implies that the ground 
state is indeed AFM. This indicates that \CFPAO\, belongs to the class of systems where the order changes from FM to AFM as \TC\ decreases, as is the
case in \CRPO\, under pressure (previous section), and \CRFPO\, (next section). This raises some interesting questions: There must be a Lifshitz point on the
phase boundary shown in Fig.~\ref{CeFeAsPO_phase_diagram_new}, and there must be a QPT from the FM phase to the AFM phase, followed by an AFM QPT.
These issues have not been investigated either experimentally or theoretically.

\paragraph{\CRFPO}
\label{CRFPO}

The fact that \CRPO\, is a low-temperature ferromagnet (see Sec.~\ref{CRPO}) and \CFPO\, is a paramagnet with very strong FM fluctuations (see Sec.~\ref{CFPAO}) 
motivated the study of the series \CRFPO. The substitution of Fe for Ru is isoelectronic and affects just the Fe(Ru)P layers without causing much disorder in the CeO
layers responsible for the magnetism. \textcite{Kitagawa_et_al_2012} have investigated polycrystalline 
(oriented powder) samples belonging to this series by $^{31}$P nuclear magnetic resonance (NMR). 
They measured the Knight shift $K$, which is proportional to the uniform magnetization, and the spin-lattice relaxation rate $(1/T_{1})$, which is a measure of the 
fluctuations perpendicular to the applied field direction, 
for fields parallel and perpendicular to the $c$-axis ($H\parallel c$ and $H \perp c$, respectively) of the tetragonal crystallographic structure. Because of the XY-type 
anisotropy, the largest signal is found for $K_{\perp}$ and $(1/T_{1})_{\parallel}$, which both show a strong $T$ dependence.
\begin{figure}[t]
\begin{center}
\includegraphics[width=0.95\columnwidth,angle=0]{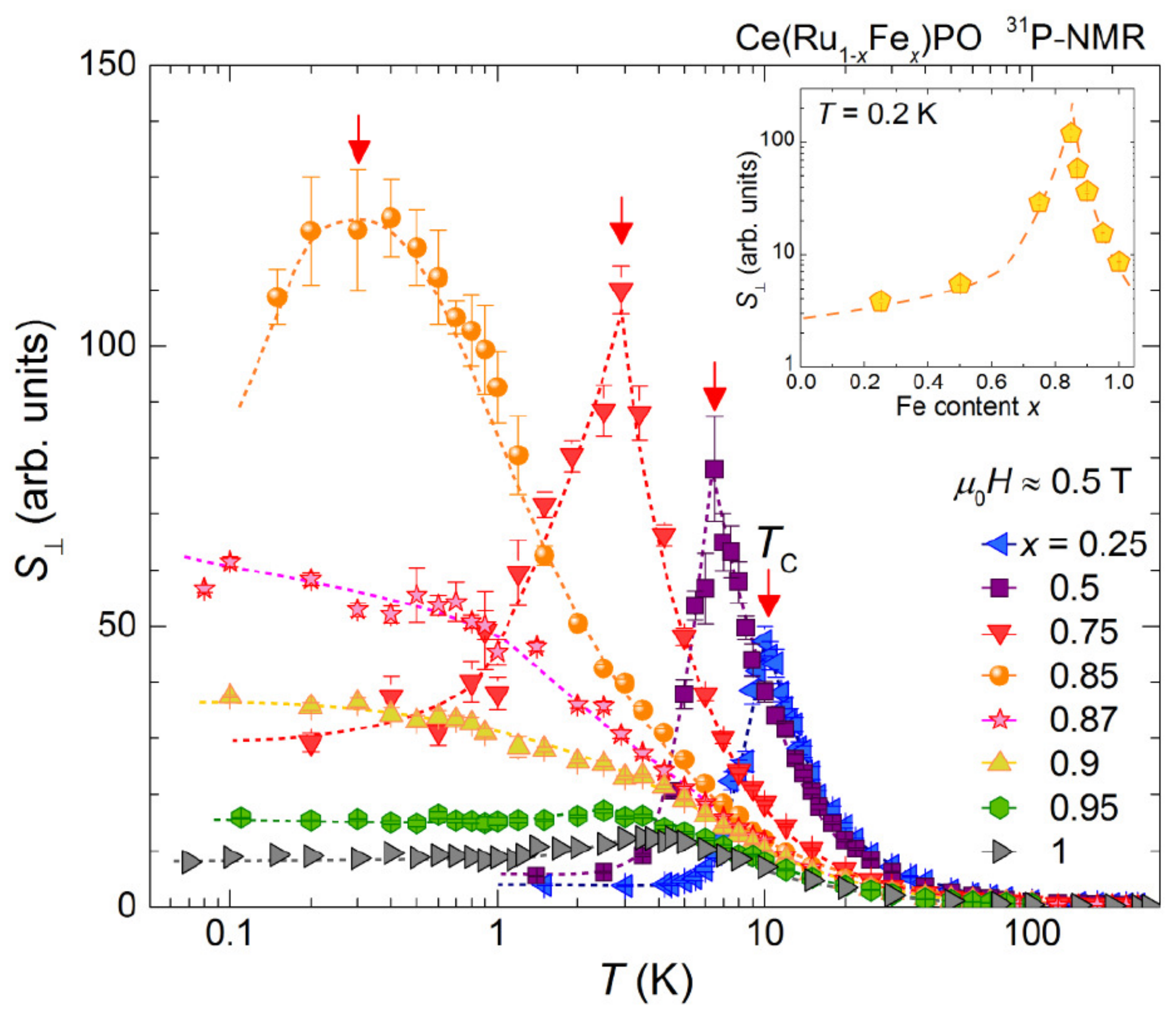}
\end{center}
\caption{Temperature dependence of the in-plane spin fluctuations $S_{\perp} = (1/2T_{1})_{H \parallel c}$ at $\mu_0H = 0.5$\,T for various values of $x$. The peak 
              indicates the FM transition temperature \TC, which is only visible for $x \le 0.85$. For $x > 0.85$, $S_{\perp}$ is constant at low $T$. The inset shows 
              the $x$-dependence of $S_{\perp}$ at a temperature of 200\,mK.  $S_{\perp}$ peaks at \xc\, indicating the presence of a QCP. 
              From~\textcite{Kitagawa_et_al_2012}.}
\label{CeRuFePO_fluctuations}
\end{figure}

\begin{figure}[t]
\begin{center}
\includegraphics[width=0.9\columnwidth,angle=0]{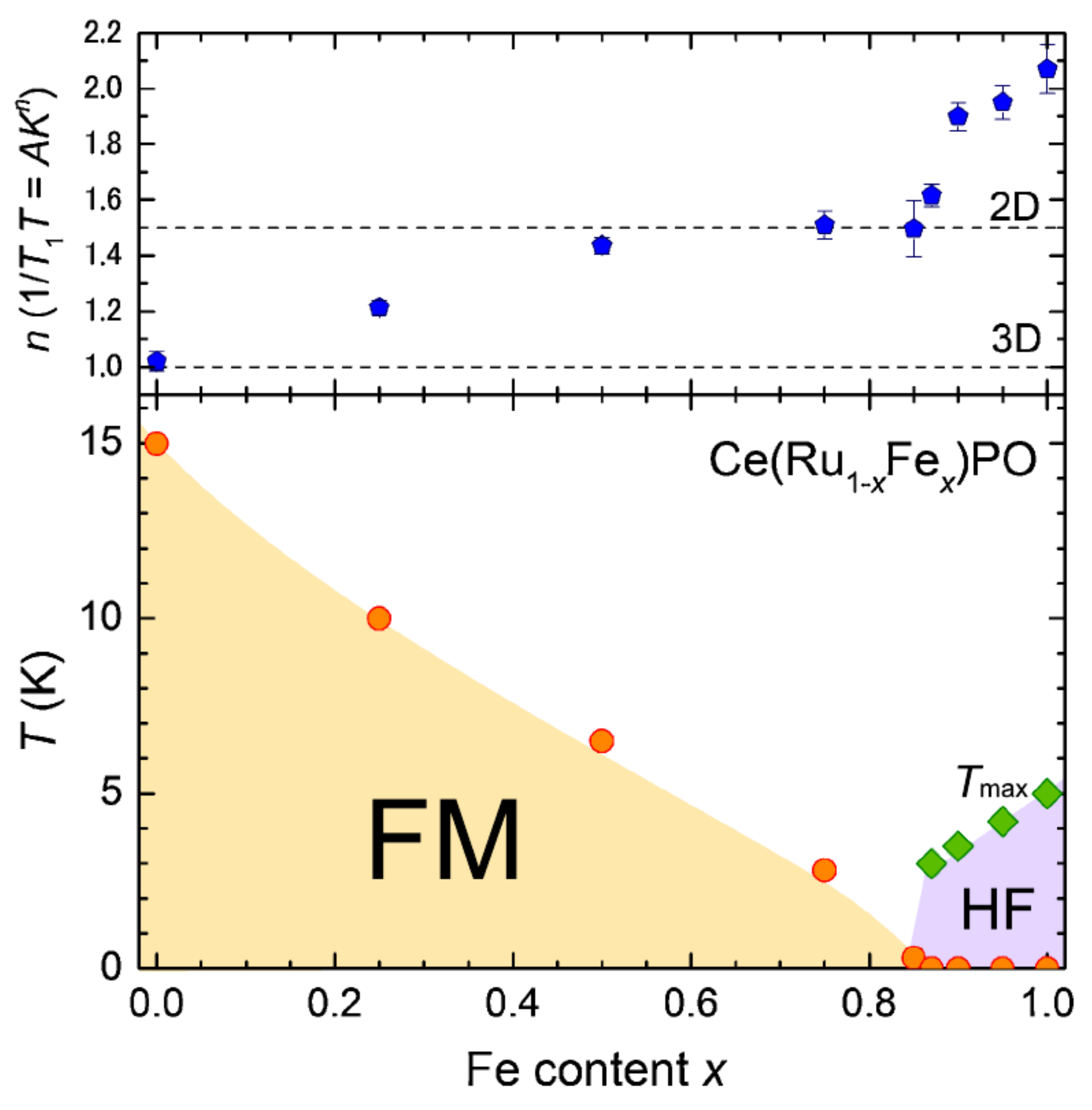}
\end{center}
\caption{The lower panel shows the $x$-$T$ phase diagram of \CRFPO\, derived from NMR studies. Strong fluctuations are indicative of a QCP at
               \xc\, $\approx 0.86$. At larger $x$ the ground state is a heavy-fermion (HF) paramagnet, bounded by the temperature $T_{\text{max}}$ where
               the Knight shift shows a peak. The exponent $n$ in the upper panel is indicative of the effective dimensionality of magnetic correlations, with
               $n=1.5$ corresponding to two-dimensional correlations. From ~\textcite{Kitagawa_et_al_2013}.
}
\label{CeRuFePO_phase_diagram_2D}
\end{figure}

The FM transition temperature was determined by the large increase of $K(T)$ at \TC\, and the peak seen in $(1/T_{1})_{\parallel}$ as shown in 
Fig.~\ref{CeRuFePO_fluctuations} for $\mu_0H = 0.5$\,T. The phase diagram extracted from these NMR measurements is shown in Fig.~\ref{CeRuFePO_phase_diagram_2D}.
With increasing Fe content, both \TC\, and the ordered moment, as determined from the Knight shift (not shown), are continuously suppressed until both
vanish at \xc\, $\approx 0.86$. The phase transition is clearly second order at $x = 0$ and remains second order for all concentrations, although a significant broadening 
of the relaxation rate and of the Knight-shift increase is seen at $x = 0.85$. This might indicate a spin-glass-like or short-range ordered state, as was observed in pure 
\CFPO~\cite{Lausberg_et_al_2012a}, see Sec.\ \ref{CeFePO}, 
but this is not quite clear. It is possible that short-range order, if it is present in these samples, is suppressed
by the nonzero applied field of about 0.5\,T for most of the NMR measurements; in \CFPO\, the short-range ordered state is suppressed at this field strength. One thus
must keep in mind that Fig.~\ref{CeRuFePO_phase_diagram_2D} does not show a zero-field phase diagram, and this is more relevant for $x \approx x_c$ than for
small $x$. 

The strong fluctuations observed near $x_c$ are a clear sign of a QCP, the nature of which is not purely ferromagnetic. The NMR data in a field $\mu_0H=0.07\,$T
for a sample with $x = 0.85$ showed that, in addition to the homogeneous FM ($q=0$) component of the fluctuations that levels off below $T = 3\,$K, there are AFM
($q\neq 0$) components that continue to increase as the temperature is lowered towards \TC\,$\approx 300\,$mK \cite{Kitagawa_et_al_2012}. Similar behavior
was found in \YRS\, near an AFM QCP~\cite{Ishida_et_al_2003}. The unusual behavior near the QCP was further investigated by \textcite{Kitagawa_et_al_2013},
who concluded that the suppression of ferromagnetism is due to a change of the effective dimensionality of the FM fluctuations from three-dimensional to 
two-dimensional near $x_c$. 
This is in contrast to what was observed in stoichiometric CeRuPO under hydrostatic pressure, where the magnetic correlations remain three-dimensional at all values of
the pressure, including $p_c$ \cite{Kitagawa_et_al_2014}. Another
interesting effect is a metamagnetic crossover (not a first-order transition) at a field $H_M$ perpendicular to the 
$c$-axis~\cite{Kitagawa_et_al_2011,Kitagawa_et_al_2012}. The tips of these ``crossover wings'' coincide with the QCP and the wings grow with increasing field;
their shape is thus very different from that of the tricritical wings observed in conjunction with the tricritical point in the systems discussed in Sec.\ \ref{subsec:II.B}.
A possibly related observation is that in pure \CFPO, NFL behavior commonly associated with a QCP has been observed around 
$\mu_0H\approx 4$T~\cite{Kitagawa_et_al_2011}. The authors suggested that $H_M$ represents the field that breaks the local Kondo singlet, and that the critical behavior
is driven by the Kondo breakdown accompanied by a Fermi-surface instability. 
\begin{figure}[t]
\begin{center}
\includegraphics[width=0.95\columnwidth,angle=0]{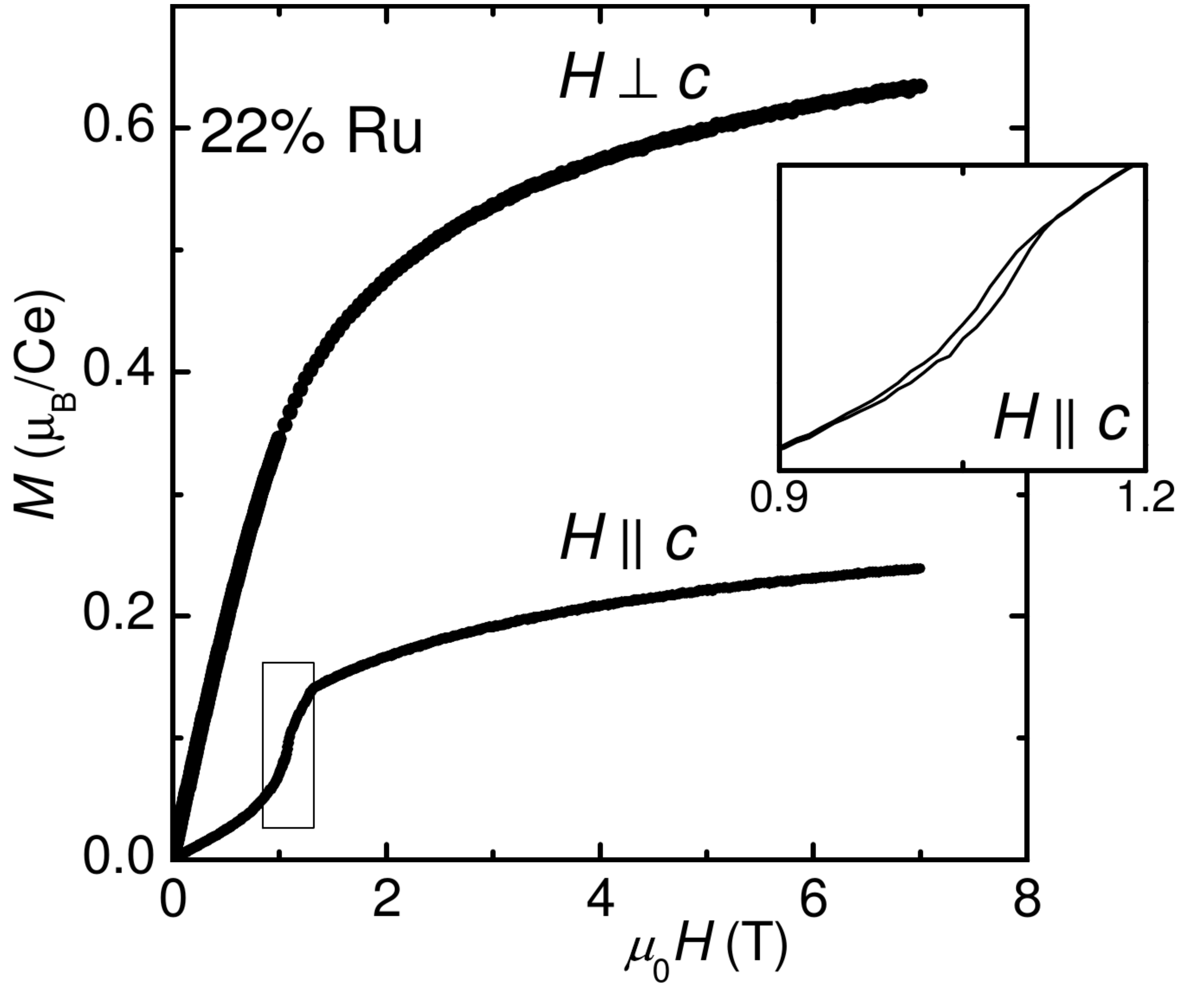}
\end{center}
\caption{Isothermal magnetization of a single crystal of CeRu$_{0.22}$Fe$_{0.78}$PO for $H\parallel c$ and $H \perp c$ at $T \approx 1.8$\,K. Note the absence of
              a remanent magnetization at $H=0$. The inset shows a weak hysteresis loop near $\mu_0H = 1.1\,$T for $H \parallel c$. After \textcite{Krellner_Jesche_2014}.
}
\label{CeRu22FePO}
\end{figure}

The NMR results described above paint a picture that is rather different from that of \CRPO\ under pressure, where there is no doubt that a phase of AFM character
intervenes before the FM QCP is reached, see Sec.~\ref{CRPO}. However, in recent work on single crystals with a Ru content close to 20\%, 
\textcite{Krellner_Jesche_2014} found an AFM ground state between the FM and the PM phases, just as in \CRPO\ under pressure. An example is shown in
Fig.~\ref{CeRu22FePO} which displays the field dependence of the magnetization for a sample with 22\% of Ru content at $T \approx 1.8$\,K, below the transition 
temperature of 2.5\,K for this concentration. There is no remanent magnetization at zero field in both field directions. In addition, for $H \parallel c$ a metamagnetic increase of the magnetization with a small hysteresis loop is found for $\mu_0H\approx 1.1$T (see the inset in Fig.~\ref{CeRu22FePO}) indicating a first-order metamagnetic transition. This is reminiscent of the situation in \CFPAO\, (cf. the inset in Fig.~\ref{CeFeAsPO_phase_diagram_new}) and \NF\,~\cite{Moroni_et_al_2009}.
\subsubsection{Other Kondo-lattice systems}
\label{subsubsec:II.D.3}

\paragraph{\CAS}
\label{CAS}

\CAS\, is a Kondo-lattice system with a planar magnetic anisotropy. It is one example of a ferromagnet where the moments are unexpectedly aligned along the 
magnetic hard direction; others include YbNiSb~\cite{Bonville_et_al_1992}, \CRPO~\cite{Krellner_et_al_2007}, \YNP~\cite{Steppke_et_al_2013} and 
\YRCSFM~\cite{Lausberg_et_al_2013}. 
Neutron scattering experiments have shown that the ordered moment of 0.41\mub/Ce aligns uniaxially along the tetragonal $c$-axis, whereas magnetization 
measurements indicate a strong magnetocrystalline anisotropy with the basal plane as the magnetic easy plane~\cite{Takeuchi_et_al_2003,Araki_et_al_2003}. 
For this reason, quantum criticality in \CAS\, has been investigated by transversal-field tuning with $H \perp c$. The critical field was found to be
$H_{c} \approx 2.8$\,T~\cite{Strydom_et_al_2008, Zou_et_al_2013, Logg_et_al_2013}. The transition is suspected to remain second order to the
lowest \TC\, measured, about 2\,K.

The FM order in \CAS\ can also be suppressed by hydrostatic pressure. 
The first experiments under pressure (up to about 50\,kbar) were performed by~\textcite{Sidorov_et_al_2003}. They measured the resistivity and 
ac heat capacity of very pure single crystals with a RRR $\approx 285 - 480$. 
\begin{figure}[t]
\begin{center}
\includegraphics[width=0.9\columnwidth,angle=0]{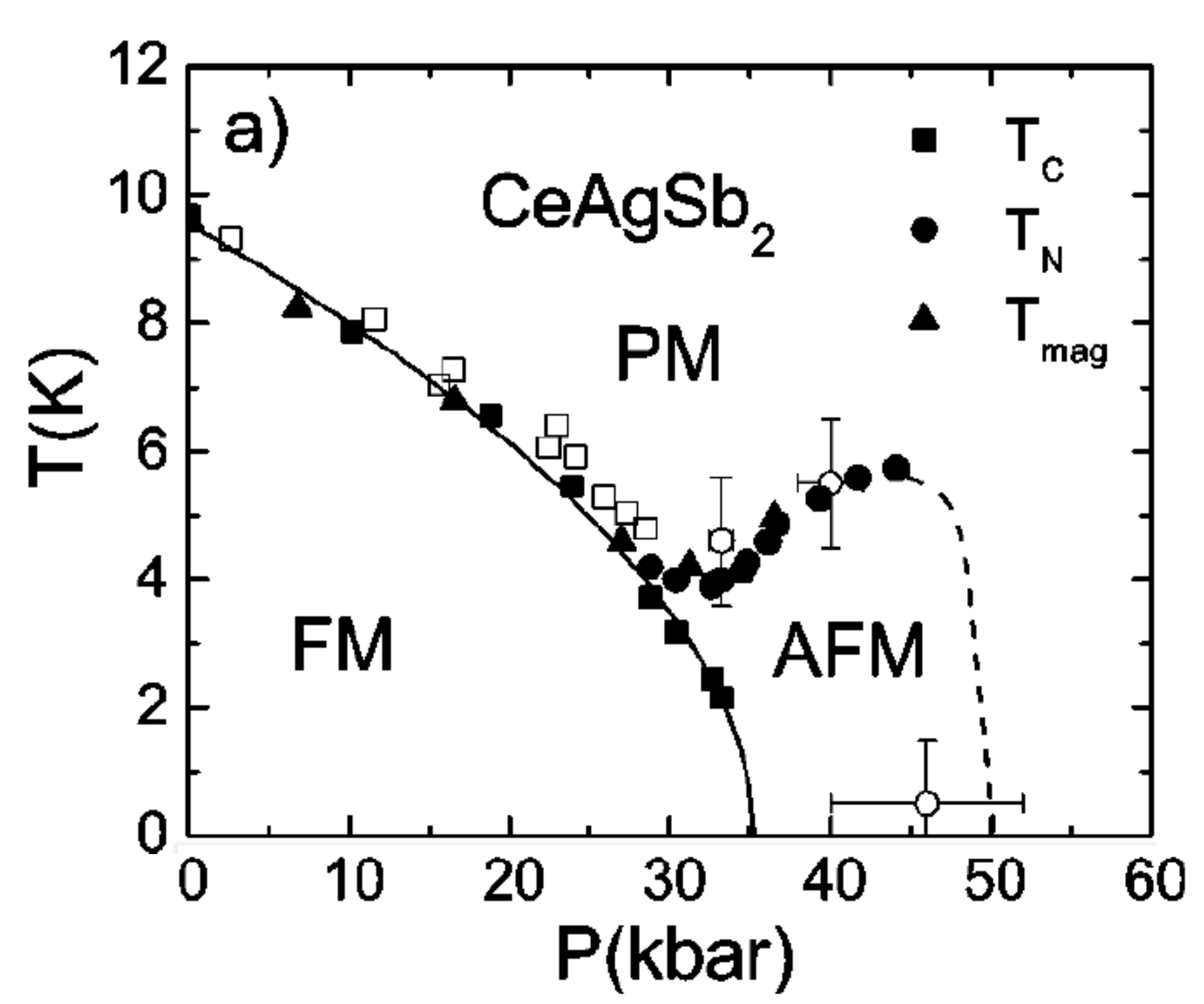}
\end{center}
\caption{Pressure-temperature phase diagram of \CAS\. \TC\, and $T_{N}$ were determined from $d\rho/dT$ while $T_{mag}$ from ac-calorimetry measurements.
               From~\textcite{Sidorov_et_al_2003}. }
\label{CeAgSb2}
\end{figure}
They found that the FM state changes into a presumably AFM state above 27\,kbar, with the new phase persisting to about 46\,kbar (see Fig.~\ref{CeAgSb2}). 
The thermal FM phase transition is second order for $p < 27$\,kbar and the entropy below \TC\, indicates that the Kondo temperature in this material is well below 
\TC~\cite{Sidorov_et_al_2003,Zou_et_al_2013}. In the pressure range 27\,kbar $ < p < 33$\,kbar the resistivity signature (specifically in $d\rho/dT$) at the FM transition 
sharpens considerably with increasing $p$, indicating that the FM-to-AFM transition is first order. This is reminiscent of the situation observed in other ferromagnets 
which show a change to AFM oder, such as \YRCS, \CRPO\, or Nb$_{1-y}$Fe$_{2+y}$ (cf. Tab.~\ref{table:3}). 
However, subsequent experiments (probing the resistivity and magnetization under pressure) by other groups could not detect the AFM phase, possibly because 
of lower sample quality (RRR $\approx 110$) or the limited resolution of the measurements~\cite{Kobayashi_et_al_2007, Logg_et_al_2013}.

\paragraph{CeRu$_{2}$Ge$_{2}$ and \CRGS}
\label{par:CRGS}

CeRu$_{2}$Ge$_{2}$ is a FM Kondo lattice system with \TC\ $\approx 8$\,K and a spontaneous magnetization of 1.96\,\mub\ along the tetragonal 
$c$-axis~\cite{Boehm_et_al_1988,Besnus_et_al_1991}. Its Kondo temperature is $T_{\rm K} \approx 2$\,K and the Sommerfeld coefficient is rather 
small, 20\,mJ/K$^{2}$mol. In some crystalline samples specific heat and magnetization measurements exhibit two transitions, a hump at $T_{\rm N} \approx 8.2$\,K 
and a sharp transition at \TC\ $\approx 7.7$\,K~\cite{Fontes_et_al_1996,Raymond_et_al_1999a}. Between \TC\ and $T_{\rm N}$, long-range order was identified to be AFM with 
an incommensurate propagation vector ${\bm q} = (0.31,0,0)$. Studies under hydrostatic 
pressure~\cite{Kobayashi_et_al_1998,Wilhelm_Jaccard_1998,Suellow_et_al_1999} have revealed a rich $p$ - $T$ phase diagram that is well reproduced by Si 
substitution for Ge~\cite{Haen_Bioud_Fukuhara_1999}. Both local-moment FM and AFM phases exit at low pressures. Above 20\,kbar the FM phase changes into a second low-$T$ AFM phase. The magnetic order is then 
rapidly suppressed near $p_{\rm c} \approx 67$\,kbar, accompained by NFL behavior with a linear-in-$T$ resistivity. For $p > p_{\rm c}$ the ground state is a Fermi 
liquid with an enhanced quasiparticle mass that decreases towards higher pressures~\cite{Suellow_et_al_1999}. A similar phase diagram was obtained by substituting Ru by Fe~\cite{Raymond_et_al_1999a}.

Although the $p$-$T$ phase diagram of CeRu$_2$Ge$_2$ and its Si-doped variety has the general shape shown in Fig. 2\,c), the underlying physics may be different from other systems. This is because the change in ground state from FM to AFM is located 
in the phase diagram where the Kondo temperature is very small, much smaller than the transition temperatures~\cite{Suellow_et_al_1999}. The 
modification of the ordered state is then very likely due to a variation of the exchange interactions between the local moments with pressure and is 
not driven by the mechanism discussed in Sec.~\ref{subsec:III.E}.

\paragraph{\YRS\ and \YRCS}
\label{par:YRCS}

The heavy-fermion metal \YRS\, is a prototypical example of a quantum critical system~\cite{Gegenwart_et_al_2002,Custers_et_al_2003}. Since its discovery 
by~\textcite{Trovarelli_et_al_2000} its properties have been intensively investigated; for reviews, see~\textcite{Gegenwart_et_al_2008, Si_Steglich_2010}. Here
we focus only on those properties that are related to ferromagnetism and FM fluctuations. 

\YRS\, crystallizes in the body-centered tetragonal ThCr$_{2}$Si$_{2}$ structure. The Yb ions are in the trivalent state as indicated by the high-$T$ Curie-Weiss behavior 
of the susceptibility $\chi(T)$ with an effective magnetic moment of 4.4\,\mub, i.e. close to what is expected for a free Yb$^{3+}$ ion. The Weiss temperatures 
$\Theta_{W}(B\parallel~c) = -180$\,K and $\Theta_{W}(B\perp~c) = -9$\,K indicate a strong magnetocrystalline anisotropy~\cite{Trovarelli_et_al_2000}. 
The crystalline electric field splits the $J = 7/2$ levels into 4 Kramers doublets, leaving the ground state 
separated from the three excited doublets by approximately 17, 25, and 43\,meV, respectively~\cite{Stockert_et_al_2006}. \YRS\, has a large Kondo temperature 
$T_{K} \approx 25$\,K~\cite{Koehler_et_al_2008}; nevertheless, a small unscreened magnetic moment of about $10^{-3}$\mub/Yb~\cite{Ishida_et_al_2003} orders 
antiferromagnetically below $T_{N} \approx 0.07$\,K \cite{Trovarelli_et_al_2000}. The exact magnetic structure is still unknown. $T_{N}$ can be suppressed by a 
magnetic field $B \approx 0.06$\,T ($B\perp~c$, with $c$ being the magnetically hard axis)~\cite{Gegenwart_et_al_2002} or negative chemical pressure 
($p \approx -0.25$\,GPa)~\cite{Mederle_et_al_2001, Macovei_et_al_2008}, which tunes the system to unconventional QCPs. 

\YRS\, shows pronounced NFL behavior in transport and thermodynamic quantities indicating the presence of strong spin fluctuations. For instance, the resistivity 
$\rho(T) \propto T$ below 10\,K and the Sommerfeld coefficient diverges as a power law $C/T \propto T^{-0.34}$ for $T \alt 0.3$K, similar to what is observed in the
low-$T$ ferromagnet \YNP~\cite{Krellner_et_al_2009}. The latter behavior has been interpreted in terms of a breakup of the heavy quasiparticles at the 
QCP~\cite{Custers_et_al_2003}. At low temperature the susceptibility also shows interesting behavior: 
$\chi_{\perp c}$ is very large for an antiferromagnet ($\approx 8.5 \cdot 10^{-6}$\,m$^{3}$/mol $\approx 0.18$\, SI) and about 20 times larger than $\chi_{\parallel c}$. 
This, and the value of the Sommerfeld-Wilson ratio of about 30, indicate the presence of strong FM fluctuations~\cite{Gegenwart_et_al_2005}, consistent with
NMR~\cite{Ishida_et_al_2003} and neutron scattering data~\cite{Stock_et_al_2012}. In an intermediate temperature range, for $0.3\,$K $\alt T \alt 4\,$K,
$C(T)/T \propto \ln(T_{0}/T)$, where $T_{0} \approx 25$\,K, can be considered the 
characteristic spin-fluctuation temperature according to \textcite{Moriya_1985}. Pressure stabilizes the magnetic order, increasing both $T_{N}$ and the value of the 
ordered moment~\cite{Mederle_et_al_2001,Knebel_et_al_2006}. Moreover, at small pressures an additional transition is observed at a lower temperature $T_{L}$,
which moves towards $T_{N}$ with increasing pressure. At about 5\,GPa, $T_{L} \approx T_{N}$ and~\textcite{Knebel_et_al_2006} proposed a ferromagnetic ground 
state above 5\,GPa in \YRS.
\begin{figure}[ht]
\begin{center}
\includegraphics[width=0.95\columnwidth,angle=0]{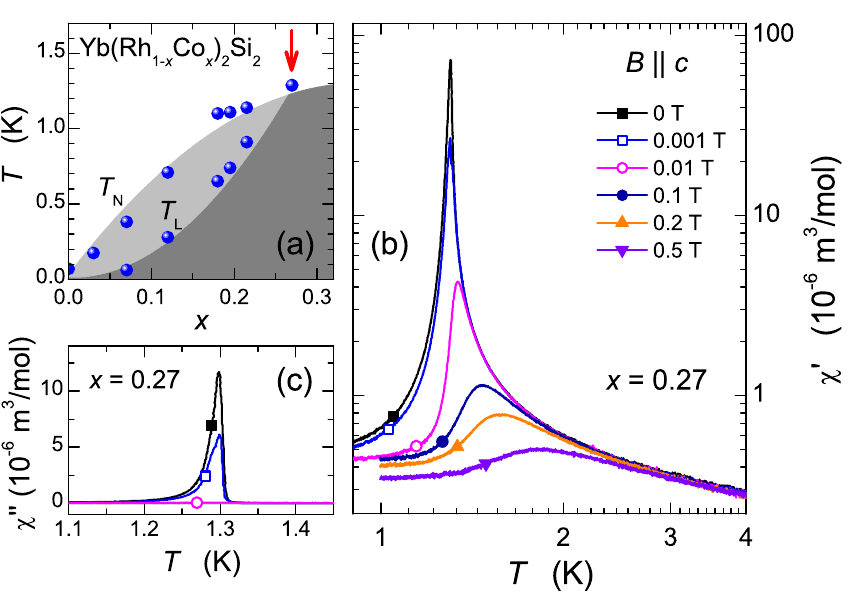}
\end{center}
\caption{(a) $T - x$ phase diagram of \YRCS\, as measured by \textcite{Klingner_et_al_2011}
                   showing the transition temperatures $T_{L}$ and $T_{N}$. The red arrow marks the sample with $x = 0.27$ which is FM.
                  (b) Real part $\chi'(T)$ of the susceptibility for \YRCSFM\, in different magnetic fields with $B \parallel c$. The sharp peak at $T_{C} = 1.30\,$K and $B = 0$ 
                  is suppressed and shifted towards higher $T$ with increasing field. (c) Temperature dependence of the imaginary part $\chi''(T)$ of the susceptibility. 
                  From~\textcite{Lausberg_et_al_2013}.}
\label{YRS_PD}
\end{figure}

Isoelectronic substitution of Rh by Co in \YRCS\, leads to a similar effect as pressure; the correspondence is excellent  for $p \leq 2.5$\,GPa~\cite{Klingner_et_al_2011}. 
The phase diagram for $x < 0.3$ is shown in Fig.~\ref{YRS_PD}. At $x = 0.27$, which corresponds to about 4.5\,GPa, $T_{L} = T_{N} = 1.3$\,K. In this sample FM order 
was indeed found by~\textcite{Lausberg_et_al_2013}. Surprisingly, the moments order along the
magnetic hard axis (i.e., the $c$-axis), similarly to what occurs in \YNP~\cite{Steppke_et_al_2013} and \CRPO~\cite{Krellner_et_al_2007}, and despite
the large magnetocrystalline anisotropy (which is about 6 in \YRCSFM). A plot of the real and imaginary parts of the susceptibility with $B \parallel c$ is 
shown in Fig.~\ref{YRS_PD} (b) and (c), respectively. A pronounced peak is seen in both quantities, with huge absolute values. 
In \YRCSFM\, this anomalous behavior can be well explained in terms of a Heisenberg model with competing FM and AFM exchange 
interactions~\cite{Andrade_et_al_2014}. This model also explains the observation that the transition from the AFM to the FM at, e.g., $x=0.21$, is first
order, as it is in Nb$_{1-y}$Fe$_{2+y}$, see Sec.\ \ref{NF}.

The discovery of ferromagnetism in  \YRCSFM\ suggests that the state below $T_{L}$ (the dark gray area in 
Fig.~\ref{YRS_PD}(a)) is also FM with a field-induced FM QCP. However, recent investigations
by~\textcite{Hamann_et_al_2014} indicate a much richer phase diagram with additional AFM phases. \YRCS\, thus appears to be one of the systems 
whose ground state changes from FM into AFM while approaching the putative FM QCP (cf. Fig~\ref{fig:schematic_phase_diagrams}\,c)). As in
CeRu$_2$Si$_2$, the physics of Yb(Rh$_{1-x}$Co$_x$)$_2$Si$_2$ is controlled by the evolution of the Kondo temperature, the magnetic anisotropy,
and the exchange interactions, and not simply by the mechanism discussed in Sec.\ \ref{subsec:III.E} that could be valid for a simpler system such as
Nb$_{1-y}$Fe$_{2+y}$. However, the strong similarity between the phase diagrams and some of the properties of these two materials is intriguing.

\subsubsection{Discussion, and comparison with theory}
\label{subsubsec:II.D.4}

All of the phase diagrams discussed in this section have the same overall structure: 
As the Curie temperature decreases as a function of some control
parameter, the ground state changes from a homogeneous FM to some modulated magnetic state that is often summarily referred to as  AFM, even if
its detailed structure is not known.%
\footnote{\label{AFM_evidence_footnote} It must be noted, however, that the evidence for an AFM phase is stronger in some materials than in others,
                         and that additional experimental work is needed in many cases, see the discussions of the individual systems above.} 
This is an important distinction from a theoretical point of view, as classic antiferromagnetism 
involves structure on an atomic scale, whereas other modulated states, such as the SDW in \NFy, Sec.\ \ref{NF}, or the helimagnetism in MnSi, Sec.\ \ref{par:II.B.1.a}, are
long-wavelength phenomena. In any case, the resulting phase diagram typically has a control-parameter range where the modulated state and the
FM state are observed subsequently as the temperature is swept, followed by a range where only the modulated phase exists, see
Figs.\ \ref{NFy}, \ref{CeRuPO_phase_diagram}, and \ref{CeAgSb2}. The general structure of these phase diagrams is shown schematically in
Fig.~\ref{fig:schematic_phase_diagrams}c). It is quite complicated, with two QPTs (one between the FM state and the modulated state, and one
between the modulated state and the nonmagnetic state), and a multicritical point where the three phases meet. 

The issues related to the nature and properties of AFM QCPs are very different from those of FM QCPs, and are not the main subject of
this review. Specifically, most of the theoretical concepts discussed in Sec.\ \ref{sec:III}, with the exception of Sec.\ \ref{subsec:III.E}, apply to FM QCPs
only, and questions related to the first-order vs. second-order nature of any AFM QPT must not be confused with the corresponding questions for FM QPTs.
Moreover, in some systems, especially Kondo-lattice systems, such a change from FM to AFM order can occur as the result of competing interactions,
i.e., frustration, as a result of pressure or chemical substitution. These mechanisms are very different from the one discussed in Sec.\ \ref{subsec:III.E}.

An interesting topic is the transition from the FM to the modulated phase, and in particular the order of this transition at zero temperature.
FM-to-AFM transitions are common even at high temperatures, and are often accompanied by a structural phase transition, which makes
them first order. The nature of the FM-to-AFM/SDW transitions discussed in this section is probably different, as indicated by the small
wave number that characterizes the modulated phase. They likely belong to the class of Lifshitz transitions, which separate a homogeneous phase 
from a phase with a modulated order parameter, see, \textcite{Hornreich_Luban_Shtrikman_1975} and references therein, and also
\textcite{Chaikin_Lubensky_1995}, and have been considered in many different contexts. 
On the basis of the mechanism discussed in Sec.\ \ref{sec:III} there
are general theoretical reasons to believe that a QPT from a metallic FM to a modulated magnet is generically of first order, although a detailed 
general theory remains to be worked out, see also
Sec.\ \ref{subsec:IV.B} point 3. Classically, this transition can be either first order or second order \cite{Chaikin_Lubensky_1995}.
A generalization of the theory reviewed in Sec.\ \ref{subsec:III.B}
that allows for a modulated order parameter has been developed by \textcite{Karahasanovic_Kruger_Green_2012}, see
Fig.~\ref{figure:nematic_spiral_uniform_phase_diagram} (the nematic phase may or may not be present) and the discussion in Sec.~\ref{subsec:III.E}.
The structure of the resulting phase diagram agrees with what is observed in, e.g., \NFy, see Fig.~\ref{NFy}. However, the theory predicts a first-order transition
from the AFM/spiral phase to the PM, whereas the SDW-PM transition in \NFy\ is observed to be continuous.
% It is not known whether in \NFy\ the wave vector of the SDW changes continuously throughout the phase, as expected from the theory.

FM Kondo lattice systems have been studied theoretically by \textcite{Perkins_et_al_2007} by means of a mean-field theory, and by
\textcite{Yamamoto_Si_2010} by means of a RG treatment. Both found a second-order QPT, i.e., the RG treatment found that the earlier generalized
Stoner theory is exact with respect to the order of the transition. Technically, this is because the calculation by \textcite{Yamamoto_Si_2010} does not
yield the nonanalytic wavenumber dependence of the spin propagator found in a related model \cite{Chubukov_Pepin_Rech_2004}, which destroys
the FM QCP and leads to a first-order transition as described in Sec.\ \ref{subsubsec:III.B.2}. While it is conceivable that the two models are different
in this respect, this seems unlikely, and more work on this topic is needed. In any case, these theories consider a QPT from a ferromagnetic metal
to a paramagnetic metal, which so far has not been observed in Kondo-lattice systems (although in some materials the experimental situation is not
quite clear yet, see, e.g., CeRu$_{1-x}$Fe$_x$PO in Sec.\
\ref{CRFPO}).

\subsection{System showing glass-like behavior, short-range order, or other strong-disorder effects}
\label{subsec:II.E}

%\textcolor{red}{[MB: Please check everything, including the \%-ed out text in the source code.]} 
This section (with the exception of Sec.\ \ref{subsubsec:II.E.4}) describes ferromagnetic metallic systems that display effects believed to be characteristic of strong 
disorder in the region where the FM order
is destroyed, and for many of them the nature and precise location of the FM QPT is not clear. Some materials display effects that have been interpreted as 
evidence for a quantum Griffiths region on the PM side of the QCP, with or without additional evidence of glassy freezing of the rare regions or clusters that 
characterize the Griffiths region. A special case is CeFePO, which displays short-range magnetic order in the absence of strong quenched disorder. The 
systems discussed here are listed in Table~\ref{table:4}.%
\footnote{We do not include diluted magnetic semiconductor, such as Fe$_{1-x}$Co$_{x}$S$_{2}$. For an example, see~\textcite{Guo_et_al_2008}.} 
They are arranged with respect to their phenomenology and/or its interpretation. 

The behavior characteristic of these materials 
can often be obtained by substituting a magnetic element by a non-magnetic one of the same series, e.g., uranium by thorium in \UTNS\ for the actinide 
series, or nickel by vanadium in \NV\ for the transition-metal series. This usually introduces substantial amounts of quenched disorder. The $x$ - $T$ phase diagram is 
often characterized by a pronounced tail (cf. Fig.~\ref{fig:schematic_phase_diagrams}\,d), % where the phase transition becomes smeared, a common feature of diluted magnetic alloys. This is because the thermodynamics is dominated by locally ordered clusters. Right above the tail, non-Fermi-liquid behavior can be observed, in form of power-laws in thermodynamic quantities, usually ac-susceptibility and specific heat. This is due to quantum Griffiths singularities, that can be observed in a temperature window above the ordering or freezing temperature (see Sec.~\ref{subsubsec:III.B.3}). 
and the region immediately above the tail is generally characterized by NFL behavior. The tail has been interpreted in terms of locally ordered clusters, and the NFL 
behavior in terms of quantum Griffiths singularities, a topic that we review in Sec.\ \ref{subsec:III.D}. This is, however, not the only possible explanation for a tail in 
the phase diagram, see the discussion in Sec.\ \ref{subsubsec:III.B.3}. Some of the theoretical and experimental results pertinent to this section
have been summarized by~\textcite{Vojta_2010}. As in the case of Sec.\ \ref{subsec:II.D}, different experiments and their interpretations are not consistent for some
materials, and it is possible that some of the systems discussed below will eventually be classified with those in Sec.\ \ref{subsec:II.C}. Conversely, some materials
discussed in Sec.\ \ref{subsec:II.C} may eventually be found to belong in the current section, especially URu$_{2-x}$Re$_x$Si$_2$, Sec.\ \ref{URuReSi}, and 
possibly Ni$_x$Pd$_{1-x}$, Sec.\ \ref{NP}, while \CFPO\ may belong into Sec.\ \ref{subsec:II.B} according to some experimental results, see Sec.\ \ref{CeFePO}.
\begin{table*}[t]
\caption{Systems showing short-range order or spin-glass (SG) freezing. \TC\ = Curie temperature, $T_{\text{g}} =$
 freezing temperature, $\rho_0 =$ residual resistivity. QC = quantum critical. QGP = quantum Griffiths phase. N/A = not applicable. n.a. = not available.}
\smallskip
\begin{ruledtabular}
\begin{tabular}{llllllll}
System 
   & Order of 
      & $T_{\text{C}}$/K 
         & $T_{\text{g}}$/K 
            & \hskip -10pt magnetic 
               & tuning  
                  & Disorder  
                     & Comments \\

   & Transition$\,^a$ 
      &        
         &                    
            & \hskip -10pt moment/$\mu_{\text{B}}\,^b$
               &  parameter  
                  & ($\rho_0/\mu\Omega$cm)$\,^c$
                     & \\
%\\
\hline\\[-7pt]
%%%%%%%%%%%%%%%%%%%%%%%%%%%%%%%
CePd$_{1-x}$Rh$_x$ 
   & n.a.
      & 6.6 -- 3 $\,^{d}$
         & 3 -- 0 $\,^{e}$
            & n.a.
               & composition$\,^{1,2}$
                  & n.a.
                     & Kondo cluster  \\
  &  &  &  &  &  &  & glass \\                    
\\[-7pt]
%%%%%%%%%%%%%%%%%%%%%%%%%%%%%%%
CePt$_{1-x}$Rh$_x$ 
   & n.a.
      & 6 -- 2$\,^f$
         & 3 -- 2$\,^g$
            & n.a.
               & composition$\,^{3}$
                  & n.a.
                     & cluster glass (?)$\,^h$\\
\\[-5pt]

%%%%%%%%%%%%%%%%%%%%%%%%%%%%%%%%
Ni$_{1-x}$V$_x$ 
   & n.a.
      & 633 -- $\approx 30\,^{4,i}$
         & $\approx 30$ -- $0.2\,^{4,j}$
            & 0.6 -- 0
               & composition$\,^{4}$
                  & n.a.
                     & cluster glass (?)$\,^h$\\
\\[-5pt]
%%%%%%%%%%%%%%%%%%%%%%%%%%%%%%%%
UNi$_{1-x}$Co$_x$Si$_2$ 
   & n.a.
      & 95 -- 8.6$\,^{5,k}$
         & 6 (?)$\,^{5,l}$
            & n.a.
               & composition$\,^5$
                  & 15\,$\mu\Omega$cm$\,^{6,m}$
                     & glassy phase ?\\[2pt]
\hline\\[-7pt]
%%%%%%%%%%%%%%%%%%%%%%%%%%%%%%%%
U$_{1-x}$Th$_x$NiSi$_2$ 
   & n.a.
      & 95 -- 29$\,^{7,n}$
         & 29 -- $\approx 4\,^{7,o}$
            & n.a.
               & composition$\,^7$
                  & n.a.
                     & possible QGP \\
\\[-5pt]
%%%%%%%%%%%%%%%%%%%%%%%%%%%%%%%%%%%%%%%%%%
CeTi$_{1-x}$V$_x$Ge$_{3}$ 
   & 2nd$\,^{8,9}$
      & 14 -- 2.8$\,^{8,9}$
         & n.a.
            & 1.5$\,^{9}$
               & composition$\,^{9}$
                  & 22$\,^{9,p}$
                     & QGP ? \\[2pt]
\hline\\[-7pt]
%%%%%%%%%%%%%%%%%%%%%%%%%%%%%%%%
Sr$_{1-x}$Ca$_x$RuO$_3$ 
   & n.a.
      & $\approx 100$ -- 40$\,^{10,q}$
         & $\approx 40$ -- $\approx 5\,^{10,r}$
            & 1 -- 0
               & composition$\,^{10}$
                  & 8$\,^{s}$
                     & thin films \\[2pt]
\hline\\[-7pt]
%%%%%%%%%%%%%%%%%%%%%%%%%%%%%%%%%%%%%%%%%%
CeFePO
   & N/A
      & N/A
         & 0.9$\,^{11}$
            & n.a.
               & none
                  & 800$^{\,12,t}$
                     & low intrinsic\\
  &  & & & & & & disorder
\\
%%%%%%%%%%%%%%%%%%%%%%%%%%%%%%%%
\hline\hline\\[-5pt]
\multicolumn{8}{l} {$^a$ For the disappearance of homogeneous FM order.\quad 
                              $^b$ Per formula unit unless otherwise noted.
                              }\\
\multicolumn{8}{l} {$^c$ For the highest-quality samples.\quad
                              $^d$ For $x = 1$ -- $0.6$~\cite{Sereni_et_al_2007}
                              }\\
\multicolumn{8}{l} {$^e$ For $x = 0.6$ -- $0.9$~\cite{Sereni_et_al_2007,Westerkamp_et_al_2009}.\quad
                              $^f$ For $x = 0$ -- 0.7\quad 
                              $^g$ For $x = 0.5$ and 0.6.\quad
                              }\\
\multicolumn{8}{l}  {$^h$ See Footnote \ref{cluster_glass_footnote}.\quad
                              $^i$ For $x = 0$ -- $0.105$~\cite{Ubaid-Kassis_et_al_2010}.\quad
                              $^j$ For $x = 0.11$ -- $0.1225$~\cite{Ubaid-Kassis_et_al_2010}.\quad
                               }.\\
\multicolumn{8}{l} { $^k$ For $x = 0$ -- $0.96$~\cite{Pikul_Kaczorowski_2012}.\quad
                               $^l$ For $x = 0.98$~\cite{Pikul_Kaczorowski_2012}.\quad
                              $^m$ For UNiCoSi.\quad
                              }\\
\multicolumn{8}{l} {$^n$ For $x = 0$ -- $0.7$~\cite{Pikul_et_al_2012a}.\quad
                              $^o$ For $x = 0.7$ -- $0.9$~\cite{Pikul_et_al_2012a}.\quad
                              $^p$ For CeTiGe$_{3}$.}\\
\multicolumn{8}{l} {
                              $^q$ For $x = 0.15$ -- $0.38$~\cite{Demko_et_al_2012}.\quad    
                              $^r$ For $x = 0.38$ -- $0.52$~\cite{Demko_et_al_2012}.}\\
\multicolumn{8}{l} {$^s$ For $x=0$ \cite{Schneider_et_al_2010}. RRR values varied from a high of 28.9 for $x=0$ to a low of 2.9 for $x=0.5$.}\\
\multicolumn{8}{l}  {$^t$ High $\rho_{0}$ not intrinsic, but due to granularity of the polycrystalline sample.}\\
\\[-5pt]
\hline\\[-5pt]
\multicolumn{2}{l} {$^1$ \textcite{Sereni_et_al_2007}}
      & \multicolumn{2}{l} {$^2$ \textcite{Westerkamp_et_al_2009}}     
            & \multicolumn{2}{l} {$^{3}$ \textcite{Kawasaki_et_al_2008, Kawasaki_et_al_2009}}
                  & \multicolumn{2}{l} {$^{4}$ \textcite{Ubaid-Kassis_et_al_2010}}\\
\multicolumn{2}{l} {$^5$ \textcite{Pikul_Kaczorowski_2012}}
      & \multicolumn{2}{l} {$^6$ \textcite{Kaczorowski_1996}}     
            & \multicolumn{2}{l} {$^{7}$ \textcite{Pikul_et_al_2012a}}
                  & \multicolumn{2}{l} {$^{8}$ \textcite{Manfrinetti_et_al_2005}}\\     
\multicolumn{2}{l} {$^{9}$ \textcite{Kittler_et_al_2013}}
      & \multicolumn{2}{l} {$^{10}$ \textcite{Demko_et_al_2012}}     
            & \multicolumn{2}{l} {$^{11}$ \textcite{Lausberg_et_al_2012a}}
                  & \multicolumn{2}{l} {$^{12}$ \textcite{Bruening_et_al_2008}}\\ 
%%%%%%%%%%%%%%%%%%%%%%%%%%%%%%%%%%%%%%%%%%%%%%%%%%%%%%%%%%
\end{tabular}
\end{ruledtabular}
\vskip -3mm
\label{table:4}
\end{table*}

\subsubsection{Systems with glass-like features}
\label{subsubsec:II.E.1}

The systems with the most obvious strong-disorder effects include \CPR, \NV, and \UNCS. Their unusual behavior has been interpreted in terms of a Griffiths
region in the PM phase, with symptoms of glassy freezing at the lowest temperatures. 

\paragraph{\CPR}
\label{CPR}

\begin{figure}[t]
\begin{center}
\includegraphics[width=0.95\columnwidth,angle=0]{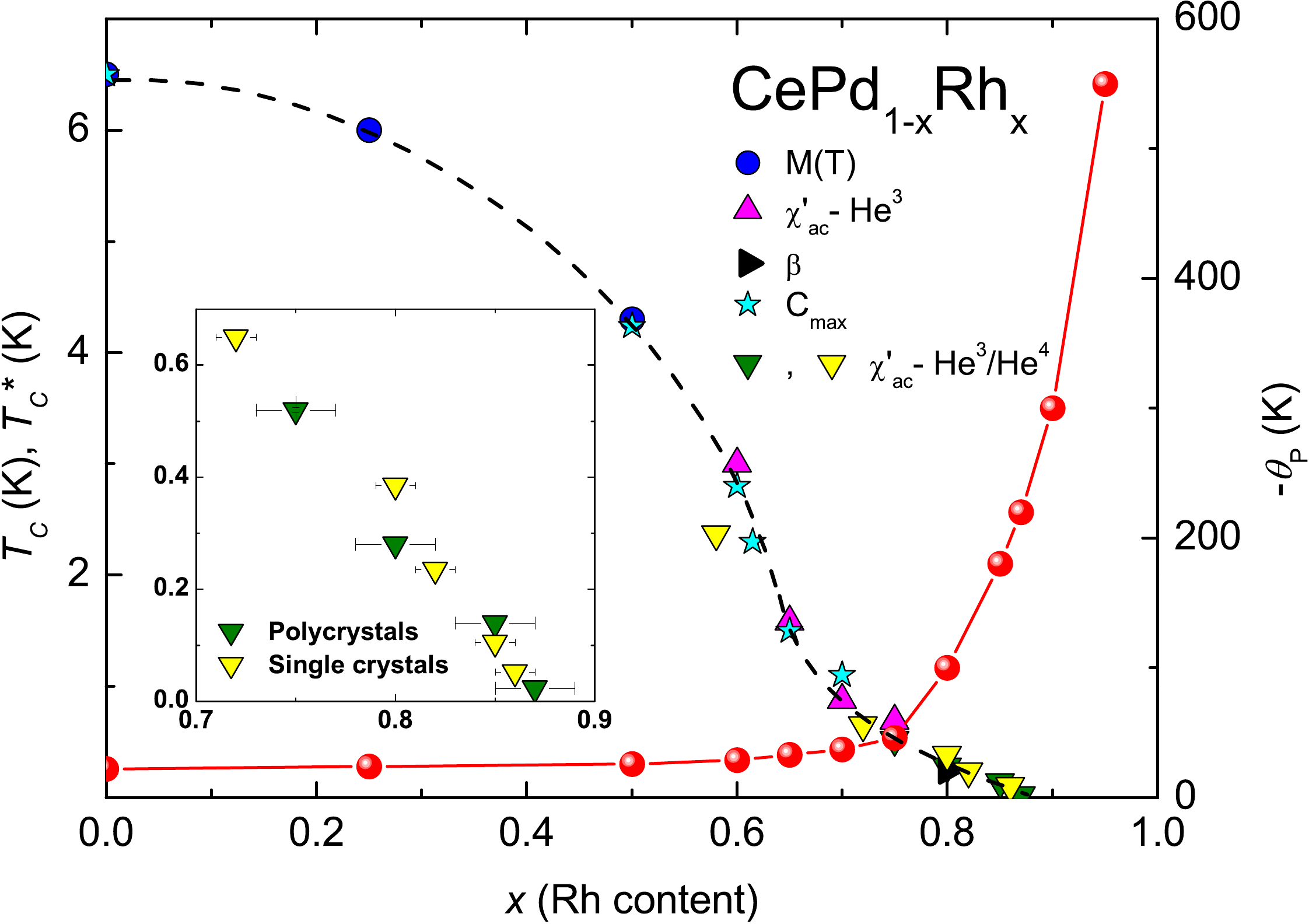}
\end{center}
\caption{Magnetic phase diagram of \CPR. Left scale: composition dependence of the ordering (freezing) temperature \TC\ ($T_{\textrm{C}}^{*}$) deduced from 
              various measurement techniques: magnetization ($M$), ac susceptibility ($\chi'_{ac}$), thermal expansion ($\beta$) and specific heat ($C$). The inset 
              shows \TC($x$) values observed in $\chi'_{ac}(T)$ for $x > 0.7$ in poly- and single crystals. Right scale: the Weiss temperature $\theta_{\textrm{P}}$. 
              Afer \textcite{Sereni_et_al_2007, Westerkamp_et_al_2009}.}
\label{CePdRh_phase_diagram}
\end{figure}

The Curie temperature of \CPR\  as a function of Rh concentration has been followed over more than two decades in temperature, down to 25\,mK at $x = 0.87$, in both
polycrystals and single crystals, and the $x$ - $T$ phase diagram shows a pronounced tail (cf. Fig.~\ref{CePdRh_phase_diagram}), as shown schematically in 
Fig.~\ref{fig:schematic_phase_diagrams}\,d)~\cite{Sereni_et_al_2007}. The unusual properties of this system have been interpreted in terms of a quantum Griffiths phase 
(QGP) \cite{Westerkamp_et_al_2009}.

The entire series crystallizes in the orthorhombic CrB structure. Earlier studies on polycrystals have shown that the system evolves from a FM ground state in CePd with 
\TC\ = 6.6\,K to a non-magnetic intermediate-valence state in CeRh~\cite{Kappler_et_al_1991,Sereni_et_al_1993}. The chemical substitution of the Ce-ligand Pd with 
Rh induces not just a volume effect (positive chemical pressure), but also increases the local hybridization strength of the cerium $4f$ electrons with the conduction 
electrons, leading to a strong enhancement of the Kondo temperature $T_{K}$~\cite{Sereni_et_al_2007}. In addition, the Rh substitution introduces disorder. 
%
%\begin{figure}[ht]
%\begin{center}
%\includegraphics[width=0.95\columnwidth,angle=0]{figures/eps/CePdRh_ac_susceptibility.eps}
%\end{center}
%\caption{(Color online) Temperature dependence of the ac susceptibility ($\chi'_{ac}(T)$). Every curve has been normalized at its maximum value. The temperature scale of the sample with $x = 0.60$ ͑only͒ is shown on the upper abscissa. Figure adapted from Ref.~\textcite{Sereni_et_al_2007}.}
%\label{CePdRh_ac_susceptibility}
%\end{figure}
%

Evidence for the FM nature of the ordered state comes from the $T$-dependence of the ac susceptibility $\chi'(T)$, which shows large and sharp maxima for all samples
ranging from $x = 0.6$ to $x = 0.87$~\cite{Sereni_et_al_2007}. No maximum was observed down to 20\,mK in a sample with $x = 0.9$; this indicates a critical 
concentration for the loss of ferromagnetism very close to $x_c = 0.87$ (other types of magnetic order can be ruled out for this sample)~\cite{Westerkamp_et_al_2009}. 
The resulting phase diagram in Fig.~\ref{CePdRh_phase_diagram} shows the transition temperature deduced from measurements of various observables
as a function of the composition $x$. 
\textcite{Westerkamp_et_al_2009} have attributed the continuous decrease of \TC\ with increasing $x$ on the competition between FM order and growing Kondo 
screening. The curvature of the phase boundary \TC\ changes from negative to positive at $x \approx 0.6$, displaying a long tail towards higher Rh contents. In this 
concentration range, the Kondo temperature $T_{K} \approx 2\vert\theta_{\textrm{P}}\vert$, with $\theta_{\textrm{P}}$ the paramagnetic Weiss temperature obtained from 
fits of the dc susceptibility at high temperatures, strongly increases with $x$.
The main mechanism governing $T_{K}$ is the hybridization of the Ce $4f$ electrons with the valence electrons of the surrounding ligands. In Ce-based compounds, 
Rh ligands are known to lead to much larger $T_{K}$ than Pd ligands~\cite{Koelling_et_al_1985}. Thus, in \CPR\ the effect of the Rh-ligands is much stronger than the effect of the Pd-ligands once 
the Rh content reaches a critical value close to 0.7. Very likely, the random distribution of Rh and Pd ligands creates regions with different local values of $T_{K}$. An 
analysis of the entropy and the slope of $\chi'(T)$ at 2\,K revealed some fraction of unscreened magnetic moments, even at high $x$ where the average $T_{K}$ is 
already above 50\,K. The pronounced maxima in $\chi'(T)$ of samples with concentrations $x \geq 0.6$ exhibit a frequency dependence similar to that observed in 
spin glasses~\cite{Westerkamp_et_al_2009}. The relative temperature shift of about 3 to 10\% per frequency decade is considerably 
larger than that in canonical metallic spin glasses (where typical values are 1 to 2\%), but well below the value of about 28\% observed in 
superparamagnets~\cite{Mydosh_1993}. This behavior, and zero-field-cooled and field-cooled magnetization measurements, have been interpreted as evidence for 
the existence of clusters of magnetic moments in the system below a certain temperature $T_{\text{cluster}}$~\cite{Westerkamp_et_al_2009}.

\begin{figure}[b]
\begin{center}
\includegraphics[width=0.9\columnwidth,angle=0]{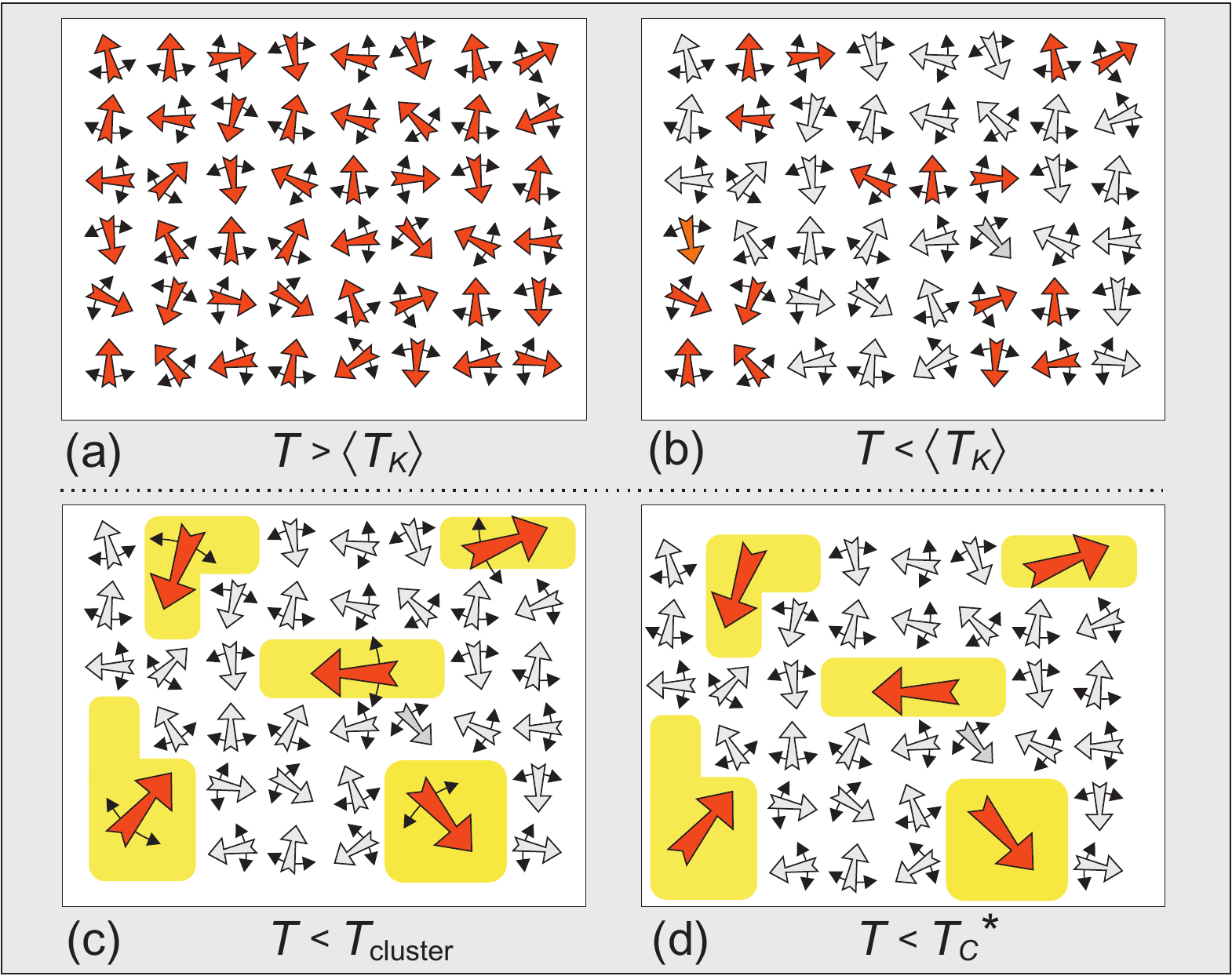}
\end{center}
\caption{Schematic representation of the formation of a Kondo-cluster glass in \CPR\, as a function of $T$. The small red and gray arrows indicate fluctuating 
             $f$-electron moments above and below their respective Kondo temperature; the big red arrows represent FM clusters. The temperature decreases from 
             frame (a) to (d). See the text for additional information. From~\textcite{Westerkamp_2009}.}
\label{CePdRh_clusters}
\end{figure}

The behavior for $x > 0.6$ can be visualized in terms of different magnetic states the system goes through upon cooling, see Fig.~\ref{CePdRh_clusters}. At 
temperatures high enough to overcome the Kondo screening, fluctuating magnetic moments exist on every Ce site, indicated by the small red arrows, panel (a). Below 
the average Kondo temperature $\langle T_{K}\rangle$ an increasing number of $f$-moments becomes screened, as represented by the gray arrows, panel (b). 
However, due to the statistical distribution of Rh dopants on the Pd site and the strong dependence of the local $T_{K}$ on the number of nearest neighbors, there 
remain regions where the Kondo scale has not yet been reached. Inside these regions, the $f$-moments are still unscreened (indicated in red). At even lower 
temperatures, $T < T_{\text{cluster}}$, these moments form clusters with predominantly FM coupling of the moments. In panel (c), the large red arrows represent the 
total magnetic moment of each cluster. Within this temperature regime, the clusters are fluctuating independently. Upon further cooling below a temperature 
$T_{\textrm{C}}^{*}$, random freezing of the cluster moments sets in, leaving a static spin configuration as displayed in panel (d). Since the broad distribution of local 
Kondo temperatures is thought to be responsible for the cluster formation, \textcite{Westerkamp_et_al_2009} called the low-$T$ state in \CPR\ a ``Kondo-cluster glass".\footnote{\label{cluster_glass_footnote} The term "cluster glass" is frequently used in a spin-glass context (see, e.g., \textcite{Itoh_et_al_1994}), and 
                                                               different authors use it for various phenomena and concepts whose underlying physics may be quite different.}
The observation of reentrant depolarization in recent neutron-depolarization imaging experiments~\cite{Schmakat_et_al_2015} on the same samples investigated by~\textcite{Westerkamp_et_al_2009} seems to confirm the presence of the Kondo-cluster-glass state. The general behavior of \CPR\ is quite different from that observed in many other disordered NFL systems, such as CeNi$_{1-x}$Cu$_x$, where the Ce valence remains nearly trivalent and where a percolative cluster scenario has been proposed~\cite{Marcano_et_al_2007}.%

\begin{figure}[t]
\begin{center}
\includegraphics[width=0.9\columnwidth,angle=0]{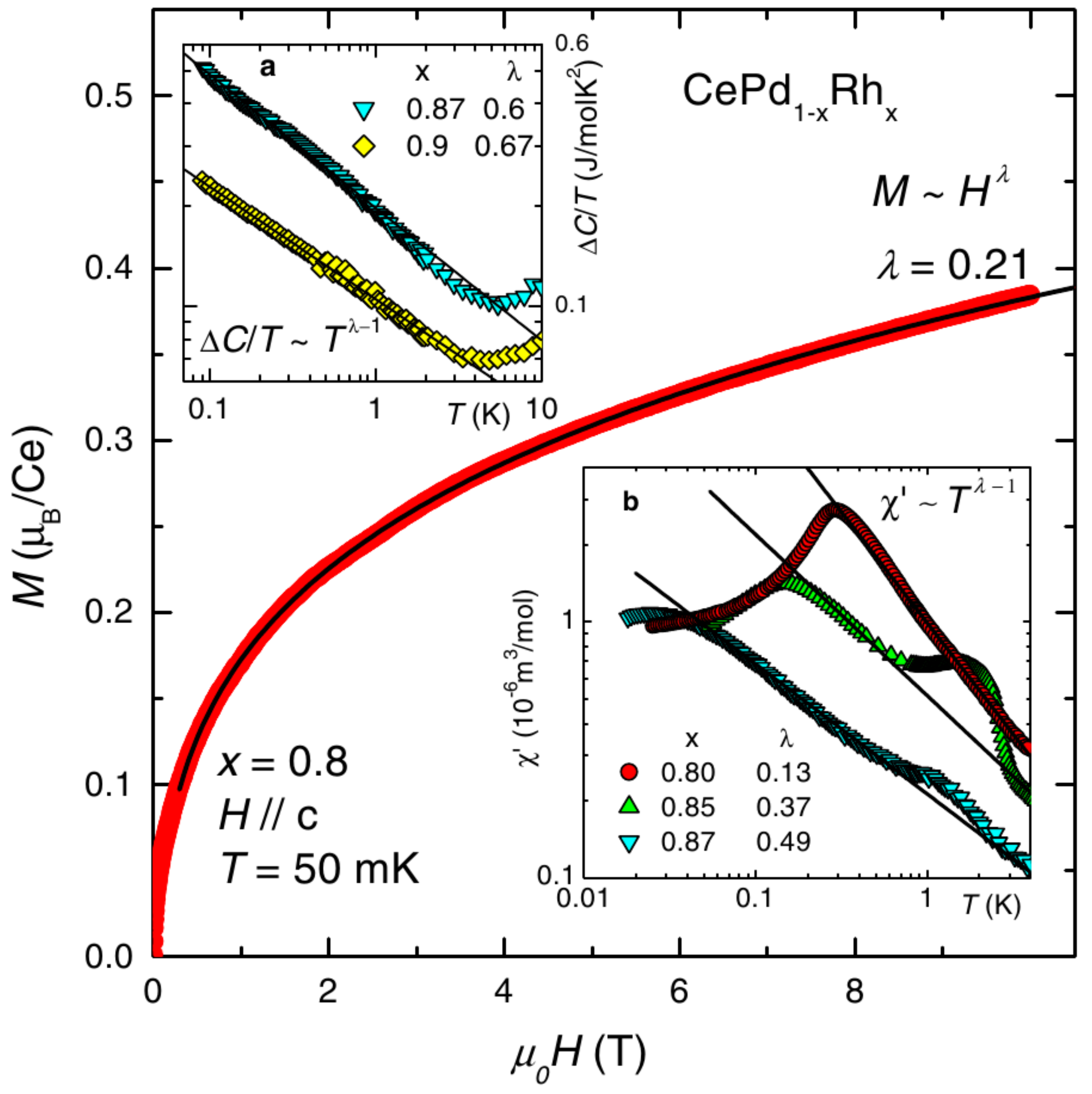}
\end{center}
\caption{Comparison of thermodynamic data for \CPR\ with the quantum Griffiths phase scenario (cf. Sec.~\ref{subsec:III.D}). Main figure: Field 
              dependence of the magnetization $M$ for a single crystal with $x = 0.8$ at 50\,mK. The data follow a power law $M\propto H^{\lambda}$ with $\lambda = 0.21$. 
              Inset a: $4f$ Sommerfeld coefficient for polycrystals with $x = 0.87$ and $0.9$ \cite{Pikul_et_al_2006} plotted on a double-logarithmic scale. Solid lines 
              indicate a $T^{\lambda - 1}$ power-law behavior. Inset b: $T$-dependence of the ac susceptibility $\chi'(T)$ of three polycrystals with $x = 0.8$, 0.85 and
              0.87. The lines are fits to a $T^{\lambda - 1}$ power law. From~\textcite{Westerkamp_et_al_2009}.}
\label{CePdRh_Griffiths}
\end{figure}

Specific-heat measurements have shown the existence of NFL behavior for concentrations $0.85 \leq x \leq 0.9$~\cite{Deppe_et_al_2006, Pikul_et_al_2006}. 
Samples in this concentration range show a power-law $T$-dependence $C(T)/T \sim T^{\lambda - 1}$, with exponents $\lambda = 0.6$ 
and 0.67 for $x = 0.87$ and 0.9, respectively (see inset a of Fig.~\ref{CePdRh_Griffiths}). Power-law behavior has also been found for the $T$-dependent ac 
susceptibility, see inset b of Fig.~\ref{CePdRh_Griffiths}. These findings suggest that there are strong fluctuations in an entire range of Rh concentrations,
which raises the question whether \TC\ going to zero near $x = 0.87$ represents a FM QCP or not.
%associated with the disappearence of ferromagnetism in \CPR\, and raise the question whether a FM QCP is present at $x \approx 0.9$ or not. 
%In several Ce-based ferromagnets the increase of the Kondo interaction with pressure tends to stabilize an AF ground state before the QCP is reached~\cite{Suellow_et_al_1999,Burghardt_et_al_2005}. In addition, it is questionable whether the FM-QCP in a Kondo-lattice system can be described by itinerant scenarios like those described in Sec.~\ref{sec:III}.
%

A clear answer was given by~\textcite{Westerkamp_et_al_2009} who measured the Gr\"uneisen ratio, defined as $\Gamma\propto \beta/C$, with $\beta$ the 
volume thermal expansion coefficient and $C$ the specific heat. $\Gamma$ must diverge as a power law as $T$ goes to zero at any QCP that obeys scaling~\cite{Zhu_et_al_2003}; this has been confirmed
experimentally for several Kondo-lattice systems exhibiting an AFM QCP~\cite{Kuechler_et_al_2003}. Close to the critical concentration where the anomaly in 
$\chi'(T)$ disappears, a weak logarithmic behavior, $\Gamma(T) \propto \ln T$, has been found, contrary to the power-law divergence expected at a FM QCP. 
This shows that there is no FM QCP in \CPR\ at this concentration. Rather, in the region $0.7 < x < 0.9$ and in the temperature range 
$T_{\textrm{C}} \leq T \leq T_{\text{cluster}}$ the observations show power laws that are consistent with the QGP scenario that predicts 
$\chi'(T) \propto C(T)/T \propto T^{\lambda - 1}$ and $M \propto H^{\lambda}$, with $0 \leq \lambda \leq 1$ an $x$-dependent exponents
\cite{Castro_Neto_et_al_1998, Dobrosavljevic_Miranda_2005}, and $\Gamma(T) \propto \log(T)$~\cite{Vojta_2009}, see Sec.~\ref{subsec:III.D}. 
This is demonstrated in Fig.~\ref{CePdRh_Griffiths}, which shows that both the specific-heat coefficient and 
the susceptibility follow a power-law behavior above \TC, where the exponent varies systematically with $x$. However, the exponents $\lambda$
obtained from the specific heat and the susceptibility are not the same, in contrast to the theoretical prediction.

The field-dependent magnetization at 50\,mK also follows a power law~\cite{Brando_et_al_2010, Westerkamp_et_al_2009}. It is worth noting that \CPR\, displays a very 
small magnetic anisotropy at $x \approx 0.8$~\cite{Deppe_et_al_2006}, suggesting Heisenberg symmetry which is needed for the realization of the QGP 
scenario~\cite{Vojta_Schmalian_2005}. According to this interpretation, the observations for $0.7 \alt x \alt 0.9$ represent quantum Griffiths behavior in
the PM phase, and the true QCP, which one expects in the vicinity of $x = 0.7$, so far has not been observed.

\paragraph{CePt$_{1-x}$Rh$_x$}
\label{CePtRh}

\textcite{Kawasaki_et_al_2008, Kawasaki_et_al_2009} have substituted Pt in CePt (Sec.\ \ref{par:II.B.3.CePt}) by Rh to reduce \TC\ and found that
CePt$_{1-x}$Rh$_x$ behaves very similarly to CePd$_{1-x}$Rh$_x$. These authors have fitted the freezing temperature, identified as the temperature
where the ac susceptibility shows a pronounced peak, to a Vogel-Fulcher law with good results.

\paragraph{\NV}
\label{par:NiV}

The alloy \NV\ is a weak itinerant ferromagnet. A small amount of vanadium (about 12\%) suppresses \TC\ to zero from \TC\ $\approx$ 630\,K in pure 
nickel \cite{Boelling_1968}. It is attractive for studying quantum Griffiths effects for several reasons:
\begin{figure}[ht]
\begin{center}
\includegraphics[width=0.9\columnwidth,angle=0]{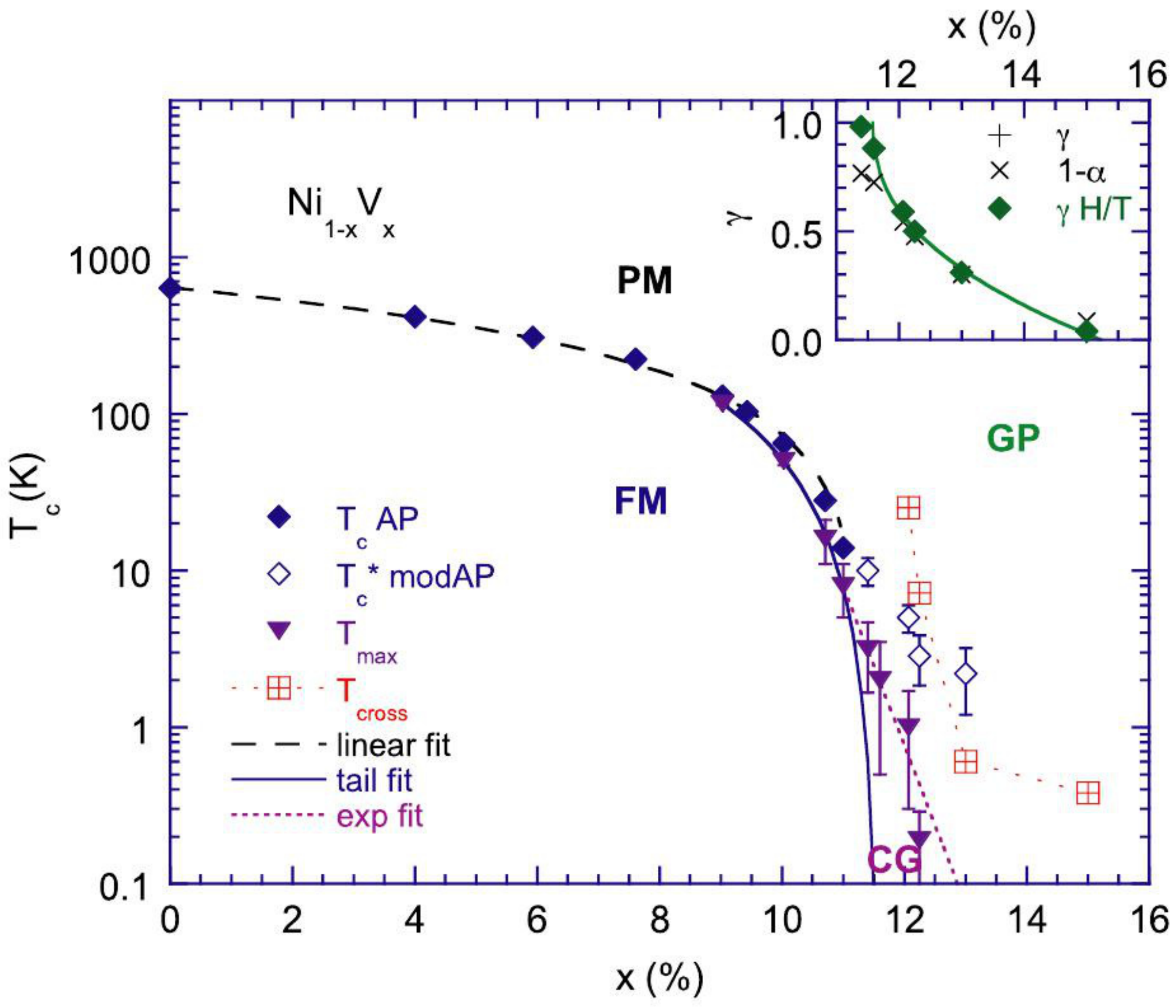}
\end{center}
\caption{$x$ - $T$ phase diagram of \NV\ plotted on a semilogarithmic scale. \TC\ was estimated from standard and modified Arrott plots. Also shown is the 
              temperature $T_{\textrm{max}}$ of the maximum in the ac susceptibility, which marks the freezing into the cluster glass (CG) state. For $0.114 \leq x \leq 0.15$
              effects consistent with a quantum Griffiths phase (GP) were observed in the susceptibility ($\chi \propto T^{-\gamma}$) and magnetization 
              ($M \propto H^{\alpha}$) , see Fig.~\ref{NiV_susceptibility}. The red squares mark the crossover from the GP to the cluster glass phase (CG).   
              The inset shows the strong $x$-dependence of the exponents.
              % in agreement with the theory which predicts $\lambda = \alpha = 1 - \gamma$. [Note:This is an AMF theory, so I took out that remark. We don't even
              % quote these theory papers. DB]
              From~\textcite{Schroeder_et_al_2011}.}
\label{NiV_phase_diagram}
\end{figure}
i) it is a simpler system than Kondo-lattice ferromagnets and has Heisenberg symmetry, ii) the high \TC\ of nickel allows the effects to be observable in a relatively large 
temperature range compared to other systems, such as \CPR\ where the maximal \TC\ is 6.6\,K, and iii) a vanadium impurity causes a strong reduction (about 90\%) of the 
magnetic moment of the neighboring Ni atoms, which creates well-defined large defects and therefore significant disorder. This is in contrast to \NP, where isoelectronic 
Pd substitution does not introduce much disorder, and a large amount of Pd (about 97.5\%) is needed to suppress $T_{\textrm{C}}$ to zero, see Sec.~\ref{NP}.
\begin{figure}[ht]
\begin{center}
\includegraphics[width=0.95\columnwidth,angle=0]{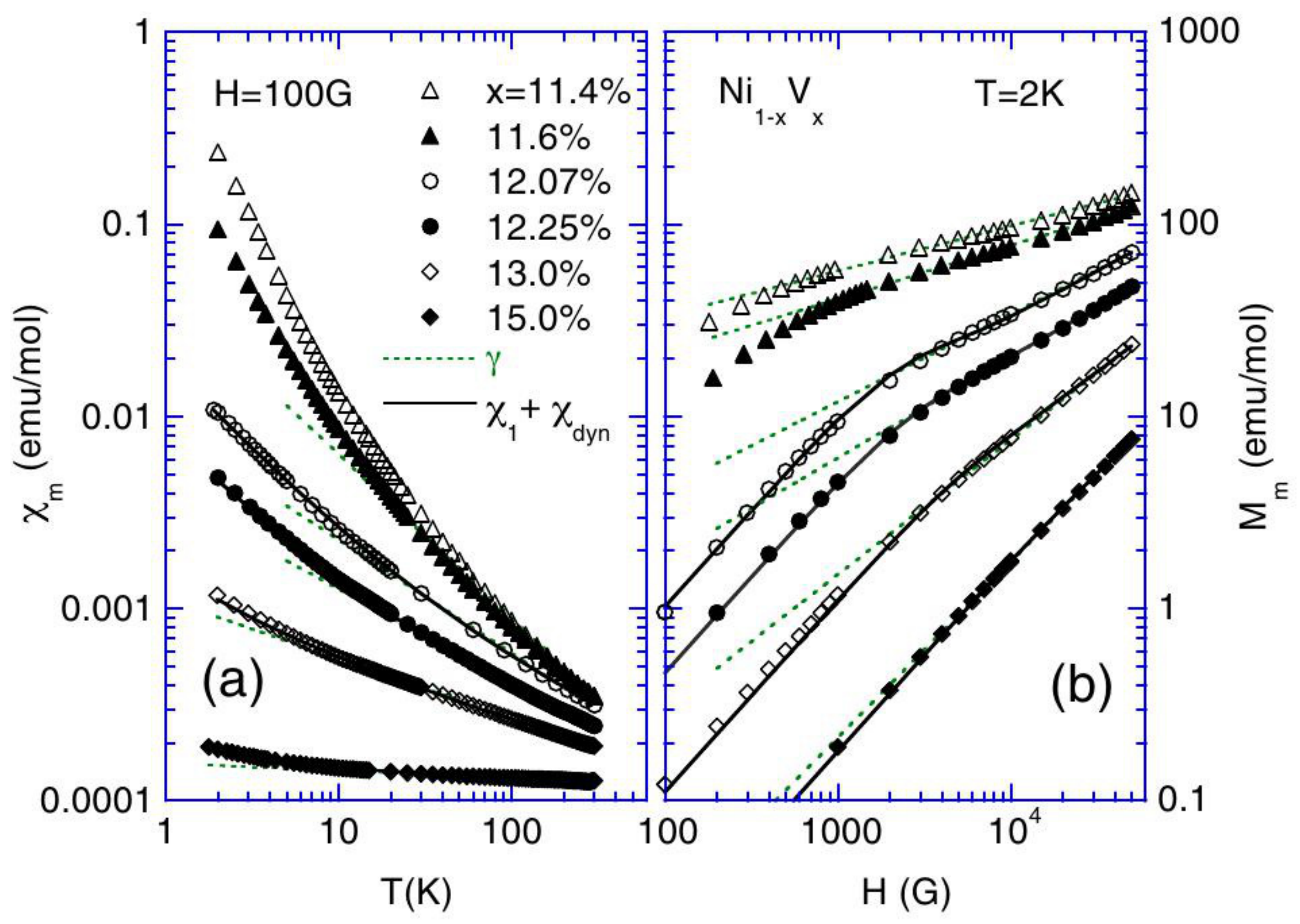}
\end{center}
\caption{Log-log plots of the low-field susceptibility $\chi_{m} = M/H - \chi_{orb}$ vs $T$, and of the magnetization $M_{m} = M - \chi_{orb}H$ at 2\,K vs $H$, for 
              samples with $0.114 \leq x \leq 0.15$, with $\chi_{orb}$ the orbital contribution to $\chi$. The dashed lines emphasize the power-law behavior 
              $\chi_{m} \propto T^{-\gamma}$ and $M_{m} \propto H^{\alpha}$. From~\textcite{Ubaid-Kassis_et_al_2010}.}
\label{NiV_susceptibility}
\end{figure}

\textcite{Ubaid-Kassis_et_al_2010} have measured the magnetization $M$ and the ac susceptibility $\chi$ of several samples with $0 \leq x \leq 0.15$. The $x$ - $T$ phase 
diagram is shown in Fig.~\ref{NiV_phase_diagram} (note the log-linear plot). For $x \leq 0.11$ the transition temperature was estimated from standard Arrott plots, 
$H/M = a + bM^{2}$. For larger values of $x$ the determination of \TC\ was model dependent; for $x \geq 0.11$ and fields $H > 0.5\,$T a modified Arrott plot, 
$M^{1/\beta} = M_{0}^{1/\beta}(T) + c(H/M)^{1/\gamma}$, was used. Recent $\mu$SR data confirmed that the long-range FM order is lost for $x$ close to 0.11
\cite{Schroeder_et_al_2014}. In addition to the Arrott plots, the  temperature $T_{\textrm{max}}$ of the maximum in the ac 
susceptibility was determined as in the case of \CPR. The field and frequency dependence of  $T_{\textrm{max}}$ for  $x \approx 0.12$ was found to be consistent
with what is expected for a cluster glass (see Sec.\ \ref{subsubsec:III.D.3}). The dashed line in Fig.~\ref{NiV_phase_diagram} is a linear extrapolation of $\ln$ \TC\ 
vs. $x$, which represents a shape of the phase diagram similar to that in Fig.~\ref{fig:schematic_phase_diagrams}\,d).

In the paramagnetic region of the phase diagram with $0.114 \leq x \leq 0.15$ the $T$-dependence of the susceptibility $\chi_{m} = M/H - \chi_{orb}$ (with a small 
orbital contribution $\chi_{orb} = 6 \times 10^{-5}$\,emu/mol) can be fitted to a power law $T^{-\gamma}$ for $10 \leq T \leq 300$\,K, see 
Fig.~\ref{NiV_susceptibility}. The magnetization $M_{m} = M - \chi_{orb}H$ also shows a power-law behavior, $H^{\alpha}$ at 2\,K, for $3000 \leq H \leq 50000$\,G. 
These power laws are typical signatures of a quantum Griffiths phase. Deviations from this behavior at low temperatures has been ascribed to the formation of a
cluster-glass phase~\cite{Ubaid-Kassis_et_al_2010}; the crossover between the two is marked by red points in Fig.~\ref{NiV_phase_diagram}. 
% It is worth mentioning that the exponents extracted from $\chi_{m}(T)$ and $M_{m}(H)$ fit the expected values given by the theory $\lambda = \alpha = 1 - \gamma$ (cf. Sec.~\ref{subsec:III.D}). [Note: No, this is an AFM theory that they should not even compare to; it is not the theory we discuss in Sec. III.D.]
It would be interesting to check whether the specific heat of these samples also follows a power law, $C(T) \propto T^{\lambda}$. 
%Moreover, the Griffiths exponent $\lambda$ was found to be proportional to $(x - x_{c})^{0.42}$ with the value 0.42 close to that expected from strong-disorder renormalization group result~\cite{Millis_1993} (see green line in the inset of Fig.~\ref{NiV_phase_diagram}). [Note: I'm not sure what this refers to, so I took it out for now. The Millis reference is not right, and Schroeder et al don't seem to say what the green line in the inset represents. DB]
%
\paragraph{\UNCS}
\label{UNiCoSi}

UNiSi$_{2}$ is a ferromagnet (see Sec.\ \ref{par:UNiSi}) in which nickel can be substituted by cobalt. Since UCoSi$_{2}$ is a paramagnet which shows strong 
spin fluctuations 
\cite{Kaczorowski_1996}, the series \UNCS\, is a promising system to look for a FM QCP. This has been investigated by \textcite{Pikul_Kaczorowski_2012}. 
Cobalt substitution leaves both the uranium lattice and the orthorhombic crystal structure intact, while reducing the unit cell volume only slightly (about 1.2\%). 
Since the $b\,$-axis stretches with increasing $x$, while the $a$ and $c$ axes shrink, Co substitution is not equivalent to hydrostatic pressure. The main 
effect of the doping seems to be the modification of the intersite coupling between the U magnetic moments:
\begin{figure}[b]
\begin{center}
\includegraphics[width=0.9\columnwidth,angle=0]{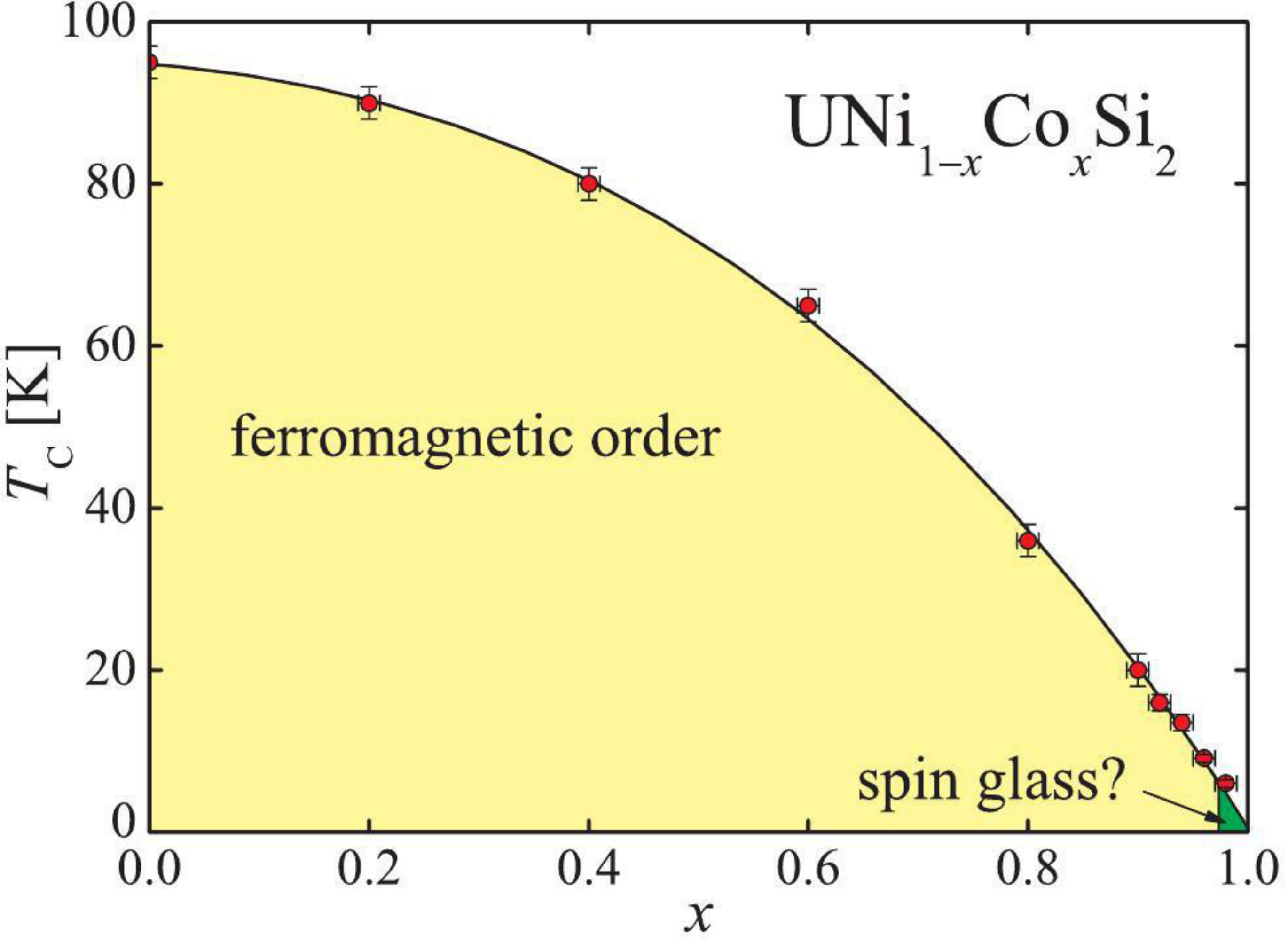}
\end{center}
\caption{$x$ - $T$ phase diagram of \UNCS\ derived from magnetization and specific-heat measurements. From \textcite{Pikul_Kaczorowski_2012}.}
\label{pikul2012b}
\end{figure}
A Curie-Weiss analysis of the susceptibility shows that the effective moment is almost unaffected by the Co substitution, whereas the Curie-Weiss temperature 
$\theta_{CW}$ varies from 95\,K in UNiSi$_{2}$ to $-70$\,K in UCoSi$_{2}$. The transition temperature as well as the remanent magnetization decrease continuously 
with increasing $x$. The clear onset of the magnetization observed at \TC\ for $0 \leq x \leq 0.96$ evolves for $x = 0.98$ into a small anomaly at $T\approx 6\,$K,
which is also seen in $C(T)/T$ as a broad hump. For $x = 1$ the anomaly is absent. The ground state of the $x = 0.98$ sample is unknown;
\textcite{Pikul_Kaczorowski_2012} have suggested that it is a spin-glass-like state with competing FM and AFM interactions (see Fig.~\ref{pikul2012b}). The
FM transition at nonzero temperature appears to be second order for all samples where it is clearly present; the order of the transition from the FM state to
the glassy state, if a glassy phase indeed exists, is not known. UCoSi$_{2}$ shows a logarithmic enhancement of the Sommerfeld coefficient, $C(T)/T \propto -\ln T$,
down to the lowest temperature measured, $T = 2\,$K, which \textcite{Pikul_Kaczorowski_2012} have interpreted as indicating vicinity to a QCP.

%This system, therefore, does not show any tendency to first-order QPT near the putative FM QCP nor a pronounced tail in the 
%$x - T$ phase diagram (although only one sample was investigated in the spin-glass-like region). Interestingly, the end compound of the series, UCoSi$_{2}$, displays 
%a remarkably strong logarithmic enhancement of the Sommerfeld coefficient, $C(T)/T \propto -\log(T)$, with decreasing $T$ down to 2\,K. It might be a rare case of a 
%stoichiometric compound positioned exactly at the FM QCP at zero pressure.    
%
%

\subsubsection{Other systems showing effects of strong disorder}
\label{subsubsec:II.E.2}

\paragraph{\UTNS}
\label{par:UThNiSi}

\begin{figure}[t]
\begin{center}
\includegraphics[width=0.9\columnwidth,angle=0]{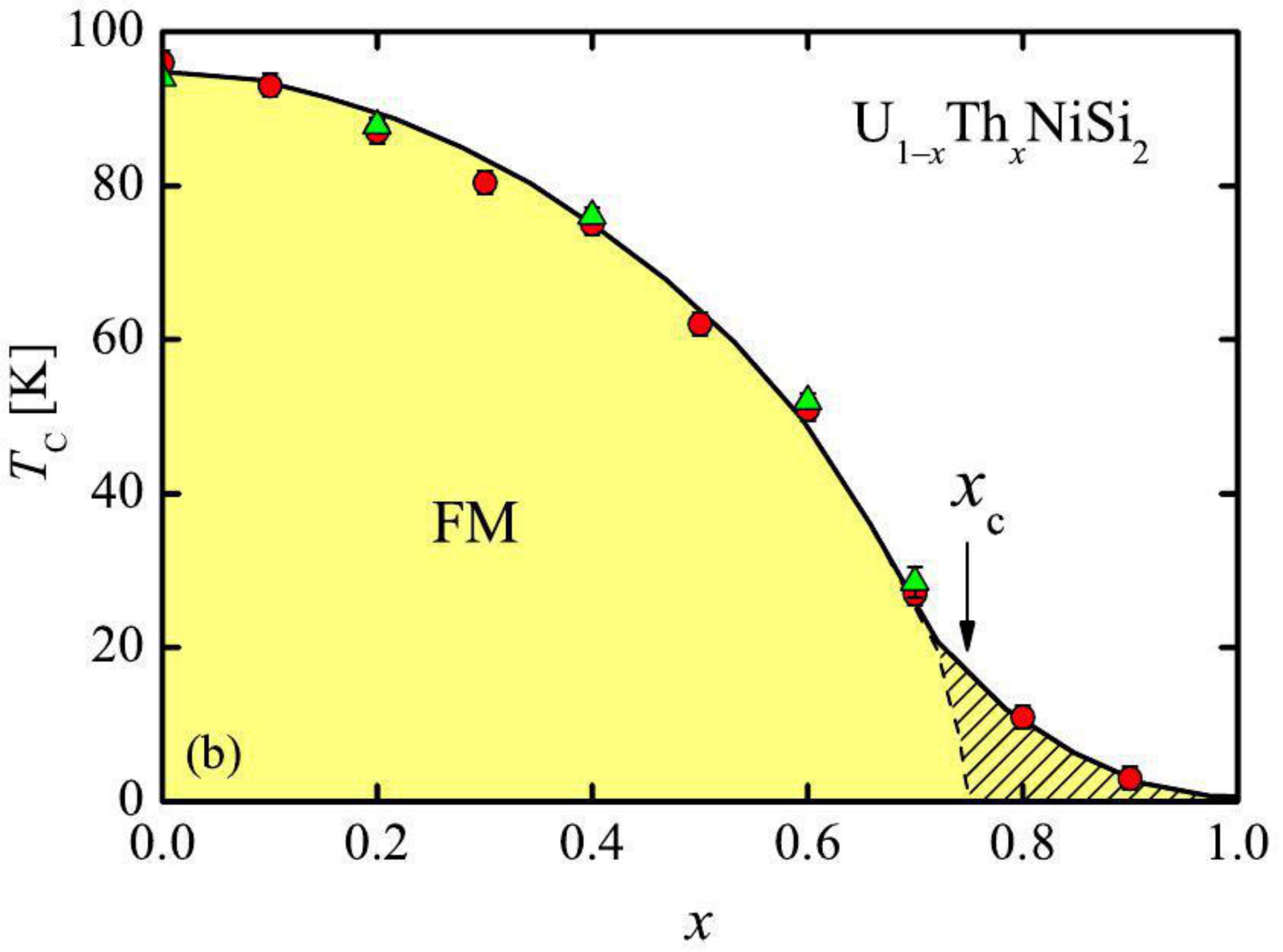}
\end{center}
\caption{$T$ - $x$ phase diagram of \UTNS\ derived from magnetization $M(T)$ and specific heat $C(T)$ measurements. The circles correspond to the maxima in 
             $\chi(T) = M(T)/B$ or minima in $\partial M(T)/\partial T$, and the triangles correspond to the minima in $\partial (C(T)/T)/\partial T$. The arrow ($x_{c}$) 
             indicates the position of the putative FM QCP obtained from extrapolating the low-$x$ curvature of the phase separation line. From~\textcite{Pikul_et_al_2012a}.}
\label{pikul2012a2}
\end{figure}

The system \UTNS\ with partial substitution of thorium at the uranium site was investigated by~\textcite{Pikul_et_al_2012a}. Alloying with non-magnetic Th causes the 
unit cell volume to expand without changing the crystal structure. This shifts the FM phase transition to lower temperatures as shown in Figs.~\ref{pikul2012a1} and
\ref{pikul2012a2}. The 
feature in the specific heat that signals the phase transition broadens with increasing $x$. At $x = 0.8$ the transition is no longer clearly visible in $C(T)/T$, while it can be still seen in $\partial M(T)/\partial T$.
At $x_{c} \approx 0.75$ the long-range FM order seen for $0 \leq x \leq 0.7$ changes smoothly into short-range or spin-glass-like order. The $x$ - $T$ phase diagram is shown in Fig.~\ref{pikul2012a2}. $x_c$ indicates the position of the putative FM QCP; around this concentration the phase 
boundary changes its curvature and develops a marked tail (cf. Fig.~\ref{fig:schematic_phase_diagrams}\,d). For the sample with 
$x = 0.8$, $C(T)/T$ does not show a maximum, as one would expect in a spin-glass-type~\cite{Mydosh_1993}, and it does not level off either as in a Fermi liquid, but 
keeps increasing with decreasing $T$. This anomalous behavior was also observed in other doped FM systems, for instance, \CPR, where the thermodynamics in the
``tail'' region of the phase diagram, are believed to be dominated by quantum Griffiths effects (see~\textcite{Westerkamp_et_al_2009} and Sec.~\ref{CPR}). 

\paragraph{\CTG}
\label{par:CeTiVGe}

CeTiGe$_{3}$ is a FM Kondo-lattice system with \TC\ = 14\,K and a hexagonal perovskite BaNiO$_{3}$-type structure which is not common in intermetallic 
systems~\cite{Manfrinetti_et_al_2005}. Recently, the system \CTG\ has been studied as a candidate for FM quantum 
criticality \cite{Kittler_et_al_2013}. So far, only polycrystalline samples with a relatively large residual resistivity $\rho_{0} \approx 22$\,$\mu\Omega$cm 
(for CeTiGe$_{3}$) have been investigated. At high temperature the susceptibility has been reported to follow a Curie-Weiss behavior with a negative Weiss temperature 
$\theta_{W} = -36.5$\,K, indicating predominantly AFM interactions. The effective moment is 2.64\,$\mu_{\textrm{B}}$, close to the value of 2.54\,$\mu_{\textrm{B}}$ for trivalent free $^{+3}$Ce ions. The resistivity increases with decreasing temperature, displaying a maximum at about 35\,K which indicates the onset of Kondo 
coherence.
\begin{figure}[b]
\begin{center}
\includegraphics[width=0.95\columnwidth,angle=0]{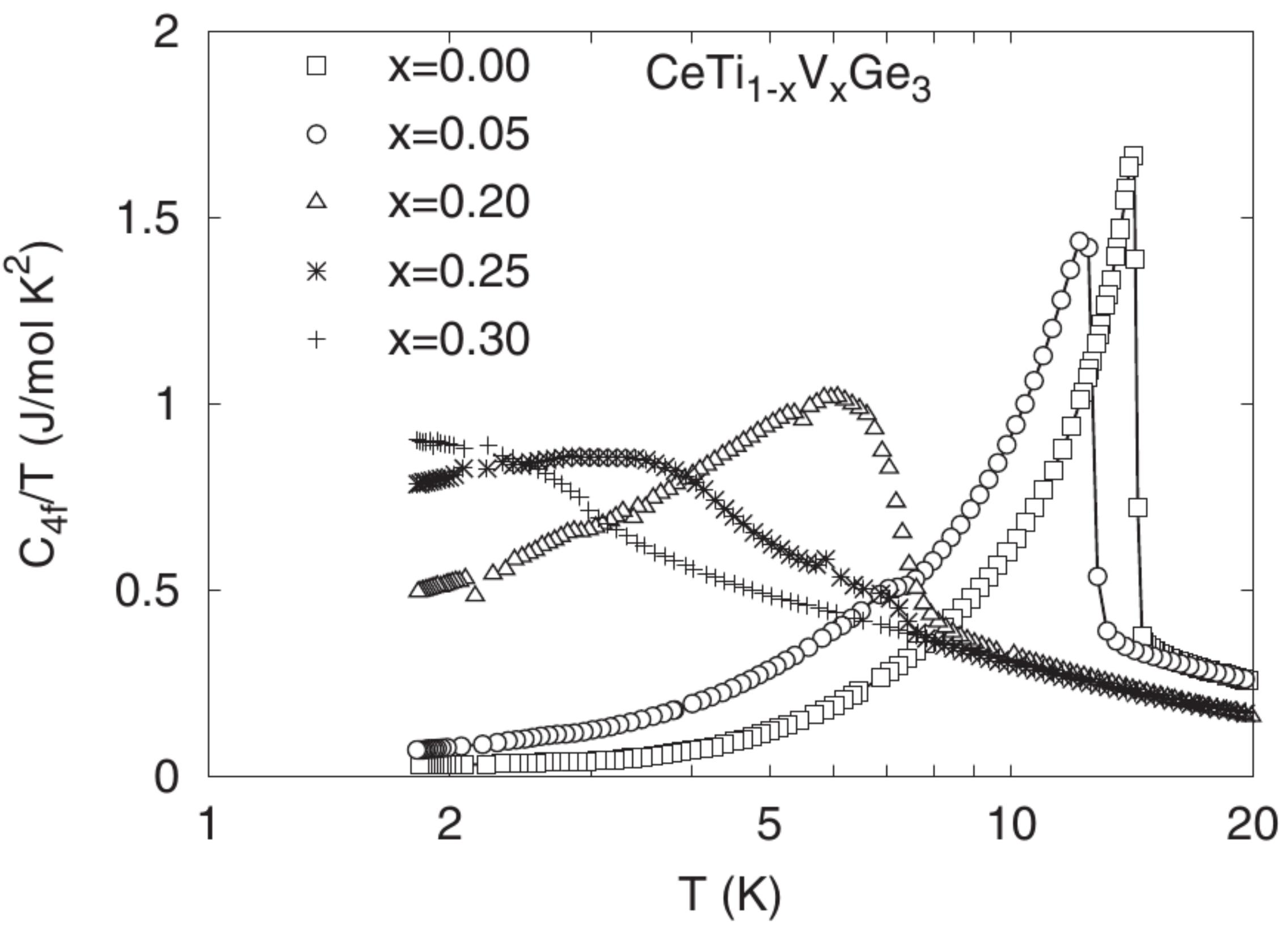}
\end{center}
\caption{$4f$-electron derived Sommerfeld coefficient $C_{4f}/T$ of \CTG. The phase transition is mean-field like for small $x$ and broadens for larger $x$. 
             For $x = 0.3$ the transition temperature was estimated to be 2.8\,K. From~\textcite{Kittler_et_al_2013}.}
\label{CeTiVGe3_specific_heat}
\end{figure}
The Ce ground state is a Kramers doublet, but the entropy just above \TC\ is larger than the $R\ln2$ expected for a doublet ground state, suggesting that the 
CEF splitting is small and that the Kondo temperature can not easily be determined from the entropy. The ordered moment measured by neutron powder diffraction 
within the FM phase is 1.5~\,$\mu_{\textrm{B}}$/Ce, and the ordering is collinear, with moments pointing along the crystallographic $c$-axis~\cite{Kittler_et_al_2013}. 
The analysis of the specific heat below \TC\ suggests the presence of a spin gap $\Delta/k_{\textrm{B}} \approx 0.8$ \TC\ in the magnetic excitation spectrum, which indicates a rather strong magnetic anisotropy. 

Non-isoelectronic vanadium substitution for titanium in \CTG\ permits to reduce \TC\ and completely suppress it at $x_{c} \approx 0.35$, while Ce retains its +3 valence in all \CTG\ samples~\cite{Kittler_et_al_2013}. Interestingly, CeVGe$_{3}$ shows AFM order below 4\,K~\cite{Bie_Mar_2009}. Magnetization measurements demonstrate that the ordered moment is also reduced with increasing $x$. The phase transition is second order down to about 3\,K, as indicated by the mean-field-like feature in the specific heat shown in Fig.~\ref{CeTiVGe3_specific_heat}. The transition 
remains second order and ferromagnetic even at higher V concentrations, but it broadens strongly for $x$ close to $x_{c}$. This is an indication of strong disorder 
effects. The logarithmic increase of $C_{4f}/T$ towards low temperatures for the $x = 0.3$ sample indicates the presence of spin fluctuations, which might arise from 
the presence of a QCP at $x_{c}$ or, more probably, from quantum Griffiths effects, similarly to what has been observed in \CPR\ and \NV.

\begin{figure}[t]
\begin{center}
\includegraphics[width=0.9\columnwidth,angle=0]{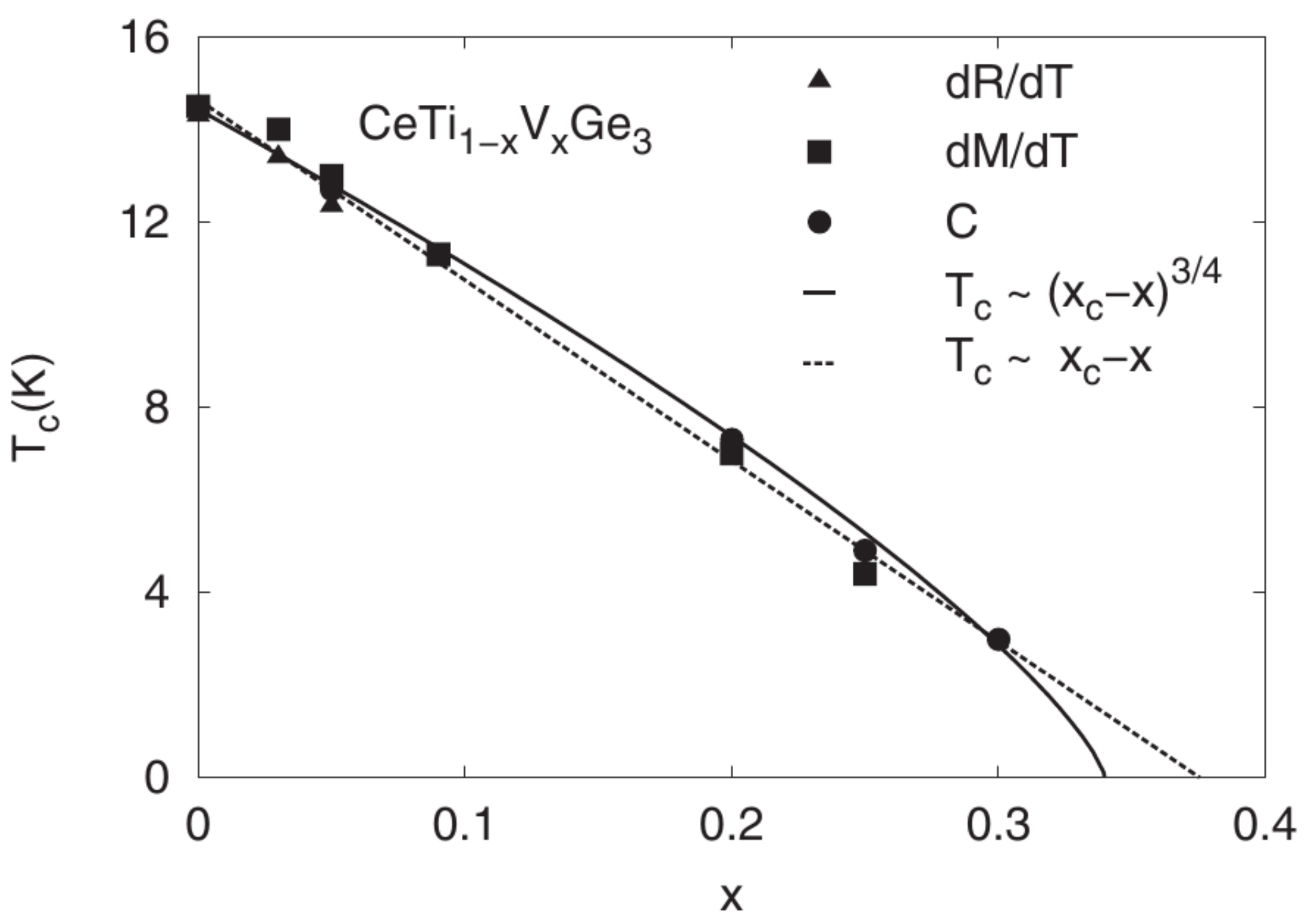}
\end{center}
\caption{$x - T$ phase diagram of \CTG. The Curie temperature was estimated by resistivity, magnetization and specific-heat measurements. \TC\ decreases 
              approximately linearly with increasing $x$, as indicated by the dotted line. Also shown is a fit by \TC\,$\propto (x_{c} - x)^{3/4}$ (solid line), which is 
              expected for a three-dimensional ferromagnet within the Herz-Millis-Moriya theory (cf. Sec.~\ref{subsubsec:III.C.2}). From~\textcite{Kittler_et_al_2013}.}
\label{CeTiVGe3_phase_diagram}
\end{figure}

The experimental phase diagram of \CTG\ is shown in Fig.~\ref{CeTiVGe3_phase_diagram}. The dotted line denotes a linear decrease of \TC\ with $x$, and the
solid line is a fit to the behavior expected from Hertz-Millis-Moriya theory, $T_{C} \sim (x_{c} - x)^{3/4}$, see Sec.~\ref{subsubsec:III.C.2}. The lowest temperature 
achieved was 2\,K. Low-temperature data in the region at $x \approx x_{c}$ would be required to determine whether or not a Griffiths region as discussed in
Sec.\ \ref{subsec:III.D} is indeed present in this system.

\subsubsection{A thin-film system: \SCRO\ (thin-film samples)}
\label{subsubsec:II.E.3}

\begin{figure}[t]
\begin{center}
\includegraphics[width=0.85\columnwidth,angle=0]{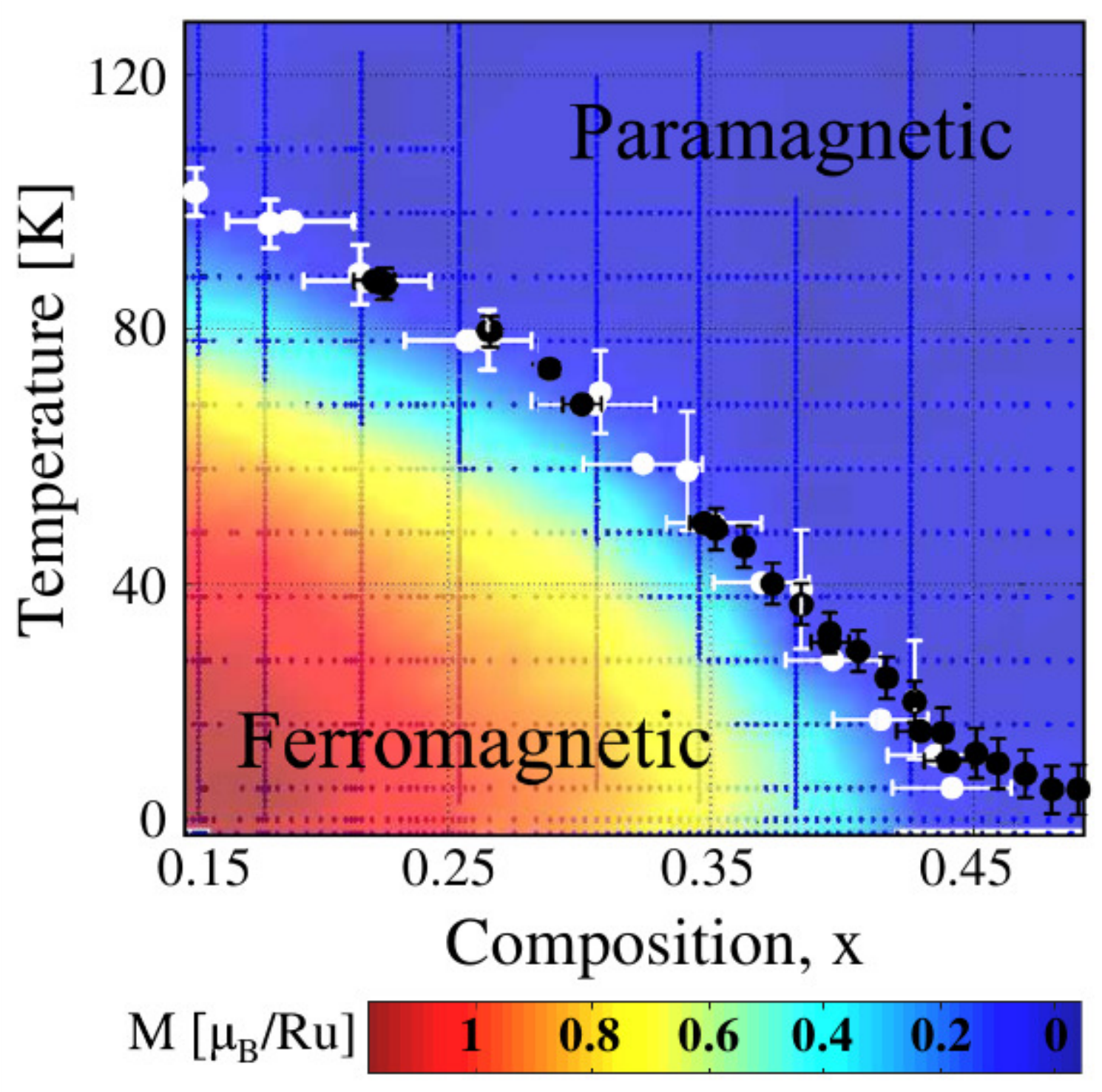}
\end{center}
\caption{Contour plot of the remanent magnetization in the $T$ - $x$ plane for an epitaxial film of \SCRO. Black and white symbols indicate the transition temperature
              determined from susceptibility and magnetization measurements, respectively. From~\textcite{Demko_et_al_2012}.}
\label{SrCaRuO3}
\end{figure}

Bulk (ceramic and powder) samples of \SCRO\ have been discussed in Secs.\ \ref{subsubsec:SrRuO} and \ref{subsubsec:II.C.1}, respectively. Thin films have
been grown epitaxially by \textcite{Schneider_et_al_2010}, \textcite{Wissinger_et_al_2011}, and \textcite{Demko_et_al_2012}. The former authors found that 
\TC\ decreases roughly linearly with $x$, with an extrapolated critical value $x_c \approx 0.7$, which is in agreement with results on powder, polycrystalline, 
and ceramic samples. However, \textcite{Wissinger_et_al_2011} found significant differences between film and bulk samples, including a higher value of
$x_c$ for films. \textcite{Demko_et_al_2012} measured the magnetization and susceptibility using a magneto-optical technique on a composition-spread epitaxial film of 200\,nm thickness. They found a phase diagram that differs
markedly from previous results, including those by \textcite{Schneider_et_al_2010}, namely, a pronounced tail with an onset around $x = 0.4$,
see Fig.~\ref{SrCaRuO3}. They interpreted their results as the FM-to-PM quantum phase transition being destroyed by the disorder. This is consistent with a 
microscopic model which considers spatial disorder correlations at smeared phase transitions \cite{Vojta_2003} with a behavior different from that at critical 
points~\cite{Demko_et_al_2012, Svoboda_et_al_2012}.
\subsubsection{A system showing short-range order: \CFPO}
\label{CeFePO}

\begin{figure}[t]
\begin{center}
\includegraphics[width=0.95\columnwidth,angle=0]{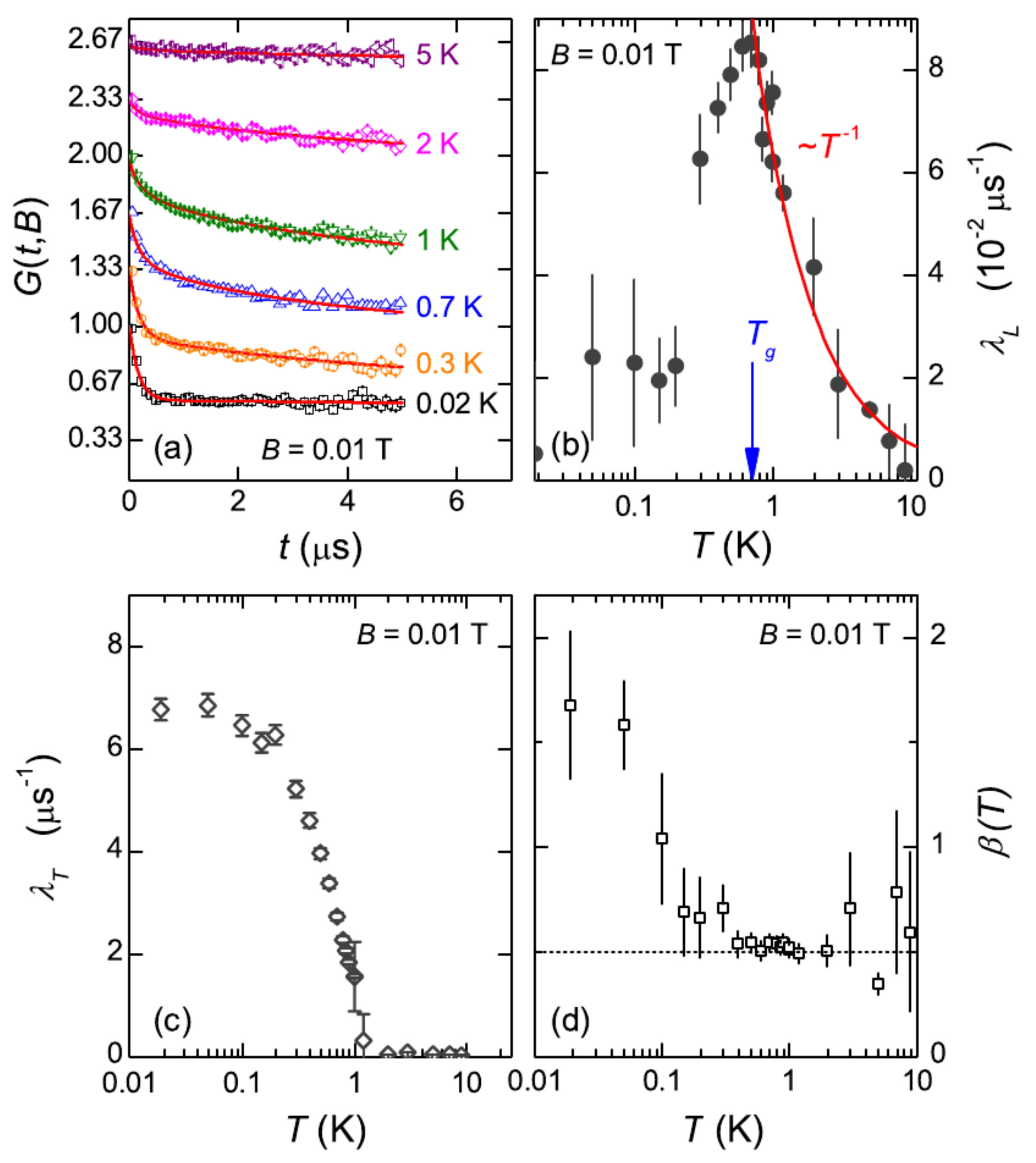}
\end{center}
\caption{Normalized muon-spin asymmetry function $G(t,B)$ at different temperatures. The red lines are fits to Eq.~\ref{asymmetry}. $\lambda_{T}$ and $\lambda_{L}$ 
              are the static and dynamic relaxation rates, and $\beta$ is the exponent in Eq.~\ref{asymmetry}. The peak in $\lambda_{L}(T)$ and the strong increase of 
              $\lambda_{L}(T)$ mark the spin-freezing temperature $T_{g}$. From~\textcite{Lausberg_et_al_2012a}.}
\label{CeFePO_uSR}
\end{figure}

\CFPO, a homologue of the quaternary pnictides, is a stoichiometric Kondo lattice system that is very close to a FM instability~\cite{Bruening_et_al_2008}. However, 
its ground state is neither ferromagnetic nor paramagnetic, but a particular short-range ordered state~\cite{Lausberg_et_al_2012a}, which is very unusual for a clean
system. The first comprehensive low-temperature study of \CFPO\, was performed by \textcite{Bruening_et_al_2008}, who investigated polycrystals by measurements of 
the uniform susceptibility, resistivity, NMR (oriented powder) down to 2\,K, and specific heat down to 0.4\,K. They found that \CFPO\, is a heavy-fermion system (iron is 
not magnetic in this compound) with a Kondo temperature $T_{\textrm{K}} \approx 10$\,K, a Sommerfeld coeffient $\gamma = 0.7$\,J/K$^{2}$mol, which corresponds
to a mass enhancement of 50, a Sommerfeld-Wilson ratio of 5.5, and a Korringa ratio $S_{0}/T_{1}TK^{2} \approx 0.065$, indicating the presence of FM correlations. 
Below 10\,K the broadening of the line shape of the NMR spectra for small fields $H \perp c$ -- but not for $H \parallel c$ -- suggests that in this temperature regime FM 
short-range correlations start to be relevant that cannot be ascribed to disorder. Thus, only the basal-plane component of the cerium $4f$ moment is FM correlated. 
This strong anisotropy reflects the quasi-two-dimensional crystal structure.     
Later, $\mu$SR experiments were performed on poly- and single crystals, together with ac susceptibility and specific heat measurements down to 
0.02\,K~\cite{Lausberg_et_al_2012a}. The ac susceptibility shows a frequency dependent peak at $T_{g} \approx 0.9$\,K, whose dependence on the modulation frequency is larger than that 
found in canonical spin glasses and smaller than that of superparamagnet. In addition, the entropy measured below the freezing maximum is just 1\% of $R\ln 2$. 
A summary of the $\mu$SR results in zero and small longitudinal field is shown in Fig.~\ref{CeFePO_uSR}. The normalized muon-spin asymmetry function $G(t,B)$, 
which gives the dynamic and static contributions to the Ce-$4f$ moments, is plotted at several temperatures. The static relaxation rate dominates at short times $t$, 
while the dynamic spin fluctuations are probed at long $t$. Both contributions can be fitted with
\be
G(t,B) = G_{1}e^{-(\lambda_{T}t)} + G_{2}e^{-(\lambda_{L}t)^{\beta}}\ , 
\label{asymmetry}
\ee
where $\lambda_{T}$ and $\lambda_{L}$ are the static (transverse) and dynamic (longitudinal) relaxation rates, respectively. The transition from dynamic to short-range 
static magnetism below $T_{g}$ is clearly seen in the evolution of these parameters with $T$ in Fig.~\ref{CeFePO_uSR}, which involves 100\% of the sample volume. 
The value of $\beta$ for $T \geq T_{g}$ indicates a broad distribution of fluctuating local fields which become static below $T_{g}$ where $\beta \approx 1.7$. 
Moreover, the observation of a time-field scaling of $G(t,B)$ strongly suggests cooperative behavior. In conclusion, there is no FM-QCP in \CFPO\, but rather some 
short-range order with a particular texture whose nature is unknown. 
The fact that in \CFPO\, strong FM fluctuations are present, and that AFM phases were observed in samples slightly doped with Ru or As (see Sec.~\ref{CRFPO} and 
Sec.~\ref{CFPAO}), suggests that a delicate interplay between FM and AFM correlations is present in the system. This might cause frustration and favor exotic states. 
There are theoretical predictions of textured states in itinerant systems close to a FM instability, see Sec.~\ref{subsec:III.E}. An
interestig proposal was put forward by \textcite{Thomson_Kruger_Green_2013}, who suggested a helical glass state. This state is the result of weak disorder which 
destabilizes the FM state and enhances the susceptibility towards incommensurate, spiral magnetic ordering. Even the best samples of \CFPO\, have a relatively small 
RRR of approximately 5, but this may be attributed to the presence of strong FM fluctuations rather than quenched disorder. Since the amount of disorder necessary to generate 
the spiral state is rather small, the mechanism proposed by \textcite{Thomson_Kruger_Green_2013} is a viable candidate for explaining the observations.

\subsubsection{Discussion, and comparison with theory}
\label{subsubsec:II.E.4}

The unusual properties observed in \CPR\  and \NV\ have been interpreted in terms of theoretical ideas discussed in Secs.\ \ref{subsubsec:III.D.1} and
\ref{subsubsec:III.D.3}, namely, a quantum Griffiths phase and interactions between rare regions that lead to glassy properties. While some aspects of the
theory agree very well with the experimental results (see, e.g., the fits in Fig.~\ref{CePdRh_Griffiths}), there also are discrepancies. For instance, the theory
predicts that the field dependence of the magnetization, $M(H) \propto H^{\lambda}$, and the temperature dependence of the specific-heat coefficient,
$C(T)/T \propto T^{1-\lambda}$, are governed by the same exponent $\lambda$, whereas the data yield different values for the two coefficients,
see Fig.~\ref{CePdRh_Griffiths}. For Kondo systems, there are other theoretical scenarios that so far have not been explored in the context of these
experiments, see Sec.\ \ref{subsubsec:III.D.2}. The evidence for a glassy phase in \UNCS\ (Sec.\ \ref{UNiCoSi}) is much weaker, and the lowest \TC\ where 
a clear FM transition has been observed is rather high at \TC\ = 8.6\,K. Experiments at lower temperatures in the region close to $x=1$ are needed
to determine the nature of the FM QCP in this system. 

The evidence for strong-disorder effects in the systems discussed in Sec.\ \ref{subsubsec:II.E.2} is weaker and largely based on the shape of the phase
diagram. As we discuss in Sec.\ \ref{subsubsec:III.B.3} there are other possible explanations for a ``tail'' in the phase diagram 
(see Fig.~\ref{figure:phase_diagram_disordered}) and further investigations are needed to ascertain whether quantum Griffiths or related effects are
indeed present in these materials. 

% The thin-film samples of \SCRO\ discussed in Sec.\ \ref{subsubsec:II.E.3} fall into a separate class, just as the quasi-one-dimensional systems in Sec.~\ref{subsubsec:II.C.3} do. The reason is that in two-dimensional systems with quenched disorder the concept of a Fermi liquid breaks down \cite{Altshuler_Aronov_1984, Lee_Ramakrishnan_1985, Belitz_Kirkpatrick_1994}. Experiments on such materials are likely to see a complicated mixture of effects due to weak localization, Althuler-Aronov effects, quantum critical behavior, and quantum Griffiths phenomena, and the details may depend strongly on the morphology of the films. The discrepancies between the results discussed in Sec.\ \ref{subsubsec:II.E.3} on one hand, and Secs.\ \ref{SCRO_ceramic} and \ref{SCRO_powder} on the other, are thus not surprising, and a complete understanding of the thin-film results may require a rather complex analysis. In this context, see also point 5. in Sec.\ \ref{subsec:IV.A}.

%\input{section_II/subsection_II.F}
%\input{section_II/subsection_II.G}
%
\section{Theoretical results}
\label{sec:III}
In this section we describe and discuss theoretical developments pertinent to the
experiments discussed in Sec.\ \ref{sec:II}. % Theoretical results
\subsection{Soft modes in metals}
\label{subsec:III.A}
In Sec.\ \ref{subsec:I.B} we discussed why, in the absence of soft modes other than the order-parameter
fluctuations, the quantum ferromagnetic transition in clean systems is expected to be generically continuous 
with mean-field static exponents and a dynamical exponent $z=3$. We also mentioned theoretical results
that showed that this expectation breaks down in general, and in Sec.\ \ref{sec:II} we discussed many
experiments that show a first-order transition rather than a continuous one. Indeed,
Tables~\ref{table:1a},~\ref{table:1b} and \ref{table:2_pt_1}, \ref{table:2_pt_2}, \ref{table:2_pt_3}
show that most of the observed low-temperature transitions into a homogeneous ferromagnetic phase are first order
unless the system is either disordered or quasi-one-dimensional. We will argue below that the systems
in Table~\ref{table:2_pt_1} can be understood in terms of a crossover from an asymptotic first-order
transition to a pre-asymptotic regime that is described by Hertz theory.
What underlies this breakdown of the 
original expectation is the existence, in metallic magnets, of soft two-particle excitations
that couple to the magnetization. These excitations are the result of two characteristic properties of metals, viz.,
(1) a sharp Fermi surface, and (2) a nonzero density of states at the Fermi level. As we will see, they can be
understood as representing a Goldstone mode that results from a spontaneously broken gauge symmetry. In disordered systems,
soft critical order-parameter fluctuations exist in addition to this fermionic soft mode and govern the critical behavior. 
Finally, the rare-region effects mentioned
in Sec.\ \ref{subsec:I.B} also can be understood as a certain class of soft excitations. In this section we explain
the importance of soft modes, give a classification, and discuss in more detail the fermionic Goldstone mode that is
of particular importance for quantum ferromagnets.

\subsubsection{Why we should care about soft modes}
\label{subsubsec:III.A.1}

The only way in which observables, be it thermodynamic quantities or time
correlation functions, can display nonanalytic behavior as a function of 
temperature, frequency, wavenumber, or an external field, is via the existence
of soft modes, i.e., correlation functions that diverge as the frequency and the
wavenumber go to zero. A useful concept in this context is to consider soft
modes as leading to a distribution of relaxation times.
Consider, for example, a diffusive process (e.g., \onlinecite{Forster_1975}), i.e., a correlation function that
behaves for small wave vectors ${\bm k}$ and long times $t$ as
\bse
\label{eqs:3.1}
\be
C({\bm k},t) \propto e^{-\nu{\bm k}^2 t}\ ,
\label{eq:3.1a}
\ee
with $\nu$ some kinetic coefficient.
This corresponds to a ${\bm k}$-dependent relaxation time 
\be
\tau({\bm k}) = 1/\nu{\bm k}^2\ .
\label{eq:3.1b}
\ee
\ese
For a fixed wave vector the decay is exponential, but the relaxation time
diverges as ${\bm k}\to 0$. As a result, the local time correlation function
in $d$ dimensions,
\be
C({\bm x}=0,t) = \frac{1}{V}\sum_{\bm k} \,C({\bm k},t) \propto 1/t^{d/2}
\label{eq:3.2}
\ee
decays algebraically with time. It is also illustrative to write 
the local time correlation function as an integral over the relaxation time,
\bse
\label{eqs:3.3}
\be
C({\bm x}=0,t) = \int_{\tau_0}^{\infty} d\tau\,P(\tau)\,e^{-t/\tau}\ .
\label{eq:3.3a}
\ee
Here $\tau_0 = 1/\nu k_0^2$, with $k_0$ the upper cutoff on the momentum
integral in Eq.\ (\ref{eq:3.2}), and $P(\tau)$ is a relaxation-time distribution function.
Comparing Eqs.\ (\ref{eq:3.3a}) and (\ref{eq:3.2}) we see that
\be
P(\tau) \propto 1/\tau^{(d+2)/2}\ .
\label{eq:3.3b}
\ee
\ese
The algebraic decay, or long-time tail, of the time correlation function is thus a
result of the power-law decay of the distribution function $P(\tau)$.
The Laplace transform $C(z)$ of $C({\bm x}=0,t)$ is a nonanalytic function of the 
complex frequency $z$ in the point $z=0$; for ${\text{Im}}\, z >0$ and
$(d-2)/2$ not integer the leading low-frequency behavior is
\bea
C(z) &=& i\int_0^{\infty} dt\,e^{i z t}\,C({\bm x}=0,t) 
\nonumber\\
       &\propto& z^{(d-2)/2} + (\text{terms analytic in $z$})\ .
\label{eq:3.4}
\eea
Observables that couple to such diffusive modes will be given in terms of integrals
whose integrands contain diffusive correlation functions. This results in a nonanalytic
dependence on, for instance, the temperature or the frequency.

Such nonanalytic behavior can be generic, i.e., exist in an entire phase,
or it can occur only at special points in the phase diagram. An example 
of the former are Goldstone modes, which arise from the spontaneous breaking
of a global continuous symmetry \cite{Forster_1975, Zinn-Justin_1996}. The prime example of the latter are critical
fluctuations, which are soft only at a critical point \cite{Ma_1976}. Other mechanisms for producing generic
soft modes include conservation laws, and gauge invariance \cite{Belitz_Kirkpatrick_Vojta_2005}.
For the purposes of this review, we are interested in four classes of soft modes in metals. The first
class consists of

\begin{enumerate}[label=\roman*)]
\item single-particle excitations. These are the well-known excitations that are represented by the 
soft single-particle Green function. They exist because of the sharpness of the Fermi surface and 
the existence of a nonvanishing quasiparticle weight. They are 
responsible for the leading behavior of observables in a
Fermi liquid, for instance, the linear temperature dependence of the specific heat.
\end{enumerate}
\par\noindent
An example of the effects of soft single-particle excitations that is relevant in the current context
is the paramagnon propagator in a metallic magnet. As a function of the wave vector ${\bm k}$ and
the imaginary Matsubara frequency $i\Omega$ it has the form \cite{Doniach_Engelsberg_1966,
Hertz_1976}
\bse
\label{eqs:3.5}
\be
P({\bm k},i\Omega) = \frac{1}{t + a{\bm k}^2 + b\vert\Omega\vert/\vert{\bm k}\vert^n}\ .
\label{eq:3.5a}
\ee
Here $t$ is the distance from the magnetic transition, and $a$ and $b$ are constants. $n$ is an
integer that depends on the physical situation. For clean and disordered metallic ferromagnets, $n=1$ and
$n=2$, respectively. For antiferromagnets, $n=0$. The spectrum or dissipative part of the corresponding
causal function, $P''({\bm k},\omega) = {\rm Im}\,P({\bm k},i\Omega\to\omega+i0)$, reads
\be
P''({\bm k},\omega) =\frac{\omega/\vert{\bm k}\vert^n}{(t + a{\bm k}^2)^2 + \omega^2/\vert{\bm k}\vert^{2n}}\ .
\label{eq:3.5b}
\ee
\ese
We see that the paramagnon excitation is damped with a damping coefficient given by $\omega/\vert{\bm k}\vert^n$.
Physically, this reflects the coupling of the magnetic collective mode to the soft single-particle
excitations in the itinerant electron system. It is usually referred to as Landau damping, in analogy to the
corresponding effect in a collisionless classical plasma \cite{Landau_Lifshitz_X_1981}. The same damping
mechanism is applicable to the plasmon mode in a charged Fermi liquid, and the zero-sound mode in a
neutral one \cite{Pines_Nozieres_1989}. For later reference we also note that the $\vert\Omega\vert$ singularity
on the imaginary-frequency axis in Eq.\ (\ref{eq:3.5a}) implies, for fixed ${\bm k}$, a $1/\tau^2$ long-time tail
for $P$ as a function of the imaginary time $\tau$ (see, e.g., \onlinecite{Belitz_Kirkpatrick_Vojta_2005}). 
We will encounter the ferromagnetic paramagnon propagator again later in this section, see Eqs.\ (\ref{eq:3.45}, \ref{eq:3.75}),
and the $1/\tau^2$ long-time tail will be important for the discussion in Sec. \ \ref{subsubsec:III.D.3}.

\medskip
The second class consists of
\begin{enumerate}[label=\roman*)]
\setcounter{enumi}{1}
\item soft two-particle excitations that are the Goldstone modes of a broken gauge symmetry
with the density of states at the Fermi level as the order parameter. The excitations were first
identified as Goldstone modes by \textcite{Wegner_1979} in the context of disordered electron
system, and we will discuss them in detail in Sec.\ \ref{subsubsec:III.A.2} below.
\end{enumerate}

The third class are
\begin{enumerate}[label=\roman*)]
\setcounter{enumi}{2}
\item Griffiths or rare-region effects in disordered systems \cite{Griffiths_1969, McCoy_1969}.
\end{enumerate}
\par\noindent
These are normally not thought of as akin to soft modes. To see the connection, let us
consider, for instance, a classical Ising system with randomly missing bonds in its disordered phase. 
In an infinite system, below the transition temperature of the clean system, but above the one
of the actual bond-disordered one, one can find arbitrarily large regions with linear dimension $L$ that happen to contain
no missing bonds. In such a region, the spins are ordered, but the probability of finding
such a region will decrease exponentially with its volume $L^d$. In order to destroy such a rare region, 
a surface free energy must be overcome. The relaxation time for a cluster of linear size $L$ will therefore 
be \cite{Randeria_Sethna_Palmer_1985, Bray_1988}
\be
\tau(L) = \tau_0\,e^{\sigma L^{d-1}}\ ,
\label{eq:3.6}
\ee
with $\tau_0$ a microscopic time scale and $\sigma$ a surface tension. This 
time scale diverges as $L\to\infty$, just as the diffusive relaxation time in Eq.\ (\ref{eq:3.1b})
diverges as ${\bm k}\to 0$, only here the divergence is exponential. In order to estimate time
correlation functions $C(t)$ we need to weigh a factor $\exp(-t/\tau(L))$ with the probability 
$P(L) \propto \exp(-cL^d)$ of finding a rare region in the first place, and integrate over all sizes $L$. We thus expect
\bse
\label{eqs:3.7}
\be
C(t) \propto \int_0^{\infty} dL\ \exp\left[-cL^d - (t/\tau_0)e^{-\sigma L^{d-1}}\right]\ ,
\label{eq:3.7a}
\ee
where $c$ is a constant. For large times $t$ the integral can be evaluated by the method of steepest descent.
The typical length scale $L$ is $L_{\rm typ} \propto [\ln(T/\tau_0)]^{1/(d-1)}$, and the
leading contribution to $C(t)$ is \cite{Randeria_Sethna_Palmer_1985}
\be
C(t\to\infty) \propto \exp\left[-b[\ln (t/\tau_0)]^{d/(d-1)}\right]\ ,
\label{eq:3.7b}
\ee
\ese
with $b = c/\sigma^{d/(d-1)}$ another constant. We see that the time-correlation function again decays
slower than exponentially, albeit faster than any power. We can again define a distribution function for
relaxation times by writing
\bse
\label{eqs:3.8}
\be
C(t) = \int_{\tau_0}^{\infty} d\tau\,P(\tau)\,e^{-t/\tau}\ .
\label{eq:3.8a}
\ee
The leading behavior for large $\tau$ is
\be
P(\tau\to\infty) \propto \exp\left[-b[\ln(\tau/\tau_0)]^{d/(d-1)}\right]\ .
\label{eq:3.8b}
\ee
\ese
The analogy to diffusive soft modes, Eqs.\ (\ref{eqs:3.1}) - (\ref{eqs:3.3}), is now obvious. 

For later reference we mention that Eq.\ (\ref{eq:3.6}) holds for classical Ising magnets only;
for other systems the exponent may be different. We will discuss Griffiths effects, and their
relevance for quantum magnets, in more detail in Sec.\ \ref{subsec:III.D}.

\medskip
All of the above are generic soft modes. In addition, we will encounter
\begin{enumerate}[label=\roman*)]
\setcounter{enumi}{3}
\item critical fluctuations at the quantum critical point in disordered ferromagnets. These are
analogous to the critical fluctuations at classical transitions \cite{Ma_1976}. However, as we
will see in Sec.\ \ref{subsubsec:III.C.2} their effects at the quantum ferromagnetic transition
are rather weak.
\end{enumerate}

Finally, another class of generic soft modes is represented by phonons, or elastic deformations
in a continuum model. These can couple to the magnetization, and this effect has been studied
extensively for classical magnets (\onlinecite{Bergman_Halperin_1976}, and references therein). For quantum
magnets no convincing treatment exists; see Sec.\ \ref{subsec:III.F}.

\subsubsection{Goldstone modes in metals}
\label{subsubsec:III.A.2}

We now illustrate in some detail the nature of the second class of soft modes listed above, which 
are of crucial importance for
the breakdown of Hertz theory. We first show that spinless noninteracting electrons, at $T=0$, 
possess soft two-particle excitations that can be understood as the Goldstone modes resulting 
from a spontaneously broken gauge symmetry. We then generalize these arguments to the case 
of interacting electrons with spin.

\paragraph{Goldstone modes in a Fermi gas}
\label{par:IIII.A.2.a}

Consider free electrons with mass $\me$ and chemical potential $\mu$
described by fermionic (i.e., Grassmann-valued) fields ${\bar\psi}_n({\bm k})$ and $\psi_n({\bm k})$ that depend on
a wave vector ${\bm k}$ and a fermionic Matsubara frequency $\omega_n = 2\pi T(n+1/2)$ ($n = 0, \pm 1, \pm2, \ldots$).
These fields are temporal Fourier transforms of fields ${\bar\psi}({\bm k},\tau)$ and $\psi({\bm k},\tau)$ that depend on 
the imaginary-time variable $\tau$. In terms of these fields, the free-fermion action reads \cite{Negele_Orland_1988}
\be
S_0[{\bar\psi},\psi] = \sum_{\bm k}\sum_n {\bar\psi}_n({\bm k})\left[i\omega_n - {\bm k}^2/2\me + \mu\right]\psi_n({\bm k})\ .
\label{eq:3.9}
\ee
Single-particle excitations are described by the Green function
\be
G_n({\bm k}) = \langle \psi_n({\bm k})\, {\bar\psi}_n({\bm k})\rangle = 1/(i\omega_n - \xi_{\bm k})
\label{eq:3.10}
\ee
with $\xi_{\bm k} = {\bm k}^2/2\me - \mu$. These are soft in the sense that $G_n({\bm k})$ diverges for
wave vectors on the Fermi surface, $\xi_{\bm k} = 0$, as the frequency approaches zero. Of greater interest in the current
context are two-particle excitations. Consider the correlation function
\bse
\label{eqs:3.11}
\bea
D_{nm}({\bm k},{\bm q}) &\equiv& \langle{\bar\psi}_n({\bm k}_+)\,\psi_m({\bm k}_-)\,{\bar\psi}_m({\bm k}_-)\,\psi_n({\bm k}_+)\rangle 
\nonumber\\
&=& \delta_{nm}\,\delta_{{\bm q},0}\,\left(G_n({\bm k})\right)^2 - G_n({\bm k}_+)\,G_m({\bm k}_-)\ ,
\nonumber\\
\label{eq:3.11a}
\eea
where ${\bm k}_{\pm} = {\bm k} \pm {\bm q}/2$, and the second line follows from Wick's theorem. 
Multiplying Eq.\ (\ref{eq:3.11a}) with the inverse of 
$G_n({\bm k}_+)$ and $G_m({\bm k}_-)$, respectively, and subtracting the resulting two
equations, we find
\be
\left(i\Omega_{n-m} - {\bm k}\cdot{\bm q}/\me\right) D_{nm}({\bm k},{\bm q}) = G_n({\bm k}_+) - G_m({\bm k}_-)\ .
\label{eq:3.11b}
\ee
\ese
Now analytically continue to real frequencies according to $i\omega_n \to \Omega + i0$, $i\omega_m \to -\Omega - i0$,
and consider the limit ${\bm q}\to 0$, $\Omega \to 0$. Eq.\ (\ref{eq:3.11b}) then becomes
\be
D^{+-}({\bm k},{\bm q}\to 0;\Omega\to 0) = \frac{i\,G''({\bm k},\Omega=0)}{\Omega - {\bm k}\cdot{\bm q}/\me}\ .
\label{eq:3.12}
\ee
Here $D^{+-}(\Omega)$ is the analytic continuation of $D_{nm}$, and $G''$ denotes the spectrum of the Green
function. We see that the correlation function $D^{+-}$ diverges in the limit of zero wave vector ${\bm q}$ and zero
frequency $\Omega$, provided the spectrum of the Green function is nonzero. For free electrons this is the case
for all values of ${\bm k}$; if we replace the free electrons by band electrons, $\xi_{\bm k} = \epsilon_{\bm k} - \mu$,
where $\epsilon_{\bm k}$ is determined by the underlying lattice structure, it remains true everywhere within the
band. An equivalent statement is that it is true wherever there is a nonzero density of states. We have thus 
identified the correlation function $D_{nm}({\bm k},{\bm q})$, Eq.\ (\ref{eq:3.11a}), as a soft mode
of noninteracting electrons. The nature of this soft mode is ballistic, i.e., the frequency $\Omega$ scales linearly with
the wave number $\vert{\bm q}\vert$ for small ${\bm q}$.

This simple result is much more general and significant than one might expect at first sight. To explain why this is
the case, the following analogy is useful. Consider a classical XY ferromagnet with magnetization
${\bm m}$ in the presence of a small homogeneous magnetic field ${\bm h}$. Let the magnitude of ${\bm m}$ and
${\bm h}$ be $m$ and $h$, respectively. In the paramagnetic phase, ${\bm m}$ is proportional to ${\bm h}$, and 
$m(h)$ is an analytic function of $h$; in particular, $m(h=0)=0$. However, in the ferromagnetic phase this is not
true. ${\bm m}$ still points in the same direction as ${\bm h}$, but $m$ is not an analytic function of $h$ at $h=0$: 
$m(h=\pm 0) = \pm m_0$, with $m_0$ the spontaneous magnetization. Now let the system be in the ferromagnetic
phase, and consider an infinitesimal rotation of the field, ${\bm h} \to {\bm h} + \delta{\bm h}$, that leaves $h$ 
unchanged. Then the magnetization simply follows the field, with $m$ also unchanged. Hence
$\vert\delta{\bm m}\vert/m = \vert\delta{\bm h}\vert/h$. But $\vert\delta{\bm m}\vert/\vert\delta{\bm h}\vert$ is the
homogeneous transverse susceptibility $\chi_{\perp}$, and hence
\be
h\,\chi_{\perp} = m\ .
\label{eq:3.13}
\ee
This simple argument \cite{Ma_1976} shows that the transverse susceptibility diverges in the limit
of zero field everywhere in the ordered phase where $m(h\to 0)\neq 0$. It can be made technically more
elaborate by proving a Ward identity that takes the form of Eq.\ (\ref{eq:3.13}) \cite{Zinn-Justin_1996}, but
the simple argument contains all physically relevant points: The soft mode (that is, the magnon or transverse
magnetization fluctuation) is a Goldstone mode that results from a spontaneously broken continuous symmetry
\cite{Goldstone_1961, Forster_1975, Zinn-Justin_1996}; in this case,
the rotational symmetry in spin space that leads to a nonzero order parameter $m$.

Now return to free fermions. Consider a local (in imaginary time) gauge transformation
\bse
\label{eqs:3.14}
\bea
{\bar\psi}({\bm k},\tau) &\to& e^{-i\alpha\tau}\, {\bar\psi}({\bm k},\tau)
\nonumber\\
\psi({\bm k},\tau) &\to& e^{i\alpha\tau}\, \psi ({\bm k},\tau)
\label{eq:3.14a}
\eea
with $\alpha$ real, or, equivalently,
\bea
{\bar\psi}_n({\bm k}) &\to& {\bar\psi}_{n-\alpha}({\bm k})
\nonumber\\
\psi_n({\bm k}) &\to& \psi_{n-\alpha}({\bm k})\ .
\label{eq:3.14b}
\eea
\ese
The second and third terms in Eq.\ (\ref{eq:3.9}) are invariant under this transformation, but the frequency
term is not; it acts analogously to a magnetic field in the classical XY model. Explicitly, we have
\be
S_0[{\bar\psi},\psi] \to \sum_{\bm k}\sum_n {\bar\psi}_n({\bm k})\left[i\omega_n - {\bm k}^2/2\me + \mu + i\alpha \right]\psi_n({\bm k})\ .
\label{eq:3.15}
\ee
If we now let $\alpha \to 0$ and consider the expectation value $\langle\psi_n({\bm k})\,{\bar\psi}_n({\bm k})\rangle$ we
see that, upon analytic continuation to real frequencies, $\alpha>0$ vs. $\alpha<0$ makes the difference between a
retarded and an advanced Green function. The latter are not the same anywhere within the band, and the U(1) gauge
symmetry expressed by Eqs.\ (\ref{eqs:3.14}) is thus {\em spontaneously} broken. Eq.\ (\ref{eq:3.12}) can now be
interpreted in perfect analogy to Eq.\ (\ref{eq:3.13}): The spectrum of the Green function is the order parameter
of a spontaneously broken continuous symmetry, the frequency acts as the field conjugate to the order parameter,
and the soft correlation function $D^{+-}$ is the Goldstone mode associated with the broken symmetry. 
This remarkable analogy was first found by \textcite{Wegner_1979} (see
also \onlinecite{Schaefer_Wegner_1980, McKane_Stone_1981}) in the context of disordered electrons, where
the soft modes have a diffusive frequency-momentum relation and are commonly referred to as
``diffusons" \cite{Akkermans_Montambaux_2011}. That the same argument holds for clean electrons, with the diffusive soft modes replaced by
ballistic ones, was first realized  by \textcite{Belitz_Kirkpatrick_1997}, and a detailed analysis was given
by \textcite{Belitz_Kirkpatrick_2012}. In these papers the symmetry considered was an SO(2) rotation in
frequency space that is isomorphic to the U(1) gauge transformation above. We stress that the broken symmetry
discussed above, and the resulting existence of the
soft modes, has nothing to do with the conservation law for the particle number.

\bigskip
\paragraph{Goldstone modes in a Fermi liquid}
\label{par:III.A.2.b}

We now turn to what happens if one considers interacting electrons and takes spin into account. Interactions have two effects. 
One is the appearance of an inelastic scattering rate, both in the Green function and in the propagator $D^{+-}$. However, this rate
vanishes at $T=0$. The second change is the appearance of an additional term on the right-hand side of Eq.\ (\ref{eq:3.11b}),
with is related to a three-particle correlation function. This term has a different functional dependence on the interaction
than the difference of Green functions in Eq.\ (\ref{eq:3.11b}) (for instance, it vanishes in the limit of vanishing interactions,
whereas the right-hand side of Eq.\ (\ref{eq:3.11b}) does not), and therefore cannot change the fact that $D^{+-}$ diverges
in the limit of vanishing frequency and wave number \cite{Belitz_Kirkpatrick_2012}. This is consistent with what one would
expect from Fermi-liquid theory, which posits that there is a one-to-one correspondence between free-electron states and
states in a Fermi liquid \cite{Landau_Lifshitz_IX_1991}. Basic properties such as the soft-mode spectrum will thus not be 
changed by interactions, only the coefficients in the soft propagator will acquire interaction dependences. 

Spin provides another complication, which is conveniently dealt with by means of introducing bosonic matrix variables $Q$ that
are isomorphic to bilinear products of fermion fields:
\begin{widetext}
\be
Q_{nm}({\bm x},{\bm y}) \cong \frac{i}{2}\left(\begin{array}{cccc}
                -\psi_{n\uparrow}({\bm x}){\bar\psi}_{m\uparrow}({\bm y}) & -\psi_{n\uparrow}({\bm x}){\bar\psi}_{m\downarrow}({\bm y})
                       & -\psi_{n\uparrow}({\bm x}) \psi_{m\downarrow}({\bm y}) & \ \ \psi_{n\uparrow}({\bm x}) \psi_{m\uparrow}({\bm y}) \\
               -\psi_{n\downarrow}({\bm x}){\bar\psi}_{m\uparrow}({\bm y}) & -\psi_{n\downarrow}({\bm x}){\bar\psi}_{m\downarrow}({\bm y})
                       & -\psi_{n\downarrow}({\bm x})\psi_{m\downarrow}({\bm y}) & \ \ \psi_{n\downarrow}({\bm x})\psi_{m\uparrow}({\bm y}) \\
                 \ \ {\bar\psi}_{n\downarrow}({\bm x}){\bar\psi}_{m\uparrow}({\bm y}) & \ \ {\bar\psi}_{n\downarrow}({\bm x}){\bar\psi}_{m\downarrow}({\bm y})
                      & \ \ {\bar\psi}_{n\downarrow}({\bm x})\psi_{m\downarrow}({\bm y}) & - {\bar\psi}_{n\downarrow}({\bm x})\psi_{m\uparrow}({\bm y})\\
                - {\bar\psi}_{n\uparrow}({\bm x}){\bar\psi}_{m\uparrow}({\bm y}) & -{\bar\psi}_{n\uparrow}({\bm x}){\bar\psi}_{m\downarrow}({\bm y})
                      & -{\bar\psi}_{n\uparrow}({\bm x})\psi_{m\downarrow}({\bm y}) & \ \ {\bar\psi}_{n\uparrow}({\bm x})\psi_{m\uparrow}({\bm y})
                  \end{array}\right)\ .
\label{eq:3.16}
\ee
\end{widetext}
We also define the Fourier transforms
\bse
\label{eqs:3.17}
\be
Q_{nm}({\bm k},{\bm p}) = \frac{1}{V} \int d{\bm x}\,d{\bm y}\ e^{-i{\bm k}\cdot{\bm x} + i{\bm p}\cdot{\bm y}}\,
                                          Q_{nm}({\bm x},{\bm y}).
\label{eq:3.17a}
\ee
and
\be
Q_{nm}({\bm k};{\bm q}) = Q_{nm}({\bm k}+{\bm q}/2,{\bm k}-{\bm q}/2)\ .
\label{eq:3.17b}
\ee
\ese
The $4\times 4$ matrix $Q_{nm}$ can be expanded in a spin-quaternion basis
\be
Q_{nm}({\bm x},{\bm y}) = \sum_{r,i=0}^3 (\tau_r\otimes s_i)\,{^i_r Q}_{nm}({\bm x},{\bm y})\ ,
\label{eq:3.18}
\ee
where $\tau_0 = s_0 = \openone_2$ is the unit $2\times 2$ matrix, and $\tau_{1,2,3} = -s_{1,2,3}
= -i\sigma^{1,2,3}$. An explicit inspection of the 16 matrix elements shows that $i=0$ and $i=1,2,3$
represent the spin-singlet and spin-triplet channels, respectively. Similarly, $r=0,3$
represents the particle-hole channel, i.e., products of the form ${\bar\psi}\psi$, whereas $r=1,2$
represents the particle-particle channel, i.e., products of the form ${\bar\psi}{\bar\psi}$ or
$\psi\psi$. For our purposes we will need only the particle-hole degrees of freedom. There is some
redundancy in the $Q_{nm}$, and all of the matrix elements are not independent. A convenient choice of the
independent elements are those with $n\geq m$.

The above considerations then show that the matrix elements ${^0_0Q}_{nm}$ with $n\geq 0$ and $m<0$ are
soft modes. It is easy to see, by using discrete symmetries that connect the various channels, that in the absence
of external fields {\em all} of the ${^i_r Q}_{nm}$ with $n\geq 0$, $m<0$ are soft \cite{Belitz_Kirkpatrick_1997, Belitz_Kirkpatrick_2012}.
Symmetry-breaking fields change this. For instance, an external magnetic field gives a mass to the particle-particle channel,
and also to two of the three particle-hole spin-triplet channels ($i=1,2$ for a magnetic field in the $z$-direction). 
A nonzero magnetization with a homogeneous component in a magnetically ordered phase has the same effect;
this will be important later. 

To summarize, of the two-particle degrees of freedom ${^i_r Q}_{nm}({\bm k};{\bm q})$ defined by Eqs.\ (\ref{eq:3.16}) - (\ref{eq:3.18}),
those with $n\geq 0$ and $m<0$ are soft modes in the sense that their two-point correlation functions diverge in the limit of
vanishing wave vector ${\bm q}$ and vanishing frequency $i\Omega_{n-m} \to \Omega + i0$. They represent the Goldstone mode of
a spontaneously broken continuous symmetry expressed by the gauge transformation in Eqs.\ (\ref{eqs:3.14}). Physically, the
broken symmetry reflects the difference between retarded and advanced degrees of freedom, and the spectrum of the
single-particle Green function is the corresponding order parameter. Notice that the ${^i_r Q}_{nm}({\bm k};{\bm q})$ are
soft for any value of ${\bm k}$ for which the spectrum of the Green function is nonzero. There thus are an infinite number
of soft two-particle modes in a Fermi liquid. This is qualitatively different from the case of electrons in the presence of quenched disorder,
for which only the zeroth moment $\sum_{\bm k} {^i_r Q}_{nm}({\bm k};{\bm q})$ is soft, see below.

\paragraph{Goldstone modes in a disordered Fermi liquid: Diffusons}
\label{par:III.A.2.c}

Historically, the notion of a spontaneously broken continuous symmetry, and the resulting Goldstone modes, in 
many-fermion systems was first developed for noninteracting electrons in the presence of quenched disorder
\cite{Wegner_1979, Schaefer_Wegner_1980, Pruisken_Schaefer_1982, McKane_Stone_1981}, and it was
instrumental for Wegner's matrix nonlinear sigma model describing the Anderson metal-insulator transition
\cite{Wegner_1979}. The derivation of \textcite{Schaefer_Wegner_1980} was later generalized to the
case of interacting electrons in the presence of disorder \cite{Belitz_Kirkpatrick_1997}. In the notation of
Eq.\ (\ref{eq:3.11b}), the two crucial differences in the disordered case are: (1) Only the zeroth moment
with respect to ${\bm k}$ of the correlation function $D$ (more precisely, the disorder average of $D$),
$\sum_{\bm k} D_{nm}({\bm k},{\bm q})$, is soft if $n$ and $m$ have
different signs, and (2) the resulting soft modes have a diffusive character, $\Omega \sim {\bm q}^2$, as opposed
to the ballistic modes in the clean case. Denoting the soft modes analogous to Eq.\ (\ref{eq:3.12}) by
$D^{+-}({\bm q},\Omega)$, one has
\be
D^{+-}({\bm q}\to 0,\Omega\to 0) = \frac{\pi N(\epsilonF)}{\Omega + D{\bm q}^2}\ ,
\label{eq:3.19}
\ee
with $N(\epsilonF)$ the density of states at the Fermi level, and $D$ a diffusion coefficient.
These diffusive soft modes are often referred to as ``diffusons'' in the
literature, and their counterparts in the particle-particle channel as ``Cooperons''. In the language of the $Q$-matrices,
Eqs.\ (\ref{eq:3.16}) - (\ref{eq:3.18}), the diffusons and Cooperons are given by the correlation functions of the 
$\sum_{\bm k} {^i_r Q}_{nm}({\bm k};{\bm q})$. Note that there are many more soft modes in a clean system
than in a disordered one, which makes the soft-mode analysis in clean systems more complicated.

\subsection{Effects of fermionic soft modes: Simple physical arguments}
\label{subsec:III.B}
In this section we give some simple arguments, both physical and structural, for
why fermionic fluctuations cause the ferromagnetic quantum phase transition
in clean two- or three-dimensional metallic systems to always be discontinuous. 
We then discuss how the presence of quenched disorder modifies
this conclusion. 

\subsubsection{Renormalized Landau theory}
\label{subsubsec:III.B.1}

We are interested in a theory that describes
the magnetization or order-parameter (OP) field, $\mathbf{m}$, the
fermionic degrees of freedom or conduction electrons described by
the Grassmann fields of Sec. \ref{subsec:III.A}, and the coupling between them.
Accordingly, the action consist of three distinct parts, 
\be
S[{\bm m};\bar\psi,\psi] = -{\cal A}_{\text{OP}}[{\bm m}] + S_{\text{F}}[\bar\psi,\psi] + S_{\text{c}}[{\bm m};\bar\psi,\psi]\ .
\label{eq:3.20}
\ee
They denote a purely bosonic part of the action that governs the order parameter, a purely fermionic one that describes
the conduction electrons, and a coupling between the two. 
\footnote{\label{action_notation} We denote actions that depend on bosonic (number-valued) fields only by ${\cal A}$, 
                actions that depend on fermionic (Grassmann-valued) fields, or a mix of bosonic and fermionic fields, by $S$, 
                and use a sign convention such that $S$ and ${\cal A}$ enter the exponential with a plus and minus sign,
                respectively.}
The partition function is given by
\be
Z = \int D[{\bm m}]\,D[\bar\psi,\psi]\ e^{S[{\bm m};\bar\psi,\psi]}
\label{eq:3.21}
\ee
Note that we do not specify the origin of the magnetization; in general it can be due to the conduction
electrons, or due to localized electrons in a different band, or a combination of the two. If one formally
integrates out the conduction electrons one obtains a effective Landau-Ginzburg-Wilson (LGW) action in terms
of the OP only,
\bse
\label{eqs:3.22}
\be
Z = \int D[{\bm m}]\ e^{-{\cal A}_{\text{eff}}[{\bm m}]}
\label{eq:3.22a}
\ee
where
\be
{\cal A}_{\text{eff}}[{\bm m}] = {\cal A}_{\text{OP}}[{\bm m}] - \ln \int D[{\bar\psi},\psi]\ e^{S_{\text{F}}[\bar\psi,\psi] 
    + S_{\text{c}}[{\bm m};\bar\psi,\psi]}.
\label{eq:3.22b}
\ee
\ese
A crucial question is how the conduction electrons couple to the
magnetization. In general, the latter will couple to both the orbital angular momentum and
to the spin of the electrons. The former poses interesting questions that have received
little attention to date, and we will not discuss it here. The latter coupling is via a Zeeman-like term 
\be
S_{\text{c}}[{\bm m};\bar\psi,\psi] = c \int dx\ {\bm m}(x)\cdot{\bm n}_{\text{s}}(x)\ .
\label{eq:3.23}
\ee
Here $c$ is a coupling constant and ${\bm n}_{\text{s}}$ is the electronic
spin-density,
\be
{\bm n}_{\text{s}}(x) = \sum_{a,b}{\bar\psi}_{a}(x){\bm\sigma}_{ab}\psi_{b}(x)\ ,
\label{eq:3.24}
\ee
with ${\bm\sigma} = (\sigma^{x},\sigma^{y},\sigma^{z})$ 
the Pauli matrices. $a,b=(\uparrow,\downarrow)$ are spin indices, and $x \equiv ({\bm x},\tau)$
comprises both the real-space position ${\bm x}$ and the imaginary-time variable $\tau$, and
$\int dx \equiv \int_V d{\bm x} \int_0^{1/T} d\tau$, where the spatial integration is over the system
volume $V$.

For simplicity we now treat the OP in a mean-field approximation, i.e., we replace the 
fluctuating magnetization ${\bm m}(x)$ by an $x$-independent magnetization $m$ that
we take to point in the 3-direction. We will discuss the validity of this approximation in
Sec. \ref{subsec:III.C}. Denoting the 3-component of ${\bm n}_{\text{s}}$ by 
$n_{\text{s}}$, the second term in Eq.\ (\ref{eq:3.22b}), which describes the effect of the 
coupling between the fermions and the OP, can be written
\bea
\delta{\cal A}[m] &=& - \ln \int D[{\bar\psi},\psi]\ e^{S_{\text{F}}[{\bar\psi},\psi] + cm\int dx\,n_{\text{s}}(x)}
\nonumber\\
        &=& -\ln \left\langle e^{cm\int dx\,n_{\text{s}}(x)} \right\rangle_{\text{F}}\ ,
\label{eq:3.25}
\eea
where in the second line we have dropped a constant contribution to the action, and 
$\langle\ldots\rangle_{\text{F}}$ denotes an average with the action $S_{\text{F}}$.

Now consider the longitudinal spin susceptibility $\chi(h)$ of fermions governed by the action $S_{\text{F}}$ and
subject to a magnetic field $h$. It is given by the correlation function
\be
\chi(h) = \frac{T}{V}\int dx\,dy\,\langle\delta n_{\text{s}}(x)\,\delta n_{\text{s}}(y)\rangle_{S_{\text{h}}}\ ,
\label{eq:3.26}
\ee
where $S_{\text{h}} = S_{\text{F}} + h\int dx\,n_{\text{s}}(x)$, and 
$\delta n_{\text{s}}(x) = n_{\text{s}}(x) - \langle n_{\text{s}}(x) \rangle_{S_{\text{m}}}$.
By differentiating Eq.\ (\ref{eq:3.25}) twice with respect to $m$ it is easy to show that
\be
\frac{d^2\,\delta{\cal A}}{dm^2} = -\frac{V}{T}\,c^2 \chi(cm)\ .
\label{eq:3.27}
\ee
Since $\delta{\cal A}[m=0] = d\,\delta{\cal A}/dm\vert_{m=0} = 0$, we now have
\be
\delta{\cal A}[m] = \frac{-V}{T}\,c^2 \int_0^m dm_1 \int_0^{m_1} dm_2\ \chi(cm_2)\ .
\label{eq:3.28}
\ee
 $S_{\text{OP}}$ will have the usual Landau form of a power series in powers of $m^2$,
and the complete renormalized Landau free-energy density $f_{\text{eft}} = -(T/V){\cal A}_{\text{eft}}$
thus is
\bse
\label{eqs:3.29}
\be
f_{\text{eff}}[m] = t\,m^2 + \delta f[m] + u\,m^4 + O(m^6)\ .
\label{eq:3.29a}
\ee
Here $t$ and $u$ are Landau parameters, and
\be
\delta f[m] = -c^2 \int_0^m dm_1 \int_0^{m_1} dm_2\ \chi(cm_2)\ .
\label{eq:3.29b}
\ee
\ese

This result expresses the correction to the usual Landau action in terms of the spin susceptibility
of nonmagnetic fermions in the presence of an effective homogeneous magnetic field given by $c\,m$.
It is a ``renormalized Landau theory'' in the sense that it includes the effects of fluctuations extraneous
to the order-parameter fluctuations. The remaining question is the behavior of the susceptibility that 
represents these fluctuations for small $m$. As we will see, this susceptibility is not an analytic function
of $m$ at $m=0$.

\subsubsection{Clean systems}
\label{subsubsec:III.B.2}

It has been known for a long time that various observables in a Fermi liquid are
nonanalytic functions of the temperature. For instance, the specific heat coefficient
has a $T^2\ln T$ term \cite{Baym_Pethick_1991}. The spin susceptibility in
a three-dimensional system was found to have no such nonanalytic behavior
\cite{Carneiro_Pethick_1977}. However, this absence of a nonanalyticity was later shown to 
be accidental, and to pertain only to the $T$-dependence in three dimensions. In dimensions
$d\neq 3$ there is a $T^{d-1}$ nonanalyticity, and even in $d=3$ at $T=0$ the inhomogeneous spin susceptibility has
a $k^2\ln k$ wave-number dependence \cite{Belitz_Kirkpatrick_Vojta_1997, Chitov_Millis_2001b, 
Galitski_Chubukov_Das_Sarma_2004}. This nonanalyticity
is a direct consequence of the soft modes discussed in Sec.\ \ref{subsec:III.A} above. 
From general scaling arguments one expects a corresponding nonanalyticity for the
homogeneous susceptibility at zero temperature as a function of a magnetic field $h$, namely,
$\chi(h) \propto \text{const.} + h^{d-1}$ in generic dimensions, and 
$\chi(h) \propto \text{const.} - h^2 \ln h$ in $d=3$. These scaling arguments have
been shown to be exact, as far as the exponent is concerned, by a renormalization-group
treatment \cite{Belitz_Kirkpatrick_2014}, and they are consistent with explicit perturbative
calculations \cite{Misawa_1971, Barnea_Edwards_1977, Betouras_Efremov_Chubukov_2005}. The sign of the effect is universal
and can be established as follows. Fluctuations suppress the tendency of a
Fermi liquid to order ferromagnetically, and therefore the fluctuation correction to the
bare zero-field susceptibility is negative, $\delta\chi(0) < 0$. A small magnetic field
suppresses the fluctuations, and therefore $\delta\chi(h) - \delta\chi(0) > 0$. This implies
that the nonanalyticity in $\chi(h\to 0)$ has a positive sign:
\be
\chi(h\to 0) = \chi(0) + \begin{cases} a_d\, h^{d-1} & \text{for $1<d<3$} \\
                                                          a_3\,h^2\ln(1/h) & \text{for $d=3$}
                                   \end{cases}\ ,
\label{eq:3.30}
\ee
where $a_d>0$. For the renormalized Landau free-energy density, Eq.\ (\ref{eq:3.29a}), we
thus obtain
\bea
f_{\text{eff}}[m]  &=& -h\,m + t\,m^2 + u\,m^4 
\nonumber\\
&&\hskip -0pt - v_d\times\begin{cases} m^{d+1} + u\,m^4 & (1<d<3)\\
                                                  m^4\ln(1/m) & (d=3)\ .
                             \end{cases} \hskip 20pt
\label{eq:3.31}
\eea
Here $v_d \propto c^{d+1} > 0$, and we have added an external magnetic field $h$. For $d=3$ this result was first derived by
\onlinecite{Belitz_Kirkpatrick_Vojta_1999}. The negative term in the free energy,
which dominates the quartic term for all $d\leq 3$, necessarily leads to a first-order
ferromagnetic transition. We stress that while this is a fluctuation-induced first-order
quantum phase transition, the relevant fluctuations are {\em not} the order-parameter
fluctuations, but are fermionic in nature. For purposes of an analogy with the well-known
classical fluctuation-induced first-order transitions \cite{Halperin_Lubensky_Ma_1974},
the latter play a role that is analogous to that of the vector potential in superconductors,
or the director fluctuations at the nematic-smectic-A transition. An important difference,
however, is that in these classical systems the order-parameter fluctuations are below
their upper critical dimension, which makes them strong enough to make the first-order
transition very weak and hard to observe at best, and destroy it altogether at worst \cite{Anisimov_et_al_1990}.
By contrast, in the case of a quantum ferromagnet the order-parameter fluctuations are
{\em above} their upper critical dimension, so the first-order transition predicted
by the renormalized Landau theory is expected to be much more robust.

\begin{figure}[t]
\begin{center}
\includegraphics[width=0.9\columnwidth,angle=0]{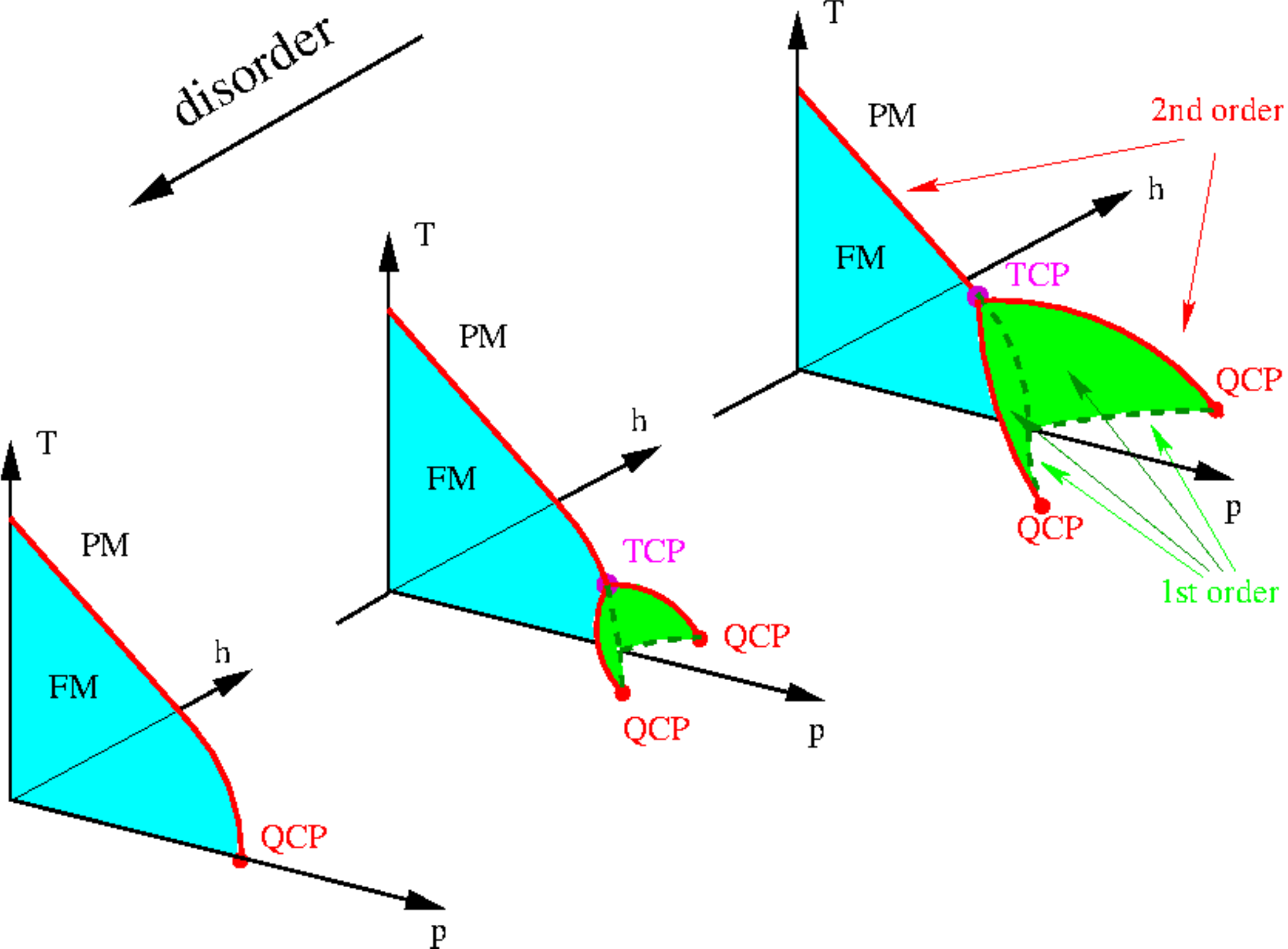}
\end{center}
\caption{Schematic phase diagram in the space spanned by temperature ($T$),
              hydrostatic pressure ($p$), and magnetic field ($h$). Shown are the ferromagnetic (FM)
              and paramagnetic (PM) phases, the tricritical point (TCP), and the quantum critical
              points (QCP). Solid and dashed lines denote second-order and first-order transitions,
              respectively. The tricritical wings emerging from the TCP are surfaces of first-order
              transitions. The three panels show the predicted evolution of the phase diagram with
              increasing disorder. After \textcite{Sang_Belitz_Kirkpatrick_2014}. }
\label{figure:wings_schematic}
\end{figure}

A nonzero temperature cuts off the magnetic-field singularity \cite{Betouras_Efremov_Chubukov_2005},
and with increasing temperature the fluctuation-induced term in the free energy becomes less and
less negative. Suppose the Landau parameter $t$ at $T=0$ is a monotonically increasing function of 
a control parameter, say, hydrostatic pressure $p$, and let $t(p=0,T=0) < 0$. Then there will be a
quantum phase transition at some nonzero pressure $p_{\text{c}}$. As the transition temperature
is increased from zero by lowering $p$, one then expects a tricritical point in the phase diagram. 
Below the tricritical temperature the transition will be discontinuous due to the mechanism described 
above, while at higher temperatures it will be continuous. In the presence of an external magnetic
field there appear surfaces of first-order transitions, or tricritical wings \cite{Belitz_Kirkpatrick_Rollbuehler_2005}, 
and the phase diagram has the schematic structure shown in the right-most panel in Fig.~\ref{figure:wings_schematic}.%
\footnote{\label{wings_footnote} The presence of tricritical wings is characteristic of any phase diagram that contains a
tricritical point \cite{Griffiths_1970, Griffiths_1973}, it is not restricted to the ferromagnetic
quantum phase transition.}
The third law of thermodynamics, in conjunction with various Clapeyron-Clausius relations, puts constraints on the shape of the wings at low 
temperatures \cite{Kirkpatrick_Belitz_2015a}. Most importantly, the wings must be perpendicular to the $T=0$ plane, and they cannot be 
perpendicular to the zero-field plane. These constraints, as well as the overall wing structure, are in excellent agreement with experimentally 
observed phase diagrams, see, for instance, Figs.\ \ref{figure:UGe_2_phase_diagram}, \ref{figure:MnSi_phase_diagram},
\ref{fig:URhGe_phase_diagram}, and \ref{fig:UCoAl_phase_diagram}.

The theory also predicts a correlation between the tricritical temperature $T_{\text{tc}}$ and the 
magnetic moment $m_1$ just on the FM side of the first-order transition. In terms of the parameters
in Eq.\ (\ref{eq:3.31}) one finds $T_{\text{tc}} = T_0\,e^{-u/v_3}$ and $m_1 = m_0\,e^{-1/2} e^{-u/v_3}$,
where $T_0$ and $m_0$ are microscopic temperature and magnetization scales, respectively
\cite{Belitz_Kirkpatrick_Rollbuehler_2005}. For given scales $T_0$ and $m_0$, which one expects
to vary little within members of a given class of materials, the theory thus predicts that $T_{\text{tc}}$
is proportional to $m_1$. As we have pointed out in Sec.\ \ref{subsubsec:II.B.5}, this is in good
agreement with experiments.

\subsubsection{Disordered systems}
\label{subsubsec:III.B.3}

In the presence of quenched disorder the general logic of the arguments given above
remains intact, but important aspects change qualitatively. First, the fermionic soft
modes are diffusive in nature, rather than ballistic, see Sec.\ \ref{par:III.A.2.c}. 
This slowing-down of the electrons favors the tendency towards ferromagnetism 
and as a result the combined disorder and interaction fluctuations increase the 
bare susceptibility, $\delta\chi(0) > 0$.%
\footnote{\label{disorder_enhancement_footnote} The notion that quenched disorder
{\em favors} ferromagnetic order is somewhat counterintuitive, given that in
classical systems, long-range order is negatively affected by it \cite{Cardy_1996}.
Indeed, the ferromagnetic $T_{\text{c}}$ usually decreases with increasing
disorder, see Sec.\ \ref{sec:II}. However, at sufficiently low temperature the
diffusive motion of the electrons leads to an increase in the effective exchange
interaction. The interplay between these two effects is discussed at the end
of this subsection.} 
A small magnetic field will again suppress the effect of the fluctuations, and the
nonanalytic contribution to $\chi(h)$ therefore has a negative sign. Second, the
changed nature of the fermionic soft modes ($\Omega \sim {\bm q}^2$ rather than
$\Omega \sim \vert{\bm q}\vert$) leads to a different exponent, namely,  
\be
\chi(h\to 0) = \chi(0) - {\tilde a}_d\,h^{(d-2)/2}\qquad \text{for $d>2$}\ ,
\label{eq:3.32}
\ee
with ${\tilde a}_d > 0$. This expectation is borne out by explicit perturbative
calculations \cite{Altshuler_Aronov_Zyuzin_1983}. Third, the entire notion of a
disordered Fermi liquid breaks down for $d\leq 2$ due to localization effects
\cite{Belitz_Kirkpatrick_1994, Lee_Ramakrishnan_1985}, so the only physical
dimension where the current discussion applies is $d=3$.

The renormalized Landau free-energy density in $d=3$ now becomes \cite{Belitz_Kirkpatrick_1996, Kirkpatrick_Belitz_1996}
\be
f_{\text{eff}}[m]  = -h\,m + t\,m^2 + v\,m^{5/2} + u\,m^4\ ,
\label{eq:3.33}
\ee
with $v>0$. We see that, for very general reasons, quenched disorder
leads to a second-order or continuous transition, but the Landau
theory for this transition is not standard because of the $m^{5/2}$ term
which leads to unusual critical exponents. In particular, the exponents$\,^{\ref{exponents_footnote}}$
$\beta$ and $\delta$ for the order parameter, and $\gamma$ for the
order-parameter susceptibility, are
\be
\beta = 2\quad,\quad \delta = 3/2\quad,\quad \gamma = 1\ .
\label{eq:3.34}
\ee
Other critical exponents will be discussed in Sec.\ \ref{subsec:III.C}.

Equations (\ref{eq:3.31}) and (\ref{eq:3.33}) are valid for the extreme cases of ultraclean and
strongly disordered systems, respectively. An equation of state that interpolates between the
two has been constructed by \textcite{Sang_Belitz_Kirkpatrick_2014}; the schematic evolution
of the phase diagram with increasing disorder is shown in Fig.~\ref{figure:wings_schematic}. The theory allows to distinguish
between three distinct disorder regimes, characterized by the residual resistivity $\rho_0$: 
\vskip 2pt
\par\noindent{\it Regime I (Clean):} $\rho_0 \alt \rho_0^{(1)}$. The transition at low temperature is first
order, and there is a tricritical point in the phase diagram. The tricritical temperature decreases with
increasing disorder.
\vskip 2pt
\par\noindent{\it Regime II (Intermediate):} $\rho_0^{(1)} \alt \rho_0 \alt \rho_0^{(2)}$. The transition is 
second order down to $T=0$. The critical
behavior is mean-field-like, as predicted by Hertz theory, except extremely close to the
critical point, where it crosses over to the exponents given by Eq.\ (\ref{eq:3.34}). 
\vskip 2pt
\par\noindent{\it Regime III (Disordered):} $\rho_0 \agt \rho_0^{(2)}$. The transition is second order, and 
the critical exponents are given
by Eq.\ (\ref{eq:3.34}). In this regime the quantum Griffiths effects discussed in Sec.\ \ref{subsec:III.D} are
expected to be present and to compete with the critical behavior. 
\vskip 2pt
\par\noindent
A rough semi-quantitative estimate (see footnote \ref{resistivity_footnote}) for
the boundaries between the three regimes yields $\rho_0^{(1)} \approx$ 1 to several $\mu\Omega$cm, and
$\rho_0^{(2)} \approx $ 100 to several hundred $\mu\Omega$cm. 

\begin{figure}[t]
\begin{center}
\includegraphics[width=0.9\columnwidth,angle=0]{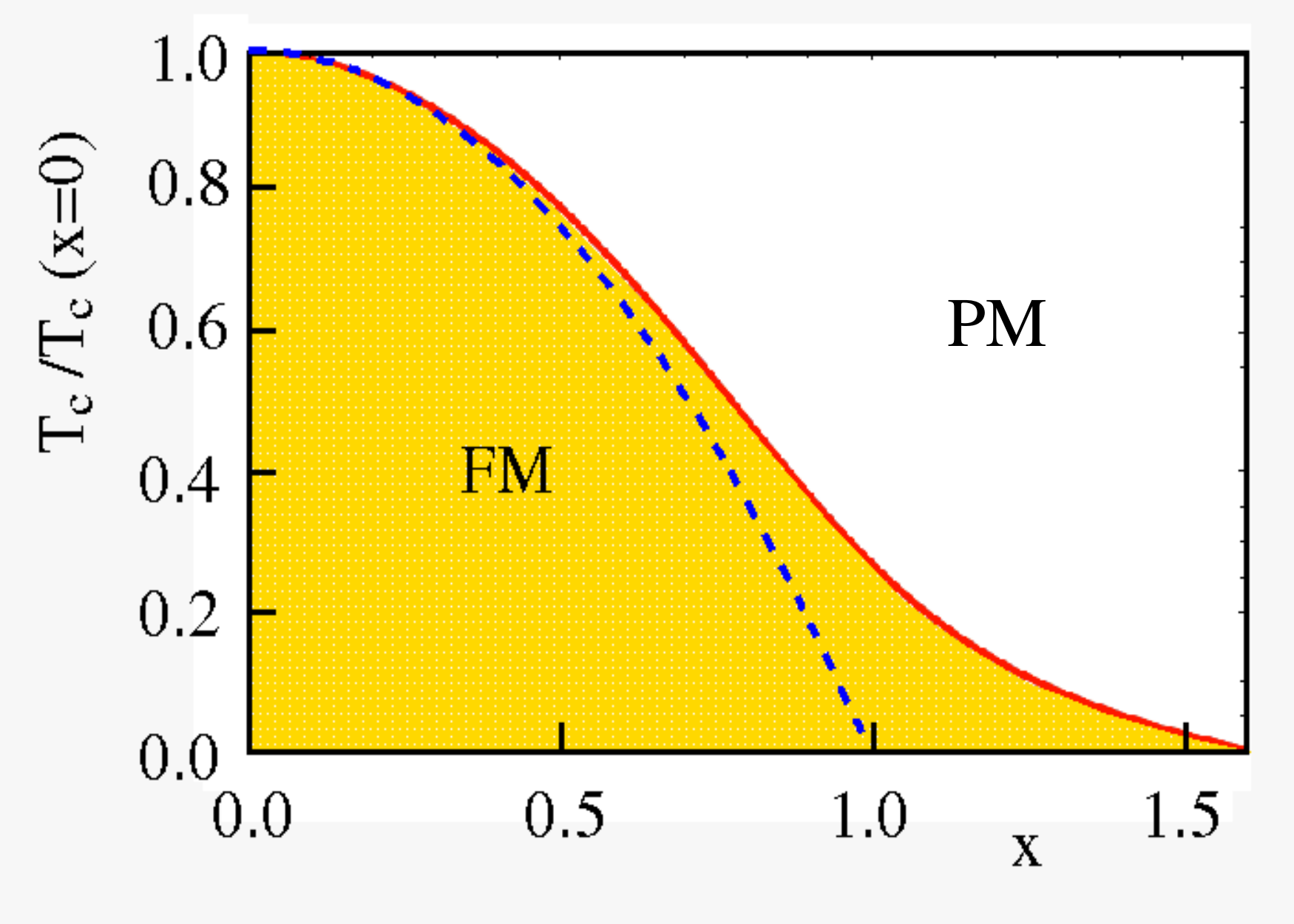}
\end{center}
\caption{Schematic phase diagram in the temperature-disorder plane as given by Eq.\ (\ref{eq:3.35}).
              The dashed line reflects the classical dilution effect of the
              quenched disorder only ($a=0$ in Eq.\ (\ref{eq:3.35})); the solid (red) line also reflects the
              quantum effects, with $a=1$, $b=10$, due to the diffusive dynamics of the electrons. See the text for additional information.
              }
\label{figure:phase_diagram_disordered}
\end{figure}

We now discuss the expected qualitative shape of the phase diagram.
Let $x$ be a dimensionless measure of the disorder, i.e., $x\propto 1/\tau$ with $\tau$
the elastic mean-free time. As mentioned above,$^{\ref{disorder_enhancement_footnote}}$ there are two
competing influences of $x$ on the critical temperature. One is a classical
dilution effect that suppresses the transition temperature to zero at sufficiently
large values of $x$ \cite{Cardy_1996}. For simplicity, let us assume that this leads to
$T_{\text{c}}(x) = 1 - x^2$, with $T_{\text{c}}$ measured in units of 
$T_{\text{c}}(x=0)$. (Adding a term linear in $x$ does not change the
qualitative discussion that follows.) The other is an increase in $T_{\text{c}}$ due to the
diffusive nature of the electron dynamics, which increases the effective spin-triplet
interaction \cite{Altshuler_Aronov_Zyuzin_1983}. Indeed, the increase in the zero-field
susceptibility mentioned above Eq.\ (\ref{eq:3.32}) is proportional to this increase in the
interaction amplitude. For small disorder, this effect is 
linear in the disorder at $T=0$, and it is cut off by the
temperature itself, i.e., it is strongest for small values of $T_{\text{c}}$. A
simple schematic way to represent these two effects is to write
\be
T_{\text{c}}(x) = 1 - x^2 + \frac{a\,x}{1 + b\,T_{\text{c}}(x)/x}\ .
\label{eq:3.35}
\ee
The resulting qualitative shape of the phase diagram is shown in
Fig.~\ref{figure:phase_diagram_disordered}.
The two competing disorder effects can lead to an inflection point in the plot
of $T_{\text{c}}$ vs. $x$ that is often seen in experiments, see Figs.\ \ref{pikul2012a2}
and \ref{CePdRh_phase_diagram}.
Another possible interpretation of this shape of the phase diagram is a smeared
transition due to quantum Griffiths effects that have been ignored in the above
arguments; this is discussed in Sec.\ \ref{subsec:III.D}. Note that, in addition to the
inflection point in the phase diagram, the unusually large value of the exponent
$\beta$, Eq.\ (\ref{eq:3.34}), can also mimic a smeared transition.

We end this section by recalling a very general result on critical fixed points in systems
with quenched disorder due to \textcite{Harris_1974} and
\textcite{Chayes_et_al_1986}. Harris investigated a necessary condition for the critical
behavior of a clean system to remain unchanged by a small amount of quenched 
disorder. The physical argument is that in order for the transition to stay sharp, the
disorder-induced fluctuations of the location of the critical point in parameter space
must be small compared to the distance from the critical point. This implies that the
correlation length must diverge sufficiently fast as the critical point is approached,
which leads to the requirement
\be
\nu \geq 2/d
\label{eq:3.36}
\ee
for the correlation-length exponent $\nu$, which is often referred to as the Harris
criterion. In Harris's original argument this was a
constraint on the exponent $\nu$ of the {\em clean} system which, if hyperscaling
holds, is equivalent to the condition $\alpha < 0$ for the specific-heat exponent
$\alpha$. It does not apply to the quantum ferromagnetic transition, since the
transition in clean systems is not continuous. However, \textcite{Chayes_et_al_1986} showed
rigorously that, for a large class of disordered systems that undergo
a continuous phase transition, Eq.\ (\ref{eq:3.36}) must hold for the exponent $\nu$
that characterizes the {\em disordered} fixed point. The value of $\nu$ corresponding
to the exponents given in Eq.\ (\ref{eq:3.34}), viz., $\nu = 1$ in $d=3$ \cite{Kirkpatrick_Belitz_1996},
satisfies this constraint, while the mean-field value $\nu = 1/2$ of Hertz theory does
not. We will come back to this point in Sec.\ \ref{subsubsec:III.C.2}.

\subsection{Effects of order-parameter fluctuations, and comparison with experiment}
\label{subsec:III.C}

In the previous subsection we discussed very simple physical arguments for the
nature of the ferromagnetic quantum phase transition. We treated the order parameter in
a mean-field approximation and integrated out all fermionic degrees of freedom. The
fermionic soft modes then lead to a Landau free energy that is not an analytic function
of the magnetization. An obvious question regards the effects of the order-parameter
fluctuations that were neglected in this procedure. In this section we discuss these
effects, and also consider the related question of the behavior not asymptotically close
to the quantum phase transition. All of these issues are important for the relation of the
theory to the experimental results discussed in Sec.\ \ref{sec:II}.

\subsubsection{Coupled field theory for soft modes}
\label{subsubsec:III.C.1}

We return to the coupled field theory in Eq.\ (\ref{eq:3.20}). For the purposes of
discussing fluctuation effects, integrating out the fermions, as in Eq.\ (\ref{eqs:3.22}), is
not desirable, as it will result in a non-local field theory for the order-parameter fluctuations.%
\footnote{\label{nonlinear_footnote} Historically, this was the route taken by
\textcite{Hertz_1976}, who integrated out the fermions in a tree approximation. For
disordered systems, it was later refined by \textcite{Kirkpatrick_Belitz_1996}, who 
showed that fermionic loops destabilize Hertz's critical fixed point. While this method works 
for power-counting purposes, the coupled local field theory developed later for disordered 
\cite{Belitz_et_al_2001a, Belitz_et_al_2001b} and clean \cite{Kirkpatrick_Belitz_2012} systems
is more versatile and easier to handle, and we use it here. For clean systems, the fermionic 
fluctuations destroy the critical fixed point altogether and change the order of the transition, as we have discussed in
Sec.\ \ref{subsubsec:III.B.2}. However, the pre-asymptotic behavior, which is governed by a
critical fixed point that ultimately is unstable, can still be important and is discussed in
Sec.\ \ref{subsubsec:III.C.2}. }
A better strategy is to separate the fermionic degrees of freedom into soft and massive
modes, and integrate out only the latter to arrive at an effective theory that treats all of the soft
modes on equal footing. This is possible using the identification of fermionic soft modes
explained in Sec.\ \ref{subsec:III.A}. The resulting effective action can then be analyzed
by means of renormalization-group methods. 

In Sec.\ \ref{subsec:III.A} we have seen that the soft fermionic degrees of freedom are
given by those matrix elements $Q_{nm}$, Eq.\ (\ref{eq:3.16}), for which the two Matsubara
frequencies have different signs. Denoting these by $q_{nm}$ (with $n>0$, $m<0$ implied, see the
remark after Eq.\ (\ref{eq:3.10})), and the massive modes
by $P_{nm}$, we first rewrite the action from Eq.\ (\ref{eq:3.20}) in terms of the $q$ and
$P$ instead of the $\bar\psi$ and $\psi$:
\footnote{\label{Lagrange_multiplier_footnote} Technically, this can be achieved by constraining all terms
                         in the action that contain the Grassmann fields to higher than bilinear order to the matrix
                         field $Q$ from Eq.\ (\ref{eq:3.16}) by means of a Lagrange multiplier field and integrating
                         out the Grassmann fields, see \textcite{Belitz_Kirkpatrick_1997, Belitz_Kirkpatrick_2012}. 
                         For simplicity we suppress the dependence of the action on the Lagrange
                         multiplier field in our notation.}
\bea
{\cal A}[{\bm m};q,P] &\equiv& -S[{\bm m};{\bar\psi},\psi]
\nonumber\\
&=& {\cal A}_{\text{OP}}[{\bm m}] + {\cal A}_{\text{F}}[q,P] + {\cal A}_{\text{c}}[{\bm m};q,P]\qquad
\label{eq:3.37}
\eea
If we now integrate out the massive modes $P$, we can formally write the partition function
\bse
\label{eqs:3.38}
\be
Z = \int D[{\bm m}]\,D[q]\ e^{-{\cal A}_{\text{eff}}[{\bm m},q]}\ ,
\label{eq:3.38a}
\ee
in terms of an effective action 
\bea
{\cal A}_{\text{eff}}[{\bm m},q] &=& {\cal A}_{\text{OP}}[{\bm m}] - \ln \int D[P]\,e^{-{\cal A}_{\text{F}}[q,P] - {\cal A}_{\text c}[{\bm m};q,P]}
\nonumber\\
&\equiv& {\cal A}_{\text{OP}}[{\bm m}] + {\cal A}_{\text{F}}[q] + {\cal A}_{\text{c}}[{\bm m},q]\ .
\label{eq:3.38b}
\eea
\ese
Integrating out the $P$ cannot be done exactly, but any approximation that respects the symmetries of the
action suffices. 

Before we discuss the various terms in this effective action in more detail, we make a few general remarks.
${\cal A}_{\text{OP}}$ is a standard LGW action, supplemented by a random-mass term in the disordered
case, see below. ${\cal A}_{\text{F}}$ has a Gaussian contribution that reflects the soft modes identified
and discussed in Sec.\ \ref{subsubsec:III.A.2}, as well as higher-order terms to all orders in $q$. The soft modes
are diffusive in disordered systems, and ballistic
in clean ones, but apart from this and the random-mass term in ${\cal A}_{\text{OP}}$ there are no
structural differences between clean and disordered systems as far as these two terms in the action are
concerned. The coupling ${\cal A}_{\text{c}}$ has a contribution that is bilinear in ${\bm m}$ and $q$, and in
addition terms of order ${\bm m}\,q^n$, where $n$ can be any integer. The bilinear term leads to the
characteristic Landau damping in the paramagnon propagator \cite{Doniach_Engelsberg_1966, Hertz_1976},
i.e., to a frequency dependence of the form $\vert\Omega\vert/\vert{\bm k}\vert$ in clean systems, and
$\vert\Omega\vert/{\bm k}^2$ in disordered ones. At the level of terms bilinear in ${\bm m}$ and $q$ there
is again no other structural difference between the clean and disordered cases. However, the terms of
order ${\bm m}\,q^2$ generate, in a renormalization procedure, a nonanalytic wave-number dependence
of the paramagnon propagator that has the form $\vert{\bm k}\vert^{d-1}$ in clean systems, and $\vert{\bm k}\vert^{d-2}$
in disordered ones. The sign of this term is different in the two cases. If one replaces the fluctuating order
parameter by its expectation value, this term leads to the renormalized Landau theory described in
Sec.\ \ref{subsubsec:III.B.1}, with a first-order transition in the clean case and a second-order one in the
disordered case.

\subsubsection{Clean systems}
\label{subsubsec:III.C.2}

In clean systems, the transition at $T=0$ in a vanishing magnetic field was found to be of first order if
order-parameter fluctuations are neglected, see Sec.\ \ref{subsubsec:III.B.2}. Order-parameter fluctuations
are thus cut off before the system reaches a critical point, remain finite, and do not change the nature
of the transition.%
\footnote{\label{OP_fluctuations} This statement needs to be interpreted with care in the
               light of footnote \ref{first_order_footnote}. The first-order transition is described
               by a strong-coupling RG fixed point \cite{Nienhuis_Nauenberg_1975, Fisher_Berker_1982},
               and the relevant fluctuations are effectively already included in the generalized Landau
               theory represented by Eq.\ (\ref{eq:3.31}).}$^,$%
\footnote{\label{fluctuation_induced_second_order} There are examples of phase transitions that are first order
              on a mean-field level, yet are driven second order by fluctuations \cite{Fucito_Parisi_1981}.
              Motivated in part by early experiments that reported a second-order transition in ZrZn$_2$
              \cite{Grosche_et_al_1995}, \textcite{Kirkpatrick_Belitz_2003b}
              proposed that a similar mechanism may be operative in quantum ferromagnets in a
              certain parameter region. However, later experiments on cleaner samples showed that
              the transition is first order after all \cite{Uhlarz_Pfleiderer_Hayden_2004}, and to date 
              there is no clear evidence for the $T=0$ transition to be second order in any clean bulk quantum 
              ferromagnet. For questionable cases that require more experimental scrutiny see the discussion
              in Sec.\ \ref{subsubsec:II.B.1}.}
However, if the transition is weakly first order, then there will be a sizable region in parameter space where
the physical behavior is controlled by the unstable fixed point that is described by Hertz theory, and only
asymptotically close to the transition will the renormalization-group flow turn away towards the strong-coupling
fixed point that describes the first-order transition. It is therefore important to fully understand the results of
Hertz theory and its predecessors, even if they ultimately do not describe the nature of the transition correctly.
Also, order-parameter fluctuations do affect the various lines of second-order transitions in the phase diagram
shown schematically in Fig.~\ref{figure:wings_schematic}. In this section we describe all of these effects.

\paragraph{Hertz's action, and relation to spin-fluctuation theory}
\label{par:III.C.2.a}

In clean systems, the relevant fermionic soft modes are the Fermi-liquid Goldstone modes discussed
in Sec.\ \ref{subsubsec:III.A.2}. The soft-mode action has not been derived in a closed form, but can
be obtained to any desired order in the soft degrees of freedom $q$. To Gaussian order it reads
\cite{Belitz_Kirkpatrick_2012}
\begin{widetext}
\be
{\cal A}_{\text{F}}[q] = -8\sum_{r=0,3}\sum_{i=0}^3\sum_{{1,2}\atop{3,4}}\frac{1}{V}\sum_{\bm k}
{^i_r q}_{12}({\bm k})\left(\delta_{13}\,\delta_{24}\,\frac{1}{\varphi_{12}({\bm k})} - \delta_{1-2,3-4}2T\gamma_i\right)
{^i_r q}_{34}(-{\bm k}) + O(q^3)
\label{eq:3.39}
\ee
\end{widetext}
Here $\gamma_0$ and $\gamma_{1,2,3}$ are the spin-singlet and spin-triplet interaction amplitudes, respectively,
$1\equiv n_1$ etc is a shorthand for Matsubara frequencies, and the function $\varphi$ is given by
\be
\varphi_{12}({\bm k}) = N_{\text{F}}\,\frac{2\pi G}{k}\,\varphi_d(Gi\Omega_{1-2}/k)\ ,
\label{eq:3.40}
\ee
where $G$ is a coupling constant whose bare value is the inverse Fermi velocity $1/\vF$. $\varphi_d$ can be expressed
in terms of Gauss's hypergeometric function:
\be
\varphi_d(z) = \frac{i}{z}\,{_2F}_1(1,1/2,d/2;1/z^2)\ .
\label{eq:3.41}
\ee
For $d=1,2,3$ it reduces to the
familiar expressions for the hydrodynamic part of the Lindhard function in these dimensions:
\bse
\label{eqs:3.42}
\bea
\varphi_{d=1}(z) &=& -iz/(1-z^2)\ ,
\label{eq:3.42a}\\
\varphi_{d=2}(z) &=& \sgn({\text{Im}} z)/\sqrt{1-z^2}
\nonumber\\
                   &\equiv& i/\sqrt{z+1}\sqrt{z-1}\ ,
\label{eq:3.42b}\\
\varphi_{d=3}(z) &=& \frac{-i}{2}\,\ln\left(\frac{1-z}{-1-z}\right)\ .
\label{eq:3.42c}
\eea
\ese
Note that the vertex $1/\varphi$ scales as a function that is linear in either the frequency or the wave number
(except in $d=1$), and that this is true also for the second term in parentheses in Eq.\ (\ref{eq:3.39}) 
due to the structure of the frequency constraint. This feature reflects the Goldstone modes. Another part
of the Gaussian action involves a soft Lagrange multiplier field $\lambda$. Its Gaussian vertex is minus
the noninteracting part of the $q$-vertex, and the only effect of the $\lambda$ is to ensure that the 
Goldstone modes affect observables only in interacting systems, as is obvious from many-body
perturbation theory. For simplicity, we do not show the $\lambda$-part
of the action.$^{\ref{Lagrange_multiplier_footnote}}$

The order-parameter field ${\bm m}$, whose expectation value is proportional to the magnetization, 
couples linearly to the electron spin density, which is linear in the matrix field $Q$, with
a dimensionless coupling constant $c = O(1)$. To linear order in the soft component $q$ the coupling reads
\bse
\label{eqs:3.43}
\be
{\cal A}_{\text{c}}[{\bm m},q] = 8c\sqrt{T}\sum_{\bm k}\sum_{12}\sum_{r,i} {^i_r q}_{12}({\bm k})\,{^i_r b}_{12}(-{\bm k})
   + O(m\,q^2)
\label{eq:3.43a}
\ee
where
\be
{^i_r b}_{12}({\bm k}) = (-)^{r/2} \sum_n \delta_{n,n_1-n_2} \left[m_n^i({\bm k}) + (-)^{r+1}\,m^i_{-n}({\bm k})\right]
\label{eq:3.43b}
\ee
\ese
is a symmetrized version of the order-parameter field ${\bm m}_n({\bm k})$ with components $m_n^{1,2,3}$.

Finally, the order-parameter action is an ordinary quantum $\phi^4$-theory,%
\footnote{\label{spin_precession_footnote} This action is missing a term of order $\Omega\,{\bm m}^3$ that
 describes the Bloch spin precession of the order parameter in the magnetic field of all the other magnetic
 moments. This term is absent in the case of an Ising order parameter, but in all other cases it is important,
 for instance, for producing the correct dynamics of the spin waves. In a field-theoretic context, it is sometimes
 referred to as a Wess-Zumino or Chern-Simons term, and its topological aspects are stressed \cite{Fradkin_1991}.
 For the purposes of our discussion it is RG irrelevant, and we drop it.}
\bea
{\cal A}_{\text{OP}}[{\bm m}] &=& -\sum_{{\bm k},n} {\bm m}_n({\bm k})\,\left[ t + a{\bm k}^2 + b(\Omega_n)^2\right]
   \cdot {\bm m}_{-n}(-{\bm k})
   \nonumber\\
&& + u \int d{\bm x}\,T \sum_{n_1,n_2,n_3} \left({\bm m}_{n_1}({\bm x})\cdot {\bm m}_{n_1}({\bm x})\right)
\nonumber\\
&& \hskip 30pt \times \left({\bm m}_{n_3}({\bm x})\cdot {\bm m}_{-n_1-n_2-n_3}({\bm x})\right)\ ,
\label{eq:3.44}
\eea
with $t$, $a$, $b$, and $u$ the coupling constants of this LGW action. 

We now have specified all parts of the effective action, Eq. \ (\ref{eq:3.38b}), to bilinear order in ${\bm m}$ and
$q$. It is important to realize that it is {\em not} a fixed-point action corresponding to a critical fixed point. The
terms of $O(m\,q^2)$ that are not shown explicitly in Eq.\ (\ref{eq:3.43a}) are relevant with respect to the
fixed point represented by this action and lead to the first-order transition described in Sec.\ \ref{subsubsec:III.B.2}.
However, depending on the strength of the first-order transition, there will be a sizable regime where the 
renormalization-group flow is dominated by the unstable fixed point. The physical behavior in this regime will
thus be given by the action as written above, before it crosses over to the first-order transition. To study this
pre-asymptotic behavior it is convenient to integrate out the fermion fields $q$, which yields the action derived
by \textcite{Hertz_1976}. In particular, the Gaussian order-parameter or paramagnon propagator reads
\bea
\left\langle m_n^i({\bm k})\,m^j_m({\bm p})\right\rangle &=& \delta_{{\bm k},-{\bm p}}\,\delta_{n,-m}\,\delta_{ij}\,\frac{1}{2}
\hskip 30pt
\nonumber\\
&&\hskip -50pt \times\frac{1}{t + a{\bm k}^2 + b(\Omega_n)^2 + Gc\vert\Omega_n\vert/\vert{\bm k}\vert}\ .
\label{eq:3.45}
\eea
Here we have replaced the vertex $1/\varphi_{12}({\bm k})$ in Eq.\ (\ref{eq:3.39}) with a schematic one
that is linear in $\Omega$ and $k$ for simplicity. We see that the coupling to the electronic Goldstone modes
generates the characteristic Landau-damping term proportional to $\vert\Omega_n\vert/\vert{\bm k}\vert$ in
the paramagnon propagator. The term quadratic in the frequency in Eq.\ (\ref{eq:3.44}) is therefore not
the leading frequency dependence and can be dropped. The approximate effective action thus becomes
\bea
{\cal A}_{\text{Hertz}} &=& -\sum_{{\bm k},n} {\bm m}_n({\bm k})\,\left[t + a{\bm k}^2 + Gc\vert\Omega_n\vert/\vert{\bm k}\vert\right]\cdot {\bm m}_{-n}({\bm k})
\nonumber\\
&& + u \int d{\bm x}\,T \sum_{n_1,n_2,n_3} \left({\bm m}_{n_1}({\bm x})\cdot {\bm m}_{n_2}({\bm x})\right)
\nonumber\\
&& \hskip 30pt \times \left({\bm m}_{n_3}({\bm x})\cdot {\bm m}_{-n_1-n_2-n_3}({\bm x})\right)\ .
\label{eq:3.46}
\eea

This action was derived and studied by \textcite{Hertz_1976}, and its finite-temperature properties were 
analyzed by \textcite{Millis_1993}. Many of the explicit results had been derived earlier by means of a
theory of spin fluctuations that one would now classify as a self-consistent one-loop theory; see
\textcite{Moriya_1985} and references therein, \textcite{Lonzarich_1997}, and \textcite{Lonzarich_Taillefer_1985}. This development
was analogous to that in the area of classical critical dynamics, where mode-coupling theories
\cite{Fixman_1962, Kadanoff_Swift_1968, Kawasaki_1967, Kawasaki_1970, Kawasaki_1976} were
followed by renormalization-group treatments \cite{Hohenberg_Halperin_1977}. In what follows, we
will derive these results by means of scaling arguments, which is analogous to a third angle of attack
on the classical dynamical scaling problem \cite{Ferrell_et_al_1967, Ferrell_et_al_1968, Halperin_Hohenberg_1967}.

\paragraph{Scaling analysis of the pre-asymptotic regime}
\label{par:III.C.2.b}

From the action, Eq.\ (\ref{eq:3.46}), we see that the frequency scales as $\Omega \sim k^3$. 
That is, the dynamical exponent is
\be
z= 3\ ,
\label{3.47}
\ee
independent of the dimensionality. The theory thus is above its upper critical dimension for all $d > 4-z = 1$. Let $t$ be the
distance from criticality at $T=0$
\footnote{\label{t_footnote} $t$ in Eqs.\ (\ref{eq:3.44}) - (\ref{eq:3.46}) denotes the bare distance
from criticality; here and in what follows we use the same symbol for its renormalized or physical counterpart.}
and define static exponents by the dependence of the observables on $t$
in the usual way, see Appendix \ref{app:B}. The static
exponents $\nu$, $\beta$, $\eta$, $\gamma$, and $\delta$ then all have their usual mean-field values for all $d>1$:
\be
\nu = \beta = 1/2\quad,\quad \eta = 0\quad,\quad \gamma = 1\quad,\quad \delta = 3\quad.
\label{eq:3.48}
\ee
The quartic coefficient $u$, with scale dimension $[u] = -(d+z-4) = -(d-1)$ is a dangerous irrelevant variable (DIV) with respect to the 
order parameter, which is why  $\beta$ and $\delta$ deviate from the naive scaling results. (For a general
discussion of the DIV concept, see, \textcite{Ma_1976} and \textcite{Fisher_1983}.) 

$u$ also plays an important role for the temperature dependence of many observables, which is not simply
determined by $z$ due to the dangerous irrelevancy of $u$. We now show how the relevant results can be
obtained from scaling arguments. Scaling functions will be denoted by $F$ with a subscript indicating the
observable in question. 
%In many cases we give the results for $d=3$ only; the generalization to arbitrary $d$ is obvious.

\medskip\par
$(i)$ {\it Correlation length.} Let $\delta(t,T)$ be the T-dependent distance from the QCP, such that $\delta(t,0) = t$. 
Then the homogeneity law for $\delta$ is
\bea
\delta (t,T) &=& b^{-\gamma/\nu}F_{\delta} (tb^{1/\nu}, Tb^z, ub^{-(d-1)})
\nonumber\\
                  &=& b^{-2}F_{\delta} (t\,b^2, T\,b^3, u\,b^{-(d-1)})\ .
\label{eq:3.49}
\eea
The temperature dependence of $\delta$ results from a one-loop contribution that is proportional to $u$.
Based on this purely structural observation we conclude that
\bse
\label{eqs:3.50}
\be
F_{\delta}(0,1,y\to 0) \propto y
\label{eq:3.50a}
\ee
This is in agreement with the perturbative result obtained by \textcite{Moriya_Kawabata_1973a} who found,
for $d=3$,
\be
\delta = t + {\rm const.}\times u\,T^{4/3}\ .
\label{eq:3.50b}
\ee
\ese
$u$ is thus dangerously irrelevant with respect to the $T$-dependence of $\delta$. In contrast, 
$F_{\delta}(1,0, y\to 0) \propto {\rm const.}$, and also $\partial F_{\delta}(x,1,y\to 0)/\partial x\vert_{x=0} = {\rm const.}$ 
Choosing $b = T^{-1/3}$ we can therefore Taylor expand in the first argument of $F_{\delta}$. 
Specializing to $d=3$ we have, for $t/T^{2/3} \ll 1$,
\bea
\delta(t,T) &=& T^{2/3} F_{\delta}(t/T^{2/3}, 1, uT^{2/3}) 
\nonumber\\
                &\propto& T^{2/3} \left[uT^{2/3} + {\rm const.}\times t/T^{2/3} + \ldots\right]\qquad.
\label{eq:3.51}
\eea
For $T_{\text{c}}$, defined by $\delta(t,T_{\text{c}})=0$, this gives $t \propto T_{\text{c}}^{2/3} \times u\,T_{\text{c}}^{2/3} 
\propto T_{\text{c}}^{4/3}$, or \cite{Moriya_Kawabata_1973a}
\be
T_{\text{c}} \propto (-t)^{3/4}\ .
\label{eq:3.52}
\ee
This result is due to the dangerous irrelevancy of $u$; in its absence one would have $T_{\text{c}} \propto (-t)^{3/2}$.
This result was confirmed by \textcite{Millis_1993}, who showed how to obtain it from an RG analysis of Hertz's action. 
The importance of the DIV was stressed by him and also by \textcite{Sachdev_1997}.

The above results also determine the behavior of the correlation length $\xi \propto 1/\sqrt{\vert\delta\vert}$: At $T=0$ we
have $\xi(t\to 0,T=0) \propto \vert t\vert^{-1/2}$, in agreement with the value of $\nu$ in Eq.\ (\ref{eq:3.48}), whereas at
$t=0$ we have $\xi(t=0,T\to 0) \propto T^{-\nu_T}$, with \cite{Millis_1993}
\be
\nu_T = 2/3\ .
\label{eq:3.53}
\ee
For general $d$, $\nu_T = (d+1)/6$.

\medskip\par
$(ii)$ {\it Magnetic susceptibility.} Now consider the magnetization susceptibility $\chi$. The scaling law is
\bea
\chi(t,T ) &=& b^{\gamma/\nu}\,F_{\chi}(t\,b^{1/\nu}, T\,b^z, u\,b^{-(d-1)})
\nonumber\\
              &=& b^{2}\,F_{\chi}(t\,b^2, T\,b^3, u\,b^{-(d-1)})\ .
\label{eq:3.54}
\eea
Since we are dealing with a Gaussian theory, $\delta(t,T) \sim 1/\chi(t,T)$, and we know from the behavior of $\delta(t,T)$
that $u$ is dangerously irrelevant, viz.
\be
F_{\chi}(0,1,y\to 0) \propto 1/y\ .
\label{eq:3.55}
\ee
At $t=0$ in $d=3$ we thus have $\chi(t=0,T\to 0) \propto T^{-\gamma_T}$ with
\be
\gamma_T = 4/3\ ,
\label{eq:3.56}
\ee
and more generally $\gamma_T = (d+1)/3$. 
At $T=0$, on the other hand, we have $F_{\chi}(1,0,y\to 0) = {\rm const.}$, and hence 
$\chi(t\to 0,T=0) \propto \vert t\vert^{-1}$ in agreement with the value of $\gamma$ in Eq.\ (\ref{eq:3.48}).

\medskip\par
$(iii)$ {\it Magnetization.}  The magnetization $m$ obeys a scaling law
\bea
m(t,T) &=& b^{-(d+z-2+\eta)/2}\,F_m(t\,b^{1/\nu}, T\,b^z, u\,b^{-(d-1)})
\nonumber\\
          &=& b^{-(d+1)/2}\,F_m(t\,b^2, T\, b^3, u\,b^{-(d-1)})\ .
\label{eq:3.57}
\eea
In general, $m$ is affected by the DIV $u$, just as $\chi$ is. However, at $t=0$
this is not the case, which can be seen as follows: Since $m \propto \sqrt{-\delta/u}$, and $\delta(t=0) \propto u$,
see Eq.\ (\ref{eq:3.50b}), $u$ drops out and hyperscaling works. We thus have
\be
m(t=0,T) = b^{-(d+1)/2}\,F_m(0,T\,b^3,0)\ ,
\label{eq:3.58}
\ee
or $m(t=0,T) \propto T^{\beta_T}$ with $\beta_T = (d+1)/6$ in general, or, for $d=3$,
\be
\beta_T = 2/3\ .
\label{eq:3.59}
\ee
In interpreting this exponent one needs to keep in mind that the magnetization is nonzero only for
$-t > u\,T^{4/3}$ (putting a constant equal to unity), see Eq.\ (\ref{eq:3.51}). For $-t \gg u\,T^{4/3}$
one observes static scaling with small temperature corrections, and for $-t < u\,T^{4/3}$ the scaling
function vanishes identically.
%, so $\beta_T$ is observable only in a narrow region where $-t \agt u\,T^{4/3}$.
The exponent $\beta_T$ therefore cannot be observed via the $T$-dependence of $m$ at $r=0$. It
does, however, determine the more general scaling form of $m$ as a function of $t$ and $T$, see
\textcite{Kirkpatrick_Belitz_2015b}.

We also note that combining $m\propto\sqrt{-\delta}$ with $\delta \propto T^{4/3} - T_{\text c}^{4/3}$, which
follows from Eq.\ (\ref{eq:3.50b}), yields \cite{Moriya_1985}
\be
m^2 \propto T_{\text c}^{4/3} - T^{4/3}
\label{eq:3.59'}
\ee

\medskip\par
$(iv)$ {\it Specific heat.} The free energy density $f$ obeys a homogeneity law
\be
f(t,T) = b^{-(d+z)}\,F_f(t\,b^{1/\nu}, T\,b^z)\ .
\label{eq:3.60}
\ee
For the specific-heat coefficient $\gamma = -\partial^2 f/\partial T^2$ this implies
\be
\gamma(t,T) = b^{3-d}\,F_{\gamma}(t\,b^2, T\,b^3)\ .
\label{eqs:3.61}
\ee
For $d=3$ scaling thus yields $\gamma = {\text{const.}}$. An explicit calculation of the Gaussian fluctuation contribution
to the free energy \cite{Brinkman_Engelsberg_1968a, Makoshi_Moriya_1975, Millis_1993, Lonzarich_1997} yields
\be
\gamma(t,T=0) \propto \ln t \quad,\quad \gamma(t=0,T) \propto \ln T\ .
\label{eq:3.62}
\ee
For general $d>1$, the exponents $\bar\alpha$ and $\bar\alpha_T$ (for a definition, see Appendix \ref{app:B}) 
are $\bar\alpha = (3-d)/2$ and $\bar\alpha_T = (3-d)/3$.%
\footnote{\label{specific_heat_scaling_footnote} These exponents describe the leading fluctuation contribution to the
specific-heat coefficient. For $d>3$ the latter is subleading compared to a constant non-scaling contribution.}

\medskip\par
$(v)$ {\it Electrical resistivity.} In order to discuss relaxation rates, we start with the single-particle rate 
$1/\tau_{\text{sp}}$. This is
dimensionally an energy, and hence has a scale dimension $[1/\tau_{\text{sp}}] = z = 3$. The relevant homogeneity
law thus is
\be
1/\tau_{\text{sp}}(t,T) = b^{-z}\,F_{\tau}(t\,b^{1/\nu}, T\,b^z) = T\,F_{\tau}(1,T/t^{3/2})\ .
\label{eqs:3.63}
\ee
At $t=0$ we thus have $1/\tau_{\text{sp}}(t=0,T\to 0) \propto T$. For $t\neq 0$ we must recover the Fermi-liquid result 
$1/\tau_{\text{sp}} \propto T^2$, 
which implies $F_{\tau}(1,x\to 0) \propto x$, or $1/\tau_{\text{sp}} \propto T^2/t^{3/2}$.

The transport rate $1/\tau_{\text{tr}}$, which determines the electrical resistivity, is also dimensionally an energy, 
but its scale dimension is {\em not} equal to $z$. The reason is the backscattering factor in the Boltzmann equation,
which provides an extra factor of $k^2 \sim b^{-2}$, with $k$ the hydrodynamic wave number.%
\footnote{\label{backscattering_footnote} This is because we consider scattering by long-wavelength magnetic
               excitations. In a Fermi liquid, both $1/\tau_{\text{sp}}$ and $1/\tau_{\text{tr}}$ scale as $T^2$.}
We thus have $[1/\tau_{\text{tr}}] = z+2$, and the homogeneity law for the resistivity $\rho$ is
\be
\rho(t,T) = b^{-(z+2)} F_{\rho}(t\,b^{1/\nu}, T\,b^z) = T^{5/3}\,F_{\rho}(1,T/t^{3/2})\ .
\label{eqs:3.64}
\ee
At $t=0$ we recover the result of \textcite{Mathon_1968}: $\rho(t=0,T\to 0) \propto T^{5/3}$. For $t\neq 0$ we
can again invoke the Fermi-liquid $T^2$ behavior to conclude $\rho \propto T^2/t^{1/2}$.

Note that $\rho$ does not obey naive scaling. While this is true for many of the observables discussed above, in this case the reason is 
not a DIV. Rather, the underlying relaxation rate does not have its naive scale dimension. As in the case of a DIV, 
this must be established by explicit calculations; it cannot be deduced from general scaling arguments.
In the disordered case, this particular complication does not occur, see Sec.\ \ref{par:III.C.3.e} (iii).

\medskip\par
To summarize, the critical exponents at Hertz's fixed point, which determine the pre-asymptotic behavior in a 
clean system before the first-order nature of the transition becomes manifest, are given for all $d>1$%
\footnote{See footnote \ref{specific_heat_scaling_footnote} for the proper interpretation of $\bar\alpha$ and
$\bar\alpha_T$ for $d>3$.}
by
\bea
z &=& 3\ ,
\nonumber\\
\nu &=& 1/2 \quad,\qquad\quad\ \   \nu_T = (d+1)/6\ ,
\nonumber\\
\beta &=& 1/2 \quad,\qquad\quad\ \  \beta_T = (d+1)/6\ ,
\nonumber\\
\delta &=& 3\ ,
\nonumber\\
\gamma &=& 1 \quad,\qquad\qquad\ \ \gamma_T = (d+1)/3\ ,
\nonumber\\
\eta &=& 0\ ,
\nonumber\\
\bar\alpha &=& (3-d)/2\quad,\quad \bar\alpha_T = (3-d)/3\ .
\label{eqs:3.65}
\eea

\paragraph{First- and second-order transitions; tricritical behavior; quantum critical points}
\label{par:III.C.2.c}

Order-parameter fluctuations do affect the second-order transition above
the tricritical temperature (the line of second-order transitions about TCP in Fig.~\ref{figure:wings_schematic}), 
where the critical behavior is in the appropriate classical universality class; Heisenberg, XY, 
or Ising, depending on the nature of the magnet. Along the wing-critical lines (between TCP and QCP
in Fig.~\ref{figure:wings_schematic}) the critical behavior is always in the classical Ising universality class,
since the presence of a magnetic field effective makes the order parameter one-dimensional.
The tricritical behavior at the tricritical point is described by the mean-field theory with
logarithmic corrections, as the upper critical dimension for a classical tricritical point is
$d_{\text{c}}^{+} = 3$ \cite{Wegner_Riedel_1973}.

\paragraph{Quantum critical points at the wing tips}
\label{par:III.C.2.d}

A magnetic field restores the quantum critical point that is suppressed in zero field: The tricritical
wings end in a pair of QCPs (QCP in Fig.~\ref{figure:wings_schematic}) in the $T=0$ plane at a point $(t_{\text{c}},h_{\text{c}})$, with
$t_{\text{c}}$ and $h_{\text{c}}$ the critical values of the control parameter and the magnetic field,
respectively. The magnetization is nonzero at this point and has a value $m_{\text{c}}$. This QCP
has the remarkable property that the quantum critical behavior can be determined exactly. The
reason is that the nonzero field and magnetization give the fermionic Goldstone modes a mass, and the
field conjugate to the order parameter therefore does not change the soft-mode structure of the
system. Under these conditions, Hertz theory is expected to be valid \cite{Belitz_Kirkpatrick_Vojta_2002}. In the 
present case, an expansion in powers of $\delta m = m - m_{\text{c}}$ about the QCP shows that the
quantity $\mathfrak{h} = 2m_{\text{c}}\delta t - \delta h$, with $\delta t = t - t_{\text{c}}$ and
$\delta h = h - h_{\text{c}}$, plays the role of the conjugate field. Switching on
an external magnetic field from $h=0$ gives certain soft modes a mass, but changing $h$ from
$h_{\text{c}} \neq 0$ does not lead to further changes in the soft-mode spectrum, and neither does
changing the value of $t$. Hertz theory thus gives the exact static quantum critical behavior, i.e.,
\be
\beta = \nu = 1/2\quad,\quad \gamma = 1\quad,\quad \eta = 0\quad,\quad \delta = 3\ .
\label{eq:3.66}
\ee
The dynamical behavior can be determined as follows. The magnetization at criticality as a function of the 
conjugate field obeys the homogeneity law (which has the effects of the DIV $u$ built in)
\be
\delta m(\mathfrak{h}) = b^{-\beta/\nu} F_{\delta m}(\mathfrak{h}\,b^{\beta\delta/\nu})\ .
\label{eq:3.67}
\ee
With mean-field values for the exponents this yields $\delta m \propto \mathfrak{h}^{1/3}$. But $\delta t$, and
therefore $\mathfrak{h}$, within Hertz theory scales as $T^{(d+1)/3}$, see Eq.\ (\ref{eq:3.50b}) and its
generalization to a general $d$. We thus find that, at the QCP as a function of $T$, the magnetization
decreases as
\be
\delta m(T) = m(T) - m(T=0) \propto - \mathfrak{h}^{1/3} \propto - T^{(d+1)/9}\ ,
\label{eq:3.68}
\ee
or $T^{4/9}$ in $d=3$. This is the result obtained by \textcite{Belitz_Kirkpatrick_Rollbuehler_2005}
using different arguments. The reasoning above has the advantage that it also immediately yields the
behavior of the magnetic susceptibility $\chi$, which is easier to measure. It obeys
\be
\chi (\mathfrak{h}) = b^{\gamma/\nu}\,F_{\chi}(\mathfrak{h}\,b^{\delta\beta/\nu})\ ,
\label{eq:3.69} 
\ee
which yields for the $T$-dependence of $\chi$ at the QCP
\be
\chi(T) \propto T^{-2(d+1)/9}\ ,
\label{eq:3.70}
\ee
or $T^{-8/9}$ in $d=3$. We stress again that this is a rare example of a QCP where the quantum
critical behavior can be determined exactly.

We also mention that in the presence of weak quenched disorder, weak enough for the tricritical wings to still be present 
(see Fig.~\ref{figure:wings_schematic} and the related discussion), the asymptotic critical behavior is unknown. In a
transient pre-asymptotic region the behavior is governed by Hertz's 
fixed point for disordered systems; for a discussion of pre-asymptotic behavior, see \textcite{Kirkpatrick_Belitz_2015b}.
However, this fixed point is ultimately unstable since it violates the Harris criterion and the true critical fixed point may be
of a strong-disorder type.

\paragraph{Comparison with experiment}
\label{par:III.C.2.e}

In order to compare these theoretical predictions with experiments, we recall that the theory
states that {\em if} there is a quantum phase transition to a homogeneous ferromagnetic state
in a clean bulk system, {\it then} the transition is first order. This qualification is
important, for reasons that we now recall: (1) The transition at low temperatures may be to
a different state, see Secs.\ \ref{subsec:II.D}, \ref{subsec:II.E}, and \ref{subsec:III.E}. (2) The presence
of quenched disorder has a qualitative effect on the transition, and sufficiently strong disorder
will always render the transition second order, see Secs.\ \ref{subsubsec:III.B.3} and \ref{subsubsec:III.C.2}.
(3) The fermionic soft modes that drive the transition first order exist only in two- and three-dimensional
systems; the theory therefore does not apply to quasi-one-dimensional materials.

With these caveats taken into account, we first consider the systems listed in Tables~\ref{table:1a},~\ref{table:1b}. 
With one exception, these are all rather clean systems that show a first-order transition, as
theoretically expected. The only questionable case is YbCu$_2$Si$_2$, which is strongly disordered; 
however, the nature of the magnetic order is not clear. In the rather clean system URhAl a tricritical point 
is suspected but so far has not been conclusively observed \cite{Combier_2013}.

The materials in Table~\ref{table:2_pt_1} comprises systems that are rather clean, with 
residual resistivities comparable to those in Tables~\ref{table:1a},~\ref{table:1b}, yet show a second-order transition.
The behavior observed in these systems is consistent with the pre-asymptotic critical behavior
governed by Hertz's fixed point that was discussed in Sec.\ \ref{par:III.C.2.b}. In particular, the
characteristic $(-t)^{3/4}$ behavior of the Curie temperature, Eq.\ (\ref{eq:3.52}), was observed 
as early as 1975 by \textcite{Sato_1975} in (Ni$_{1-x}$Pd$_x$)$_3$Al, and the behavior of the specific-heat
coefficient is consistent with (\ref{eq:3.62}). For a more recent observation of the scaling of $T_{\text{C}}$,
see Fig.~\ref{fig:NiPd_phase_diagram}. The most obvious interpretation of these observations
is that these experiments indeed probe the pre-asymptotic region, and following the critical
temperature to lower values would reveal a tricritical point and a first-order transition at the
lowest temperatures. This expectation is supported by the fact that the lowest $T_c$ observed
so far in these systems is relatively high, and by the observation that $T_c$ at intermediate
temperatures also follows the $(-t)^{3/4}$ law in systems where the transition at asymptotically
low temperatures is known to be first order, for instance, in MnSi \cite{Pfleiderer_et_al_1997}.
An experimental confirmation or otherwise of this expectation would be very important. Another
important experimental check of the theory would be the critical behavior at the wing tips,
Eqs.\ (\ref{eq:3.66}), (\ref{eq:3.68}), (\ref{eq:3.70}), which has not been studied so far.

\subsubsection{Disordered systems}
\label{subsubsec:III.C.3}

For disordered systems, the situation is qualitatively different since the transition is
continuous at the mean-field level. While the development of the effective action proceeds in exact
analogy to Sec.\ \ref{subsubsec:III.C.2}, the final result is a stable critical fixed point where the
asymptotic critical behavior is not given by power laws due to the existence of marginal operators
\cite{Belitz_et_al_2001a, Belitz_et_al_2001b}.

\paragraph{Effective soft-mode action}
\label{par:III.C.3.a}

In a disordered
system, the relevant fermionic soft modes are the diffusons discussed in Sec.\ \ref{subsubsec:III.A.2}. 
Their effective action can be written in a closed form, namely, the
matrix nonlinear sigma model developed by \textcite{Finkelstein_1983}
for studying the Anderson-Mott metal-insulator transition problem (for reviews, see
\textcite{Belitz_Kirkpatrick_1994} and \textcite{Finkelstein_2010}). The quenched disorder is handled technically by means
of the replica trick \cite{Edwards_Anderson_1975, Grinstein_1985} If one denotes the soft 
modes by ${^i_r}q_{nm}^{\alpha\beta}({\bm q}) = \Theta(-nm)
\sum_{\bm k} {^i_r}Q_{nm}^{\alpha\beta}({\bm k};{\bm q})$, with $Q_{nm}$ from Eqs.\ (\ref{eq:3.16})
and $\alpha$, $\beta$ replica indices, it can be written
\bse
\label{eqs:3.71}
\bea
{\cal A}_{\text{F}}[q] &=& {\cal A}_{{\rm NL}\sigma{\rm M}}[q]
\nonumber\\
&=&\frac{-1}{2G}\int d{\bf x}\,
     \tr\!\!\left(\nabla\hat Q ({\bf x})\right)^2 + 2H\!\! \int d{\bf x}\,\tr\!\!\left(\Omega\,{\hat Q}({\bf x})\right)
     \nonumber\\
&&\hskip 0pt  + {\cal A}_{\rm int}^{(s)}[\pi\,N_{\rm F}{\hat Q}/2] + {\cal A}_{\rm int}^{(t)}[\pi\,N_{\rm F}{\hat Q}/2]\ ,
\label{eq:3.71a}
\eea
where
\be
{\hat Q} = \left( \begin{array}{cc}
                 \sqrt{1-qq^{\dagger}} & q   \\
                    q^{\dagger}        & -\sqrt{1-q^{\dagger} q} \\
           \end{array} \right)\quad,
\label{eq:3.71b}
\ee
is a nonlinear function of $q$, and $\Omega$ is a frequency matrix with elements
\be
\Omega_{12} = \left(\tau_0\otimes s_0\right)\,\delta_{12}\,\omega_{n_1}\quad.
\label{eq:3.71c}
\ee
\ese
Here $1 \equiv (n_1,\alpha_1)$, etc., and $\tr$ denotes a trace over all discrete degrees of freedom (frequency,
spin, particle-hole, and replica).
The coupling constant $G$ is proportional to the bare (i.e., Boltzmann) resistivity, and thus is a measure
of the disorder strength. $H$ is proportional to the specific heat coefficient. The first two terms in Eq.\ (\ref{eq:3.71a}) 
describe noninteracting electrons. They are the fermionic version \cite{Efetov_Larkin_Khmelnitskii_1980} of Wegner's
nonlinear sigma model for the Anderson localization problem \cite{Wegner_1979}. Note the diffusive structure of
these two terms once they are expanded to $O(q^2)$, with the gradient squared scaling as a frequency. The last
two terms reflect the electron-electron interactions in the spin-singlet and spin-triplet channels, respectively. They
are quadratic in ${\hat Q}$ with coupling constants $K_s$ and $K_t$, respectively, and are effectively linear in the frequency \cite{Finkelstein_1983}. 
They therefore do
not spoil the soft-mode structure of the nonlinear sigma model but just renormalize the prefactor of the frequency
in the diffusion pole.

The magnetization again couples linearly to the electron spin density, the soft part of which is linear in ${\hat Q}$.
The coupling term reads
\bea
{\cal A}_{\text{c}} &=& \sqrt{\pi T c}\int d{\bf x}
                              \sum_{\alpha}\sum_n\sum_{i=1}^{3}
          m_n^{i,\alpha}({\bf x})\sum_{r=0,3} (\sqrt{-1})^r \sum_m
\nonumber\\
&&\times\tr\left[\left(\tau_r\otimes s_i\right)\,
      {\hat Q}^{\alpha\alpha}_{m,m+n}({\bf x}) \right]\quad,
\label{eq:3.72}
\eea
with $c$ a coupling constant. $m^i$ ($i=1,2,3$) denotes again the three components of the order-parameter field
${\bm m}$, which now also carries a replica index $\alpha$. It determines the physical magnetization $m$ via the relation
\be
m = \mu_{\text{B}}\sqrt{\pi N_{\text{F}}^2 T/K_t} \langle m^{i,\alpha}_{n=0}({\bm x})\rangle\ .
\label{eq:3.73}
\ee

The order-parameter action is very similar to the one in the clean case, Eq.\ (\ref{eq:3.44}),
except that there is an additional quartic term that arises from the quenched disorder,
\bea
{\cal A}_{\text{OP}} &=&-\sum_{{\bf k},n}\sum_{\alpha}
   {\bm m}^{\alpha}_n({\bf k}) \left[t + a\,{\bf k}^2 + b\,(\Omega_n)^2\right] \cdot {\bm m}^{\alpha}_{-n}(-{\bf k}) 
   \nonumber\\
   && + u\int\hskip -2pt d{\bf x}\ T\hskip -5pt
   \sum_{n_1,n_2,n_3}\hskip -2pt \sum_{\alpha}
   \left({\bf m}_{n_1}^{\alpha}({\bf x})
        \cdot {\bf m}_{n_2}^{\alpha}({\bf x})\right)
\nonumber\\
&&\hskip 30pt\times   \left({\bf m}_{n_3}^{\alpha}({\bf x})\cdot
     {\bf m}_{-n_1-n_2-n_3}^{\alpha}({\bf x})\right)
      \nonumber\\
      && + v\int d{\bf x}\sum_{n_1,n_2}\sum_{\alpha,\beta}
   \left\vert{\bf m}_{n_1}^{\alpha}({\bf x})\right\vert^2\,
                \left\vert{\bf m}_{n_2}^{\beta}({\bf x})\right
    \vert^2\ .
\label{eq:3.74}
\eea   
The last term, with coupling constant $v$, is a random-mass or random-temperature term that arises 
from the disorder dependence of the bare distance from criticality whose disorder average is
given by $t$. There also is a term cubic in $m$, which carries at least one gradient or frequency
and is less relevant for the critical behavior than the terms shown.

The soft-mode action given by Eqs.\ (\ref{eqs:3.71}) - (\ref{eq:3.74}) was motivated and derived
by \textcite{Belitz_et_al_2001a} from an underlying microscopic fermionic action.%
\footnote{\label{Omega_squared_footnote} The term $b(\Omega_n)^2$ in Eq.\ (\ref{eq:3.74})
was erroneously written as $b\vert\Omega_n\vert$ in Ref.\ \textcite{Belitz_et_al_2001a}. ${\cal A}_{\text{OP}}$
is constructed such that it contains no effects of the fermionic soft modes, and therefore does
not contain any Landau damping, which results from the soft fermionic single-particle excitations,
see Sec.\ \ref{subsubsec:III.A.1}. This difference is of no consequence, as the term in question is
RG irrelevant in either case.}
However, such
a derivation is not necessary. All parts of the effective action written above can be obtained from
more general considerations namely, (1) the existence of an effective soft-mode theory for
disordered interacting electrons, (2) symmetry considerations for a quantum $\phi^4$-theory with
a vector order parameter, and (3) the Zeeman coupling between the order parameter and the
electron spin density. In particular, the order-parameter part of the action can either be written down
based on symmetry considerations, or derived by means of a Hubbard-Stratonovich decoupling of
the particle-hole spin-triplet interaction term in the underlying fermionic action. In the latter case,
a spin-triplet interaction will be generated again by renormalization in the fermionic sector as long 
as a spin-singlet interaction is present. The presence of the last term in Eq.\ (\ref{eq:3.71a}) therefore
does not constitute any double counting. 

\paragraph{Hertz's action}
\label{par:III.C.3.b}      

As in the clean case, if we keep only the term of $O(Mq)$ in Eq.\ (\ref{eq:3.72}) and integrate
out the fermions, we recover
Hertz's action \cite{Hertz_1976} (plus the random-mass term, which was not considered by Hertz). 
%In particular, the Gaussian order-parameter or paramagnon propagator reads
%\bea
%\langle{^i}M_{n}^{\alpha}({\bm k})\,{^j}M_{m}^{\beta}({\bm p})\rangle &=& \delta_{{\bm k},-{\bm p}}\,\delta_{n,-m}\,
%   \delta_{ij}\,\delta_{\alpha\beta}\,\frac{1}{2}
%   \nonumber\\
%   &&\hskip -40pt\times\frac{1}{t + a{\bm k}^2 + b\,(\Omega_n)^2 + \frac{Gc\vert\Omega_n\vert}
%        {{\bm k^2}+GH\vert\Omega_n\vert}}\ .\qquad
%\label{eq:3.75}
%\eea   
%We see that the coupling to the diffusive electrons generates the characteristic Landau-damping term
%proportional to $\vert\Omega_n\vert/{\bm q}^2$ in the magnetization propagator. The term quadratic in
%the frequency in Eq.\ (\ref{eq:3.74}) is therefore not the leading frequency dependence, and can be
%dropped. Neglecting all terms of higher then linear order in $q$ in Eq.\ (\ref{eq:3.72}), as well as the
%random-mass term, the action becomes
The only difference is that the Landau-damping term now has the form $\vert\Omega\vert/{\bm k^2}$
due to the diffusive nature of the fermionic soft modes. The paramagnon propagator thus reads
\bea
\langle{^i}M_{n}^{\alpha}({\bm k})\,{^j}M_{m}^{\beta}({\bm p})\rangle &=& \delta_{{\bm k},-{\bm p}}\,\delta_{n,-m}\,
   \delta_{ij}\,\delta_{\alpha\beta}\,\frac{1}{2}
   \nonumber\\
   &&\hskip -40pt\times\frac{1}{t + a{\bm k}^2 + b\,(\Omega_n)^2 + \frac{Gc\vert\Omega_n\vert}
        {{\bm k^2}+GH\vert\Omega_n\vert}}\ .\qquad
\label{eq:3.75}
\eea   
Dropping the random-mass terms, the action
becomes
\bea
{\cal A}_{\text{Hertz}} &=&-\sum_{{\bf k},n}\sum_{\alpha}
   {\bm m}^{\alpha}_n({\bf k}) \left[t + a\,{\bf k}^2 + Gc\vert\Omega_n\vert/{\bm k}^2\right] 
   \nonumber\\
   &&\hskip 100pt \times {\bm m}^{\alpha}_{-n}(-{\bf k}) 
   \nonumber\\
   && \hskip -30pt + u_4\int\hskip -2pt d{\bf x}\ T\hskip -5pt
   \sum_{n_1,n_2,n_3}\hskip -2pt \sum_{\alpha}
   \left({\bf m}_{n_1}^{\alpha}({\bf x})
        \cdot {\bf m}_{n_2}^{\alpha}({\bf x})\right)
        \nonumber\\
        &&\hskip 0pt \times \left({\bf m}_{n_3}^{\alpha}({\bm x})\cdot{\bm m}_{-n_1-n_2-n_3}^{\alpha}({\bm x})\right)\ .
\label{eq:3.76}
\eea
Power counting again suggests a continuous phase transition with mean-field static
critical exponents, only now the upper critical dimension is $d_c^+=0$, and the dynamical 
critical exponent is $z=4$. 
%The problem with this reasoning is again that the same physics that leads to the Landau damping term also leads to the terms of higher order in $q$ in Eq.\ (\ref{eq:3.72}), which make the fixed point unstable. 
This fixed point is ultimately unstable, since the same physics that leads to the Landau damping term also leads to the terms of higher order in $q$ in 
Eq.~(\ref{eq:3.72}). Nevertheless, as in the clean case (see Sec.~\ref{subsubsec:III.C.2}) it is important to study his fixed point since it is experimentally
relevant in a pre-asymptotic crossover region \cite{Kirkpatrick_Belitz_2014}. In the disordered case this is true {\em a fortiori} since the effects that 
destabilize Hertz's fixed point still result in a continuous transition, albeit with different exponents.

%Nevertheless, the critical behavior at this fixed point is interesting, since it illustrates some fundamental scaling concepts, and also since it may be relevant experimentally in some pre-asymptotic crossover region. 
The homogeneity relations and exponents for Hertz's action are obtained by a straightforward
modification of the development in Sec.\ \ref{par:III.C.2.b}. The dynamical critical exponent is
now
\bse
\label{eqs:3.77}
\be
z = 4\ ,
\label{eq:3.77a}
\ee
which makes the upper critical dimensionality $d_{\text{c}}^+ = 0$, and the DIV $u$ has a
scale dimension $[u] = -d$. For all $d>0$ the exponents are given by (cf. (\ref{eqs:3.65}) for the clean case)
\bea
\nu &=& 1/2 \quad,\qquad\quad\ \   \nu_T = (d+2)/8\ ,
\nonumber\\
\beta &=& 1/2 \quad,\qquad\quad\ \  \beta_T = (d+2)/8\ ,
\nonumber\\
\delta &=& 3\ ,
\nonumber\\
\gamma &=& 1 \quad,\qquad\qquad\ \ \gamma_T = (d+2)/4\ ,
\nonumber\\
\eta &=& 0\ ,
\nonumber\\
\bar\alpha &=& (4-d)/2\quad,\quad \bar\alpha_T = (4-d)/4\ .
\label{eqs:3.77b}
\eea
\ese
For $d>4$, $\bar\alpha$ and $\bar\alpha_T$ describe the leading fluctuation contribution to the
specific-heat coefficient, see footnote \ref{specific_heat_scaling_footnote} for the analogous
statement in the clean case. 

%We stress again that these exponents do not describe any physical asymptotic critical behavior since the same class of terms are missing from the fixed-point action that make Hertz's fixed point unstable in the clean case. Here, they lead to a continuous transition with non-mean-field exponents, as opposed to the first-order transition in the clean case. Another problem with Eq.\ (\ref{eq:3.76}) is that the value of the correlation-length exponent, $\nu = 1/2$, violates the Harris criterion (in the formulation by \textcite{Chayes_et_al_1986}) for all $d<4$, see Eq.\ (\ref{eq:3.36}). This also indicates that the action as written cannot be complete.

As mentioned above, these exponents do not describe the physical asymptotic critical behavior. Another indication of this is the value of the
correlation-length exponent, $\nu = 1/2$, which violates the requirement $\nu \geq 2/d$, Eq.~(\ref{eq:3.36}), for all $d<4$. For the purpose of
finding the true asymptotic critical behavior it is preferable  to not integrate out the fermions, but rather deal with the
coupled soft-mode field theory for analyzing the fixed-point structure.

\paragraph{Fixed-point action}
\label{par:III.C.3.c}      

The lowest-order term that was neglected in Eq.\ (\ref{eq:3.76}) is the 
term of $O(mq^2)$ in Eq.\ (\ref{eq:3.72}). It is easy to see that this generates a
renormalization of the 2-point $m$-vertex that is proportional to $\vert{\bm k}\vert^{d-2}$. For dimensions
$d<4$, the gradient-squared term in Eq.\ (\ref{eq:3.74}) is therefore not the leading wave-number
dependence, and it is convenient to add the generated term to the bare action. In a schematic form
that suppresses everything not necessary for power counting, the effective action then 
reads \cite{Belitz_et_al_2001a}
\bea
{\cal A}_{\rm eff}[m,q]&=&-\int d{\bm x}\ m\left[t + a_{d-2}\,
   \partial_{\bm x}^{d-2} + a_2\,\partial_{\bm x}^2\right] m
\nonumber\\
  && \hskip 60pt + O(\partial_{\bm x}^4\, m^2, m^4)
\nonumber\\
&&\hskip -60pt -\frac{1}{G}\int d{\bm x}\,(\partial_{\bm x} q)^2 
     + H\int d{\bm x}\,\Omega\,q^2 + (K_s+K_t)\,T\int d{\bm x}\, q^2
\nonumber\\
&& \hskip -60pt - \frac{1}{G_4}\int d{\bm x}\,\partial_{\bm x}^2\, q^4
   + H_4\int d{\bm x}\,\Omega\,q^4 + O(T q^3, \partial_{\bm x}^2\, q^6,
      \Omega\,q^6)
\nonumber\\
&& \hskip -60pt + \sqrt{T}\,c_1 \int d{\bm x}\,m\,q + \sqrt{T}\,c_2 \int d{\bm x}\,m\,q^2 + O(\sqrt{T}mq^4)\ .
\nonumber\\
\label{eq:3.78}
\eea
Here the fields are understood to be functions of position and frequency,
and only quantities that carry a scale dimension are shown. The bare values
of $G_4$ and $H_4$ are proportional to those of $G$ and 
$H$. $K_s$ and $K_t$ are the coupling constants of the terms of $O(q^2)$ in the interacting parts
of ${\cal A}_{{\rm NL}\sigma{\rm M}}$. A term of order $Tq^3$ that arises
from the same part of the action is not important for the problem of magnetic criticality. It therefore
is not shown although its coupling constant squared has 
the same scale dimension as $1/G_4$ and $H_4$. $c_1$ and
$c_2$ are the coupling constants of the terms that result from expanding 
${\cal A}_{\rm c}$ in powers of $q$. Their bare values are proportional to $c$.

\paragraph{Fixed points, and their stability}
\label{par:III.C.3.d}

The action shown schematically in Eq.\ (\ref{eq:3.78}) can be analyzed for critical fixed points
by means of standard renormalization-group (RG) techniques \cite{Ma_1976}. We assign a scale
dimension $[L] = -1$ a length $L$, and $[\tau] = -z$ to the imaginary time $\tau$ (with $z$ to be
determined). Under a RG transformation that involves the rescaling of lengths by a factor $b$, all
scaling quantities $A$ will transform according to $A \to A]\,b^{[A]}$. In particular, temperature $T$
and frequency $\Omega$ have scale dimensions $[T] = [\Omega] = z$. 

It is illustrative to first again look for a fixed point that describes the mean-field critical behavior of Hertz
theory. To this end, let us look for a fixed point where the coupling constants $a_2$ and $c_1$ 
are marginal. This results in standard mean-field static critical behavior, and a dynamical exponent 
$z=4$, all of which is consistent with the action given in Eq.\ (\ref{eq:3.76}) and with the 
paramagnon propagator, Eq.\ (\ref{eq:3.75}). The requirement that the
action be dimensionless leads to $[m] = (d-2)/2$, which makes $t$ relevant with $[t]=2$. The critical
exponents $\eta$ and $\nu$ are thus $\eta = 0$, and $\nu = 1/2$. This fixed point is unstable for $d<4$, since
$[a_{d-2}] = 4-d$, and $a_{d-2}$ is thus relevant for all $d<4$. This is obvious if one adds the
term with coupling constant $a_{d-2}$ to the bare action, as we have done above, but less so if
one chooses the bare value of $a_{d-2}$ to be zero and have the physics related to $a_{d-2}$
be generated by the term with coupling constant $c_2$. In that case, a careful analysis of the
time scales involved leads to the same conclusion \cite{Belitz_et_al_2001a}. All of this in consistent
with the fact that the mean-field value $\nu=1/2$ violates the Harris criterion discussed in 
Sec.\ \ref{subsubsec:III.B.3}, see Eq.\ (\ref{eq:3.27}),
and therefore cannot represent the correct critical behavior in a disordered system.

The above discussion suggests that one should look for a fixed point where only $c_1$ is
required to be marginal, which implies $[m] = 1 + (d-z)/2$. The diffusons will be unaffected by the
magnetic transition, and hence the scale dimension of the soft fermion field $q$ is $[q] = (d-2)/2$.
This also implies that there is a diffusive time scale characterized by a dynamical exponent
\be
z_{\text{diff}} = 2
\label{eq:3.79}
\ee
{\em in addition} to the critical dynamical time scale whose exponent we denote by $z_c$. This 
presence of more than one time scale complicates the power counting arguments, as the scale
dimensions of various coupling constants can depend on the context they appear in. That is, the
scale dimension $z$ of the various factors of temperature or frequency in the effective action can
be equal to $z_{\text{diff}}$ or $z_c$, depending on the context. In particular,
$a_{d-2}$ can be irrelevant if the paramagnon propagator appears as an internal propagator in the
loop expansion, while it will be marginal in the critical paramagnon propagator, which implies $[m]=1$.
This in turn leads to $z_c = d$ and $\eta = 4-d$. This makes $a_2$ irrelevant, while $t$ is relevant
with $[t] = d-2$. The three independent critical exponents thus are
\bse
\label{eqs:3.80}
\be
\nu = \frac{1}{d-2}\quad,\quad \eta = 4-d \quad,\quad z_c = d\ .
\label{eq:3.80}
\ee
Note that for this fixed point, $\nu$ satisfies the Harris criterion. The remaining static exponents are
given by $d$-dependent generalizations of Eq.\ (\ref{eq:3.25}) \cite{Belitz_et_al_2001b}:
\be
\beta = \frac{2}{d-2}\quad,\quad \gamma = 1\quad,\quad \delta = d/2\ .
\label{eq:3.80b}
\ee
Equation (\ref{eq:3.80b}) is valid for $2<d<6$. For $d \geq 6$, $\beta$ and $\delta$ lock into their mean-field values;
for $\nu$ and $\eta$ this happens for $d \geq 4$. The $T$-dependence of the observables at criticality, $t=0$, is
determined by \cite{Belitz_et_al_2001b}
\be
\beta_T = \beta/2\nu\quad,\quad\gamma_T = \gamma/2\nu .
\label{eq:3.80c}
\ee
\ese

To discuss the stability of this fixed point we now consider the remaining coupling constants in the
effective action, Eq.\ (\ref{eq:3.78}). $c_2$ has a scale dimension $[c_2] = 1 - z/2$, and thus is
irrelevant if $z = z_c$, but marginal if $z = z_{\text{diff}}$. Moreover, a careful analysis shows
that due to the existence of two different time scales even some operators that are irrelevant
by power counting may effectively act as marginal operators \cite{Belitz_et_al_2001b}. The reason is that naive power counting
is based on a length scale argument, which can be modified if the scale factor $b$ represents a frequency
rather than a length. Since the difference between the two dynamical exponents $z_{\text{diff}}$ and $z_c$
is equal to $d-2$, this implies that coupling constants with a naive scale dimension given by $-(d-2)$
can act as marginal operators under certain conditions. A detailed analysis \cite{Belitz_et_al_2001b} shows
that this is actually the case, and as a consequence all terms that are shown explicitly in Eq.\ (\ref{eq:3.78})
are important for determining the leading critical behavior and constitute the fixed-point action.

\paragraph{Asymptotic critical behavior}
\label{par:III.C.3.e}

The conclusion so far is that the fixed-point action represented by Eq.\ (\ref{eq:3.78}) contains
marginal operators that result, order by order in a loop expansion, in logarithmic corrections to the
fixed point with critical exponents given by Eqs.\ (\ref{eqs:3.80}). The remaining question is what the
result is if the loop expansion is summed to all orders. 

$(i)$ {\it Integral equations for the diffusion coefficients}.
Surprisingly, the above question can be answered
exactly without resorting to a small parameter (such as an expansion in $\epsilon = d-4$) 
\cite{Kirkpatrick_Belitz_1996, Belitz_et_al_2001a, Belitz_et_al_2001b}. 
This hinges on various properties of the loop expansion that are
revealed by a detailed analysis: First, at the fixed point of interest the fermionic dynamics remain
diffusive. The coupling constants $K_s$ and $K_t$ do not change this, and therefore can be
ignored. Second, $G$ and $c_2$ are not singularly renormalized. Third, the renormalized versions
of $G_4$ and $H_4$ are proportional to those of $G$ and $H$, as are their bare values. Finally,
$c_1$ is held fixed by definition of the fixed point. This leaves the renormalizations of $H$ and the
two-point order-parameter vertex $u_2 = t + a_{d-2}\vert{\bm k}\vert^{d-2} + a_2{\bm k}^2$ to be
determined. The resummation of the loop expansion to all orders can be expressed in terms of
two coupled integral equations for $H$ and $u_2$ or, equivalently, for the thermal diffusion 
coefficient $D(\Omega) = 1/GH(\Omega)$ and the spin diffusion coefficient 
$D_s({\bm k},\Omega) = 16\pi\,u_2({\bm k},\Omega)/Gc_1^2$, both of which acquire a frequency
dependence under renormalization. These integral equations take the form \cite{Belitz_et_al_2001b}%
\footnote{\label{integral_equations_footnote} Historically, these equations were first derived
and solved as a solution of Finkelstein's (1983) generalized nonlinear sigma model for
disordered interacting electrons \cite{Belitz_Kirkpatrick_1991, Kirkpatrick_Belitz_1992b}.
The nature of the transition they describe was unclear at the time. The
connection with ferromagnetism was made by \textcite{Kirkpatrick_Belitz_1996}, and
elaborated upon by \textcite{Belitz_et_al_2001a, Belitz_et_al_2001b}.}
\bse
\label{eqs:3.81}
\bea
D_{s}({\bf k},\Omega)&=&D_{s}^{0} + \frac{iG}{2V}\sum_{\bf p}
   \int_{0}^{\infty }d\omega \frac{1}{{\bf p}^{2} - i\omega/D(\omega)}\qquad\qquad
\nonumber\\
&&\hskip -30pt\times\frac{1}{({\bf p+k)}^{2}-i(\omega +\Omega)/D(\omega +\Omega )}\quad,
\label{eq:3.81a}
\eea
\be
\frac{1}{D(\Omega)} = \frac{1}{D^{0}} + \frac{3G}{8V}\sum_{{\bf p}}
   \frac{1}{\Omega} \int_{0}^{\Omega}d\omega \frac{1}{-i\omega 
   + {\bf p}^{2}D_{s}({\bf p},\omega )}\quad.
\label{eq:3.81b}
\ee
\ese
Here $D^0$ and $D_s^0$ are the bare diffusion coefficients at zero wave number. $D$ and $D_s$ simultaneously
go to zero at a critical value of $G$, and in the vicinity of that critical point the integral equations can be solved
analytically. It turns out that the logarithmic corrections obtained in perturbation theory do not change the power
laws given in Eqs.\ (\ref{eqs:3.80}), but rather result in log-normal corrections to power-law scaling. For instance,
the magnetization $m$ at $T=0$ in $d=3$ as a function of $t$ has an asymptotic behavior \cite{Belitz_et_al_2001b}
\bse
\label{eqs:3.82}
\be
m(t\to 0) \propto \vert t\,g(\ln(1/\vert t\vert))\vert^{\beta}\ ,
\label{eq:3.82a}
\ee
with $\beta = 2$ from Eqs.\ (\ref{eqs:3.80}) and
\be
g(x\to\infty) \propto e^{(\ln x)^2/2\ln(3/2)}\ .
\label{eq:3.82b}
\ee
Similarly, at criticality as a function of a magnetic field $h$,
\be
m(t=0,h\to 0) \propto \left[h\,g\left(\frac{1}{3}\ln(1/h)\right)\right]^{1/\delta}\ ,
\label{eq:3.82c}
\ee
with $\delta = 3/2$ from Eqs.\ (\ref{eqs:3.80}). The specific heat also has a log-normal critical behavior.
However, the critical exponent $\gamma$ comes without logarithmic corrections; the magnetic susceptibility
diverges as
\be
\chi(t\to 0) \propto 1/\vert t\vert\ .
\label{eq:3.82d}
\ee
\ese

$(ii)$ {\it Scaling considerations: Thermodynamic quantities.}
All of the above results are conveniently summarized in the following generalized homogeneity law for the
free energy density \cite{Belitz_et_al_2001b}:
\bea
f(t,T,h) &=& b^{-(d+z_c)}\,f_1(t\,b^{1/\nu}, T\,b^{z_c}, h\,b^{z_c})
\nonumber\\
&& + b^{-(d+z_g)}\,f_2(t\,b^{1/\nu}, T\,b^{z_g}, h\,b^{z_c})\ .\quad
\label{eq:3.83}
\eea
Here $z_c$ is the critical dynamical exponent, which determines the temperature dependence of the
specific heat, and $z_g$ is the dynamical exponent due to the generic soft modes, which determines
the temperature dependence of the order parameter and its susceptibility. If $z_c \geq z_g$ (this {\em has}
to be the case, see Sec.\ \ref{subsubsec:III.C.4} below), we obtain homogeneity laws for the 
order parameter $\partial f/\partial h$, the order-parameter susceptibility $\chi = \partial^2 f/\partial h^2$,
and the specific-heat coefficient $\gamma_C = -\partial^2 f/\partial T^2$,
\bse
\label{eqs:3.84}
\bea
m(t,T,h) &=& b^{-(d+z_g-z_c)}\,F_m(t\,b^{1/\nu}, T\,b^{z_g}, h\,b^{z_c})\ ,
\nonumber\\
\label{eq:3.84a}\\
\chi(t,T;k) &=& b^{-(d+z_g-2z_c)}\,F_{\chi}(t\,b^{1/\nu}, T\,b^{z_g};k\,b)\ ,\hskip 20pt
\label{eq:3.84b}\\
\gamma_C(t,T) &=& b^{-(d-z_c)}\,F_{\gamma}(t\,b^{1/\nu}, T\,b^{z_c})\ .
\label{eq:3.84c}
\eea
In Eq.\ (\ref{eq:3.84b}) we have added the wave-number dependence of $\chi$. 
Also of interest is the scaling of the critical temperature $\Tc$ with $t$. $\Tc$ is the
temperature where the order parameter vanishes, or the susceptibility diverges, and
from Eq.\ (\ref{eq:3.84a}) or (\ref{eq:3.84b}) we obtain
\be
\Tc \propto (-t)^{\nu z_g}\ .
\label{eq:3.84d}
\ee
\ese

All critical exponents can now be expressed in terms of $z_c$, $z_g$, and $\nu$.$^{\ref{exponents_footnote}}$
We have
\bse
\label{eqs:3.85}
\bea
\bar\alpha &=& \nu(z_c-d)\quad,\qquad \ \ \ \bar\alpha_T = (z_c-d)/z_c\ ,
\label{eq:3.85a}\\
\beta &=& \nu(d+z_g-z_c)\quad,\quad \beta_T = (d+z_g-z_c)/z_g\ ,
\nonumber\\
\label{eq:3.85b}\\
\gamma &=& \nu(2z_c-d-z_g)\quad,\  \, \gamma_T = (2z_c-d-z_g)/z_g\ ,
\nonumber\\
\label{eq:3.85c}\\
\delta &=& z_c/(d+z_g-z_c)\ .
\label{eq:3.85d}\\
\eta &=& d+2+z_g-2z_c\ .
\label{eq:3.85e}
\eea
Finally, $\nu_T$ follows from the requirement that $\chi(t=0,T\to 0;k)$ must be proportional to $T^{-\gamma_T}$
times a function of $k\xi$, which yields
\be
\nu_T = 1/z_g\ .
\label{eq:3.85f}
\ee
\ese
The log-normal terms multiplying the power laws can be expressed in terms of a scale dependence of the
independent exponents $z_c$, $z_g$, and $\nu$. It is convenient to write, for $2<d<4$,
\bse
\label{eqs:3.86}
\bea
z_c &=& d + \lambda\ ,
\label{eq:3.86a}\\
z_g &=& 2 + \lambda\ ,
\label{eq:3.86b}\\
1/\nu &=& d-2 + \lambda\ ,
\label{eq:3.86c}
\eea
\ese
where $\lambda$ is defined as
\be
\lambda = \ln g(\ln b)/\ln b\ ,
\label{eq:3.87}
\ee
with $g(\ln b)$ from Eq.\ (\ref{eq:3.82b}).

This critical behavior is expected to be exact provided a continuous transition into a homogeneous
ferromagnetic phase occurs. However, rare-region effects may mask this critical behavior. Theories
that deal with such effects are discussed in Secs.\ \ref{subsec:III.D} and \ref{subsec:III.E}. 

$(iii)$ {\it Scaling considerations: Electrical resistivity.} We finally mention the electrical resistivity $\rho$. The
transport relaxation rate is dominated by the disorder, which is unaffected by the magnetic ordering. 
The scale dimension of $\rho$ at a ferromagnetic QCP is therefore zero. However, $\rho$ does depend
on the critical dynamics, since the paramagnon propagator enters the calculation of $\rho$ in perturbation
theory. Going back to Eq.\ (\ref{eq:3.78}) we see that one-loop corrections to $\rho$ can be constructed, 
for instance, from one vertex with coupling constant $1/G_4$, or from two vertices with coupling constant
$c_2$. These terms belong to the class of least irrelevant variables with respect to the critical fixed point;
their scale dimension is $-(d-2)$. Denoting the least irrelevant variables collectively by $u$, we thus have
the following homogeneity law for the resistivity:
\bea
\rho(t,T) &=& F_{\rho}(t\,b^{1/\nu}, T\,b^{z_c}, u\,b^{-(d-2)})
\nonumber\\
             &=& \text{const.} + b^{-(d-2)} {\tilde F}_{\rho}(t\,b^{1/\nu}, T\,b^{z_c})\ ,
\label{eq:3.88}
\eea
where we have used that fact that the leading correction to $\rho$ is linear in $u$. At criticality, this
yields
\be
\rho(t=0,T) \propto T^{(d-2)/z_c}
\label{eq:3.89}
\ee
For the $t$-dependence at $T=0$ there are additional logarithmic complications due to a resonance
between the scale dimensions of $u$ and $t$, see \textcite{Belitz_et_al_2001b}.

Alternatively, one can argue that $\rho$ consists of a background contribution that does not scale,
and a singular one contribution $\delta\rho$ that does. Since $\rho$ is dimensionally a length to 
the power $d-2$, one expects
\be
\delta\rho(t,T) = b^{-(d-2)} F_{\delta\rho}(t\,b^{1/\nu}, T\,b^{z_c})\ ,
\label{eq:3.90}
\ee
which again yields (\ref{eq:3.89}). Note that this argument builds in the DIV $u$, so naive scaling
works.

\paragraph{Pre-asymptotic behavior}
\label{par:III.C.3.f}

The logarithmic nature of the asymptotic critical behavior described above suggests that it is valid only
in an exponentially small region around the critical point. Indeed, a numerical solution of the integral
equations, Eqs.\ (\ref{eqs:3.81}), shows that the behavior in a reasonably observable region around
criticality is given by effective power laws that correspond to the quantity $\lambda$ defined in
Sec.\ \ref{par:III.C.3.e} being approximately $\lambda \approx 2/3$ in a large range of scales \cite{Kirkpatrick_Belitz_2014}.
For instance, the specific-heat coefficient follows effective power laws with exponents$^{\ref{exponents_footnote}}$
\bse
\label{eqs:3.91}
\be
\bar\alpha^{\text{eff}} \approx 0.4\quad,\quad \bar\alpha_T^{\text{eff}} \approx 0.18
\label{eq:3.91a}
\ee
over almost three decades. Similarly, the critical temperature dependence of the spin susceptibility and the
magnetization is given by effective exponents
\be
\gamma_T^{\text{eff}} \approx 0.625\quad,\quad\beta_T^{\text{eff}}\approx 0.75\ ,
\label{eq:3.91b}
\ee
and the effective static exponents relevant for these two observables are
\be
\gamma = 1\quad,\quad \beta^{\text{eff}} \approx 1.2\quad,\quad \delta^{\text{eff}} \approx 1.83\ .
\label{eq:3.91c}
\ee
For the exponent that determines the shape of the phase diagram in the $T$-$t$ plane, Eq.\ (\ref{eq:3.84d}),
we have
\be
(\nu z_g)^{\text{eff}} \approx 1.6\ .
\label{eq:3.91d}
\ee
\ese
Only the value of $\gamma$ is the same in the pre-asymptotic and asymptotic regions, respectively. 
This is important for the interpretation of experiments.

\paragraph{Summary of critical exponents in the disordered case}
\label{par:III.C.3.e1}

In summary, the critical exponents for the disordered case in $2<d<4$ dimensions in both the asymptotic and the pre-asymptotic
regions are give by Eqs.\ (\ref{eqs:3.85}, \ref{eqs:3.86}). In the asymptotic regime they do not represent
pure power-law behavior since $\lambda$ is the scale-dependent object given in Eq.\ (\ref{eq:3.87}). In
the pre-asymptotic regime, $\lambda \approx 2/3$, and the exponents represent effective power laws.

\paragraph{Relation to experiment}
\label{par:III.C.3.g}

The interpretation of experiments on strongly disordered systems is difficult for various reasons. First,
the control parameter tends to be the chemical composition, which necessitates the preparation of a
separate sample for each data point. This makes the precise determination of the critical point very
difficult, and neither the precision nor the absolute values of the distance from criticality are anywhere
near the values that in classical systems are known to be necessary for a reliable determination of
critical exponents. Second, the Griffiths-region effects discussed in Sec.\ \ref{subsec:III.D} below
are expected to be pronounced in strongly disordered systems and coexist with critical phenomena.

A well-studied strongly disordered system is URu$_{2-x}$Re$_x$Si$_2$, see Sec.\ \ref{subsubsec:II.C.2}. 
\textcite{Bauer_et_al_2005} found a quantum critical point at $x\approx 0.3$; scaling plots yielded
exponent values $\delta = 1.56$ and $\beta_T = 0.9$. $\gamma_T$ was inferred from the Widom relation
(which does hold in this context, see Sec.\ \ref{par:III.C.2.b}(vi)), $\gamma_T = \beta_T (\delta-1) = 0.5$.
The specific-heat coefficient showed a $\ln T$ behavior over a wide range of $x$ values. 
A later analysis \cite{Butch_Maple_2009} put the critical concentration at $x \approx 0.15$ and found
continuously varying exponents in the range $0.6 \geq x \geq 0.2$, with $\delta \to 1$, $\gamma_T \to 0$,
and $\beta_T \approx 0.8$ roughly constant. These data were taken in what the experimentalists interpreted
as the ordered phase, so Griffiths-region effects, which might explain the continuously varying exponents,
should not be present. If the data represent critical phenomena, then continuously varying exponents are
hard to understand. Also, an exponent $\gamma_T = 0$, which must signal a divergence of the order-parameter
susceptibility that is only logarithmic, would be very unusual. 

Another relevant material is Ni$_{1-x}$V$_x$, see Sec.\ \ref{par:NiV}. \textcite{Ubaid-Kassis_Schroeder_2008}
found a critical point at $x_c \approx 0.11$ with $\gamma_T = 0.37\pm 0.07$, $\beta_T \approx 0.5$, and
$\delta = 1.8 \pm 0.2$. The value of $\delta$ agrees very well with Eqs.\ \ref{eqs:3.91}, the agreement
for $\gamma_T$ and $\beta_T$ is less satisfactory. These data were reinterpreted by 
\textcite{Ubaid-Kassis_et_al_2010} in terms of a Griffiths phase for $x<x_c$. 

We finally mention that the exponent that governs the scaling of the critical temperature $\Tc$ with the
control parameter is equal to $2$ asymptotically, and about $1.6$ in the pre-asymptotic region, see
Eqs.\ (\ref{eq:3.84d}) and (\ref{eq:3.91d}). This is in contrast to the result from Hertz theory in the
clean case, where the corresponding value is $3/4$, see Eq. (\ref{eq:3.52}). An exponent well greater than
$1$ is qualitatively consistent with the ``tail'' in the phase diagram observed in many disordered systems,
see Figs. \ref{pikul2012a2} and \ref{SrCaRuO3}, and also with the schematic phase diagram shown in
Fig.~\ref{figure:phase_diagram_disordered}. As discussed in Sec.\ \ref{subsec:II.E}, these tails are often
interpreted as signalizing quantum Griffiths effects. We note that these two interpretations are not
mutually contradictory; more detailed experimental investigations will be needed to distinguish between
them.

\subsubsection{Exponent relations}
\label{subsubsec:III.C.4}

At a classical critical point, only two static critical exponents are independent. This implies that there must exists
relations between various exponents.%
\footnote{\label{exponent_relations_footnote} These exponent relations are also often referred to as ``scaling relations'',
or ``scaling laws'', the latter not to be confused with the homogeneity laws that are often referred to by the same term.}
These relations have a complicated history, and some of them were initially found empirically \cite{Stanley_1971}. 
Several of them take the form of a rigorous inequality that turns into an equality if certain conditions are fulfilled.
Well-known examples are Widom's equality%
\footnote{\label{primed_exponents_footnote} The early literature distinguished between the exponents 
 ${\alpha}'$, $\gamma'$, $\nu'$ in the ordered phase, and ${\alpha}$, $\gamma$, $\nu$ in the disordered one. 
 The RG later showed that the primed and unprimed exponents have the same values in most cases, and 
 for our purposes we do not distinguish between them.}
\cite{Widom_1964}
\be
\gamma = \beta (\delta -1)\ ,
\label{eq:3.92}
\ee
Fisher's equality \cite{Fisher_1964}
\be
\gamma = (2-\eta)\nu\ ,
\label{eq:3.93}
\ee
and the Essam-Fisher relation \cite{Essam_Fisher_1963}%
\footnote{\label{alpha_footnote} In Eqs.\ (\ref{eqs:3.94}) and (\ref{eqs:3.95}) $\alpha$ denotes the
usual specific-heat exponent, which at a thermal phase transition is identical with the
exponent $\bar\alpha$ defined in Appendix \ref{app:B}.}
\bse
\label{eqs:3.94}
\be
\alpha + 2\beta + \gamma = 2\ .
\label{eq:3.94a}
\ee
The latter was deduced heuristically as an equality; \textcite{Rushbrooke_1963b} showed that the related
inequality
\be
\alpha + 2\beta + \gamma \geq 2
\label{eq:3.94b}
\ee
\ese
holds rigorously, and that it fails to hold as an equality if and only if the ratio $C_m/C_h$ approaches unity
as the critical point is approached, $C_m$ and $C_h$ being the specific heat at constant magnetization
and constant field, respectively \cite{Rushbrooke_1965}. 

Another class of exponent relations involves the spatial dimension $d$ explicitly. An example is
\bse
\label{eqs:3.95}
\be
d\nu = 2 - \alpha\ ,
\label{eq:3.95a}
\ee
which follows from a scaling assumption for the free energy, and which under weaker assumptions turns into the 
inequality \cite{Josephson_1967}
\be
d\nu \geq 2 - \alpha\ ,
\label{eq:3.95b}
\ee
\ese
These relations tend to hold only below an upper critical dimensionality, when no DIVs are
present, and the conditions that make them hold are often referred to as ``hyperscaling''. More generally, we will
refer to scaling that is unaffected (affected) by DIVs as ``strong'' (``weak'') scaling,
respectively. Examples of strong and weak scaling in the current context are the disordered fixed point in $2<d<4$ dimensions
and Hertz's fixed point in $d>1$ dimensions that were discussed in Sec.\ \ref{par:III.C.3.e} and \ref{par:III.C.2.b}, respectively. 

Systems at a QCP tend to be above their upper critical dimension, and hyperscaling is in general violated. Furthermore,
we need to distinguish between the exponents that describe the behavior of observables as a function
of the distance $t$ from criticality at $T=0$, and the exponents that describe the behavior at $t=0$ as a function of
$T$. We denote the former by $\bar\alpha$, $\beta$, $\gamma$, etc., and the latter by $\bar\alpha_T$, $\beta_T$,
$\gamma_T$, etc., as we have done throughout this review.$^{\ref{exponents_footnote}}$ An obvious question then is what, if any, exponent
relations hold for the $T$-exponents. This question has recently been explored by \textcite{Kirkpatrick_Belitz_2015b}; here we summarize
the main results.

%\paragraph{Exponent relations at a QCP}
%\label{par:III.C.4.a}

We first discuss what one would expect in the presence of strong scaling. 
Let $E$ be an observable with scale dimension $[E] = \epsilon/\nu$ with $\nu$ the correlation-length exponent 
(this defines the critical exponent $\epsilon$), and consider its dependence on $t$ and $T$. The homogeneity law 
that governs the universal critical behavior of $E$ is
\be
E(t,T) = b^{-\epsilon/\nu} E(t\,b^{1/\nu}, T\,b^z)\ ,
\label{eq:3.96}
\ee
Here $z$ is the dynamical exponent that governs the temperature scale dominant for $E$. From this relation
we obtain $E(t,T=0) \propto t^{\epsilon}$, and $E(t=0,T) \propto T^{\epsilon/\nu z} \equiv T^{\epsilon_T}$. We
see that, quite generally, if $\epsilon$ is an exponent that governs the $t$ dependence of an observable, then
the exponent $\epsilon_T$ that governs the $T$ dependence at criticality is
\be
\epsilon_T = \epsilon/\nu z\ .
\label{eq:3.97}
\ee
This argument assumes that $E$ is not affected by a DIV, i.e., it assumes strong scaling. For
instance, Eq.\ (\ref{eq:3.97}) does not hold for the exponents $\beta$ and $\beta_T$ in Sec.\ \ref{par:III.C.2.b}
because of the DIV $u$. To make a point that will be useful later, we now use (\ref{eq:3.97}) to rewrite (\ref{eq:3.96}) as
\be
E(t,T) = b^{-\epsilon/\nu} E(t\,b^{1/\nu}, T\,b^{\epsilon/\nu\epsilon_T})\ .
\label{eq:3.98}
\ee
If strong scaling holds, this contains the same information as (\ref{eq:3.96}). However, if one interprets $\epsilon_T$
as independent of $\epsilon$, unconstrained by (\ref{eq:3.97}), then (\ref{eq:3.98}) correctly describes the
temperature scaling of $E$ even in the absence of strong scaling. The value of $\epsilon_T$ in this case is 
determined by the DIV that breaks strong scaling and thus requires additional information that can not be
obtained by scaling arguments alone. Scaling thus becomes less general and less powerful if DIVs are present.

Also note that (\ref{eq:3.97}), if used in Eqs.\ (\ref{eq:3.92}, \ref{eq:3.93}), immediately yields the Widom and Fisher
equalities for the $T$-exponents,
\bea
\gamma_T &=& \beta_T (\delta - 1)\ ,
\label{eq:3.99}\\
\gamma_T &=& (2 - \eta)\nu_T\ .
\label{eq:3.100}
\eea
This derivation assumes strong scaling. However, as we will now show, these equalities can remain valid
even in the absence of strong scaling; strong scaling is sufficient, but not necessary. 

\medskip
\par
$(i)$ {\it Widom's equality:}
Consider the general homogeneity law for the magnetization, e.g., as expressed by the first line in Eq.\ (\ref{eq:3.57}). 
The effects of DIVs, if any, can be encoded in an effective homogeneity law
\be
m(t,h,T) = b^{-\beta/\nu}\,{\tilde F}_m(t\,b^{1/\nu}, h\,b^{\beta\delta/\nu}, T\,b^{\beta/\nu\beta_T})\ ,
\label{eq:3.101}
\ee
where $\beta$, $\delta$, and $\beta_T$ are the physical exponents. For Hertz's fixed point, they are
given by Eqs.\ (\ref{eq:3.48}, \ref{eq:3.59}), and they include the effects of the DIV $u$. For the marginally
stable fixed point in the disordered case, they are given by Eqs.\ (\ref{eqs:3.80}). 
Note that (\ref{eq:3.101}) does {\em not} imply the relation (\ref{eq:3.97}) between $\beta$
and $\beta_T$, which holds only in the presence of strong scaling; rather, it is an example of (\ref{eq:3.98})
with the general interpretation given after that equation. Differentiating with respect to $h$ we obtain a corresponding 
homogeneity law for the susceptibility,
\be
\chi(t,T) = b^{(\delta - 1)\beta/\nu} {\tilde F}_{\chi}(t\,b^{1/\nu}, T\,b^{\beta/\nu\beta_T})\ ,
\label{eq:3.102}
\ee
which likewise includes the effects of any DIVs. The latter can be quite complicated, as we have seen in Sec.\ \ref{par:III.C.2.b}:
At Hertz's fixed point, $m(t,T=0)$ depends on $u$ while $m(t=0,T)$ does not, whereas for $\chi$ the converse is true. Nevertheless, 
Eq.\ (\ref{eq:3.102}) implies both (\ref{eq:3.92}) and the analogous equality (\ref{eq:3.99}). 
An explicit check via Eqs.\ (\ref{eqs:3.65}, \ref{eqs:3.77}) shows that this does indeed hold for Hertz's fixed
point in both the clean and disordered cases. It also holds at the physical disordered QCP, as can be seen from 
Eqs.\ (\ref{eqs:3.80}) (with the exponents
properly interpreted to account for the log-log-normal terms), and even for the
pre-asymptotic behavior discussed in Sec.\ \ref{par:III.C.3.f}, see the exponent values
given there. This illustrates that Eq.\ (\ref{eq:3.102}) is general; it depends only on the homogeneity law for the magnetization.
The Widom equality, both for the exponents $\beta$, $\gamma$ and $\beta_T$, $\gamma_T$, is
thus an example of a scaling relation that remains valid even in the absence of strong scaling.%
\footnote{There is no guarantee that even ordinary scaling (as opposed to hyperscaling) relations
 remain valid in the absence of strong scaling, and no general criteria seem to exist. For an example where
 they do not, see \textcite{Fisher_Sompolinsky_1985}.}
The only thing affected by the DIV that invalidates strong scaling is the relation between the various exponents
and their $T$-counterparts, i.e., (\ref{eq:3.97}) does not hold.

Analogous statements holds for
\medskip
\par
$(ii)$ {\it Fisher's equality:} The spin-spin pair correlation function $G(r)$, in complete generality, has the form
\be
G(r) = \frac{e^{-r/\xi}}{r^{d-2+\eta}}\ ,
\label{eq:3.103}
\ee
which defines both the correlation length $\xi$ and the critical exponent $\eta$. Integrating over space we obtain the
homogeneous spin susceptibility
\be
\chi = \int d{\bm x}\ G(r) \propto \xi^{2-\eta}\ .
\label{eq:3.104}
\ee
Since $\xi(t,T=0) \propto t^{-\nu}$ and $\xi(t=0,T) \propto T^{-\nu_T}$, see Sec.\ \ref{par:III.C.2.b}{\it (i)}, we
obtain Fisher's equality both in the form of (\ref{eq:3.93}) and in the analogous form (\ref{eq:3.100}).
Checking this again for Hertz's fixed point we see, from Eqs.\ (\ref{eqs:3.65}, \ref{eqs:3.77}), that it does indeed hold
for all $d>1$. Similarly, (\ref{eq:3.92}) holds for the mean-field exponents from Eq.\ (\ref{eq:3.48}).

We now discuss
\medskip\par
$(iii)$ {\it Rushbrooke's inequality:} The inequality (\ref{eq:3.94b}) is based on the general thermodynamic
relation
\be
1 - \gamma_m/\gamma_h = \left[\left(\partial m/\partial T\right)_h\right]^2/\chi_T\,\gamma_h\ ,
\label{eq:3.105}
\ee
where $\gamma_m$ and $\gamma_h$ denote the specific-heat coefficient at constant magnetization
and constant magnetic field, respectively, and $\chi_T$ is the isothermal magnetic susceptibility. By
substituting the critical behavior of the various quantities, see Appendix \ref{app:B}, we obtain
Rushbrooke's inequality for the $T$-exponents at a QPT:
\be
\bar\alpha_T + 2\beta_T + \gamma_T \geq 2\ .
\label{eq:3.106}
\ee
The corresponding relation between $\bar\alpha$, $\beta$, and $\gamma$ at a QPT is obtained by using
the homogeneity law for the magnetization, Eq.\ (\ref{eq:3.101}), to show that $(\partial m/\partial h)_{T=0} \propto (-t)^{\beta(1-1/\beta_T)}$.
This yields
\be
\bar\alpha + 2\beta + \gamma \geq 2\beta/\beta_T\ .
\label{eq:3.107}
\ee
Note that this is different from the classical Rushbrooke inequality, Eq.\ (\ref{eq:3.94b}). As in the classical case, these
relations hold as inequalities if $\gamma_m/\gamma_h \to 1$ for $T\to 0$ at $t=0$ and for $t\to 0$ at $T=0$,
respectively, and as equalities otherwise. A check of Hertz's fixed point for $d>1$ in the clean case, Eq.\ (\ref{eqs:3.65}), and
for $d>0$ in the disordered case, Eqs.\ (\ref{eqs:3.77}), as well as the physical
disordered fixed point for $2<d<4$, Sec.\ \ref{par:III.C.3.d}, shows that in all cases both relations hold as inequalities.

If one assumes strong scaling, one can find larger lower bounds for the Rushbrooke inequalities and turn them into
Essam-Fisher-type equalities. For instance, consider the homogeneity relations (\ref{eqs:3.84}), which hold at the
ferromagnetic QCP in the disordered case. From Eqs.\ (\ref{eqs:3.85}, \ref{eqs:3.86}) we find
\be
\bar\alpha_T + 2\beta_T + \gamma_T = 2 + d(1/z_g - 1/z_c)\ .
\label{eq:3.108}
\ee
The rigorous thermodynamic inequality (\ref{eq:3.106}) requires $z_g < z_c$, which is indeed the case at the disordered
fixed point, see Eqs.\ (\ref{eqs:3.86}). Analogously, we have
\be
\bar\alpha + 2\beta + \gamma = \nu(z_c + z_g)\ .
\label{eq:3.109}
\ee
This again is consistent with the rigorous inequality (\ref{eq:3.107}). Eqs.\ (\ref{eq:3.108}, \ref{eq:3.109}) both hold
at the disordered transition, see Eqs.\ (\ref{eqs:3.85} - \ref{eqs:3.86}), consistent with the fact that this is a strong-scaling
fixed point. For the same reason they hold at Hertz's fixed point at the upper critical dimension ($d=1$, $z_c = z_g \equiv z = 3$
in the clean case, and $d=0$. $z_c = z_g \equiv 4$ in the disordered one), see Eqs.\ (\ref{eqs:3.65}) and (\ref{eqs:3.77}).

We finally mention the
\medskip\par
$(iv)$ {\it Hyperscaling relation} between $\bar\alpha$ and $\nu$, the classical version of which is given by Eq.\ (\ref{eq:3.95a}). 
With a strong scaling assumption, i.e., Eq.\ (\ref{eq:3.83}) with $z_c = z_g \equiv z$, we have
\be
\bar\alpha = \nu(z - d) \quad,\quad \bar\alpha_T = 1 - d/z\ .
\label{eq:3.110}
\ee
Both of these relations hold at Hertz's fixed point in both the clean and the disordered case, see Eqs.\ (\ref{eqs:3.65}, \ref{eqs:3.77}).%
\footnote{The reason why they hold even above the upper critical dimension is that the $d$-dependent exponents $\bar\alpha$ and
$\bar\alpha_T$ describe the leading fluctuation contribution to the specific-heat coefficient, which does obey strong scaling; see
footnote \ref{specific_heat_scaling_footnote}.}
They also hold at the disordered fixed point (with $z = z_c$) discussed in Sec.\ \ref{par:III.C.3.e}, see Eq.\ (\ref{eq:3.85a}).

\subsection{Rare-region effects in disordered systems}
\label{subsec:III.D}

\subsubsection{Quantum Griffiths effects}
\label{subsubsec:III.D.1}

The notion of a Griffiths phase is well established in both classical
and quantum disordered systems \cite{Griffiths_1969, McCoy_1969, Randeria_Sethna_Palmer_1985, Bray_1987,
Millis_Morr_Schmalian_2002, Vojta_2010}.%
\footnote{\label{Griffiths_history_footnote} The notion that in a random Ising system there is a whole region in
what one would naively consider the disordered phase where the free energy is not an analytic function
of the magnetic field was put forth simultaneously by \textcite{Griffiths_1969} and \textcite{McCoy_1969}.  
McCoy considered a strip-random two-dimensional classical model \cite{McCoy_Wu_1968} that is closely related to the quantum-mechanical
problem of a random transverse-field Ising spin chain \cite{Fisher_1995}. This observation, and
phenomena deriving from it, are now often referred to as (quantum) Griffiths-phase effects.}
The basic idea can be illustrated
by considering a classical randomly diluted Ising ferromagnet
in $d$-dimensions. In this model some of the ferromagnetic bonds are randomly
missing with a probability $p$. As a result, the critical temperature
$T_{\text{c}}$ in the random system is lower than the corresponding critical temperature
in the pure or non-random system. In random systems, the latter
is often denoted by $T_{\text{G}}$ and referred to as the Griffiths temperature. In general,
interesting effects occur both in the `paramagnetic' phase, $T > T_{\text{G}}$, and in the
`Griffiths phase', $T_{\text{c}} < T < T_{\text{G}}$. Here we focus on the latter.

Griffiths argued that in such a system there always exist regions of linear
size $L$ that happen to contain no missing bonds, and thus behave
as a region of the same size in the corresponding pure system. This is
true even for arbitrarily large $L$, but the probability of finding a large
region devoid of missing bonds is exponentially small,
\begin{equation}
P(L)\propto\exp(-c\,L^d)\ .
\label{eq:III.D.1.1}
\end{equation}
Here $d$ is the spatial dimensionality of the system, and $c$ is a constant.
If the size of these rare regions is large compared to the local correlation
length, then it is meaningful to speak of them as being magnetically ordered.
At criticality in a mean-field theory, the correlation length $\xi$ as a function
of an applied magnetic field $h$ scales as $\xi\sim1/h^{1/3}$. This
in turn suggests that in the entire Griffiths phase there is a
contribution $\delta F_{\text{G}}$ to the free energy that reflects both the
exponentially small probability of rare regions and the scaling of the correlation
length with the magnetic field \cite{Dotsenko_2006}:
\begin{equation}
\delta F_{G}\propto\exp[-c\,'\,h^{-d/3}]\ ,
\label{eq:III.D.1.2}
\end{equation}
with $c\,'$ another constant. That is, in the entire Griffiths phase the free energy
is a nonanalytic function of the field $h$ at $h=0$. However, the singularity is
only a very weak essential one.

The weak singularities in the thermodynamic properties in the classical
Griffiths phase are very difficult to detect experimentally. However,
the existence of ordered rare regions has a qualitative effect on the
dynamics of the equilibrium time correlation functions. This
is physically obvious since overturning large clusters of ordered
spins takes a relaxation time that grows exponentially with the size of
the cluster, and time correlation functions in the
Griffiths phase will depend on such dynamical processes. We have given
the qualitative argument in Sec. \ \ref{subsubsec:III.A.1}. The result,
Eq.\ (\ref{eq:3.7b}), was that time correlation functions are expected to
decay slower than any exponential.

The conclusion is that the static effects in the classical Griffiths phase
are very weak, but dynamic Griffiths effects are quite profound,
changing exponential decay of time correlation functions into non-exponential decay. As
we have stressed in Sec.\ \ref{sec:I}, in quantum statistical mechanics the statics and the
dynamics are coupled. This implies that Griffith-phase effects are expected to be important 
for both the statics and the dynamics near quantum phase transitions in disordered systems
in general, and in disordered quantum ferromagnets in particular. In fact,
it turns out that in the quantum case the dynamical singularities are even stronger
than suggested by the classical arguments above.

In the context of quantum mechanics, this goes back to the model proposed and
studied by \textcite{McCoy_Wu_1968} \cite{McCoy_Wu_1969, McCoy_1969}, which
is closely related to a one-dimensional quantum problem. This model was later generalized
\cite{McCoy_1970, Shankar_Murthy_1987}, and its quantum mechanical interpretation
was studied in detail by \textcite{Fisher_1992, Fisher_1995} and others 
\cite{Rieger_Young_1996, Young_1997, Pich_et_al_1998}.
The crucial thing to keep in mind is that the slow dynamics associated
with the Griffiths phase greatly affects the zero-temperature behavior.
To see this, consider a local magnetized rare region of linear
size $L$, separated by a domain wall from the rest of the system
as in the classical case. 
\footnote{\label{ordered_Griffiths_footnote} Griffiths effects also exist in the ordered phase.
                However, they are weaker than the corresponding effects in the disordered phase
                except in certain special models \cite{Motrunich_et_al_2000, Senthil_Sachdev_1996}.
                Here we focus on the disordered phase.}
Its imaginary-time local dynamic susceptibility
will decay exponentially by a quantum tunneling process. For long
imaginary times we have,
\begin{equation}
\chi_{\rm{loc}}(\tau\rightarrow\infty)\propto\exp[-\tau/\bar{\tau}(L)]\ ,
\label{eq:III.D.1.3}
\end{equation}
where $\bar{\tau}(L)$ is the characteristic relaxation time for
the tunneling process. To estimate $\bar{\tau}(L)$ we imagine a domain
wall in imaginary time space for a cluster of size $L^{d}$ in real
space. This has been considered for Ising systems \cite{Millis_Morr_Schmalian_2002, Motrunich_et_al_2000,
Pich_et_al_1998, Rieger_Young_1996, Guo_Bhatt_Huse_1996, Thill_Huse_1995}
and for Heisenberg magnets \cite{Vojta_Schmalian_2005}; for a review, see \textcite{Vojta_2010}.%
\footnote{Strictly speaking the considerations presented here are valid only
for non-Ising metallic magnets, i.e. systems with an order-parameter
dimensionality $n>1$. The reason is that the $\vert\Omega_{n}\vert$
term in the Gaussian order-parameter action corresponds, at $T=0$,
to a long-ranged $1/\tau^{2}$ decay in imaginary-time space.
Because a one-dimensional Ising model ($n=1$) can have a phase transition
with such an interaction, this implies that there might be a freezing
phase transition in imaginary-time space that is not included in the simple Griffiths
arguments given here. One-dimensional models with $n>1$ do not have
such a phase transition because they are below their lower critical
dimension even with this long-ranged interaction.}
Most of the work on this topic has been done for antiferromagnets, i.e., the case of a nonconserved
order parameter. One finds
\begin{equation}
\bar{\tau}(L)\sim\tau_0\exp({\bar\sigma} L^{d})\ .
\label{eq:III.D.1.4}
\end{equation}
Here $\tau_0$ is a microscopic time scale and ${\bar\sigma}$ is a constant,
and the overbars distinguish $\bar\tau$ and $\bar\sigma$ from the corresponding
quantities in the classical case, Eq.\ (\ref{eq:3.6}).
Effectively, in the quantum case the volume of the region is $L^{d+z}$, with
$z$ the dynamical exponent, 
and the domain wall is a hypersurface with area $L^{d+z-z}=L^{d}$. Physically,
the decay of the rare region in the quantum case is much slower
than its classical counterpart, Eq.\ (\ref{eq:3.6}), since at $T>0$ the cluster can
flip via thermal activation in addition to quantum tunneling. Equations (\ref{eq:III.D.1.1}),
(\ref{eq:III.D.1.3}) and (\ref{eq:III.D.1.4}) imply for the average local dynamic susceptibility 
\bse
\label{eqs:III.D.1.5}
\begin{equation}
\chi_{\rm{loc}}^{av}(\tau) = \int_0^{\infty} dL\exp[-c\,L^{d}-(\tau/\tau_0)e^{-\bar\sigma L^{d}}]
\label{eq:III.D.1.5a}
\end{equation}
In this case the typical length scale is $L_{\rm typ} \propto [\ln(\tau/\tau_0)]^{1/d}$, and
in the limit of large imaginary time the method of steepest descent yields
\be
\chi_{\rm{loc}}^{\rm{av}}(\tau\to\infty) \propto (\tau/\tau_0)^{-c/{\bar\sigma}}\ .
\label{eq:III.D.1.5b}
\ee
\ese
We see that quantum mechanics leads to a power-law decay. This is in contrast to the 
classical case, see Sec. \ref{subsubsec:III.A.1}. The temperature dependence of the static 
susceptibility is
\begin{equation}
\chi_{{\rm loc}}^{{\rm av}}(T\rightarrow0)=\int_{0}^{1/T}d\tau\chi_{\rm loc}^{\rm av}(\tau)\sim T^{c/{\bar\sigma}-1}
\label{eq:III.D.1.6}
\end{equation}
The remarkable conclusion is that Griffiths-phase dynamical singularities
lead to low-temperature singularities in static quantities. Similarly, the specific
heat contribution from these processes is $C_{\rm loc}^{\rm av}(T)\sim T^{c/{\bar\sigma}}$.
Note that if $c/{\bar\sigma}<1$, then these local rare-region contributions
dominate the usual Fermi liquid ones.

The conserved disordered ferromagnetic case is even more dramatic.
Physically this is because a conservation law is equivalent to a long-ranged
interaction \cite{Hoyos_Vojta_2007}, and hence slows down relaxation even more. 
\textcite{Nozadze_Vojta_2012} have argued that in this case the relaxation time in three dimensions
scales as
\begin{equation}
\bar{\tau}(L)\propto \exp[{\bar\sigma} L^{3+n}]
\label{eq:III.D.1.7}
\end{equation}
where $n=1$ if the itinerant electrons are ballistic, and $n=2$ if
they are diffusive. Technically, the extra factor of $L^n$ compared
to the antiferromagnetic or non-conserved case is a result of the $1/\vert{\bm k}\vert^n$ in
the paramagnon propagator, Eq.\ (\ref{eq:3.5a}). Following the same steps as above, one can
determine the physical observables. The local susceptibility behaves as
\begin{equation}
\chi_{\rm loc}^{\rm av}(T\rightarrow0)\propto\frac{1}{T}\exp[-A\{\ln(T_{0}/T)\}^{3/(3+n)}]\ ,
\label{eq:III.D.1.8}
\end{equation}
with $A$ a constant and $T_0$ is a microscopic temperature scale, and the specific heat
is proportional to
\begin{equation}
C_{\rm loc}^{\rm av}(T\rightarrow0)\propto\exp[-A\{\ln(T_{0}/T)\}^{3/(3+n)}]\ .
\label{eq:III.D.1.9}
\end{equation}
Finally, the magnetization $m$ at zero temperature as a function of an
applied field $H$ is
\begin{equation}
m(H\rightarrow0)\propto\exp[-A\{\ln(H_{0}/H)\}^{3/(3+n)}]\ ,
\label{eq:III.D.1.10}
\end{equation}
where $H_0$ is a microscopic magnetic field scale. Note that these exponentials
go to zero slower than any power law.

\subsubsection{Disordered local moments}
\label{subsubsec:III.D.2}

A related concept in the presence of quenched disorder is that of local
magnetic moments.%
\footnote{\label{Griffiths_rare_regions_footnote} Although Griffiths-region effects 
and local-moment effects are obviously related, this relation has, to the authors's knowledge,
not been considered in a comprehensive way.}
Importantly, disorder can both facilitate the formation of
local moments \cite{Milovanovich_Sachdev_Bhatt_1989} and weaken the
Kondo effect, which quenches local moments, to the point where the effective
Kondo temperature is zero \cite{Bhatt_Fisher_1992}. If these effects are important
in metallic ferromagnets, they will result in an absence of spin diffusion at zero
temperature and the considerations outlined in Secs.\ \ref{subsubsec:III.B.3},
\ref{subsubsec:III.C.2}, and \ref{subsubsec:III.D.1} are incomplete.

Let us recall in what sense the Kondo temperature $T_{\rm K}$ vanishes \cite{Bhatt_Fisher_1992}.
These arguments will also illustrate the importance of the long-ranged RKKY interaction in
disordered systems. First, consider a clean system.
The single-site Kondo problem has a single spin-1/2 magnetic impurity, located at the
origin, which interacts with the conduction-electron spin via an antiferromagnetic point-contact
amplitude $J>0$. Perturbation theory and RG calculations (see, e.g., \textcite{Gruner_Zawadowski_1974},
\textcite{Nozieres_1978}) indicate that the renormalized AFM coupling
becomes very large below a Kondo temperature given by
\begin{equation}
T_{\rm K} = E \exp[-1/JN_{\rm F}]\ ,
\label{eq:Kondo temperature}
\end{equation}
where $E$ is an energy scale on the order of the bandwidth or the
Fermi energy, and $N_{\rm F}$ is the density of states at the Fermi surface. 
For $T>T_{\rm K}$ there are local moments that give a Curie-like contribution to the
magnetic susceptibility,
\begin{equation}
\chi_{\rm LM}(T>T_{K})\propto\frac{\mu^{2}}{T}\,n_{\rm LM}
\end{equation}
with $\mu$ the magnetic moment and $n_{\rm LM}$
the density of local moments, respectively. For $T<T_{\rm K}$ the renormalized antiferromagnetic
coupling scales to infinity, and it is energetically favorable for the impurity spin to form a nonmagnetic
spin singlet with a conduction electron. The Curie contribution to $\chi_{\rm LM}$ is thus cut off at $T_{\rm K}$, and for
lower temperatures it is given by
\begin{equation}
\chi_{\rm LM}(T<T_{K})\propto\frac{\mu^{2}}{T_{\rm K}}\,n_{\rm LM}\ .
\end{equation}
The crucial physical point is that for $T<T_{\rm K}$ the local moment
is effectively screened, and for $T\rightarrow 0$ the system is a
conventional Fermi liquid. Also note that for $T<T_{\rm K}$ the dynamics
of the conduction electrons will be diffusive in the absence of spin-orbit
coupling effects.

Next consider a generic system with nonmagnetic disorder and with
local moments. Again consider a local moment at the origin. With a
probability
\begin{equation}
P(r)\propto\exp[-cr^{d}]
\label{eq:probaility}
\end{equation}
the local moment is located in a cavity of radius $r$ and isolated
from the rest of the system. Next assume that the exchange coupling
between the local moment and the rest of the system, with nearest
neighbors a distance $r$ away, goes as
\begin{equation}
J(r)\sim\exp[-c'r]
\label{eq:J(r)}
\end{equation}
In these equations $c$ and $c'$ are positive constants. In the absence
of $J$ the local moment would result in a Curie contribution to the magnetic susceptibility,
$\chi\propto\mu^{2}/T$. However, as we have seen above, the coupling to the conduction electrons
can quench or screen the local moment. 

If $J$ is too small, i.e., if $r$ in Eq. (\ref{eq:J(r)}) is too large, then the local moment
cannot be Kondo screened. A critical value
for $r$ can be determined by using Eq. (\ref{eq:Kondo temperature})
with $J$ replaced by $J(r)$ and $T_{\rm K}$ by $T_{\rm K}(r)$. That is,
in a disordered system there will be a range or distribution of Kondo
temperatures. Combining all of this, one concludes that a local moment will not be Kondo screened
at temperature $T$ if $r>r_{c}$ with
\begin{equation}
T=E\exp[-e^{c'r_{c}}/N_{\rm F}]\ ,
\end{equation}
or
\begin{equation}
r_{c}\propto\ln\left[\ln\frac{E}{T}\right]\ .
\label{eq:critical radius}
\end{equation}
The density of unquenched local moments, $n_{\rm LM}(T)$, is then proportional
to $P(r_{c})$ given by Eqs. (\ref{eq:probaility}) and (\ref{eq:critical radius}):
\begin{equation}
n_{\rm LM}(r_{c}) = n_{\rm LM}(T) \propto \exp\left\{ -c''\left[\ln\left[\ln\frac{E}{T}\right]\right]^{d}\right\} 
\end{equation}
with $c''$ a constant. This leads to a local moment magnetic susceptivity
given by
\begin{equation}
\chi_{\rm LM}(T)\propto\frac{\mu^{2}}{T}\exp\left\{ -c''\left[\ln\left[\ln\frac{E}{T}\right]\right]^{d}\right\} \ .
\end{equation}
We see that even though the density of local moments vanishes
as $T\rightarrow0$, $\chi_{\rm LM}$ still diverges in this limit. The
conclusion reached by \textcite{Bhatt_Fisher_1992} is that a disordered Fermi system with local moments is
never a Fermi liquid, even for arbitrarily weak disorder. An alternative way to avoid
Kondo screening in disordered systems has been discussed by \textcite{Dobrosavljevic_Kirkpatrick_Kotliar_1992}.

So far we have not considered any interactions between the local moments. This can be estimated as
follows \cite{Bhatt_Fisher_1992}.
In the metallic phase the interaction is an RKKY interaction mediated by the conduction electrons. The
interaction between spins at position $\bm{R}_{i}$ and $\bm{R}_{j}$,
separated by a distance $R_{ij}$, and coupled to the conduction electrons
via exchange interaction amplitudes $J_{i}$ and $J_{j}$, is
\begin{equation}
K_{ij}\propto\frac{J_{i}J_{j}}{R_{ij}^{d}}\,N_{\rm F}\,g_{d}(k_{\rm F}\bm{R}_{i},k_{\rm F}\bm{R}_{j})
\end{equation}
with $g_{d}$ an oscillating function of order one. For spin-$1/2$
local moments, the antiferromagnetic interactions have a stronger effect than the
ferromagnetic ones due to the Pauli principle, so we can replace $g_{d}$ by unity. To estimate the effects
of the local moment interaction we consider two local moments in cavities
separated by a distance $R$. The typical separation $R$ can be related
to $n_{\rm LM}(R)$ by $R\propto n_{\rm LM}^{-1/d}$. To relate $R$ to the temperature
it is assumed that the two spins will form a nonmagnetic singlet state
if the energy gain is greater than $T$. This gives
\begin{equation}
T\propto\frac{J^{2}(r)}{R^{d}}N_{\rm F}
\end{equation}
or 
\begin{equation}
n_{\rm LM}\propto T/N_{\rm F}J^{2}(r)\ .
\label{eq:local moment density}
\end{equation}
Using this and $n_{\rm LM}\propto\exp[-cr^{d}]$ gives
\begin{equation}
n_{\rm LM}(T)\propto\frac{T}{N_{\rm F}}\exp\{c[\ln(1/T)]^{1/d}\}
\end{equation}
and
\begin{equation}
\chi_{\rm LM}(T)\propto\exp\{c[\ln(1/T)]^{1/d}\}\ .
\end{equation}
This estimate should be considered as a lower bound. The conclusion is again that there is
a vanishing local-moment density, but a divergent magnetic susceptibility
as $T\rightarrow 0$.

The conclusions from these arguments are three-fold. First, interactions between local moments,
or rare regions in this case, are important. Second, a system with local moments does not behave
as a Fermi liquid. Third, strictly speaking there is an absence of spin diffusion at zero temperature
because of the divergent susceptibility.

\subsubsection{Interacting rare regions}
\label{subsubsec:III.D.3}

One conclusion of the previous subsection is that long-ranged RKKY interactions between local moments,
in conjunction with rare-region effects, can have qualitative effects. This suggests that similar interactions
between rare regions in a quantum Griffith phase might also be important. This question has been studied
by Dobrosavljevi{\'c} and others \cite{Dobrosavljevic_Miranda_2005, Case_Dobrosavljevic_2007a} for the
case of a Heisenberg antiferromagnet. The applicability of these ideas, with suitable modifications, to ferromagnets
remains to be studied.

These authors considered rare regions centered at points ${\bm R}_{i}$ ($i=1,2,3,\ldots$) that are characterized
by local $N$-component ($N>1$) order parameters $\bm\phi_{i}(\tau)$, with $\tau$ the imaginary-time variable. The Gaussian part of the action has the form
\begin{equation}
S^{(2)} = S^{(2)}_0 + S^{(2)}_{\rm int}\ .
\end{equation}
Here $S^{(2)}_0$ is the noninteracting part, 
\bea
S^{(2)}_0 &=& \sum_{i}\int_{0}^{\beta}d\tau d\tau'\,\bm\phi_{i}(\tau)\,\Gamma_{0}(\tau-\tau')\,\cdot\bm\phi_{i}(\tau')
\nonumber\\
&=&\sum_{n,i}\bm\phi_{i}(\Omega_{n})\,\Gamma_{0}(\Omega_{n})\,\cdot\bm\phi_{i}(-\Omega_{n})
\label{eq:noninteracting action}
\eea
with $\Omega_{n}$ a bosonic Matsubara frequency. Let us assume for simplicity that the order parameter
is not conserved, so that the noninteracting vertex is given by
\begin{equation}
\Gamma_{0}(\Omega_{n}) = \Gamma_{0}(0) + \vert\Omega_{n}\vert \ .
\label{eq:bare_vertex}
\end{equation}
The $\vert\Omega_n\vert$ nonanalyticity is the Landau damping mechanism due to the coupling of the
magnetic order parameter to the conduction electrons that was discussed in Sec.\ \ref{subsubsec:III.A.1}.
In imaginary-time space, it corresponds to a power-law decay $\Gamma_0(\tau\to\infty) \propto 1/\tau^2$.
This puts the rare region or droplet, now considered a one-dimensional classical system in $\tau$-space
with a $1/\tau^2$ interaction, at its lower critical dimension \cite{Joyce_1969}. This means that the
noninteracting rare regions cannot develop long-range order.

The interacting part of the Gaussian action is given by
\begin{equation}
S^{(2)}_{\rm int} = \frac{1}{2}\sum_{i\neq j}\int_{0}^{\beta}d\tau d\tau'\,\bm\phi_{i}(\tau)\,V(R_{ij},\tau-\tau')\,\cdot\bm\phi_{j}(\tau')
\end{equation}
The interaction between two rare regions is assumed to be a static
RKKY interaction given by
\begin{equation}
V(R_{ij},\tau) = \frac{J_{ij}}{(R_{ij})^d}\,\delta(\tau)\ .
\end{equation}
$J_{ij}$ is assumed to be a random amplitude of zero mean and variance
$\langle J_{ij}^{2}\rangle = J^{2}$. Using replica methods, \textcite{Dobrosavljevic_Miranda_2005} conclude
that the effective contribution to the total action from rare-region interactions is
\bea
\delta S &=& -\frac{1}{2}\sum_{i,j}(1-\delta_{ij})\sum_{\alpha\beta}\frac{J^{2}}{(R_{ij})^{2d}}
   \nonumber\\
   &&\hskip 0pt \times\int d\tau d\tau'\, 
   \Bigl(\bm\phi_{i}^{\alpha}(\tau)\cdot\bm\phi_{j}^{\alpha}(\tau)\Bigr)\Bigl(\bm\phi_{i}^{\beta}(\tau')\cdot\bm\phi_{j}^{\beta}(\tau')\Bigr)\ .
\nonumber\\
\eea
Here $(\alpha,\beta) = 1,2,\ldots n$ are replica labels, and the replica limit $n\rightarrow 0)$ is to be taken
at the end of any calculation. Treating this interaction in a standard mean-field approximation then gives
\begin{equation}
\delta S = -\sum_{\alpha\beta}\sum_{n,i}\bm\phi_{i}^{\alpha}(\Omega_{n})\, \Delta_{i}^{\alpha\beta}(\Omega_{n})\,\cdot\bm\phi_{i}^{\beta}(-\Omega_{n})\label{eq:renormalization}
\end{equation}
where $\Delta_i^{\alpha\beta}(\Omega_{n})$ is proportional to a weighted spatial
average of a local rare-region susceptibility,
\begin{equation}
\Delta_{i}^{\alpha\beta}(\Omega_{n}) = \frac{1}{N}\sum_{j\neq i}\frac{J^{2}}{(R_{ij})^{2d}}\, 
     \langle\bm\phi_{j}^{\alpha}(\Omega_{n})\cdot\bm\phi_{j}^{\beta}(-\Omega_{n})\rangle\ .
\label{eq:weighted_local_average}
\end{equation}
Within a self-consistent mean-field theory, the average in Eq. (\ref{eq:weighted_local_average}) is to be taken with
respect to the complete action, including the rare-region interaction term. 

Comparing Eq. (\ref{eq:noninteracting action}) and Eq. (\ref{eq:renormalization})
we see that the rare-region interactions have renormalized the Gaussian
part of the noninteracting action $S_{0}$. This is analogous
to the effects of the fermionic soft modes that was discussed in Sections
\ref{subsec:III.B} and \ref{subsec:III.C}. The importance of this term depends on its behavior
for long times or low frequencies. \textcite{Dobrosavljevic_Miranda_2005} concluded that
effectively the noninteracting vertex $\Gamma_0$, Eq.\ (\ref{eq:bare_vertex}), gets augmented
by an additive term of the form
\be
\delta\Gamma(\Omega_n) \propto {\rm const.} + \vert\Omega_{n}\vert^{\alpha-1}\ .
\end{equation}
Here $\alpha = c/{\bar\sigma}$ is the same exponent that appears in Eqs.\ (\ref{eq:III.D.1.5b})
and (\ref{eq:III.D.1.6}). It is nonuniversal and is expected to decrease
as the magnetically ordered phase is approached. Once $\alpha < 2$, the nonanalyticity
coming from the rare-region interaction is stronger than the one due to Landau damping
in the bare action, Eq.\ (\ref{eq:bare_vertex}). The order-parameter correlation function then
falls off more slowly than $1/\tau^2$ for large imaginary times. The rare region thus is above
its lower critical dimension, and thus can develop long-range order. This in turn implies that
sufficiently large droplets will freeze and form a ``cluster glass'' phase.$^{\ref{cluster_glass_footnote}}$
This concept has been used to analyze and interpret experiments on \CPR, see the discussion
in Sec.~\ref{CPR}.

Based on these considerations, which suggest that the Griffiths phase is unstable, 
\textcite{Dobrosavljevic_Miranda_2005} have proposed a
phase diagram where a cluster-glass phase appears between the paramagnetic phase and
the magnetically ordered phase. This phase diagram has been further discussed by
\textcite{Case_Dobrosavljevic_2007a}, who have argued that the transition from the paramagnet
to the cluster-glass phase is a fluctuation-induced first-order transition at low temperature, while
it is continuous at higher temperatures, with a tricritical point in the phase diagram. This mechanism
is analogous to the one described in Sec.\ \ref{subsec:III.B} for clean ferromagnets, with the
ordinary ferromagnetic order parameter replaced by the droplet order parameter, and the fermionic
soft modes replaced by the Griffiths fluctuations that were discussed in Sec.\ \ref{subsubsec:III.A.1}.

\subsubsection{The size of Griffiths effects}
\label{subsubsec:III.D.4}

The arguments for Griffiths-phase effects reviewed above are all asymptotic in nature, a characteristic
they share with other problems involving rare events, e.g., Lifshitz tails in the density of states of
disordered solid \cite{Lifshitz_1964}. A natural question for all of these phenomena is the range of
their validity. For instance, we need to ask how far from its initial value a time correlation function
has to decay before the asymptotic behavior becomes realized, or how low a temperature one has
to consider in order for the effects described in Sec.\ \ref{subsubsec:III.D.1} to becomes observable.
These and related questions have a long history. They were initially investigated for classical systems,
see below, where the predicted effects were not always observed. However, for quantum systems more recent
numerical evidence indicates substantial effects \cite{Rieger_Young_1996, Guo_Bhatt_Huse_1996, 
Pich_et_al_1998, Vojta_2010} and indeed experimental observations in many strongly disordered systems 
have been interpreted as due to quantum Griffiths effects, see the discussions in Secs.\ \ref{subsec:II.C} and \ref{subsec:II.E}. 

For classical systems, several rigorous results are available.
One interesting example is the problem of a random walk with a random distribution of static traps
that immobilize the diffusing particle if it hits one. This problem has been studied with mathematically rigorous methods
as well as by physically appealing Lifshitz-Griffiths-type arguments and extensive
numerical studies. It has been shown rigorously \cite{Donsker_Varadhan_1975, Donsker_Varadhan_1979}
that the survival probability $P(c,t)$, with $c$ the concentration of traps, decays for asymptotically long
times as
\be
\ln P(c,t\rightarrow\infty)\propto -\lambda^{\frac{2}{d+2}}\,t^{\frac{d}{d+2}}
\label{eq:survival_probability}
\ee
with $\lambda=-\ln(1-c)$. The same result was obtained by means of Griffiths-Lifshitz
arguments by \textcite{Grassberger_Procaccia_1982} and by \textcite{Kayser_Hubbard_1983}, 
who showed that the asymptotic long-time behavior
is dominated by the existence of arbitrarily large, but exponentially rare, trap-free regions. This work left
open the size of the asymptotic region. After many earlier studies, \textcite{Barkema_Biswas_van_Beijeren_2001}
(see also references therein)
showed conclusively that the asymptotic result is valid only when $P(c,t)$ is exceedingly small. For
instance, in $d=3$ the asymptotic behavior sets in only when $P(c,t) \approx 10^{-30}$ and $10^{-80}$ for
$c = 0.1$ and $c = 0.01$, respectively. For shorter times, $P(c,t)$ decays exponentially.

The Griffiths phase of the classical randomly bond-diluted Ising model mentioned in Sec.\ \ref{subsubsec:III.D.1} 
has also been studied extensively. The time dependent local spin-spin
correlation function $C(t)$ is predicted to decay as \cite{Bray_1988, Bray_1989} 
\be
C(t\rightarrow\infty)\sim\exp[-{\rm const.}\times(\ln t)^{d/(d-1)}]
\label{eq:C(t)_Ising}
\ee
Monte Carlo simulations for $d=3$ \cite{Colborne_Bray_1989} showed poor agreement with Eq.\ (\ref{eq:C(t)_Ising}). 
Plotting $\ln C(t)$ against $(\ln t)^{3/2}$ yielded substantial curvature. A better fit was obtained by plotting $\ln C(t)$ 
against $[\ln(t/\tau)]^{3/2}$, with $\tau(T)$ an adjustable parameter. A
still better fit was found using a stretched exponential or Kohlrausch form $C(t)\sim\exp[-(t/\tau)^{\beta}]$, with
$\beta$ is an increasing function of temperature that is
on the order of $0.4$. Various simulations \cite{Colborne_Bray_1989, Jain_1995, Cao_et_al_2006} suggest that
$C(t)$ must be smaller than $10^{-4}$ of its initial value before the asymptotic behavior sets in.

The situation is different for classical $n$-dimensional spins  with $n\geq2$. In this case, the Griffiths arguments 
predict \cite{Bray_1987, Bray_1988, Bray_1989}
\be
C(t\rightarrow\infty)\sim\exp[-{\rm const.}\times t^{1/2}]\ .
\label{eq:C(t)_Heisenberg}
\ee
Monte Carlo data are entirely consistent with this prediction for all but short times \cite{Colborne_Bray_1989}.

For quantum systems, the increasing power of numerical methods has yielded interesting results.
For a transverse-field Ising spin glass, Monte-Carlo simulations on two-dimensional and three-dimensional systems by \textcite{Rieger_Young_1996}
and \textcite{Guo_Bhatt_Huse_1996} found clear evidence of Griffiths-phase effects. The size of the effects 
decreased by about a factor of 4 from $d=2$ to $d=3$. The strength of the effects, compared with classical systems, 
is sometimes attributed to the fact that in the quantum case the Griffiths clusters occur as line defects, as opposed to 
point defects in classical models. Perhaps more importantly,
because quantum tunneling of a rare region is a slower process ($\tau \propto \exp{(L^d)}$) than thermally activated 
dynamics ($\tau \leq \exp{L^{(d-1)}}$) of the same rare region, the Griffiths singularities in the quantum case lead to power-law 
decays in time, or power-law singularities at low temperatures. These power-law effects in temperature can dominate the usual 
Fermi-liquid power laws in metals. In general, various (possibly nonlinear) susceptibilities will even diverge as $T \to 0$.

Collectively, these results imply that the importance of the Griffiths effects is not a-priori clear and may strongly
depend on the nature of the system. For instance, in the classical case there is a qualitative difference between
Ising and XY or Heisenberg models, see Eqs.\ (\ref{eq:C(t)_Ising}) and ({\ref{eq:C(t)_Heisenberg}). The quantum
ferromagnetic case, for both Ising and Heisenberg symmetry, is similar to the classical Ising model in the sense
that there is a activation barrier to transport, unlike the classical Heisenberg case. On the other hand, there is
numerical evidence for quantum mechanics enhancing the Griffiths effects.

\subsection{Textured phases as a way to avoid a quantum critical point}
\label{subsec:III.E}

It was realized early on that the instability of Hertz theory can signalize either a first-order
transition, or a transition into a non-homogeneous magnetic phase \cite{Belitz_Kirkpatrick_Vojta_1997,
Chubukov_Pepin_Rech_2004, Rech_Pepin_Chubukov_2006}. The conditions under which a first-order 
transition to a paramagnetic phase, or a transition to an intermediate spiral phase, respectively, occurs 
have been investigated by several authors \cite{Efremov_Betouras_Chubukov_2008, Maslov_Chubukov_Saha_2006}.
\textcite{Maslov_Chubukov_2009} concluded that in a model with a long-ranged exchange interaction
the first-order transition always pre-empts the formation of a spiral phase.

A similar physical scenario was
discussed by \textcite{Conduit_Green_Simons_2009}, who used a self-consistent many-body approach 
supplemented by a numerical evaluation of fluctuation corrections to the free energy to argue that a
spiral state can pre-empt the first-order transition as the ferromagnetic state is approached from the
paramagnetic phase. This textured magnetic phase is analogous to the Fulde-Ferrell-Larkin-Ovchinnikov state
in superconductors \cite{Fulde_Ferrell_1964, Larkin_Ovchinnikov_1964}. \textcite{Karahasanovic_Kruger_Green_2012} 
expanded this approach to a purely analytical theory that allows for instabilities towards spin-nematic phases in addition 
to a spiral one. They concluded that a complex phase diagram is possible, where the first phase encountered as
the ordered state is approached from the paramagnetic one at low temperature is a spin-nematic phase, followed by a transition to a
spiral phase, and finally another transition to a uniform ferromagnet. The proposed phase diagram is shown in
Fig.~\ref{figure:nematic_spiral_uniform_phase_diagram}.
\begin{figure}[t]
\begin{center}
\includegraphics[width=0.95\columnwidth,angle=0]{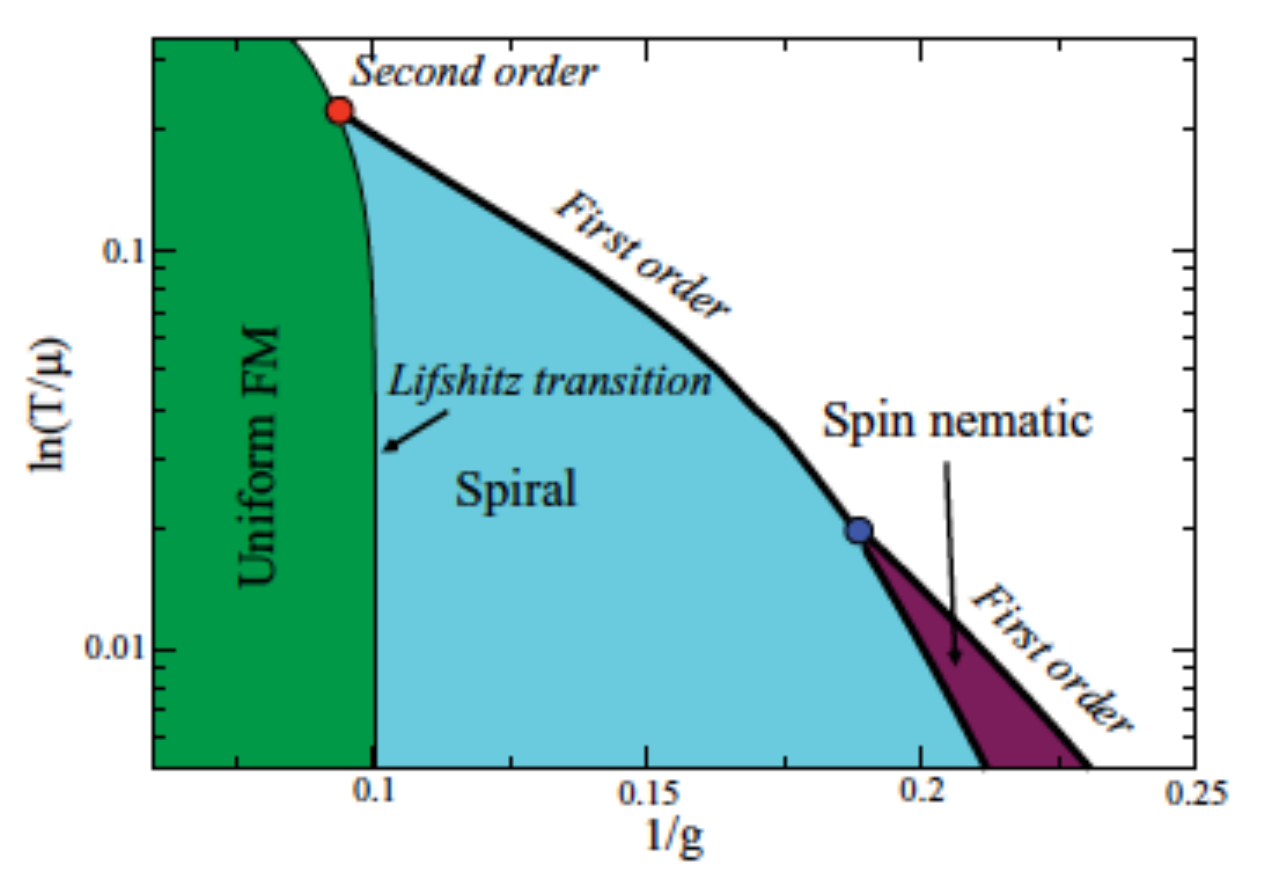}
\end{center}
\caption{Proposed phase diagram for a model allowing for spiral and spin-nematic order. $\mu$ is the chemical potential,
              and $g$ is the exchange coupling. From \textcite{Karahasanovic_Kruger_Green_2012}.}
\label{figure:nematic_spiral_uniform_phase_diagram}
\end{figure}
The possibility of a Pomeranchuk instability
towards a non-s-wave ferromagnet or magnetic nematic had also been discussed earlier by \textcite{Chubukov_Maslov_2009}.
Later work concluded that an infinite resummation of fluctuation contributions to the free energy
results in the spiral phase occupying a substantially smaller part of the phase diagram (within about 1\% of
the transition point at $T=0$) than the original theory predicted \cite{Pedder_Kruger_Green_2013}, 
but the topology of the phase diagram remained the same. Such a narrow slice of spiral order would
be easy to overlook experimentally and has so far not been observed. In two-dimensional systems, the theory
predicts a much larger spiral phase. One must keep in mind, however, that no true long-range ferromagnetic
order is possible in two dimensions at $T>0$.

The influence of nonmagnetic quenched disorder on the proposed spiral phase has been investigated
by \textcite{Thomson_Kruger_Green_2013}. These authors found that disorder is unfavorable to
long-range ferromagnetism and increases the size of the spiral or helical phase. While taken at face value this is not
consistent with the discussion in Sec.\ \ref{subsubsec:III.B.3}, there likely are many competing effects of
quenched disorder on ferromagnetism, in analogy to what is known to be the case in disordered 
superconductors, that yet have to be investigated comprehensively, and in a common context. The theory
predicts a helical spin-glass phase induced by the disorder leading to random preferred directions of the
helical axis. This is a realization of the schematic phase diagram shown in Fig.~\ref{fig:schematic_phase_diagrams}\,d), and has
been discussed by \textcite{Thomson_Kruger_Green_2013} as a possible explanation of the observations
in CeFePO by \textcite{Lausberg_et_al_2012a}. An interesting aspect of such a phase is that fluctuations
of line defects is expected to lead to a $T^{3/2}$ behavior of the electrical resistivity by the same mechanism
that has been proposed to be operative in the helical magnet MnSi \cite{Kirkpatrick_Belitz_2010}.

Another scenario for a new phase in between the paramagnetic and ferromagnetic phases was the
suggestion that the electron effective mass diverges at some distance from
criticality, leading to a new phase on the paramagnetic side of the 
transition \cite{Khodel_Shaginyan_Zverev_1997, Shaginyan_2003}.
It was shown by \textcite{Chubukov_2005} that this effect is an artifact of a purely static
electron-electron interaction, and that no such phase exists if the retardation of the interaction 
is taken into account.

\subsection{Other mechanisms for a first-order transition}
\label{subsec:III.F}
The mechanism for driving the quantum ferromagnetic transition in clean systems first order
that was discussed in Sec.\ \ref{subsubsec:III.B.2} is remarkable because of its universality.
However, in any given material less universal mechanism may be present that by themselves
would suffice to drive the transition first order. Here we briefly discuss two such mechanisms.

\subsubsection{Band structure effects}
\label{subsubsec:III.F.1}

The coefficients in the Landau free energy (cf. Eq.\ (\ref{eq:3.29a}))
\be
f_{\rm L}[m] = t\,m^2 + u_4\,m^4 + u_6\,m^6 + O(m^8)
\label{eq:3.F.1}
\ee
depend in complicated ways on the microscopic details of the system, and in particular on 
the details of the band structure. In any given material it is possible that band-structure
effects lead to a negative value of $u_4$. If $u_6>0$, this leads to a first-order transition
at some positive value of $t$, which pre-empts the second-order transition at $t=0$. Under
certain conditions, correlations can have the same effect \cite{Yamada_1993}. However,
this cannot explain the universality of the observed effect in clean low-temperature ferromagnets
that is displayed by Tables~\ref{table:1a},~\ref{table:1b}.

It is interesting to note that UGe$_2$, in addition to the pressure-induced first-order paramagnet-to-ferromagnet
transition at $p\approx 16\,$kbar, shows a metamagnetic transition at a lower pressure that is also of first order. This transition, as
well as the superconductivity that coexists with the ferromagnetism at low temperatures and intermediate
pressures, has been attributed to a special feature in the density of states of UGe$_2$ 
\cite{Sandeman_Lonzarich_Schofield_2003, Pfleiderer_Huxley_2002, Shick_et_al_2004a, Shick_et_al_2004b}.

\subsubsection{Magnetoelastic effects}
\label{subsubsec:III.F.2}

Phonons are generic soft modes in the sense of Sec.\ \ref{subsubsec:III.A.1} that couple to the
magnetization. It has been known for a long time that this can lead, under certain conditions, to a weakly first-order transition in
classical magnets \cite{Rice_1954, Bean_Rodbell_1962, Sak_1974, Bergman_Halperin_1976, deMoura_et_al_1976, 
Wegner_1974, Larkin_Pikin_1969}. 
We briefly review the conclusions for classical magnets, and then discuss the relevance of these results for quantum ferromagnets.

\paragraph{Classical magnets}
\label{par:III.F.2.a}

Consider an LGW theory for a ferromagnet with order parameter ${\bm M}$ that couples to harmonic elastic degrees
of freedom. In the simplest case of an isotropic three-dimensional system in the continuum limit the action reads \cite{Aharony_1976}
\bea
S &=& \int d{\bm x} \Bigl[t{\bm M}^2 + (\nabla{\bm M})^2 + u_4{\bm M}^4 + (\frac{K}{2} - \frac{\mu}{3})(\sum_{\alpha=1}^3
   u_{\alpha\alpha})^2 
\nonumber\\
&& + \mu \sum_{\alpha,\beta} u_{\alpha\beta}^2 + g\,{\bm M}^2\sum_{\alpha} u_{\alpha\alpha}\Bigr]\ .
\label{eq:3.F.2}
\eea
Here $K$ and $\mu$ are elastic coefficients, and 
\be
u_{\alpha\beta} = \frac{1}{2}\left(\partial_{\beta}u_{\alpha} + \partial_{\alpha}u_{\beta} + \sum_{\gamma} \partial_{\alpha}u_{\gamma}
   \partial_{\beta}u_{\gamma}\right)
\label{eq:3.F.3}
\ee
is the strain tensor in terms of derivatives of the displacement vector ${\bm u}({\bm x})$. $g$ is the magnetoelastic coupling
constant. In systems on a lattice there are additional terms \cite{Bergman_Halperin_1976, deMoura_et_al_1976}, but the
general structure of the action is the same. For systems at constant pressure, additional terms coupling the pressure to the
strain tensor need to be added \cite{Imry_1974}.

There are several important features of this action. First, the coupling is to the square of the order parameter. Second,
the coupling is to the divergence of the soft mode, i.e., the displacement vector. This is in sharp contrast to the case
of the smectic order parameter coupling to the nematic Goldstone modes at a nematic-to-smectic-A transition, or 
the superconducting order parameter coupling to the electromagnetic vector potential \cite{Halperin_Lubensky_Ma_1974}.
In both of these cases, the coupling is directly to a soft mode, which leads to a nonanalytic dependence of the free energy
on the order parameter in a renormalized Landau theory. Here, by contrast, the coupling is much weaker due to the
additional gradient, and the net effect of the elastic modes are additional terms of quartic order in the order parameter. 
Schematically, one can see this by replacing the strain tensor $u_{\alpha\beta}$ by a scalar $\epsilon$ and considering
a Landau free energy
\be
f[m,\epsilon] = t\,m^2 + u_4 m^4 + K\epsilon^2 + gm^2\epsilon\ .
\label{eq:3.F.4}
\ee
By decouplng $m$ and $\epsilon$ we see that the magnetic transition in mean-field approximation is first order
if $g^2 > 4Ku_4$. A detailed study of the nature of the phase transition as described by the LGW action (\ref{eq:3.F.2}) and
its generalizations has been done by \textcite{deMoura_et_al_1976}, who integrated out the elastic degrees of freedom, and by
\textcite{Bergman_Halperin_1976}, who performed an RG analysis of the full coupled theory. The conclusion is consistent
with the simple argument above: For a sufficiently large magnetoelastic coupling the transition may become first order, 
but whether or not this occurs depends on the bare values of the parameters in the LGW theory, i.e., on microscopic details, 
as well as on the dimensionality of the order parameter \cite{Nattermann_1977}. Magnetoelastic effects are {\em a} route to a 
first-order transition; however, they do not provide a universal route.

\paragraph{Quantum magnets}
\label{par:III.F.2.b}

\textcite{Gehring_2008} (see also \textcite{Gehring_Ahmed_2010}) and \textcite{Mineev_2011} have proposed to apply
the above results for classical magnets to the quantum ferromagnetic transition in pressure-driven systems by 
generalizing the Landau free energy (\ref{eq:3.F.4}) to
\be
f[m,\epsilon] = t(\epsilon) m^2 + u_4 m^4 + K\epsilon^2\ ,
\label{eq:3.F.5}
\ee
with $t(\epsilon) = T - T_{\text{c}}(\epsilon)$ representing the dependence of the transition temperature on the strain
(or, equivalently, on the pressure $p$). Expanding $T_{\text{c}}(\epsilon)$ for small $\epsilon$ then leads to the coupling
given in Eq.\ (\ref{eq:3.F.4}) with $g \propto dT_{\text{c}}/dp$. Since experimentally one finds $dT_{\text{c}}/dp \to \infty$
as $T_{\text{c}} \to 0$, these authors have argued that effectively the magnetoelastic coupling $g$ increases without
bounds as $T_{\text{c}}$ decreases, necessarily leading to a first-order transition at sufficiently low transition temperatures.
This line of reasoning is problematic for various reasons. First, a singular dependence of a coefficient on a field must
not be built into a Landau theory if the theory is to have any predictive value. Such a singular dependence may 
result from integrating out soft modes, such as in the treatment of classical liquid crystals or superconductors 
mentioned above \cite{Halperin_Lubensky_Ma_1974}, or in the renormalized Landau theory reviewed in Sec.\ \ref{subsec:III.B}.
In the case of compressible magnets, such a result is implausible. The coupling between the magnetic order parameter 
and the elastic deformations is weak even in the classical case, as explained above, and in the quantum case it will be
even weaker due to an additional frequency integral, which amounts to an effective extra gradient by power counting.
Second, a diverging effective magnetoelastic coupling results in a diverging volume change \cite{Bean_Rodbell_1962}.
This means that even if one accepts the substitution of the observed $T_{\text{c}}(p)$ into the Landau theory, it predicts
that a structural phase transition must necessarily accompany the first-order magnetic transition. There is no experimental
evidence for this. We conclude that currently no convincing theory for magnetoelastic effects in the quantum regime exists.
\section{Summary, Discussion, and Outlook}
\label{sec:IV}

In this section we conclude with a summary of the topics covered, some additional discussion points,
and a list of open problems. % Summary, Discussion and Outlook
\subsection{Summary, and Discussion}
\label{subsec:IV.A}
We have given an overview of the quantum phase transition problem
in metallic ferromagnets. Experimentally, a number of interesting
phase diagrams are observed, the structure of which is summarized
in Fig.~\ref{fig:schematic_phase_diagrams}. Apart from discontinuous
(first-order) and continuous (second-order) quantum phase transitions from a ferromagnet
to a paramagnet, a QPT from a FM state to an AFM or spin-wave state is observed in
some systems, while in others
the low-temperature phase near the onset of ferromagnetism is some
sort of a magnetic glass. In many systems with quenched disorder there
is evidence for quantum Griffiths effects on the paramagnetic side of the
transition. The experimental results are described in Sec.\ \ref{sec:II},
organized with respect to the structure of the phase diagram.

Theoretically, the transition from a paramagnet to a homogeneous
quantum ferromagnet is expected to be discontinuous in clean systems,
and continuous in disordered ones. In either case the behavior at the
quantum phase transition is very different from the one expected from
conventional Hertz theory. This is because of a coupling between the
magnetization and soft fermionic excitations in metals that was included
in Hertz theory in too simple an approximation and treated more thoroughly
in the theory originally developed by two of us and Thomas Vojta that is
reviewed in Sec.\ \ref{sec:III}. The results obtained by
Moriya, Hertz, and Millis are still expected to be observable in certain
pre-asymptotic regimes. The agreement between these theoretical
predictions and experimental results are generally very good for clean
systems. Strongly disordered systems are much more complicated.
Although the critical singularities at the continuous quantum ferromagnetic
transition have been calculated exactly, Griffiths-phase effects coexist
with the critical singularities and complicate the experimental analysis.

Before we conclude with a list of open problems, we add some further
remarks to the discussion in the main body of the review and mention
some related topics that we did not cover. The references in this
section are intended to be illustrative, rather than exhaustive.

%
%\smallskip\par\noindent
\begin{enumerate}[leftmargin=*, labelindent = 0pt, itemindent=11pt]
\item
Nematic phases in a Fermi fluid, in particular the nematic Goldstone modes, as well as the 
associated phase transitions to a Fermi liquid, have been investigated theoretically by
\textcite{Oganesyan_Kivelson_Fradkin_2001}.
They used a Hertz-type theory, which yields a continuous transition with mean-field
critical behavior in spatial dimensions $d=2$ and $d=3$ for all types of nematics considered. The case of a 
metallic spin-nematic, or non-s-wave ferromagnet, is theoretically closely related to the 
metallic ferromagnetic one. For such systems in the absence of quenched disorder 
it was later shown that the same mechanism operative in ferromagnets generically 
causes the spin-nematic transition to be of first order
\cite{Kirkpatrick_Belitz_2011a}. It is likely that the transition in disordered magnetic nematics is also
related to the corresponding one in ferromagnets, but this has not
been investigated so far.

For charge nematics the situation is different, since the mechanism leading to a 
first-order transition does not apply \cite{Belitz_Kirkpatrick_Vojta_2002}. Still, later work showed that the Hertz approach
breaks down close to the transition even in this case, but the breakdown is less
dramatic than in the spin channel and the transition is believed to remain continuous \cite{Metlitski_Sachdev_2010, DellAnna_Metzner_2006,
Lee_2009}. 

There is experimental evidence of charge Ising-nematic order in a number
of systems including the pnictides \cite{Chuang_et_al_2010}, Sr$_{3}$Ru$_{2}$O$_{7}$ \cite{Borzi_et_al_2007}, 
and the normal state of the cuprate superconductors, in particular YBa$_{2}$Cu$_{3}$O$_{y}$ \cite{Daou_et_al_2009}.
For a review, see \textcite{Fradkin_et_al_2010}.

\item
A point that sometimes leads to confusion is related to the models 
used by theoretical studies of the ferromagnetic QPT. \textcite{Hertz_1976} considered a simple continuum
model of free electrons that interact via a point-like spin-triplet interaction.
There are good reasons to believe that such a model does not actually
have a FM phase, see point 4. in Sec.\ \ref{subsec:IV.B}. However, the
point of an effective field theory such as Hertz's is {\em not} to
establish whether or not there is a phase transition in this, or any, model. 
Rather, it is to describe the properties of the transition, provided 
one actually occurs. The complicated band structure and other microscopic 
details that may well be necessary to produce a transition in the first place do
not affect the universal properties at the transition, and therefore can
safely be omitted from the effective theory.

More recent theories, such as \textcite{Kirkpatrick_Belitz_2012b}, consider
an effective order-parameter theory that has the existence of a magnetic
transition encoded in the parameters of the effective LGW functional. All
details of the solid-state structure that are necessary for ferromagnetism to
occur are thus hidden in these parameters. The order parameter is then
coupled to fermions, and for capturing the qualitative effects of the latter
on the FM transition again a simple continuum model suffices.

\item The experimentally observed first-order QPT in clean metallic FMs
is surprisingly robust in the light of its theoretical interpretation as a
fluctuation-induced first-order transition. For classical phase transitions,
fluctuation-induced first-order transitions have proven hard to realize.
As discussed in Sec.\ \ref{subsubsec:III.B.2} this robustness is likely the result
of the order-parameter fluctuations in the quantum case being above their
upper critical dimension, in contrast to the classical systems where a
similar mechanism is predicted to occur. This implies that the renormalized
Landau theory described in Sec.\ \ref{subsubsec:III.B.1} has a much wider
range of validity in the quantum case than analogous theories for classical
systems. 

Even in the light of the above remarks, the near-universal observation of a
first-order QPT in clean ferromagnets is surprising, given that the term in the
renormalized Landau theory that is presumed responsible for it is
logarithmic, which results in an exponential dependence of observables on
parameters. It is possible that, perhaps as a result of strong electron
correlations, an analog of van der Waals's law of corresponding states for
classical liquids holds for strongly correlated Fermi liquids, making the relevant
parameters, measured in natural units, roughly the same in different materials. This notion is
consistent with the discussion in Sec.~\ref{subsubsec:II.B.5}, and especially
with the fact that the tricritical temperature scales roughly with the magnetic
moment.
%
%\item There are rigorous results to the effect that, strictly speaking, any amount 
%of disorder will destroy a first-order transition, classical or quantum, see
%\textcite{Imry_Ma_1975, Aizenman_Greenblatt_Lebowitz_2012} and
%references therein. This can be understood in the context of the scaling description of
%first-order transitions by \textcite{Fisher_Berker_1982} (see footnote
%\ref{first_order_footnote}), which says that the correlation-length exponent $\nu$ at a 
%classical first-order transition is equal to $1/d$, with $d$ the spatial dimensionality. In 
%the quantum case, this gets generalized to $\nu = 1/(d + z) > 1/d$,
%with $z$ an appropriate dynamical exponent. In either case, this
%violates the rigorous bound $\nu > 2/d$ for a system with quenched disorder 
%\cite{Harris_1974, Chayes_et_al_1986}, which implies that the fixed point describing 
%the first-order transition is unstable with respect to disorder. In practice, one expects 
%the resulting rounding of the first-order transition to be very weak for weak disorder.
%These considerations notwithstanding, it is possible that in some disordered systems a 
%random first-order transition may be realized where the finite-size scaling exponent 
%$\nu$ considered by \textcite{Fisher_Berker_1982} is not equal to $1/d$.
%
\item
In Sections \ref{sec:II} and \ref{sec:III} we emphasized that experimental
observations of continuous ferromagnetic QPTs in strongly disordered systems
are often difficult to interpret, and the critical exponents that characterize these 
transitions are hard to measure. There are, however, qualitative features of
both theoretical results and experimental observations that indicate that
several exponents are drastically different from both the mean-field exponents
expected in a pre-asymptotic regime in weakly disordered systems, and 
classical exponents in common universality classes. 

For instance, the order-parameter exponent $\beta$ is predicted to be larger
than unity (about 1.2) in the pre-asymptotic regime where an effective power-law behavior
is expected, whereas the exponent $\delta$ is unusually small (about 1.8), see
Eq.\ (\ref{eq:3.91c}). In contrast, the mean-field values are $\beta = 1/2$ and
$\delta = 3$, and the corresponding classical values for three-dimensional Heisenberg
ferromagnets are about 0.37 and 4.8, respectively. Experiments do indeed
tend to find values of $\beta$ and $\delta$ that are larger and smaller,
respectively, than their respective mean-field values, see Secs.\ \ref{subsubsec:II.C.2}
and \ref{par:III.C.3.g}. A related issue is the shape of the phase boundary near
the QPT, with both theory and experiments concluding that there is a ``tail''
in the phase diagram, see the discussion in Sec. \ref{par:III.C.3.g}. Griffiths
effects may also contribute to the observed properties in this region, which
makes more detailed investigations desirable.

This superposition of critical phenomena and additional disorder effects
notwithstanding, the results reviewed in Sec.\ \ref{par:III.C.3.e} for the
critical behavior of an FM order parameter coupled to diffusive fermions
are believed to be exact. This type of problem also appears elsewhere. 
For instance, \textcite{Savary_Moon_Balents_2014} have considered a model
for pyrochlore iridates that couples a quantum $\varphi^4$ theory to
(in this case exotic) fermions, which results in a phase transition
with similarly unusual critical properties. 
\item
Even far away from any QPT ferromagnetic metals at low
temperatures have very interesting properties. This is generally not as
well appreciated as the problems posed by AFMs, or by FMs near a QPTs. For 
instance, in many clean FMs a {\em generic} (i.e., existing
in an entire phase) non-Fermi liquid $T^{3/2}$ resistivity is observed over a large 
low-temperature range in both the FM phase and the
paramagnetic phase \cite{Brando_et_al_2008, Sato_1975, Pfleiderer_et_al_2001, 
Niklowitz_et_al_2005, Takashima_et_al_2007}. This behavior is not well understood,
see Sec.\ \ref{subsec:IV.B}.

In disordered systems, Griffith effects
lead to generic non-Fermi-liquid behavior on the paramagnetic side
of the phase boundary as was discussed in Section III.D. In either
phase, weak-localization \cite{Lee_Ramakrishnan_1985} and Altshuler-Aronov \cite{Altshuler_Aronov_1984},
effects are expected in disordered systems. The resulting superimposed temperature
dependences of observables can be quite intricate \cite{Butenko_Bolshutkin_Pecherskaya_1990}
but in general little attention has been paid to them.

\item
There has been interesting work on magnetic phase transitions in
ferromagnetic metals under non-equilibrium conditions \cite{Mitra_et_al_2006, Mitra_Millis_2008}. 
In these systems the fermionic soft modes that play a central role
in the theory discussed in much of Sec.\ \ref{sec:III} are suppressed by boundary effects.
As a consequence a Hertz-type non-equilibrium transition has been
predicted. Non-equilibrium systems are in general very interesting
because correlations are generically greatly enhanced compared to
equilibrium systems \cite{Belitz_Kirkpatrick_Vojta_2005}.

\item
In some metallic FM materials, glassy behavior reminiscent of supercooled
viscous liquids has been observed. In particular, in Sr$_{1-x}$La$_{x}$RuO$_{3}$ a cluster-glass state with a
very broad distribution of relaxation times has been found and the time scale
of the freezing of magnetic clusters is well described by a Vogel-Fulcher
law \cite{Kawasaki_et_al_2014}. It would be interesting to examine aging in this and
similar systems, as well as search for dynamic heterogeneity in them.
Both of these concepts are central in the current understanding of
glassy behavior in supercooled liquids \cite{Parisi_Zamponi_2010, Berthier_Biroli_2011}.

\item
Very unusual phases are expected in systems where both 
electronic correlations and a strong spin-orbit interaction are present \cite{Wan_et_al_2011}.
In particular, topological semi-metal phases can occur which may be
realized in, for example, pyrochlore iridates such as
Y$_2$Ir$_2$O$_7$ \cite{Wan_et_al_2011}, Bi$_2$Se$_3$ \cite{Zhang_et_al_2009},
HgCr$_2$Se$_4$ \cite{Xu_et_al_2011}, or hetrostructures of topological
and normal insulators \cite{Burkov_Balents_2011}. This semi-metal state
is a three-dimensional analog of graphene and provides a condensed-matter 
realization of Weyl fermions that obey a two-component Dirac equation. 
Calculations for these systems based on the $LSDA+U+SO$ method (local spin-density approximation plus
correlations plus spin-orbit coupling) have suggested
a very rich phase diagram, including a QPT between a ferromagnetic
metal and a Weyl semi-metal \cite{Wan_et_al_2011}. The nature of this transition has not
been investigated.

Weyl semi-metals also have a number of interesting properties apart
from any QPT. For instance, ideas associated with them
have been used to understand the intrinsic anomalous Hall effect in
metallic ferromagnets \cite{Chen_Bergman_Burkov_2013}. These authors argue that
even Weyl nodes that do not coincide with the Fermi energy, as is believed to be the
case in SrRuO$_3$, contribute significantly to the intrinsic anomalous
Hall conductivity in ferromagnetic metals. This in turn implies that
this conductivity in ferromagnets is not purely a Fermi-surface property, which
contradicts earlier conclusions \cite{Haldane_2004}.

\item
Ferromagnetic phase transitions have been observed in a variety
of quantum Hall systems. For instance, in a gallium arsenide system in a perpendicular
magnetic field, \textcite{Piazza_et_al_1999} observed a first-order phase transition
in the $\nu=2$ and $\nu=4$ quantum Hall states. They suggested that
the source of the observed hysteresis effects was not exotic, but
was due to the expected domain structure in an easy-axis
ferromagnet. Similar behavior was observed by \textcite{DePoortere_Tutuc_Shayegan_2003} in aluminum
arsenide quantum wells. \textcite{Drichko_et_al_2012} measured magnetoresistance
properties in two p-Si/SiGe/Si quantum-well samples in a tilted
magnetic field. In a sample with p = $2\times 10^{11}$ cm$^{-2}$ they concluded there was a first-order
ferromagnetic-paramagnetic phase transition and observed phase coexistence. However, in the second
sample with p = $7.2\times 10^{10}$ cm$^{-2}$ no transition was observed. 
Stoner or RPA-like theories have been used to discuss ferromagnetic
phase transitions in quantum Hall systems \cite{Burkov_MacDonald_2002, Lopatnikova_Simon_Demler_2004},
and for the pseudospin ferromagnet realized in bilayer Quantum Hall systems
there is evidence for a first-order transition \cite{Lee_et_al_2014, Schliemann_Girvin_MacDonald_2001, Zou_et_al_2010}.

\end{enumerate}

\subsection{Open problems}
\label{subsec:IV.B}
There are a number of interesting open problems concerning the QPT
in metallic ferromagnets as well as the generic properties of low
temperature ferromagnetic metals. Here we mention and discuss some
of them.
\begin{enumerate}[leftmargin=*, labelindent = 0pt, itemindent=11pt]
\item
Additional experimental and theoretical work is needed to disentangle
Griffiths singularities and critical singularities near the ferromagnetic
QPT in disordered metals. Since Griffiths singularities generically are
stronger on the paramagnetic side of the transition \cite{Motrunich_et_al_2000}, the QPT should, if possible,
be studied from the ferromagnetically ordered side of the transition. Specifically,
although Griffiths singularities exist in the thermal and magnetic response functions on
the paramagnetic side of the transition, and in weaker forms also on the ferromagnetic side, the existence of a zero-field magnetization
uniquely implies long ranged FM order, so the singular behavior the zero-field
magnetization itself can distinguish between Griffiths singularities and critical singularities.
The relation between Griffiths physics and the Harris criterion has been discussed by
\textcite{Vojta_Hoyos_2014, Vojta_Igo_Hoyos_2014}.

We also mention that the non-Fermi-liquid behavior observed in many
clean materials in large parts of the phase diagram is very interesting.
This topic has been reviewed by \textcite{Stewart_2001}, and it remains
incompletely understood. One manifestation of non-Fermi liquid behavior
that is observed in many materials is the $T^{3/2}$ behavior of the
resistivity that was mentioned in Sec.\ \ref{subsec:IV.A}. An explanations in terms of
columnar fluctuations, which is applicable to MnSi, has been proposed by
\textcite{Kirkpatrick_Belitz_2010}. However, because of the large variety
of materials where a $T^{3/2}$ resistivity is observed, it is likely that there
is more than one mechanism that can lead to this behavior. For a related
discussion of ZrZn$_2$, see \textcite{Smith_et_al_2008}.

Similarly, weak-localization and Altshuler-Aronov
effects in weakly disordered ferromagnets deserve more attention than they
have received. The $T$-dependence of the resistivity can be very complicated, 
with many contributions from very different sources, see, e.g.,  
\cite{Butenko_Bolshutkin_Pecherskaya_1990, Mizutani_et_al_1988,Yildiz_et_al_2009}.
\item
In the presence of magnetic impurities, or impurities with a large spin-orbit
coupling, the relevant soft fermionic modes in the disordered case
will be suppressed \cite{Lee_Ramakrishnan_1985, Belitz_Kirkpatrick_1994}, 
and the nature of the ferromagnetic QPT is unclear. It is possible that, once
the generic soft modes have been eliminated, the transition will resemble
the one in disordered AFM metals, but not much is known about this case.
\item
There are materials in which no FM transition has been observed, but that nonetheless display very interesting properties that warrant further investigations. One of these is \YFA. It crystallizes in the eponymous orthorhombic structure with a single Fe site~\cite{Kerkau_et_al_2012}. Initial experiments % have 
identified correlated FM behavior~\cite{Strydom_Peratheepan_2010}. Further detailed studies on single crystals found anomalies in the magnetic susceptibility and the specific heat which obey a peculiar NFL field-temperature scaling~\cite{Park_et_al_2011,Wu_et_al_2014}. In addition, FM correlations have been found in NMR experiments~\cite{Khuntia_et_al_2012}. These observations have been interpreted as indicating that the material is close to a FM quantum phase transition. 
%, despite the FM phase transition having escaped detection thus far down to 50\,mK, and even when doped with a small surplus of Fe~\cite{Strydom_et_al_2013}. 
However, no FM transition has been detected thus far at temperatures down to 50\,mK, even upon doping with a small surplus of Fe~\cite{Strydom_et_al_2013}.
On the contrary, Fe excess or deficiency drive \YFA\ away from the critical behavior. The low-$T$ resistivity shows a Kondo-like logarithmic increase below 30\,K with a high $\rho_{0} \approx 75$\,$\mu\Omega$cm (RRR $\approx 2$), which puts \YFA\ in the group of strongly disordered systems (cf. Sec.~\ref{subsubsec:II.C.2}). Nonetheless, single-crystal structure refinement did not find any deviation from the ideal composition~\cite{Kerkau_et_al_2012}, so the origin of the large resistivity is not clear. The observed scaling behavior and the lack of a magnetically ordered phase in \YFA\ need further investigations. We also mention that in the closely related system YbFe$_2$Al$_{10}$ strong FM correlations have been observed at low temperatures \cite{Khuntia_et_al_2014}. In this material, the Yb-derived electrons at low temperature form a nonmagnetic intermediate-valent state and therefore the Fe atoms alone are responsible for the FM correlations, as is the case in \YFA.

\item
There are materials that display a transition from a metallic AFM state to a FM
at low temperatures. Two examples are CeRu$_2$Ge$_2$ \cite{Raymond_et_al_1999},
and CeRu$_2$Al$_2$B \cite{Baumbach_et_al_2012}. In both cases, the AFM-FM
transition is first order, whereas the transition from a paramagnet to an AFM at a
higher Neel temperature is second order. It is plausible that the QPT from a metallic
AFM to a FM in clean systems is first order for the same reasons as that from a
metallic paramagnet to a FM, but no theory is available for this case.

A related, and even more interesting, issue is the detailed structure
of the phase diagrams discussed in Sec.\ \ref{subsec:II.D}. These systems all must display a Lifshitz point,
and at least two QPTs. In clean systems, the QPT from the FM phase to
the modulated phase is expected to be first order, but this needs experimental confirmation.
In disordered systems, it may well be a novel type of QCP. Similarly,
the Lifshitz point may be a multicritical point with very interesting properties. These questions
have not received the attention they deserve, either experimentally or
theoretically.
\item
The question of what are necessary ingredients in a model to produce
itinerant ferromagnetism is a very old one (see, e.g., the discussion by \textcite{Varma_2010},
or \textcite{Shimizu_1964} and references therein). It has long been suspected
that in simple electron-fluid models there is no ferromagnetic phase in 
the phase diagram \cite{Ceperley_Alder_1980, Chang_Zhang_Ceperley_2010},
although some recent Quantum Monte Carlo studies suggest otherwise
\cite{Pilati_et_al_2010, Pilati_Zitchenko_Troyer_2014}. This topic has 
received much attention recently in the context of optical lattices, especially an
experiment that reported itinerant ferromagnetism in a Fermi gas of
ultra cold atoms \cite{Jo_et_al_2009}. However, subsequent experiments
by the same group cast doubt on the original interpretation of the 
data \cite{Sanner_et_al_2012}. Ferromagnetic solid-state systems
typically have  fairly complicated band structure with multiple
conduction bands. Whether or not ferromagnetism can occur in 
simpler systems, such as optical lattices, is still an open question.
If it does, the transition is expected to be first order for the same
reasons as in clean solid-state systems \cite{Duine_MacDonald_2005},
and a quantum Monte
Carlo study of a two-dimensional Stoner Hamiltonian suggests
that the strength of the first-order transition may depend on the
range of the interaction \cite{Conduit_2013}.
\item
Quenches, i.e., rapid changes of external parameter values, at zero
temperature in both clean and disordered metallic ferromagnets are
an interesting topic. \textcite{Belitz_Kirkpatrick_Saha_2007b} have shown 
that the coupling of the order parameter to the fermionic soft modes leads 
to qualitatively new effects for the late-stage coarsening. \textcite{Gagel_Orth_Schmalian_2014}
have pointed out that there is universal pre-asymptotic behavior in general
quantum quench problems due to long-range boundary effects. In FM metals this
effect is expected to be even more interesting because of the coupling
to the fermionic soft modes.
\item
There is no satisfactory theory of magneto-elastic effects in metallic
quantum ferromagnets. While the effects are expected to be rather
weak, see Sec. \ref{subsubsec:III.F.2}, they may be relevant, for instance, for
a complete understanding of the metamagnetic transition in UGe$_2$,
where a small volume change of the unit cell accompanies the magnetic
transition \cite{Shick_et_al_2004b}. Magnetoelastic effects may also be
important for stabilizing the superconductivity observed in UGe$_2$ \cite{Sokolov_et_al_2011}.
While magnetostriction effects have
been discussed as a possible reason for the tricritical point in UGe$_2$
\cite{Tateiwa_et_al_2014}, there are strong theoretical reasons against
such an interpretation, see the discussion in Sec. \ref{subsubsec:III.F.2}.
\item
 An interesting theoretical problem is to study entanglement
in quantum ferromagnets, in particular the effects on the entanglement
from the additional fermionic soft modes. Entanglement in systems
with Goldstone modes has been discussed by \textcite{Metlitski_Grover_2011}.
%
%\item
%A RG description of the fluctuation-induced first-order transition in quantum
%ferromagnets, in analogy to the theory for compressible classical magnets
%by \textcite{Bergman_Halperin_1976}, would be an interesting complement
%to the equation-of-state approach discussed in Sec.\ \ref{sec:III}. In an RG analysis
%of the coupled field theory discussed in Sec.\ \ref{subsubsec:III.C.1} one would
%expect, in the clean case, marginal operators to drive negative the coefficient of the
%quartic term in the order-parameter sector of the section, but this has never been
%checked.
%
\item
The detailed nature of quantum first-order transitions has
not been as thoroughly investigated, either theoretically or experimentally,
as the nature of classical first-order transitions. 
For example, the experimental coexistence curve
appears to be extremely steep in many FM systems, see, e.g., Figs.
\ref{figure:UGe_2_p-T_phase_diagram}, \ref{fig:UCoAl_phase_diagram},
and the phase diagrams for ZrZn$_2$ measured by \textcite{Uhlarz_Pfleiderer_Hayden_2004}
and \textcite{Takashima_et_al_2007}, but determining the coexistence curve from
different observables can lead to different detailed shapes \cite{Kabeya_et_al_2010}. 
Studies of the detailed shape,
by pressure-cycling in the $p\,$-$T$ plane, or field-cycling in the $p\,$-$H$ plane,
would be interesting. Theoretically, the shape of the coexistence curve can
be determined from the Clapeyron-Clausius equation, which has been discussed
for quantum Hall systems by \textcite{Zou_et_al_2010} and for QPTs in general and FMs in particular by \textcite{Kirkpatrick_Belitz_2015a}.

%The detailed nature of quantum first-order transitions has not been as thoroughly investigated, either theoretically or experimentally, as the nature of classical first-order transitions. 
%
%For example, a detailed study of the precise shape of the coexistence curve in the $p\,$-$T$ phase diagram for a pressure-tuned first-order phase transition would be interesting. Experimentally, the coexistence curve appears to be extremely steep in many FM systems, see, e.g., Figs. \ref{figure:UGe_2_p-T_phase_diagram}, \ref{fig:UCoAl_phase_diagram}, and the phase diagrams for ZrZn$_2$ measured by \textcite{Uhlarz_Pfleiderer_Hayden_2004} and \textcite{Takashima_et_al_2007}. A confirmation of this very steep shape, by pressure-cycling in the $p\,$-$T$ plane, or field-cycling in the $p\,$-$H$ plane, would be interesting. Theoretically, the shape of the coexistence curve can be determined from the Clapeyron-Clausius equation, which has been discussed for a first-order transition in quantum Hall systems by \textcite{Zou_et_al_2010}.
%
\item
Without trying to be exhaustive, we mention a few other FM materials that may be interesting candidates for suppressing \TC\ via pressure or chemical substitution: NpNiSi$_{2}$, a Kondo-lattice system with \TC\,= 51.5\,K \cite{Colineau_et_al_2008}; Sr$_4$Ru$_3$O$_{10}$, a layered ferromagnet with \TC\ = 148\,K (\onlinecite{Cao_et_al_1997,Crawford_et_al_2002}, see also Sec.\ \ref{subsubsec:SrRuO}); the enhanced paramagnet TiBe$_{2}$ which shows metamagnetism at 5\,T~\cite{Wohlfarth_1980}, and the series TiBe$_{2-x}$Cu$_{x}$ which shows a transition to a FM ordered state~\cite{Giorgi_et_al_1979,Acker_et_al_1981}. The latter system was intensively investiaged in the early 1980s, but a detailed and conclusive phase diagram does not exists. Since recent band-structure calculations~\cite{Jeong_et_al_2006} suggest that TiBe$_{2}$ is close to an AFM instability, it would be interesting to revisit the phase diagram of TiBe$_{2-x}$Cu$_{x}$.  
% Recently neutron depolarization
% imagining has been used to give detailed real-space information on
%the existence and nature of phase separation in $Pd_{1-x}Ni_{x}$,
%$CePd_{1-x}Rh_{x}$ and $NbFe_{2}$ at their ferromagnetic quantum
%phase transitions. 

\end{enumerate}

%**************************************************************************************************************************
%
%**************************************************************************************************************************
%
\appendix
\section{List of acronyms}
\label{app:A}

\begin{tabular}{ll}
AFM & antiferromagnet, or antiferromagnetic\\
CDW & charge-density wave\\
CEF & crystalline electric field \\
CEP & critical end point \\
DIV & dangerous irrelevant variable \\
FM & ferromagnet, or ferromagnetic\\
LGW & Landau-Ginzburg-Wilson \\
PM & paramagnet, or paramagnetic\\
RG   & renormalization group \\
QCP & quantum critical point \\
QCEP & quantum critical end point \\
QGP & Quantum Griffiths phase \\
QPT & quantum phase transition\\
RRR & residual resistance ratio\\
SC & superconductivity
\end{tabular}

\section{Definitions of critical exponents}
\label{app:B}

Let $T$ be the temperature, $t$ the dimensionless distance from criticality at $T=0$, and $h$ the magnetic field.
Consider the correlation length $\xi$, the magnetization $m$, the magnetic susceptibility $\chi$, and the
specific-heat coefficient $\gamma = C/T$ as functions of $t$, $T$, and $h$, and the susceptibility also as a function
of the wave number $k$. We define critical exponents as
follows.
\medskip\par\noindent
{\it Correlation length:} 
\be
\xi(t\to 0,T=0) \propto \vert t\vert^{-\nu}\quad,\quad \xi(t=0,T\to 0) \propto T^{-\nu_T}\ .
\label{eq:B.1}
\ee
\par\noindent
{\it Order parameter:}
\bea
m(t\to 0,T=0,h=0) &\propto& (-t)^{\beta}\ ,
\nonumber\\
m(t=0,T\to 0,h=0) &\propto& T^{\beta_T}\ ,
\nonumber\\
m(t=0,T=0,h\to 0) &\propto& h^{1/\delta}\ .
\label{eq:B.2}
\eea
\par\noindent
{\it Order-parameter susceptibility:}
\bea
\chi(t\to 0,T=0;k=0) &\propto& \vert t\vert^{-\gamma}\ ,
\nonumber\\
\chi(t=0,T\to 0;k=0) &\propto& T^{-\gamma_T}\ ,
\nonumber\\
\chi(t=0,T=0,k\to 0) &\propto& 1/k^{2-\eta}\ .
\label{eq:B.3}
\eea
\par\noindent
{\it Specific-heat coefficient:}
\be
\gamma(t\to 0,T=0) \propto \vert t\vert^{-{\bar\alpha}}\quad,\quad \gamma(t=0,T\to 0) \propto T^{-{\bar\alpha}_T}\ .
\label{eq:B.4}
\ee

$\nu$, $\beta$, $\gamma$, $\delta$, and $\eta$ are defined in analogy to the corresponding exponents
at a classical phase transition. The definition of ${\bar\alpha}$ deviates from the one of the classical
exponent customarily denoted by $\alpha$, which is defined in terms of the specific heat rather than
the specific-heat coefficient.
This is necessary in order to factor out the factor of $T$ in the relation between the specific heat
and the specific-heat coefficient, which makes no difference at a thermal phase transition, but goes to
zero at a QCP. For instance, the thermodynamic identity that underlies the Rushbrooke inequality,
Eq.\ (\ref{eq:3.100}), has no explicit $T$-dependence only of it is formulated in terms of specific-heat
coefficients rather than the specific heats. At a classical phase transition, $\bar\alpha$ 
coincides with $\alpha$.
${\bar\alpha}_T$, $\nu_T$, $\beta_T$, and $\gamma_T$ reflect the fact that a QPT can
be approached either in the $T=0$ plane, or from $T>0$. The definition of $\beta_T$ in Eq.~(\ref{eq:B.2}) 
is purely formal; $\beta_T$ cannot be observed via the $T$-dependence of $m$ at $t=0$. It does, 
however, determine the scaling behavior of the temperature derivative of $m$, which is observable, 
see \textcite{Kirkpatrick_Belitz_2015b}.

\noindent 
\acknowledgments
We have benefitted from collaborations, discussions, and correspondence with many colleagues, including 
%%%%%%%%%%%%%%%%%%%%%%%%%%%%%%%%%%%%%%%%%%%%%%%%%%%%%%%%%%
Dai Aoki, Megan Aronson,
Eric Bauer, Michael Baenitz, Ryan Baumbach, Nick Butch,
Roberto Caciuffo, Bob Cava, Andrey Chubukov, Piers Coleman, Tristan Combier, Gareth Conduit,
William Duncan, Ramzy Daou,
David Edwards,
Sven Friedemann, Veronika Fritsch, Tobias F\"orster,
Markus Garst, Philipp Gegenwart, Christoph Geibel, Andrew Green,
Sandra Hamann, Clifford Hicks, Andrew Huxley,
Donjing Jang, Marc Janoschek, Anton Jesche, 
Georg Knebel, Guido Kreiner, Cornelius Krellner, Frank Kr{\"u}ger, Robert K\"uchler,
Stefan Lausberg, Gilbert Lonzarich, Jeff Lynn,
Andy MacKenzie, Dmitrii Maslov, Maria Teresa Mercaldo, Andy Millis,
Satoru Nakatsuji, Andreas Neubauer, Michael Nicklas, 
Luis Pedrero, Heike Pfau, Adam Pikul, Christian Pfleiderer, 
J{\"o}rg Rollb{\"u}hler, Achim Rosch,
Burkhard Schmidt, Andy Schofield, Julian Sereni, Sharon Sessions, Qimiao Si, J\"org Sichelschmidt, Dmitry Sokolov, Frank Steglich, Alexander Steppke, Greg Stewart, Andre Strydom, Stefan S\"ullow,
Tomo Uemura, 
Chandra Varma, Matthias Vojta, Thomas Vojta, and
Tanja Westerkamp.
%%%%%%%%%%%%%%%%%%%%%%%%%%%%%%%%%%%%%%%%%%%%%%%%%%%%%%%%%%

This work has been supported by the National Science Foundation under grant numbers NSF DMR-09-29966, DMR-09-01907, 
DMR-1401410, and DMR-1401449, and by the Deutsche Forschungsgemeinschaft under grant number FOR-960. Part of this work 
has been supported by the National Science Foundation under Grant. No. PHYS-1066293 and the hospitality of the Aspen Center 
for Physics.
\bibliography{rmp_qfm}
\bibliographystyle{apsrmp4-1}
\end{document}